\documentclass[11pt,a4paper]{article}

\usepackage[T1]{fontenc}
\usepackage[utf8]{inputenc}
\usepackage{lmodern}
\usepackage{amsmath,amssymb,amsthm,mathtools}
\usepackage{mathrsfs}
\usepackage{microtype}
\usepackage[a4paper,margin=2.65cm]{geometry}
\usepackage{enumitem}
\usepackage{float}
\usepackage{xcolor}
\usepackage{hyperref}
\usepackage{tikz}
\usetikzlibrary{arrows.meta}

\hypersetup{
 bookmarksdepth=2,
 colorlinks=true,
 linkcolor=blue!55!black,
 citecolor=blue!55!black,
 urlcolor=blue!55!black,
 pdftitle={Quantum Dobrushin comparison and locality of fermionic equilibrium states},
 pdfauthor={Lennart Becker, Harman Preet Singh, Stefan Teufel}
}

\setlist[itemize]{topsep=4pt,itemsep=2pt,parsep=1pt}
\setlist[enumerate]{topsep=4pt,itemsep=2pt,parsep=1pt}
\allowdisplaybreaks
\newtheorem{theorem}{Theorem}[section]
\newtheorem{proposition}[theorem]{Proposition}
\newtheorem{lemma}[theorem]{Lemma}
\newtheorem{corollary}[theorem]{Corollary}
\theoremstyle{definition}
\newtheorem{definition}[theorem]{Definition}
\newtheorem{hypothesis}[theorem]{Hypothesis}
\theoremstyle{remark}
\newtheorem{remark}[theorem]{Remark}

\newcommand{\statementqedsymbol}{\ensuremath{\diamond}}
\let\proofqedsymbol\qedsymbol
\newcommand{\pushStatementQED}{\let\qedsymbol\statementqedsymbol
  \pushQED{\qed}}
\AddToHook{env/definition/begin}{\pushStatementQED}
\AddToHook{env/definition/end}{\popQED}
\AddToHook{env/hypothesis/begin}{\pushStatementQED}
\AddToHook{env/hypothesis/end}{\popQED}
\AddToHook{env/remark/begin}{\pushStatementQED}
\AddToHook{env/remark/end}{\popQED}
\AddToHook{env/proof/begin}{\let\qedsymbol\proofqedsymbol}

\newcommand{\A}{\mathcal A}
\newcommand{\Hsp}{\mathfrak h}
\newcommand{\C}{\mathbb C}
\newcommand{\R}{\mathbb R}
\newcommand{\Z}{\mathbb Z}
\newcommand{\1}{\mathbf 1}
\newcommand{\I}{\mathrm{i}}
\newcommand{\E}{\mathrm{e}}
\newcommand{\dd}{\mathrm d}
\newcommand{\supp}{\operatorname{supp}}
\newcommand{\diam}{\operatorname{diam}}
\newcommand{\dist}{\operatorname{dist}}
\newcommand{\spec}{\operatorname{spec}}
\newcommand{\GS}{\mathrm{GS}}
\newcommand{\ball}{\mathrm B}
\newcommand{\at}{\mathrm{at}}
\newcommand{\st}{\mathrm{st}}
\newcommand{\sgnh}{\mathsf S}
\newcommand{\loc}{\mathrm{loc}}
\newcommand{\pol}{\mathrm{pol}}
\newcommand{\aux}{\mathrm{aux}}
\newcommand{\cC}{\mathcal C}
\newcommand{\cI}{\mathcal I}
\newcommand{\cJ}{\mathcal J}
\newcommand{\cL}{\mathcal L}
\newcommand{\cV}{\mathcal V}
\newcommand{\ft}{\mathfrak t}
\newcommand{\KMS}{\operatorname{KMS}}
\newcommand{\Cov}{\operatorname{Cov}}
\DeclareMathOperator{\RealPart}{Re}
\newcommand{\id}{\mathrm{id}}
\newcommand{\Tr}{\operatorname{Tr}}
\newcommand{\tr}{\operatorname{tr}}
\newcommand{\an}{\mathrm{an}}
\newcommand{\Ftr}[1]{\mathbb F^{\rm tr}_{#1}}
\newcommand{\Rb}{\mathsf R_\beta}
\newcommand{\sa}{\mathsf a}
\newcommand{\cS}{\mathcal S_G}
\newcommand{\cSbeta}{\mathcal S_{G_\beta}}
\newcommand{\sG}{s_{\mathrm G}}

\renewcommand{\Gamma}{\varGamma}
\renewcommand{\Psi}{\varPsi}
\renewcommand{\Phi}{\varPhi}
\renewcommand{\Lambda}{\varLambda}
\renewcommand{\Xi}{\varXi}
\renewcommand{\Upsilon}{\varUpsilon}
\renewcommand{\Sigma}{\varSigma}

\definecolor{lpplThermal}{RGB}{38,84,140}
\definecolor{lpplProduct}{RGB}{166,103,22}

\tikzset{
  lppl/.style={
    font=\small, >={Stealth[length=5pt]}, x=1cm, y=1cm,
    box/.style={
      rounded corners=2pt, line width=0.6pt,
      align=flush center, inner sep=5pt
    },
    thermalbox/.style={box, draw=lpplThermal!80, fill=lpplThermal!6},
    productbox/.style={box, draw=lpplProduct!85, fill=lpplProduct!8},
    neutralbox/.style={box, draw=black!45, fill=white},
    notebox/.style={box, draw=black!35, dashed, fill=black!2},
    construct/.style={->, line width=0.6pt, draw=black!70},
    limit/.style={->, dashed, line width=0.7pt, draw=black!70},
    lbl/.style={
      font=\footnotesize, align=flush center,
      inner sep=2pt, text=black!80
    }
  }
}

\providecommand{\lpplsub}[1]{{\footnotesize\color{black!75}#1}}

\title{Quantum Dobrushin comparison\\ and locality of fermionic equilibrium states}
\author{Lennart Becker\thanks{\unexpanded{%
\begin{tabular}[t]{@{}l@{\ }l@{}}
\multicolumn{2}{@{}l@{}}{Fachbereich Mathematik, Universit\"at T\"ubingen,
Auf der Morgenstelle 10, 72076 T\"ubingen, Germany.}\\
Email: & \href{mailto:lennart.becker@uni-tuebingen.de}{\nolinkurl{lennart.becker@uni-tuebingen.de}}\\
       & \href{mailto:harman-preet.singh@uni-tuebingen.de}{\nolinkurl{harman-preet.singh@uni-tuebingen.de}}\\
       & \href{mailto:stefan.teufel@uni-tuebingen.de}{\nolinkurl{stefan.teufel@uni-tuebingen.de}}
\end{tabular}}}
\and Harman Preet Singh\footnotemark[1]
\and Stefan Teufel\footnotemark[1]}
\date{November 27, 2026}

\begin{document}

\maketitle

\begin{abstract}
Dobrushin's comparison theorem is one of the central tools of classical statistical mechanics: from one condition on the conditional distributions at single sites it yields uniqueness of the Gibbs measure, decay of its correlations, and stability of local expectations under distant perturbations.
For quantum lattice systems no comparison theorem of this type has been available.
We prove one for lattice fermions, with the same three consequences.
We apply it to fermions with weak many-body interactions added to a quadratic Hamiltonian, as in the standard models of electrons in insulators and superconductors.
If the quadratic Hamiltonian is gapped, there is a coupling strength below which the thermal equilibrium (KMS) state is unique at every positive temperature, its correlations decay uniformly in the temperature, and local perturbations perturb locally in a strong sense (strong LPPL): every equilibrium state of a system with a distant defect, however strong or extended, is locally close to the equilibrium state without it.
Analogous results hold without a gap at each positive temperature, and at high temperature for general interactions.
All results follow from the one comparison theorem, applied to annihilation operators constructed from the dynamics.
They include new uniqueness and stability theorems and strengthen and extend known bounds on correlations.
\end{abstract}

{\small
\par}

\setcounter{tocdepth}{2}
{\small\linespread{0.97}\selectfont\tableofcontents}

\section{Introduction}
\label{sec:introduction}

A basic problem of statistical mechanics is to decide when the interactions of a system on an infinite lattice determine its equilibrium state uniquely, and how a change of the interactions in one region affects the state in another.
For classical lattice systems, Dobrushin's comparison theorem answers both questions with one argument, and it has been among the central tools of the rigorous theory ever since \cite{Dobrushin70,Georgii88}.
An influence matrix measures how the conditional distribution of the spin at one site depends on the spin at another; if it is a contraction, the difference of two Gibbs measures on a local observable is bounded by the differences of their single-site conditional distributions, propagated through the lattice by the resolvent of the influence matrix.
Uniqueness of the Gibbs measure, decay of its correlations \cite{Foellmer82,Kuensch82}, and bounds on the effect of distant perturbations and boundary conditions all follow from this one inequality.

For quantum lattice systems, no comparison theorem of this kind has been available.
A quantum form of the DLR equations exists \cite{ArakiIon74}, but the operators entering it depend in general on the state, and quantum Gibbs states are not characterized by conditional expectations, so that Dobrushin's argument has no direct counterpart; see the discussion in \cite{Matsui94}.
Matsui proposed a way around this obstacle for quantum spin systems: he characterized the purifications of KMS states by annihilation conditions and deduced uniqueness from the ergodicity of a Markov semigroup built from operators that annihilate the purification.
His aim was a quantum analogue of Dobrushin's uniqueness condition; the condition he obtained is perturbative around infinite temperature and gives uniqueness only.
More recent conditions of Dobrushin type yield a global comparison of states in Wasserstein distance and transportation inequalities \cite{DePalmaRouze22}, and, at high temperature, rapid mixing of quantum Markov chains and decay of conditional mutual information \cite{BLMT25}.
Results on uniqueness, decay of correlations and stability under perturbations have mostly been obtained by other methods, and beyond high temperature and one dimension they are incomplete; we review them at the end of this introduction.

In this paper we prove a comparison theorem of Dobrushin's type for lattice fermions (Theorem~\ref{thm:comparison}).
As in Matsui's approach, annihilation conditions take the place of conditional distributions.
The Fock vacuum is the only state annihilated by all annihilation operators, and the equilibrium state of a quadratic Hamiltonian has, after purification, the same characterization by a family of operators that are linear in the creation and annihilation operators.
Our theorem concerns families \((F_u)\), where \(u\) specifies a lattice site and an internal label, that are small, spatially localized perturbations of such a linear family.
A state \(\omega\) that is annihilated by the family, \(\omega(F_u^*F_u)=0\) for all \(u\), is the only such state, and for every other state \(\rho\) and every local observable \(A\) the difference \(\rho(A)-\omega(A)\) is bounded by the errors \(\rho(F_u^*F_u)^{1/2}\), propagated from the site of \(u\) to the support of \(A\) by a spatially decaying kernel that plays the role of Dobrushin's resolvent.
Uniqueness, decay of correlations and bounds on the effect of distant perturbations follow as in the classical theory.
Unlike there, the unperturbed family need not describe independent sites: it can belong to a free-fermion system at any positive temperature, or at zero temperature if that system is gapped, so that the theorem is not confined to high temperature.

For an equilibrium state the family \((F_u)\) is not given in advance, as a classical specification is; it has to be constructed from the dynamics.
Matsui's operators involve the dynamics at imaginary time and require an analyticity property that he could verify only at high temperature.
Bouch's nearest-neighbor spin example on \(\Z^2\) \cite{Bouch15} has a finite imaginary-time singularity that obstructs Matsui's construction at sufficiently low temperature.
We construct the family by filtering in real time, which needs no such property.
We do this for fermions with weak many-body interactions added to a quadratic Hamiltonian, a class that contains the standard models of weakly interacting electrons in insulators and, through pairing terms, in superconductors.
Neither part needs to be translation invariant, and both may be of long range.
If the quadratic Hamiltonian is gapped, the comparison theorem then yields three results on one coupling interval that does not depend on the temperature.
First, the thermal equilibrium (KMS) state is unique at every positive temperature; for interacting quantum spin and fermion systems, uniqueness had to our knowledge been proved only at high temperature, in one dimension, and for special Hamiltonians whose local terms commute.
Second, its correlations decay uniformly in the temperature; this was known from convergent expansions \cite{DRS19,GMP17} and is obtained here also for polynomially decaying interactions, with a prefactor that is linear in the size of the smaller support.
Third, local perturbations perturb locally in a strong form (strong LPPL): every equilibrium state of a system with a distant defect is locally close to the unperturbed equilibrium state, with a bound independent of the strength, extent and range of the defect, whereas earlier bounds at positive temperature depend on the strength of the perturbation or require it to be small.
Without a gap, the same conclusions hold at each positive temperature on a temperature-dependent coupling interval, and at high temperature they hold for general spatially decaying interactions.
At zero temperature, the theorem gives new proofs of the uniqueness, the spectral gap and the strong LPPL property of the ground state \cite{DRS19,StabilityBdG26,LPPLV3}.

We now describe the comparison theorem and these results in more detail.

\subsection*{The comparison theorem}

Let \(\cC=\operatorname{CAR}(\ell^2(\Gamma,\C^n))\) be the algebra of fermionic creation and annihilation operators on a discrete metric space \(\Gamma\).
Write \(\cI=\Gamma\times\{1,\ldots,n\}\) for the labels of the local creation and annihilation operators and \(\cJ=\Gamma\times\{1,\ldots,2n\}\) for the labels of the associated Majorana operators \(\gamma_v\).
The theorem applies to families \(F_u=F_u^0+\Delta F_u\), where \(F_u^0=\frac12\sum_{v\in\cJ}\sa_{uv}\gamma_v\) is linear in the Majorana operators and \(\Delta F_u\) is a small, spatially localized correction.
The linear family satisfies the Parseval condition of Definition~\ref{def:parseval-family}, a normalization ensuring that every Majorana \(\gamma_v\) is a norm-convergent linear combination of the \(F_u^0\) and their adjoints.
The principal quantitative assumptions control the spatial decay of the coefficients \(\sa_{uv}\) and the size and spatial decay of the local terms comprising \(\Delta F_u\).
Both are measured with a common spatial weight \(W\), such as \(W(R)=\E^{\mu R^\theta}\), \(0<\theta\le1\), or \(W(R)=(1+R)^p\); the precise bounds are stated in Hypothesis~\ref{hyp:perturbed-family}.
The key conclusion is that every local observable~\(A\) admits an exact decomposition in terms of the perturbed operators \(F_u\), with spatial bounds on the coefficients (Theorem~\ref{thm:comparison}(i)):
\[
 A=\omega_F(A)\1+\sum_{u\in\cI}(F_u^*X_u+Y_uF_u),
 \qquad X_u,Y_u\in\cC.
\]
The norms of the coefficients are summable in \(u\), and their decay away from \(\supp A\) is controlled by~\(W\).
Here \(\omega_F\) is a state on \(\cC\) that is determined by the family \((F_u)\) serving as the reference state of the comparison.
Every term of the sum begins with some \(F_u^*\) or ends with some \(F_u\), so its expectation vanishes in every state \(\omega\) annihilated by the family, that is, with \(\omega(F_u^*F_u)=0\) for all \(u\).
Such a state therefore equals \(\omega_F\): the family can annihilate no other state.
In classical terms, the family corresponds to a specification and \(\omega_F\) to the Gibbs measure it specifies.
In our applications, \(\omega_F\) is the extension \(\widehat\omega\) of the equilibrium state.

For an arbitrary state \(\rho\), the decomposition gives
\[
 \rho(A)-\omega_F(A)
 =\sum_{u\in\cI}\bigl(\rho(F_u^*X_u)+\rho(Y_uF_u)\bigr),
\]
and by the Cauchy--Schwarz inequality the \(u\)-th term is at most \(\rho(F_u^*F_u)^{1/2}(\|X_u\|+\|Y_u\|)\).
Thus \(\rho\) is close to \(\omega_F\) on \(A\) if it is nearly annihilated by the operators \(F_u\) near \(\supp A\).
Theorem~\ref{thm:comparison}(ii) makes this quantitative. 
It is stated for bounded linear functionals \(\varphi\) with \(\varphi(\1)=0\), which include differences of two states.

To describe how an observable \(A\) depends on each Majorana operator, define the Majorana derivative \(\partial_vA:=\gamma_vA-\Theta(A)\gamma_v\), where \(\Theta\) is the parity automorphism.
This is a commutator for even \(A\) and an anticommutator for odd \(A\), and it vanishes when the site of \(v\) lies outside \(\supp A\).
The quantitative comparison theorem states that if 
\[
 |\varphi(F_u^*E)|+|\varphi(GF_u)|
 \le \mathfrak s_u\max\{\|E\|,\|G\|\}
 \]
 holds for some nonnegative family \((\mathfrak s_u)_{u\in\cI}\) and all \(u\in\cI,\ E,G\in\cC\), then
 \[
 |\varphi(A)|
 \le \sum_{\substack{u\in\cI\\v\in\cJ}}
       \mathfrak s_u\mathsf T_{uv}\|\partial_vA\|
\]
for all local \(A\).
Here \(\mathfrak s_u\) bounds the error associated with \(F_u\), while \(\|\partial_vA\|\) measures the dependence of \(A\) on~\(\gamma_v\).
The nonnegative kernel \(\mathsf T\) connects these two labels and depends only on the family \((F_u)\).
Let \(x_u\) and \(x_v\) denote the lattice sites of the labels \(u\) and \(v\), and let \(d\) be the metric on \(\Gamma\).
The kernel \(\mathsf T\) is small between distant sites: if the corrections \(\Delta F_u\) are sufficiently small, then by Lemma~\ref{lem:comparison-kernel-bounds},
\[
 \sum_{u\in\cI}W(d(x_u,x_v))\,\mathsf T_{uv}\le1+C_0
 \qquad\text{for every }v\in\cJ,
\]
and the same bound holds for the sum over \(v\) at fixed \(u\).
Here \(C_0\) is the constant that bounds the coefficients \(\sa_{uv}\) in Hypothesis~\ref{hyp:perturbed-family}.
Since \(\partial_vA\) vanishes unless \(x_v\in\supp A\), the errors \(\mathfrak s_u\) at sites with distance at least \(R\) from \(\supp A\) contribute at most \((1+C_0)W(R)^{-1}\sup_u\mathfrak s_u\sum_v\|\partial_vA\|\) to the bound.

The same estimate also controls correlations.
For a state \(\omega\) satisfying \(\omega(F_u^*F_u)=0\) and a local observable \(B\), apply it to \(\varphi_B(A)=\omega(AB)-\omega(A)\omega(B)\).
The errors for this functional are bounded by the norms of \(F_uB-\Theta(B)F_u\), which decay as \(x_u\) moves away from \(\supp B\).
The comparison bound then gives decay of correlations between separated observables.
Section~\ref{sec:comparison-corollaries} derives these consequences, and Subsection~\ref{sec:comparison-remarks} explains the classical correspondence.

The proof of the abstract comparison theorem constructs an auxiliary contraction semigroup \((T_t)_{t\ge0}\) from \((F_u)\).\footnote{%
For constructions of quantum Markov semigroups from modular data, see \cite{Matsui94} and \cite{Park00,BahnKoPark03,CiprianiZegarlinski24}.}
Its unique invariant state is \(\omega_F\), and \(T_tA\to\omega_F(A)\1\) in norm for every local observable \(A\). 
This defines \(\omega_F\) even when no state is annihilated by the family.
For the reference family, the Parseval condition makes every Majorana derivative decay at unit rate, \(\partial_v(T_tA)=\E^{-t}T_t(\partial_vA)\).
The corrections \(\Delta F_u\) couple the derivatives at different labels: the derivative at \(v\) is fed by those at nearby labels \(w\), with strengths \(\mathsf K_{vw}\ge0\) that satisfy the same kind of weighted bound as \(\mathsf T\) above, with constant \(\tfrac12\); for the theorem it suffices that some power of \(\mathsf K\) satisfies such a bound with a constant less than one.
Hence \(\|\partial_v(T_tA)\|\le\sum_{w\in\cJ}(\E^{t(\mathsf K-\1)})_{vw}\|\partial_wA\|\).
The derivatives therefore still decay exponentially, and their time integrals are bounded by \(\sum_{w\in\cJ}((\1-\mathsf K)^{-1})_{vw}\|\partial_wA\|\).

\subsection*{Equilibrium states as annihilated states}

To apply the comparison theorem to equilibrium states, we pass to a doubled CAR algebra, on which every KMS state has a thermal extension; for the Gibbs state of a finite system this is the fermionic form of its purification.
This extension is annihilated by operators \(F_u\) that we construct by filtering Majorana operators in time along the dynamics, and for free fermions these form a complete family of annihilation operators of a quasi-free state.
With a weak many-body interaction, propagation bounds show that they are small, spatially localized perturbations of the free family, so that the comparison theorem applies.
For polynomially decaying interactions, this step is made possible by the lossless propagation bounds of \cite{FermionicLR}: time-evolved interactions and local operators have local decompositions with the full decay exponent of the Hamiltonian, and with bounds that grow only polynomially in time.
The time filters can therefore be applied without loss of spatial decay, so that the operators \(F_u\), and with them all our spatial bounds, inherit the decay exponent of the interaction.

If the free Hamiltonian is gapped, the filter can be chosen with time decay independent of the temperature, which leads to a coupling interval and bounds that are uniform in the temperature.
A defect changes the operators \(F_u\) by an amount that decays with the distance from the defect and does not depend on its strength; this is the source of the strong LPPL estimates below.
Sections~\ref{sec:thermofield} and~\ref{sec:main-proofs} give the constructions and applications.

\subsection*{Results for weakly interacting fermions}

We consider fermionic systems with interaction \(\Phi_\lambda=\Phi_h+\lambda\cV\), where \(\Phi_h\) is quadratic and \(\cV\) is an even many-body interaction.
The quadratic part may contain spatially inhomogeneous hopping and pairing.
Both parts may be long-range: Hypothesis~\ref{hyp:locality} allows stretched-exponential decay and polynomial decay with any exponent \(p_\Phi>d_{\rm vol}\), where \(p_\Phi\) is the exponent of the polynomial weight in the interaction norms and \(d_{\rm vol}\) is the volume-growth dimension of \(\Gamma\).
In the polynomial case, the decay profile in the bounds below is \((1+D)^{-p_\Phi}\), with no loss in the exponent.
For example, for finite-range hopping and two-body interactions bounded by \((1+|x-y|)^{-\alpha}\) on \(\Z^d\), every \(\alpha>2d\) is admitted and \(p_\Phi\) can be any number in \((d,\alpha-d)\); when the site sum over the support of an observable is estimated by the size of its boundary, the exponent becomes \(p_\Phi-d\) (Lemma~\ref{lem:site-sum-bounds}).
In the stretched-exponential case, the temperature-uniform bounds under a free gap decay like \(\exp(-mD^\theta)\) for every \(\theta\) below the exponent \(\theta_{\rm in}\) of the interactions, so that for finite-range or exponentially decaying interactions they are stretched-exponential with any exponent below one; at fixed temperature without a gap the exponent \(\theta_{\rm in}\) is kept.

If the free Hamiltonian is gapped, then on a temperature-independent interval \(|\lambda|\le\lambda_{\mathrm g}\) there is a unique KMS state \(\omega_{\beta,\lambda}\) at every positive temperature (Theorem~\ref{thm:uniqueness}).
Correlations decay uniformly in temperature, with a prefactor linear in the size of the smaller observable support, allowing arbitrary finite and disconnected supports and both parities (Corollary~\ref{cor:background-correlations}).
The same coupling interval gives a unique gapped ground state and exponentially fast local convergence to it as the inverse temperature tends to infinity (Proposition~\ref{prop:ground-gap} and Corollary~\ref{cor:low-temperature}).

For the strong LPPL estimate, consider a modified Hamiltonian whose interaction \(\widetilde\Phi\) agrees with \(\Phi_\lambda\) on every interaction term meeting a region \(U\subset\Gamma\).
Thus \(\widetilde\Phi(Z)=\Phi_\lambda(Z)\) whenever \(Z\cap U\ne\varnothing\), while terms supported outside \(U\) may change arbitrarily.
In the gapped case, for \(|\lambda|\le\lambda_{\mathrm g}\), every \(\beta\)-KMS state \(\rho\) of \(\widetilde\Phi\) and every observable \(A\in\A_Y\), \(Y\Subset U\), satisfy the bound of Theorem~\ref{thm:kms-lppl}(i):
\[
 |\rho(A)-\omega_{\beta,\lambda}(A)|
 \le C\|A\|\sum_{y\in Y}G(d(y,U^c)).
\]
Here \(G\) is the stretched-exponential or polynomial decay profile of Definition~\ref{def:decay-profile}.
Both \(G\) and \(C\) are independent of temperature, the modified interaction \(\widetilde\Phi\), and the choice of KMS state \(\rho\).
Theorem~\ref{thm:gibbs} gives corresponding estimates for finite-volume Gibbs states with arbitrary even boundary operators.
Theorem~\ref{thm:continuity} also controls the effects of perturbations acting throughout the lattice by their interaction strength per site.

Without a free gap, the conclusions hold for \(|\lambda|\le\lambda_{\mathrm{ng}}(\beta)\), with the explicit sufficient threshold of Theorem~\ref{thm:uniqueness}(ii).
The locality estimates then use the temperature-dependent prefactor \(C(\beta)\) and decay profile \(G_\beta\) in place of \(C\) and \(G\).
At high temperature, the method gives uniqueness, clustering and strong LPPL for general even interactions satisfying the locality hypotheses (Theorem~\ref{thm:high-temperature}).

\subsection*{Relation to previous work}

\subsubsection*{Comparison principles and proof methods}

Theorem~\ref{thm:comparison} is a direct quantum analogue of classical Dobrushin comparison and the related covariance estimates \cite{Dobrushin70,Georgii88,Foellmer82,Kuensch82}.
Errors in annihilation conditions replace differences between conditional distributions, and one spatial comparison inequality gives uniqueness, perturbation bounds and decay of correlations; Subsection~\ref{sec:comparison-remarks} details the correspondence.
Apart from the work of Matsui discussed below, earlier quantum results that refer to Dobrushin's condition are of a different kind: a Dobrushin condition for quantum channels yields rapid mixing of a Lindbladian evolution and decay of the conditional mutual information for Gibbs states of finite systems at high temperature \cite{BLMT25}; Dobrushin's classical criterion has been applied to the Euclidean Gibbs measures of quantum anharmonic crystals on path space \cite{AKRT97a,AKRT97b}; and a non-commutative Dobrushin condition yields transportation inequalities \cite{DePalmaRouze22}.
The comparison of a one-dimensional quantum Markov state of a finite system with an arbitrary state in the proof of \cite[Theorem~2]{DePalmaRouze22} is global: all discrepancies enter the bound on the Wasserstein distance with the same weight, wherever they are located.
In Theorem~\ref{thm:comparison}, by contrast, the discrepancies enter the comparison of local expectations with weights that decay with the distance to the observable, for infinite systems and without a Markov assumption; this spatial resolution is the source of the local stability and decay of correlations obtained here.

Our proof uses relaxation of an auxiliary evolution to reconstruct observables, with classical antecedents in \cite{HolleyStroock76,AizenmanHolley87}.
For the uniqueness of KMS states, our strategy is that of Matsui \cite{Matsui94}: purification of the KMS states on a doubled algebra \cite[Definition~1]{Matsui94}, characterization of the purifications by annihilation conditions \cite[Proposition~1]{Matsui94}, and uniqueness from the ergodicity of a Markov semigroup built from the annihilating operators \cite[Proposition~2]{Matsui94}.
He treats spin-\(\tfrac12\) systems on \(\Z^d\); his operators involve the evolution of the Pauli matrices to the imaginary time \(\I\beta/2\), which requires an analyticity assumption on the dynamics, and his criterion asks the generator to be close to its form at infinite temperature, where it acts independently on the sites.
At sufficiently low temperature, the norm-analytic continuation to the imaginary time \(\I\beta/2\) fails for a single-site observable in a periodic one-dimensional fermion model with exponentially decaying interaction within our weak-coupling regime~\cite{BeckerTeufelWesleInPrep}.
We differ from his work in three respects: the unperturbed family is that of a free-fermion system at the given temperature; the annihilating operators are obtained by filtering in real time, which requires no analyticity of the dynamics; and the relaxation of the auxiliary evolution is used to reconstruct local observables with spatial bounds on their coefficients, which yields local comparison inequalities for general states, and hence local stability and decay of correlations in addition to uniqueness.
Bertini, De Sole, Posta and Presilla \cite{BDSP25} prove uniqueness and relaxation for weakly interacting dissipative lattice systems by perturbing an evolution of independent sites, adapting Liggett's construction of classical interacting particle systems \cite{Liggett85}.
Our construction of the auxiliary evolution follows theirs.
The ground-state comparison of~\cite{LPPLV3} uses a closely related mechanism, as explained in Remark~\ref{rem:ground-state-comparison}.
After quasi-local transport to a product vacuum, positive sums of squares of filtered annihilation operators control local excitation densities and hence local expectations.
\subsubsection*{Uniqueness and decay of correlations}

Low-temperature results for weakly interacting fermions have been obtained by convergent expansions.
With a gap assumption for the free system, De Roeck and Salmhofer \cite{DRS19} prove truncated imaginary-time correlation bounds uniform in temperature and volume, and stability of a simple gapped finite-volume ground state.
Giuliani, Mastropietro and Porta \cite{GMP17} construct rapidly decaying equilibrium correlations and their thermodynamic and zero-temperature limits for periodic two-dimensional band insulators. 
Related low-temperature adiabatic and response results are obtained in \cite{GLMP24}.
For small quantum perturbations of classical lattice systems with finitely many ground states, including fermionic ones, quantum Pirogov--Sinai theory constructs, at low temperature and in dimension \(d\ge2\), the infinite-volume limits of the Gibbs states with boundary conditions given by the classical ground states \cite{BKU96,DFF96}.
Regarding gapless, translation-invariant fermions, Benfatto, Falco and Mastropietro \cite{BFM14}, and more recently \cite{PSS25}, use renormalization group methods to obtain power-law decay estimates for zero-temperature response functions in one dimension, uniformly in the volume of the system. Furthermore, \cite{BGM03,BGM06} establish polynomial decay of the two-point function for two-dimensional metals away from half-filling, up to temperatures exponentially small in the coupling, while \cite{GM10} treats the Hubbard model at half-filling.

For spin systems, uniqueness and clustering at every positive temperature are known in one dimension \cite{Araki69,Araki75,Kishimoto76,KK25}.
In higher dimensions, uniqueness at every positive temperature is known for special Hamiltonians whose local terms commute: for classical interactions the KMS states are the Gibbs measures of the classical system \cite{Brascamp70}, so that the classical uniqueness theorems apply, and Kitaev's toric code and the other abelian quantum double models in two dimensions have a unique KMS state at every positive temperature \cite{AFH07,PoloOjitoProdan26}.
For quantum anharmonic crystals with small particle mass, the Euclidean Gibbs measure on path space is unique among tempered measures at all temperatures \cite{AKKR02}.
Our comparison strategy also applies to the spin systems considered in \cite{Yarotsky05,HTW22,AHLW26,HLW26}; Subsection~\ref{sec:spin-systems} gives the necessary dictionary without detailed proofs.
In addition to the temperature-uniform clustering and local indistinguishability established in \cite{AHLW26} and the exponential convergence to the ground-state limit proved in \cite{HLW26}, it gives uniqueness among all infinite-volume KMS states, linear instead of exponential dependence on observables' support sizes, and local comparison estimates uniform in the strength of distant perturbations.

At fixed positive temperature, earlier fermionic results include correlation estimates and constructive expansions \cite{Kashima10,ABPM22,ChenEtAl26}, and efficient Gibbs sampling and mixing for several weakly interacting models \cite{TZ25,SMBB,RapidMixing25}.
Correlations between odd observables decay at every positive temperature without any smallness assumption on the interaction: exponentially for finite-range interactions, with a decay length of order \(\beta\) at low temperature \cite{Hastings04}, and as a power law for two-site interactions with power-law decay \cite{HGCA17}.
For even observables no such general statement can hold, since they may exhibit long-range order.
Mixing estimates for local quantum Markov evolutions also yield clustering of stationary states \cite{KastoryanoEisert13} and stability of local expectations under small perturbations of the generator \cite{LuciaEtAl15}.
Our results without a free gap give infinite-volume KMS uniqueness and quantitative locality with an explicit sufficient coupling threshold whose inverse-temperature dependence is polynomial.

At high temperature, uniqueness and clustering are established for broad interaction classes \cite{BR97,FU15,KGKRE14,KKS25}.
Our clustering bound for stretched-exponential and polynomial interactions is linear in the smaller observable support, improving the exponential support dependence in general estimates such as \cite{KKS25}.
Some earlier results allow slower spatial decay of the interaction; Section~\ref{sec:results-high-temperature} compares the decay assumptions and support dependence.

For long-range interactions, clustering and stability of local expectations are known at high temperature \cite{KKS25,MSAC26}, clustering is known in one dimension at arbitrary temperature \cite{KK25}, and correlations between odd observables decay at every positive temperature \cite{HGCA17}.
Our weak-coupling results give uniqueness of the KMS state, decay of correlations between arbitrary local observables and strong LPPL for polynomially decaying interactions in any dimension, uniformly in the temperature when the free Hamiltonian is gapped.
To our knowledge, for general observables no such results were available at positive temperature for noncommuting long-range interactions beyond the high-temperature regime and one dimension.
\subsubsection*{Local stability and boundary effects}

Local stability asks how changes in the interaction or boundary conditions affect observations far from the modified region.
Capel, Moscolari, Teufel and Wessel \cite{CMTW25}, and M\"obus and collaborators \cite{MSAC26}, derive local comparison estimates from correlation decay.
Becker \cite{Becker26} develops the relations between clustering, local indistinguishability and LPPL for infinite fermionic systems, including situations with phase coexistence.
These comparison bounds depend on the perturbation strength.
Our strong LPPL estimate applies to every defect KMS state independently of the defect's strength, extent and range, even when the modified system has several KMS states.
This gives a new form of local stability already at high temperature; under a free gap it holds uniformly in temperature on the same weak-coupling interval as uniqueness and clustering.
The finite-volume comparison likewise allows arbitrary even boundary operators, with constants independent of their norm.

Teufel and Wessel~\cite{TeufelWessel25} established Lieb--Robinson bounds with polynomial growth in time for long-range fermionic dynamics and applied them to spectral flow and LPPL along uniformly gapped Hamiltonian paths.
Our analytic estimates use the propagation bounds of \cite{FermionicLR}, the free-fermion functional calculus of \cite{StabilityBdG26}, and the infinite-volume and restricted-dynamics results of \cite{LPPLV3}.
The same comparison theorem also gives direct proofs of ground-state uniqueness, gap stability and strong LPPL, providing an alternative to the perturbative and spectral-flow methods of \cite{DRS19,StabilityBdG26,LPPLV3}.

\subsection*{Organization}

The detailed presentation begins with the setting and equilibrium results in Sections~\ref{sec:setting} and~\ref{sec:results}.
Section~\ref{sec:results} can be read independently of the comparison theorem, which is the subject of Section~\ref{sec:comparison}.
The comparison theorem and equilibrium-state constructions appear in Sections~\ref{sec:comparison} and~\ref{sec:thermofield}, and their application in Section~\ref{sec:main-proofs}.
Section~\ref{sec:roadmap-comparison} outlines the proof of the comparison theorem. The full proofs of the comparison theorem and the equilibrium-state constructions are given in Sections~\ref{sec:comparison-proof} and~\ref{sec:thermal-family}, respectively.
The appendices collect the CAR, dynamics and filtering tools.

\section{Setting}
\label{sec:setting}

We first collect the relevant notions and definitions needed to formulate our results. Further standard facts about quasi-local fermionic algebras are collected in Appendices~\ref{app:graded} and~\ref{app:dynamics}.

\subsection{The CAR algebra and interactions}
\label{sec:geometry}

We take the position space to be a countable metric space \((\Gamma,d)\) with polynomial volume growth of dimension $d_{\rm vol} \ge 1$: there is a constant $C_\Gamma > 0$ such that, for all $x \in \Gamma$ and $R \ge 0$,
\begin{equation}
 |\ball_R(x)| \coloneq |\{ y \in \Gamma ~:~ d(x, y ) \le R \}|\le C_\Gamma(1+R)^{d_{\rm vol}}.
 \label{eq:growth}
\end{equation}
We also impose the \emph{cell-partition condition}: there are constants \(C_{\rm cell}\ge1\) and \(D_{\rm cell}\in\mathbb N_0\) such that, for every \(\ell\ge1\), there is a partition \(\Gamma=\bigsqcup_{b\in\mathcal B_\ell}Q_b^{(\ell)}\) into nonempty finite cells satisfying
\[
 \sup_b\diam Q_b^{(\ell)}\le C_{\rm cell}\ell,
\]
and each cell has at most $D_{\rm cell}$ other cells within distance $\ell$:
\[
 \sup_b\bigl|\{c\ne b:d(Q_b^{(\ell)},Q_c^{(\ell)})\le\ell\}\bigr|
 \le D_{\rm cell}.
\]
For example, any uniformly discrete set \(\Gamma\subset\R^n\) equipped with the Euclidean metric satisfies all the above conditions with \(d_{\rm vol}=n\). We write \(X\Subset\Gamma\) for finite nonempty subsets of $\Gamma$. For nonempty subsets \(X\subset\Gamma\) we define their diameter by \(\diam X:=\sup\{d(x,y):x,y\in X\}\) and for $R \ge 0$ the \emph{fattening} of $X$ is
\[
X^{(R)}:=\{z\in\Gamma:d(z,X)\le R\}.
\]
 We define the \emph{metric boundary} of a set \(X\subset\Gamma\) by
\begin{equation*}
 \partial X:=\{x\in X:\text{there exists }u\in \Gamma\setminus X
                   \text{ with }d(x,u)=d(u,X)\}.
\end{equation*}
The \emph{fermionic Fock space} over $\Gamma$, with $r \in \mathbb N$ modes per site, is
\[
   \mathcal{F}(\Gamma, \C^r) \coloneq \bigoplus_{N = 0}^{\infty} \ell^2(\Gamma ,\C^r)^{\wedge N} \, ,
\]
where $\ell^2(\Gamma ,\C^r)^{\wedge N}$ denotes the $N$-fold antisymmetric tensor product of $\ell^2(\Gamma ,\C^r)$. We write \(\Hsp:=\ell^2(\Gamma,\C^r)\) for the one-particle Hilbert space. For any $M \subset \Gamma$  the CAR algebra $\A_M \coloneq \operatorname{CAR} (\ell^2 (M, \C^r))$ is the $C^*$-subalgebra of the algebra of bounded operators on $\mathcal{F} (\Gamma, \C^r)$ that is generated by the fermionic creation and annihilation operators $a^*_{x,i}$ and $a_{x,i}$ for $x \in M$, $i \in \{1, \dots ,r\}$. These creation and annihilation operators satisfy the canonical anticommutation relations (CAR):
\[
    \{a_{x, i}, \, a_{y, j}\} = \{a_{x, i}^*, \, a_{y, j}^*\} = 0 ~ \text{and} ~ \{a_{x, i}, \, a_{y, j}^*\} = \delta_{x, y} \, \delta_{i, j}\1 \, , 
\]
for all $x$,  $y \in M$ and  $i$, $j \in \{1, \ldots, r\}$. Here, $\{A, B\} = AB + BA$ denotes the anticommutator of $A$ and $B$. We call the $C^*$-algebra $\A:=\A_\Gamma$ the \emph{CAR algebra} and
\[
    \A_\loc \coloneq \bigcup_{M \Subset \Gamma} \A_M \subset \A
\]
the local algebra. An operator is called \emph{quasi-local} if it belongs to $\A$ and \emph{local} if it belongs to $\A_\loc$. For a quasi-local operator $A$, we call the smallest set $M\subset\Gamma$ such that $A\in\A_M$ the \emph{support} of $A$.

For any $X \Subset \Gamma$,  the \emph{parity automorphism} $\Theta_X$ is defined by
\[
    \Theta_X(A) :=  (-1)^{N_X} A (-1)^{N_X}, \quad \text{for all} ~~ A \in \A_X,
\]
where $N_X \coloneq \sum_{x \in X} \sum_{i = 1}^r a_{x, i}^* a_{x, i}$ denotes the local number operator. By the quasi-local structure of $\A$ there is a unique automorphism $\Theta$ on $\A$ such that $\Theta |_{\A_X} = \Theta_X$ for all $X \Subset \Gamma$. 
The sets of \emph{even} and \emph{odd} quasi-local operators are
\[
    \A^{\pm} \coloneq \{A\in \A ~|~   \Theta (A) = \pm A \} \, .
\]
We denote the subspaces of even and odd operators supported in $M \subset \Gamma$ by $\A^{\pm}_M \coloneq  \A^{\pm}\cap \A_M$. Operators $A \in \A_{M_1}^+$ and $B\in \A_{M_2}$ supported in disjoint regions $M_1, M_2 \subset \Gamma$ commute: $[A,B] = 0$.

The physical dynamics on $\A$ is generated by interactions. An \emph{interaction} $\Phi$ is a map from the finite subsets of $\Gamma$ to the even local operators that satisfies $\Phi (M) \in \A_M$ and $\Phi(M)^* = \Phi(M)$ for all $M \Subset \Gamma$, and   \(\Phi(\varnothing)=0\). We introduce the \emph{first support moment norm}
\begin{equation*}
 \|\Phi\|_1
 :=\sup_{x\in\Gamma}\sum_{Z\ni x}|Z|\,\|\Phi(Z)\|.
\end{equation*}
We write \(\mathfrak J_1\) for the set of interactions \(\Phi\) with \( \|\Phi\|_1<\infty\). All interactions considered below belong to $\mathfrak J_1$. We will also sometimes impose spatial decay assumptions on interactions that yield suitable propagation bounds.

\subsection{Dynamics and equilibrium states}
\label{sec:equilibrium}

Each interaction \(\Phi\in\mathfrak J_1\) generates a dynamics on $\A$. To construct this dynamics, we first define the corresponding derivation $\delta_\Phi$ by the absolutely convergent sum
\begin{equation*}
 \delta_\Phi(A):=\sum_{Z}[\Phi(Z),A],
 \qquad A\in\A_\loc.
\end{equation*}
It follows from \cite[Propositions~3.2.22 and~3.1.15]{BR87} that this derivation is closable, and we denote its closure by \(\overline\delta_\Phi\). In Proposition~\ref{prop:weak-dynamics}, which relies on \cite[Proposition~2.1]{LPPLV3}, we show that \(\I\overline\delta_\Phi\) generates a strongly continuous, parity-preserving one-parameter group of $*$-automorphisms \(\tau^\Phi\) of \(\A\) with core \(\A_\loc\). That is,
\begin{equation}
 \frac{\dd}{\dd t}\tau_t^\Phi(A)
 =\tau_t^\Phi\left(\I\overline\delta_\Phi(A)\right),
 \label{eq:dynamics-convention}
\end{equation}
for any $A\in\operatorname{Dom}(\overline\delta_\Phi)$. 

We study thermal equilibrium states for these dynamics, characterized by the KMS condition. To state this condition, define the set of entire analytic elements of a strongly continuous dynamics \(\tau\) on a \(C^*\)-algebra \(\cC\) by
\[
 \cC_\an(\tau)
 :=\{B\in\cC:\ t\mapsto\tau_t(B)
       \text{ extends to a norm-entire function on }\C\}.
\]
The set \(\cC_\an(\tau)\) is a norm-dense \(*\)-subalgebra of $\cC$. For every \(z\in\C\), the map \(\tau_z\) is an algebra automorphism of \(\cC_\an(\tau)\), with \(\tau_w\tau_z=\tau_{w+z}\) for all \(w,z\in\C\). For \(A\in\cC_\an(\tau)\) it holds that
\[
 \tau_z(A^*)=\tau_{\bar z}(A)^*.
\]
For nonreal \(z\), this algebra automorphism need not extend to a bounded map on \(\cC\). See \cite[Section~5.1]{Pillet06} for these properties of analytic elements.

We now define equilibrium states of \(\tau^\Phi\) for interactions \(\Phi\in\mathfrak J_1\) on the CAR algebra \(\A\) at temperature \(T\in[0,\infty]\), resp.\ inverse temperature \(\beta:=1/T\in[0,\infty]\).

\begin{definition}[KMS states and ground states]
\label{def:equilibrium}
Let \(\Phi\in\mathfrak J_1\).
\begin{enumerate}[label=(\roman*)]
\item For \(\beta\in[0,\infty)\), the set   \(\KMS_\beta(\tau^\Phi)\) of \emph{\(\beta\)-KMS states} is the set of states \(\rho\) on \(\A\) that satisfy 
\[
 \rho\bigl(A\,\tau^\Phi_{\I\beta}(B)\bigr)=\rho(BA)
 \qquad\text{for all }A,B\in\A_\an(\tau^\Phi).
\]
\item The set of \emph{algebraic ground states} \(\GS(\tau^\Phi)\) is the set of states \(\rho\) on $\A$ that satisfy
\[
\RealPart\rho(A^*\delta_\Phi(A))\ge0 \qquad \text{for all }A\in\A_\loc.
\]
\item We say that an algebraic ground state $\rho$ has a gap $g > 0$ if
\[
 \RealPart\rho(A^*\delta_{\Phi}(A))
 \ge g\bigl[\rho(A^*A)-|\rho(A)|^2\bigr] \qquad \text{for all } A \in \A_\loc. \qedhere
\]
\end{enumerate}
\end{definition}
Ground states are the equilibrium states at temperature \(T=0\) resp.\ \(\beta=\infty\). We use the same definition for any CAR algebra over \(\Gamma\) that has finitely many modes per site, in particular for the algebras \(\A^\st\) and \(\A_\aux\) of Section~\ref{sec:thermofield}. We show in Proposition~\ref{prop:kms-nonempty} that for every \(\Phi\in\mathfrak J_1\), the set \(\KMS_\beta(\tau^\Phi)\) is nonempty and consists of even states. Likewise, we show in Lemma~\ref{lem:ground-state-existence} that the set \(\GS(\tau^\Phi)\) is nonempty.

\subsection{The free fermion system}
\label{sec:model}

We describe weakly interacting fermions using the Majorana formalism. For each \(x\in\Gamma\) and every mode \(1\le i\le r\), we define the two Majorana operators by
\begin{equation*}
 \begin{aligned}
  \gamma_{x,2i-1}&:=a_{x,i}+a_{x,i}^*,\\
  \gamma_{x,2i}&:=\I(a_{x,i}^*-a_{x,i}).
 \end{aligned}
\end{equation*}
These operators are odd, self-adjoint unitaries and generate \(\A\) as a \(C^*\)-algebra. We use the label \(u=(x,j)\in\Gamma\times \{1,\ldots,2r\}\) and write \(x_u:=x\) for the site associated with the label \(u\). They satisfy the anticommutation relations \(\{\gamma_u,\gamma_v\}=2\delta_{uv}\1\), which imply their linear independence. To describe linear combinations of Majorana generators, we use the coefficient space \(\Hsp_{\rm M}:=\ell^2(\Gamma,\C^{2r})\).

We describe free background Hamiltonians by the second quantization of one-particle operators. The free Hamiltonian is specified by a \emph{Majorana coefficient operator} \(h\in\mathcal B(\Hsp_{\rm M})\) satisfying
\begin{equation*}
 h=h^*,\qquad \overline h=-h.
\end{equation*}
Here \(\overline h\) is entrywise complex conjugation in the Majorana coefficient basis. The corresponding quadratic interaction is defined by
\begin{equation}\label{eq:quadratic-Majorana-interaction}
 \Phi_h(X):=\frac14\sum_{\substack{(u,v)\in(\Gamma\times\{1,\ldots,2r\})^2\\\{x_u,x_v\}=X}}h_{uv}\gamma_u\gamma_v,
\end{equation}
for \(1\le |X|\le2\), and we set \(\Phi_h(X):=0\) otherwise. 

We use weighted Schur norms to quantify the locality of the one-particle operator $h$. These norms also control the corresponding quadratic interaction $\Phi_{h}$.  Let \(k\) be a bounded operator on \(\ell^2(\Gamma,\C^m)\), where \(m\in\mathbb N\). For \(x,y\in\Gamma\),  denote the \((x,y)\)-matrix block of \(k\) by \(k_{xy}\in\mathcal B(\C^m)\). For \(\mu\ge0\) and \(0<\theta\le1\), we define the stretched-exponential Schur norm 
\begin{equation}
 \|k\|^{\rm Sch}_{\mu,\theta}
 :=\max\Bigl\{\sup_x\sum_y\E^{\mu d(x,y)^\theta}\|k_{xy}\|,
               \sup_y\sum_x\E^{\mu d(x,y)^\theta}\|k_{xy}\|\Bigr\},
 \label{eq:Schur-norm}
\end{equation}
and, for \(p\ge0\), the polynomial Schur norm 
\begin{equation}
 \|k\|^{\rm Sch,pol}_p
 :=\max\Bigl\{\sup_x\sum_y(1+d(x,y))^p\|k_{xy}\|,
               \sup_y\sum_x(1+d(x,y))^p\|k_{xy}\|\Bigr\}.
 \label{eq:Schur-polynomial-norm}
\end{equation}
Here \(\|\cdot\|\) denotes the operator norm on \(\mathcal B(\C^m)\).

The CAR identity implies \([\gamma_v\gamma_w,\gamma_u] =2\delta_{wu}\gamma_v-2\delta_{vu}\gamma_w\), and hence, for every label \(u\) and \(t\in\R\),
\begin{equation}\label{eq:Majorana-free-dynamics}
 \delta_{\Phi_h}(\gamma_u)  =-\sum_{v\in\Gamma\times\{1,\ldots,2r\}}h_{uv}\gamma_v\qquad\mbox{and} \qquad\tau_t^{\Phi_h}(\gamma_u)=\sum_{v\in\Gamma\times\{1,\ldots,2r\}}(\E^{-\I th})_{uv}\gamma_v.
\end{equation}
If $h$ has a finite Schur norm, these sums converge absolutely in norm.

For every \(\beta>0\), the free dynamics \(\tau^{\Phi_h}\) has a unique KMS state \(\omega_{\beta,0}\). This state is even and quasifree, with two-point function
\[
 \omega_{\beta,0}(\gamma_u\gamma_v)
 =\bigl[\1+\tanh(\beta h/2)\bigr]_{uv}.
\]
See Araki~\cite[Theorem~3]{Araki70}.

Our main theorems treat one-particle operators both with and without a spectral gap. In the gapped case, we assume that there is a \(g>0\) such that
\begin{equation}\label{eq:free-band}
 \spec(h)\cap(-g,g)=\varnothing.
\end{equation}
In this case we define the operators
\begin{equation}\label{eq:PQE}
 \sgnh:=\operatorname{sgn}(h),\qquad P_\pm:=\tfrac12(\1\pm\sgnh), \qquad |h|:=(h^2)^{1/2}.
\end{equation}
Lemma~\ref{lem:uniform-free-calculus} (see also \cite[Lemmas~A.7--A.8]{StabilityBdG26}) shows that the spectral gap allows us to transfer the locality of \(h\) to these operators. Functional calculus gives \(\overline\sgnh=-\sgnh\), \(\overline{P_+}=P_-\), and it shows that \(|h|\) is real symmetric. In the gapped case we consider the even quasifree state \(\omega_0\) determined by
\begin{equation*}
 \omega_0(\gamma_u\gamma_v)=[\1+\sgnh]_{uv}.
\end{equation*}
By \cite[Lemma~4.3 and Theorem~2]{Araki70}, this state is pure and is the unique ground state of \(\Phi_h\). In \cite[Proposition~4.3]{StabilityBdG26} it is shown that this state has a gap of at least \(g\). 

We note that without a gap assumption, the interaction \(\Phi_h\) need not have a unique ground state.

\subsection{Locality regimes, model data, and decay profiles}
\label{sec:regimes}

We will add many-body interactions $\cV$ to the free interaction \(\Phi_h\). To quantify their locality, we introduce two interaction norms. For stretched-exponential decay, let \(\mu\ge0\) and \(0<\theta\le1\), and define
\[
       \|\Phi\|_{\mu,\theta} := \sup_x\sum_{Z\ni x}|Z|\E^{\mu\diam(Z)^\theta}\|\Phi(Z)\|.
\]
For polynomial decay, let $p\ge0$ and $\kappa>0$, and define
\begin{equation}
       \|\Phi\|^{\#}_{p,\kappa} :=\sup_x\sum_{Z\ni x}\E^{\kappa|Z|}(1+\diam Z)^p\|\Phi(Z)\|. \label{eq:sharp-interaction-norm}
\end{equation}
The exponential support-size weight in \eqref{eq:sharp-interaction-norm} allows us to use lossless propagation bounds for time-evolved interactions \cite[Theorem~2.6]{FermionicLR}. Note that for $k$-body interactions,  i.e.\ $\Phi(Z)=0$ whenever $|Z|>k$, the weight \(\E^{\kappa|Z|}\leq \E^{\kappa k}\) is irrelevant. 

For \(\lambda\in\R\) and \(\cV\in\mathfrak J_1\) let
\[
 \Phi_\lambda:=\Phi_h+\lambda\cV.
\]
We call \(\Phi_\lambda\) the \emph{bulk interaction} and the dynamics \(\tau^{\Phi_\lambda}\) it generates the \emph{physical bulk dynamics}. We impose one of the following two locality assumptions on the free part and the many-body interaction. 

\begin{hypothesis}\label{hyp:locality}
One of the following two conditions holds. In either case, \(\|h\|\le C_h\).
\begin{enumerate}[label=(\Alph*),ref=(\Alph*),leftmargin=*]
\item\label{ass:exponential} \emph{Stretched-exponential locality.}
Assume that \(0<\theta_{\rm in}\le1\), \(\mu_h,\mu_{\cV}>0\) and
\[
 C_h:=\|h\|^{\rm Sch}_{\mu_h,\theta_{\rm in}}<\infty,\qquad \|\cV\|_{\mu_{\cV},\theta_{\rm in}}<\infty.
\]
\item\label{ass:polynomial-weight} \emph{Polynomial locality.}
Let \(\kappa_0=\kappa_0(D_{\rm cell})>0\) be as required in \cite[Theorem~2.6]{FermionicLR} and set $p_\Phi:=\min\{p_h,p_{\cV}\}$. Assume that \(p_\Phi>d_{\rm vol}\) and
\[
 C_h:=\|h\|^{\rm Sch,pol}_{p_h}<\infty,\qquad \|\cV\|^{\#}_{p_{\cV},\kappa_0}<\infty. \qedhere
\] 
\end{enumerate}
\end{hypothesis}

Since \(\Phi_h\) contains only terms supported on at most two sites, its interaction norms are bounded by the corresponding Schur norms of \(h\). Indeed, summing over the finitely many Majorana labels at each site gives
\begin{equation}\label{eq:quadratic-interaction-norm-bound}
 \begin{aligned}
  \|\Phi_h\|_{\mu_h,\theta_{\rm in}}&\le 2r\|h\|^{\rm Sch}_{\mu_h,\theta_{\rm in}}
  &&\text{in \ref{ass:exponential}},\\
  \|\Phi_h\|^{\#}_{p_h,\kappa_0}&\le 2r\E^{2\kappa_0}\|h\|^{\rm Sch,pol}_{p_h}
  &&\text{in \ref{ass:polynomial-weight}}.
 \end{aligned}
\end{equation}
We collect the parameters entering our main theorems in the following definition.

\begin{definition}\label{def:model-data}
The \emph{model data} consist of the volume-growth constants \(C_\Gamma\), \(d_{\rm vol}\), the number \(r\) of modes per site, the bound $C_h$ on the Schur norm of \(h\) and the interaction norm of \(\cV\). In regime \ref{ass:exponential} the model data also include \(\theta_{\rm in}\), \(\mu_h\) and \(\mu_{\cV}\). In regime \ref{ass:polynomial-weight} they include \(p_h\), \(p_{\cV}\), \(C_{\rm cell}\) and \(D_{\rm cell}\). When a result uses the free spectral-gap assumption \eqref{eq:free-band}, its model data additionally include \(g\). We call constants that only depend on the model data \emph{model constants}.
\end{definition} 

Next, we introduce defects that perturb the bulk interaction \(\Phi_\lambda\). We require that the perturbation \(\Psi\) belongs to \(\mathfrak J_1\), so  \(\Phi_\lambda+\Psi\) also belongs to \(\mathfrak J_1\) and defines a dynamics \(\tau^{\Phi_\lambda+\Psi}\), the \emph{perturbed physical dynamics}, with associated equilibrium states. We will impose no bounds on the strength and the support of \(\Psi\).

\begin{definition}\label{def:defect}
An \emph{admissible defect} is an interaction \(\Psi\in\mathfrak J_1\).
Its \emph{support} is
\begin{equation*}
 \Sigma(\Psi):=\bigcup_{\Psi(Z)\ne0}Z.
 \qedhere
\end{equation*}
\end{definition}

The bounds in our main theorems decay with the distance between two sets. We quantify this decay as follows.

\begin{definition}\label{def:decay-profile}
Define the \emph{decay profiles} \(G,G_\beta:[0,\infty]\to[0,1]\) by \(G(\infty)=G_\beta(\infty)=0\) and the following formulas for \(D\ge0\).
\begin{enumerate}[label=(\roman*),leftmargin=*]
\item \emph{With a free spectral gap.}
For a decay order \(0<\theta\le1\) and a decay rate \(m>0\), set
\[
 G(D):=\begin{cases}
 \E^{-mD^\theta},&\text{in \ref{ass:exponential}},\\[4pt]
 (1+D)^{-p_\Phi},&\text{in \ref{ass:polynomial-weight}}
 \end{cases}.
\]
\item \emph{Without a gap assumption.}
For \(m_0>0\) and \(\beta>0\), set
\[
 G_\beta(D):=\begin{cases}
 \displaystyle\exp\!\left(-\frac{m_0D^{\theta_{\rm in}}}{1+\beta}\right),
 &\text{in \ref{ass:exponential}},\\[4pt]
 (1+D)^{-p_\Phi},&\text{in \ref{ass:polynomial-weight}}
 \end{cases}.
\]
\end{enumerate}
For \(Y\Subset\Gamma\) and \(S\subset\Gamma\), define the site sums
\[
 \cS(Y,S):=\sum_{y\in Y}G(d(y,S)),
 \qquad
 \cSbeta(Y,S):=\sum_{y\in Y}G_\beta(d(y,S)). \qedhere
\]
\end{definition}
The choices of \(\theta\), \(m\) and \(m_0\) used in the estimates are specified in Section~\ref{sec:results}. 
The normalization \(G(0)=G_\beta(0)=1\) separates spatial decay from the prefactors in our estimates.
Without a gap assumption, these estimates also contain a temperature-dependent prefactor \(C(\beta)\), specified in Section~\ref{sec:results}.

The following lemma gives elementary bounds on $\cS$, including a bound in terms of the size of the boundary of $Y$. We defer its proof to Appendix~\ref{app:decay-profile-bounds}. 

\begin{lemma}\label{lem:site-sum-bounds}
Fix one of the profiles \(G\) from Definition~\ref{def:decay-profile}. For \(Y\Subset\Gamma\) and \(S\subset\Gamma\), we have
\[
 \cS(Y,S)\le |Y|G(d(Y,S)).
\]
If \(Y\cap S=\varnothing\), then there is, for every \(0<m'<m\) in the stretched-exponential case, a constant $C > 0$ that is independent of $Y$ and $S$ such that
\[
 \cS(Y,S)\le
 \begin{cases}
 C|\partial Y| \E^{-m'd(Y,S)^\theta},&G(D)=\E^{-mD^\theta},\\
 C|\partial Y| (1+d(Y,S))^{-(p_\Phi-d_{\rm vol})},&G(D)=(1+D)^{-p_\Phi}.
 \end{cases}
\]
The same estimates hold with \(G,\cS\) replaced by \(G_\beta,\cSbeta\).
In regime~\ref{ass:exponential}, fix \(0<m_0'<m_0\) and take \(\theta=\theta_{\rm in}\), \(m=m_0/(1+\beta)\) and \(m'=m_0'/(1+\beta)\); the constant \(C\) in the boundary estimate may then depend on \(\beta\).
In regime~\ref{ass:polynomial-weight}, \(G_\beta=G\), so \(C\) is independent of \(\beta\).
\end{lemma}

\section{Main results}
\label{sec:results}

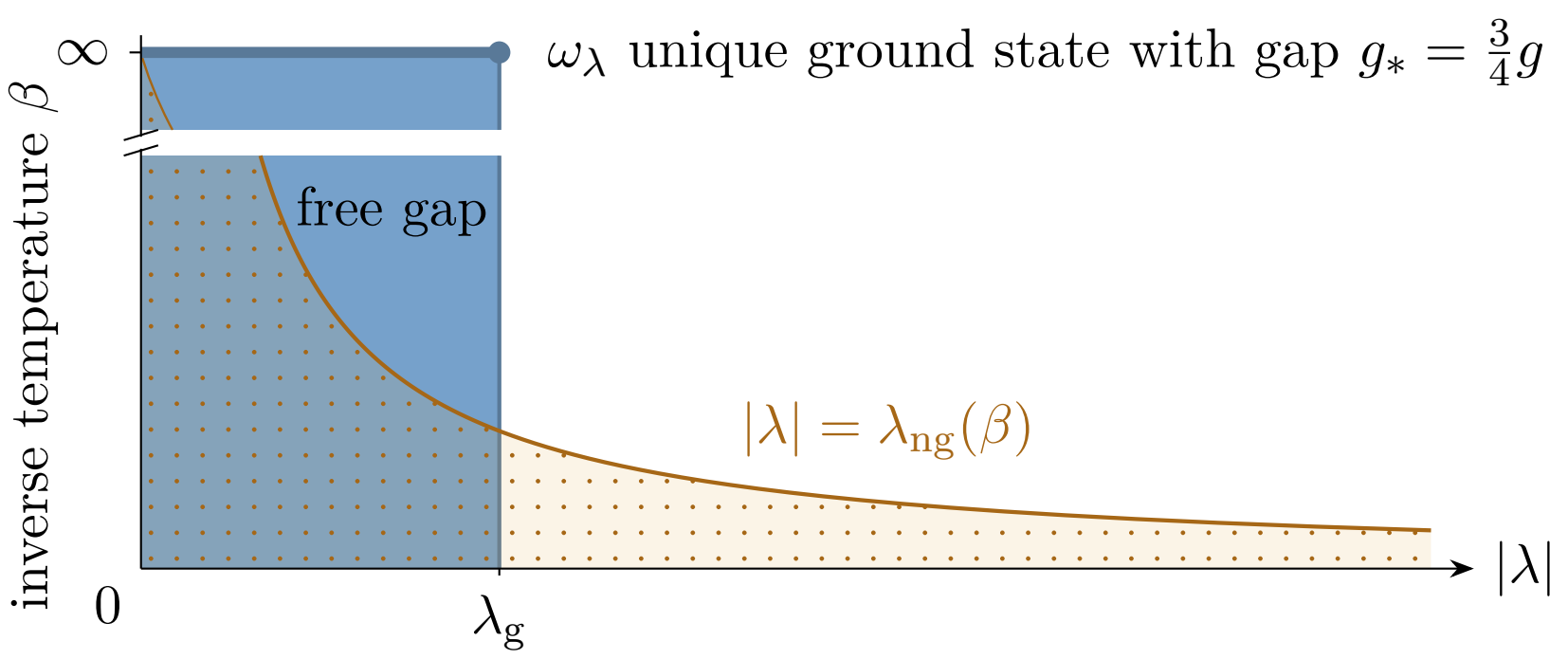
\begin{figure}[H]
  \centering
  \normalsize
\definecolor{lpplAmber}{RGB}{231,177,84}
\definecolor{lpplAmberInk}{RGB}{166,103,22}
\definecolor{lpplObs}{RGB}{118,161,203}
\begin{tikzpicture}[
    >=Stealth, x=1cm, y=0.6cm, font=\normalsize,
    gapfill/.style={fill=lpplObs},
    gapline/.style={draw=lpplObs!75!black, thick},
    ngfill/.style={fill=lpplAmber, fill opacity=0.14, path picture={
      \foreach \px in {0.07,0.25,...,9.1}
        \foreach \py in {0.07,0.25,...,3.1}
          \fill[lpplAmberInk, fill opacity=1]
            ([xshift=\px cm,yshift=\py cm]path picture bounding box.south west)
            circle (0.45pt);}},
    ngline/.style={draw=lpplAmberInk, thick},
  ]
  \def\lamg{2.5}      %
  \def\lngA{4.0}      %
  \def\xmax{9.0}      %
  \def\bmax{4.8}      %
  \def\bk{4.95}       %
  \def\bt{5.1}        %
  \def\yinf{6.0}      %
  \tikzset{declare function={lng(\b)=\lngA/\b;}}
  \pgfmathsetmacro{\bzero}{\lngA/\xmax}   %

  \fill[gapfill] (0,0) rectangle (\lamg,\bmax);
  \fill[gapfill] (0,\bt) rectangle (\lamg,\yinf);
  \draw[gapline] (\lamg,0) -- (\lamg,\bmax);
  \draw[gapline] (\lamg,\bt) -- (\lamg,\yinf);

  \path[ngfill] (0,0) -- (\xmax,0) -- (\xmax,\bzero)
                -- plot[domain=\bzero:\bmax,samples=100] ({lng(\x)},\x)
                -- (0,\bmax) -- cycle;
  \draw[ngline] plot[domain=\bzero:\bmax,samples=100] ({lng(\x)},\x);
  \path[ngfill] (0,\bt) -- (0.22,\bt)
                .. controls (0.12,{\bt+0.30}) and (0.025,{\yinf-0.18}) .. (0,\yinf)
                -- cycle;
  \draw[ngline, line width=0.45pt] (0.22,\bt)
                .. controls (0.12,{\bt+0.30}) and (0.025,{\yinf-0.18}) .. (0,\yinf);

  \node[text=lpplAmberInk, anchor=south west] at ({lng(1.0)+0.05},1.05)
       {$|\lambda|=\lambda_{\mathrm{ng}}(\beta)$};
  \node[align=center, text=black] at (1.75,4.15)
       {free gap};

  \draw[lpplObs!75!black, line width=2.2pt] (0,\yinf) -- (\lamg,\yinf);
  \fill[lpplObs!75!black] (\lamg,\yinf) circle (2.2pt);
  \node[anchor=west] at (\lamg+0.2,\yinf)
       {$\omega_\lambda$ unique ground state with gap $g_*=\tfrac34 g$};

  \draw[->] (0,0) -- (\xmax+0.3,0) node[right] {$|\lambda|$};
  \draw (0,0) -- (0,\bk-0.07);
  \draw (0,\bk+0.07) -- (0,\yinf+0.2);
  \draw (-0.12,\bk-0.15) -- (0.12,\bk-0.03);
  \draw (-0.12,\bk+0.03) -- (0.12,\bk+0.15);
  \node[below left] at (0,0) {$0$};
  \draw (\lamg,0) -- ++(0,-0.08) node[below] {$\lambda_{\mathrm g}$};
  \draw (0,\yinf) -- ++(-0.08,0) node[left] {$\infty$};
  \node[rotate=90] at (-0.75,2.6) {inverse temperature $\beta$};

\end{tikzpicture}
  \caption[Coupling regimes of the equilibrium results.]{%
    \begin{tabular}[t]{@{}p{0.6cm}@{}p{\dimexpr\linewidth-5em-0.6cm\relax}@{}}
      \tikz[baseline=-0.4ex]{%
        \fill[lpplObs] (0,0) rectangle (0.45,0.22);
        \draw[lpplObs!75!black, thick] (0.45,0) -- (0.45,0.22);}
      & \textbf{with free spectral gap:} unique KMS state, unique gapped
        ground state, low-temperature limit, strong LPPL for KMS
        states, and decay of correlations, with all bounds uniform in the entire blue region\\[0.35em]
      \tikz[baseline=-0.4ex]{%
        \fill[lpplAmber!14] (0,0) rectangle (0.45,0.22);
        \foreach \x in {0.06,0.22,0.38}
          \foreach \y in {0.05,0.17}
            \fill[lpplAmberInk] (\x,\y) circle (0.45pt);
        \draw[lpplAmberInk, thick] (0.45,0) -- (0.45,0.22);}
      & \textbf{without gap assumption:} unique KMS state, strong LPPL, and decay of correlations, with constants explicitly depending on \(\beta\)
    \end{tabular}%
  }
  \label{fig:coupling-regimes}
\end{figure}

The weak-coupling results below all use the same common intervals \(|\lambda|\le\lambda_{\mathrm g}\) and \(|\lambda|\le\lambda_{\mathrm{ng}}(\beta)\) supplied by Theorem~\ref{thm:uniqueness}, see Figure~\ref{fig:coupling-regimes}, and the decay profiles and site sums of Definition~\ref{def:decay-profile}.
When assuming the free gap condition~\eqref{eq:free-band}, all prefactors and decay rates are independent of temperature.
Without a gap assumption, the estimates use \(G_\beta\), \(\cSbeta\) and a common prefactor \(C(\beta)=C_*(1+\beta)^M\), where the theorems below provide model constants \(C_*,M\ge1\) and, in regime~\ref{ass:exponential}, \(m_0>0\).

\subsection{Uniqueness of equilibrium states and the low-temperature limit}
\label{sec:results-uniqueness}

Our first theorem shows that sufficiently weak interactions preserve uniqueness of the KMS state. If the free Hamiltonian is gapped, the interacting ground state is also unique, and the KMS states converge to it as the temperature tends to zero.

\begin{theorem}[Uniqueness of equilibrium states]\label{thm:uniqueness}
Assume Hypothesis~\ref{hyp:locality} with \(\cV\not=0\) and set 
\begin{equation}\label{eq:interaction-strength-bound}
      J:= 
 \begin{cases}
 \|\cV\|_{\mu_{\cV},\theta_{\rm in}},&\ref{ass:exponential},\\
 \|\cV\|^{\#}_{p_{\cV},\kappa_0},&\ref{ass:polynomial-weight}.
 \end{cases}
\end{equation}
\begin{enumerate}[label=(\roman*),leftmargin=*]
\item \emph{With a free spectral gap.}
Assume the free gap condition~\eqref{eq:free-band} with \(g>0\).
There exists a model constant \(c_{\mathrm g}>0\), independent of \(J\), that defines the coupling threshold   \(\lambda_{\mathrm g}:=c_{\mathrm g}/J\).
For every \(|\lambda|\le\lambda_{\mathrm g}\) and every \(\beta>0\),
\[
 \begin{aligned}
 \KMS_\beta(\tau^{\Phi_\lambda})
   &=\{\omega_{\beta,\lambda}\},\\
 \GS(\tau^{\Phi_\lambda})&=\{\omega_\lambda\}.
 \end{aligned}
\]
For each such \(\lambda\), the KMS states converge to the ground state as the temperature tends to zero:
\[
 \omega_{\beta,\lambda}
 \xrightarrow[\beta\to\infty]{\mathrm{weak}\text{-}*}
 \omega_\lambda.
\]

\item \emph{Without a gap assumption.}
There exist model constants \(c_0>0\) and \(N\ge1\), independent of \(J\), that define the coupling threshold 
\[
 \lambda_{\mathrm{ng}}(\beta)
 :=\frac{c_0}{J\beta(1+\beta)^{N-1}}.
\]
For every \(\beta>0\) and every \(|\lambda|\le\lambda_{\mathrm{ng}}(\beta)\),
\[
 \KMS_\beta(\tau^{\Phi_\lambda})
 =\{\omega_{\beta,\lambda}\}.
\]
\end{enumerate}
\end{theorem}

The low-temperature convergence in (i) follows from Corollary~\ref{cor:low-temperature} below and the norm density of the local algebra in the quasi-local algebra. We also write $\omega_{\infty, \lambda} \coloneq \omega_\lambda$ for the ground state. Throughout the permitted coupling interval, this ground state is unique and has a gap $g_* := 3g/4$.

\begin{proposition}[Ground-state gap]\label{prop:ground-gap}
Assume the hypotheses of Theorem~\ref{thm:uniqueness}(i), with \(\lambda_{\mathrm g}\) as chosen there. For every \(\lambda\) with \(|\lambda|\le\lambda_{\mathrm g}\), the unique ground state \(\omega_\lambda\) has a gap of at least \(g_*\).
\end{proposition}

\begin{remark}[Spectral flow]
Uniqueness implies that \(\omega_\lambda\) coincides for \(|\lambda|\leq \lambda_{\mathrm g}\) with the ground state obtained from \(\omega_0\) by the spectral flow constructed in \cite{StabilityBdG26}.
\end{remark}

To our knowledge, Theorem~\ref{thm:uniqueness} is the first uniqueness result beyond high temperature and one dimension for KMS states of an interacting quantum spin or fermion system whose local terms do not commute; the earlier results are reviewed in the introduction, and for weak finite-range perturbations of gapped on-site spin Hamiltonians such uniqueness is conjectured in \cite[after Corollary~2.5]{AHLW26}.
Part~(i) contains a classical case: for a non-zero on-site energy and a sufficiently weak density--density interaction, all terms of the Hamiltonian commute, the KMS states are the Gibbs measures of a classical lattice gas \cite{Brascamp70}, and Dobrushin's condition holds uniformly in the temperature.
Uniqueness and the gap of the ground state were obtained before by perturbative and spectral-flow methods \cite{DRS19,StabilityBdG26,LPPLV3}; here they follow from the comparison theorem.

\paragraph{Zero temperature limit.} 
We now compare low-temperature KMS states with ground states more quantitatively, assuming a free spectral gap. For later applications it will be useful to do this comparison locally and allow for arbitrary defects outside the relevant comparison region: we compare a \(\beta\)-KMS state \(\rho\) of \(\Phi_\lambda+\Psi\) with a ground state \(\omega\) of \(\Phi_\lambda+\Upsilon\) on the set \(\Gamma\setminus\Sigma\). Here the admissible defects \(\Psi\) and \(\Upsilon\) may differ and have arbitrary strength, but both are supported in a finite or infinite region \(\Sigma\subset\Gamma\).

\begin{theorem}[Low-temperature comparison under defects]\label{thm:ground}
Assume the hypotheses of Theorem~\ref{thm:uniqueness}(i), with \(\lambda_{\mathrm g}\) as chosen there. In regime~\ref{ass:exponential}, fix \(0<\theta<\theta_{\rm in}\).

There are constants \(C>0\) and, in regime \ref{ass:exponential}, \(m>0\), depending only on the model data and, in regime \ref{ass:exponential}, on \(\theta\), with the following property. For every \(\lambda\) with \(|\lambda|\le\lambda_{\mathrm g}\), every \(\beta>0\), \(\Sigma\subset\Gamma\), \(Y\Subset\Gamma\) and \(A\in\A_Y\),
\begin{equation*}
 \sup_{\substack{\Psi,\Upsilon\in\mathfrak J_1\\\Sigma(\Psi)\cup\Sigma(\Upsilon)=\Sigma}} \,
 \sup_{\substack{\rho\in\KMS_\beta(\tau^{\Phi_\lambda+\Psi})\\\omega\in\GS(\tau^{\Phi_\lambda+\Upsilon})}}
 |\rho(A)-\omega(A)|
 \le C\|A\|\Bigl(|Y|\E^{-\beta g_*}+\cS(Y,\Sigma)\Bigr).
\end{equation*}
Here \(G\) and \(\cS\) are as in Definition~\ref{def:decay-profile}.
\end{theorem}
The thermal contribution \(|Y|e^{-\beta g_*}\) vanishes as \(\beta\to\infty\), while the defect contribution satisfies
\[
\mathcal S_G(Y,\Sigma)\le |Y|G(d(Y,\Sigma))
\]
uniformly in temperature.
Thus, for a fixed observation region, the two states have close local expectations at low temperature and far from the defects, uniformly in their strength and in the choice of states.
Setting \(\Psi=\Upsilon=0\) removes the spatial error and gives the following quantitative convergence to the ground state.

\begin{corollary}[Low-temperature limit]\label{cor:low-temperature}
Assume the hypotheses of Theorem~\ref{thm:ground} and use the same constant \(C\).
For every \(\lambda\) with \(|\lambda|\le\lambda_{\mathrm g}\), every \(\beta>0\), \(Y\Subset\Gamma\) and \(A\in\A_Y\),
\begin{equation*}
 |\omega_{\beta,\lambda}(A)-\omega_\lambda(A)|
 \le C|Y|\|A\|\E^{-\beta g_*}.
\end{equation*}
\end{corollary}
To our knowledge, a comparable result was previously available only for weakly interacting quantum spin systems.
For weak finite-range perturbations of gapped on-site Hamiltonians, as in Subsection~\ref{sec:spin-systems}, local expectations in finite-volume Gibbs states converge to their ground-state values exponentially fast in \(\beta\), uniformly in the volume~\cite[Theorem~4.1]{HLW26}.
The bound there is stated for connected \(Y\) and has a prefactor that grows exponentially in \(|Y|\), whereas the prefactor in Corollary~\ref{cor:low-temperature} is linear in \(|Y|\) for arbitrary finite \(Y\).

The following strong LPPL principle for ground states was established in \cite[Corollary 3.4]{LPPLV3}, which also treats low-energy states.
Here it follows by setting \(\Psi=0\) in Theorem~\ref{thm:ground} and taking \(\beta\to\infty\).
Every ground state of \(\Phi_\lambda+\Upsilon\) is therefore locally close to \(\omega_\lambda\) far from the support of \(\Upsilon\), with constants independent of the defect’s strength and extent.

\begin{corollary}[Strong LPPL for ground states]\label{cor:ground-lppl}
Assume the hypotheses of Theorem~\ref{thm:ground} and use the same constant \(C\) and decay profile \(G\).
For every \(\lambda\) with \(|\lambda|\le\lambda_{\mathrm g}\), \(\Sigma\subset\Gamma\), \(Y\Subset\Gamma\) and \(A\in\A_Y\),
\[
 \sup_{\substack{\Upsilon\in\mathfrak J_1\\\Sigma(\Upsilon)=\Sigma}}
 \sup_{\rho\in\GS(\tau^{\Phi_\lambda+\Upsilon})}
 |\rho(A)-\omega_\lambda(A)|
 \le C\|A\|\,\cS(Y,\Sigma).
\]
\end{corollary}

\subsection{Locality of equilibrium states}
\label{sec:results-lppl}

We now establish locality estimates for thermal equilibrium states. We prove strong LPPL and decay of correlations, and compare finite-volume Gibbs states with the infinite-volume KMS state.

\paragraph{Strong LPPL for KMS states.}

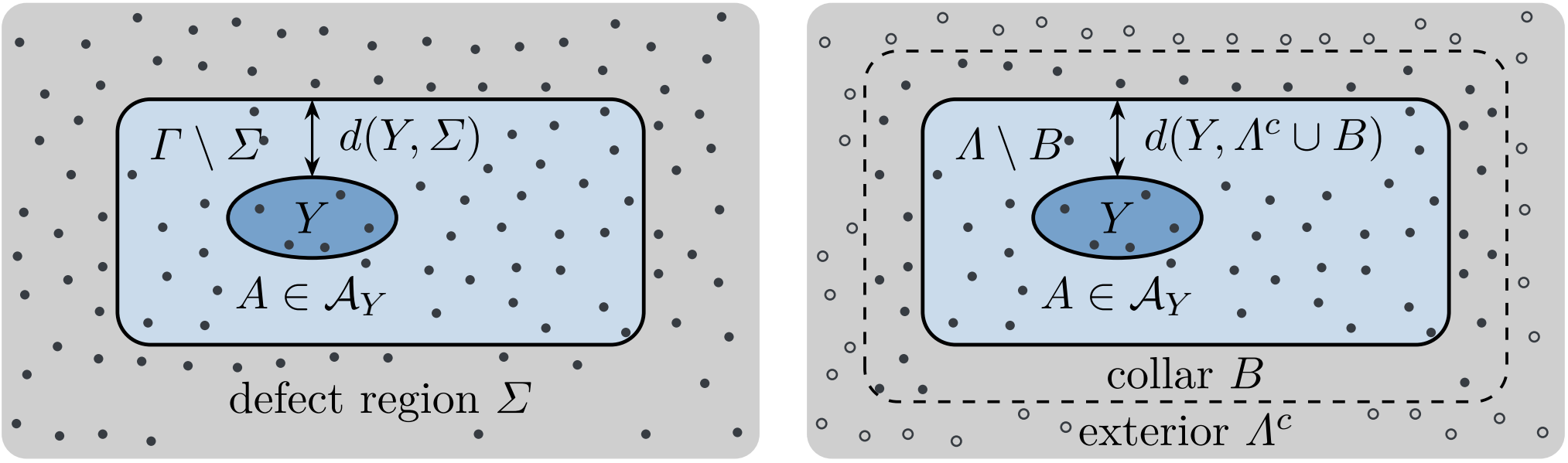
\begin{figure}[H]
\centering
\normalsize
\definecolor{lpplGamma}{RGB}{202,219,235}
\definecolor{lpplDefect}{RGB}{207,207,207}
\definecolor{lpplObs}{RGB}{118,161,203}
\definecolor{lpplSite}{RGB}{55,60,66}
\begin{tikzpicture}[>={Stealth[length=5pt]},x=0.9cm,y=0.75cm,font=\normalsize]
  \fill[lpplDefect, rounded corners=6pt] (0.00,0.00) rectangle (7.20,5.20);
  \filldraw[fill=lpplGamma, line width=0.9pt, rounded corners=8pt] (1.10,1.30) rectangle (6.10,4.10);
  \filldraw[fill=lpplObs, line width=0.9pt] (2.95,2.75) ellipse [x radius=0.8, y radius=0.46];
  \draw[<->, line width=0.6pt] (2.95,3.21) -- (2.95,4.10);
  \foreach \x/\y in {%
    4.50/4.67,4.57/4.24,4.04/4.76,3.58/4.32,3.52/4.71,5.17/4.24,2.98/4.28,5.34/4.75,4.92/4.70,%
    4.08/4.24,5.83/4.96,6.17/4.70,5.73/3.96,5.82/3.52,3.06/4.88,4.85/3.70,5.71/4.43,2.66/4.85,%
    6.24/3.80,6.63/3.93,2.40/3.96,5.53/3.14,6.30/4.21,6.70/4.56,5.25/3.80,1.93/2.91,2.38/4.42,%
    5.96/2.94,1.92/2.35,1.53/2.64,3.98/3.11,2.23/4.98,0.96/2.76,1.82/4.89,1.24/3.24,6.41/3.29,%
    3.46/2.23,4.62/3.31,0.54/2.57,1.91/4.48,2.05/1.92,0.66/3.24,6.74/3.54,6.82/2.84,5.73/2.58,%
    6.56/2.33,0.96/2.19,4.06/2.15,2.49/3.63,4.94/3.00,1.57/2.08,5.31/2.15,1.48/4.54,3.16/1.16,%
    0.15/2.31,5.12/3.36,0.36/3.63,1.39/1.55,5.30/2.56,4.40/2.95,0.93/1.68,1.29/5.03,6.24/2.11,%
    2.24/1.04,1.93/1.52,0.94/4.26,4.75/2.64,0.53/1.28,4.27/2.54,0.63/2.04,0.24/0.96,4.86/1.78,%
    3.67/1.15,0.41/4.16,0.80/4.73,0.17/4.75,4.45/2.04,6.41/1.55,0.92/1.14,6.24/2.56,4.13/1.66,%
    0.21/2.81,0.22/1.87,6.91/1.39,1.77/1.06,6.68/0.86,2.65/1.09,6.84/5.04,0.14/0.40,5.72/1.73,%
    0.73/3.86,5.17/1.49,4.77/1.16,1.33/1.11,5.93/1.44,0.96/0.28,6.81/2.01,6.12/0.28,1.42/0.20,%
    5.47/1.07,0.55/0.26,4.89/2.18,4.53/0.28,6.99/0.30,2.45/2.85,2.73/2.44,3.22/3.01,3.49/2.63,%
    3.07/2.41%
  } { \fill[lpplSite] (\x,\y) circle (1.15pt); }
  \node (aY) at (2.95,2.75) {$Y$};
  \node (aA) at (2.95,1.85) {$A\in\A_Y$};
  \node[anchor=west] (aD) at (3.05,3.65) {$d(Y,\Sigma)$};
  \node[anchor=north west] (aI) at (1.25,3.98) {$\Gamma\setminus\Sigma$};
  \node (aR) at (3.60,0.65) {defect region $\Sigma$};
  \node (aT) at (3.60,-0.38) {(a) KMS states with a defect};

  \fill[lpplDefect, rounded corners=6pt] (7.65,0.00) rectangle (14.85,5.20);
  \draw[dashed, line width=0.7pt, rounded corners=8pt] (8.20,0.65) rectangle (14.30,4.65);
  \filldraw[fill=lpplGamma, line width=0.9pt, rounded corners=8pt] (8.75,1.30) rectangle (13.75,4.10);
  \filldraw[fill=lpplObs, line width=0.9pt] (10.60,2.75) ellipse [x radius=0.8, y radius=0.46];
  \draw[<->, line width=0.6pt] (10.60,3.21) -- (10.60,4.10);
  \foreach \x/\y in {%
    12.22/4.24,11.23/4.32,12.82/4.24,10.63/4.28,11.73/4.24,13.38/3.96,13.47/3.52,13.36/4.43,13.89/3.80,%
    14.16/3.95,13.18/3.14,13.95/4.21,9.58/2.91,10.03/4.42,13.61/2.94,9.57/2.35,9.18/2.64,11.63/3.11,%
    8.61/2.76,8.89/3.24,14.06/3.29,11.11/2.23,9.56/4.48,9.70/1.92,8.34/3.24,13.38/2.58,14.16/2.33,%
    8.61/2.19,11.71/2.15,10.14/3.63,12.59/3.00,9.22/2.08,12.96/2.15,9.13/4.51,9.04/1.55,12.95/2.56,%
    12.05/2.95,8.58/1.68,13.89/2.11,9.58/1.52,8.59/4.26,12.40/2.64,11.92/2.54,8.34/2.04,12.51/1.78,%
    12.10/2.04,14.06/1.55,8.57/1.14,13.89/2.56,11.78/1.66,13.37/1.73,8.38/3.86,12.82/1.49,13.58/1.44,%
    8.34/0.80,8.75/0.79,13.90/0.87,12.54/2.18,10.10/2.85,10.38/2.44,10.87/3.01,11.14/2.63,10.72/2.41%
  } { \fill[lpplSite] (\x,\y) circle (1.15pt); }
  \foreach \x/\y in {%
    12.20/4.78,11.69/4.79,11.18/4.79,12.99/4.79,12.57/4.79,13.48/4.96,13.81/4.79,10.71/4.88,10.31/4.85,%
    14.44/4.57,9.88/4.98,9.47/4.89,8.08/2.63,14.44/3.54,14.47/2.84,7.80/2.31,8.01/3.63,8.94/5.03,%
    8.06/1.27,7.89/0.96,8.06/4.16,8.45/4.79,7.82/4.75,7.87/1.87,14.56/1.39,14.49/5.04,7.79/0.40,%
    9.71/0.51,13.03/0.51,10.11/0.36,8.61/0.28,14.46/2.01,13.77/0.28,9.07/0.20,13.43/0.51,8.20/0.26,%
    14.64/0.30,14.23/0.46%
  } { \draw[lpplSite, line width=0.5pt] (\x,\y) circle (1.15pt); }
  \node (bY) at (10.60,2.75) {$Y$};
  \node (bA) at (10.60,1.85) {$A\in\A_Y$};
  \node[anchor=west] (bD) at (10.70,3.65) {$d(Y,\Lambda^c\cup B)$};
  \node[anchor=north west] (bI) at (8.90,3.98) {$\Lambda\setminus B$};
  \node (bR) at (11.25,0.98) {collar $B$};
  \node (bE) at (11.25,0.33) {exterior $\Lambda^c$};
  \node (bT) at (11.25,-0.38) {(b) finite-volume Gibbs states};
\end{tikzpicture}
\caption[Strong LPPL and finite-volume comparison.]{Geometry of strong LPPL and finite-volume comparison. (a) A defect supported in the gray region \(\Sigma\) affects an observable supported in \(Y\) by an error controlled by \(d(Y,\Sigma)\) (Theorem~\ref{thm:kms-lppl}). (b) For finite-volume Gibbs states, the relevant distance is \(d(Y,\Lambda^c\cup B)\), where \(B\) supports the boundary operator \(K\) (Theorem~\ref{thm:gibbs}). The dashed line marks the boundary of the finite lattice \(\Lambda\). Open circles lie outside it.}
\label{fig:lppl-and-gibbs}
\end{figure}

We first compare KMS states of the background interaction with those obtained by adding an admissible defect. The constants in the bounds are independent of the strength and geometry of the defect. Figure~\ref{fig:lppl-and-gibbs}(a) illustrates the setting.

\begin{theorem}[Strong LPPL for KMS states]\label{thm:kms-lppl}\leavevmode
\begin{enumerate}[label=(\roman*),leftmargin=*]
\item \emph{With a free spectral gap.}
Assume the hypotheses of Theorem~\ref{thm:uniqueness}(i), with \(\lambda_{\mathrm g}\) as chosen there.
In regime~\ref{ass:exponential}, fix \(0<\theta<\theta_{\rm in}\).
There are constants \(C>0\) and, in regime \ref{ass:exponential}, \(m>0\), depending only on the model data and, in regime \ref{ass:exponential}, on \(\theta\), with the following property. For every \(\lambda\) with \(|\lambda|\le\lambda_{\mathrm g}\), every \(\beta>0\), \(\Sigma\subset\Gamma\), \(Y\Subset\Gamma\) and \(A\in\A_Y\),
\[
 \sup_{\substack{\Psi\in\mathfrak J_1\\\Sigma(\Psi)=\Sigma}}
 \sup_{\rho\in\KMS_\beta(\tau^{\Phi_\lambda+\Psi})}
 |\rho(A)-\omega_{\beta,\lambda}(A)|
 \le C\|A\|\,\cS(Y,\Sigma).
\]
\item \emph{Without a gap assumption.}
Assume the hypotheses of Theorem~\ref{thm:uniqueness}(ii), with \(\lambda_{\mathrm{ng}}(\beta)\) as chosen there.
In regime \ref{ass:exponential}, take \(\theta=\theta_{\rm in}\). Use \(C(\beta)\) and \(G_\beta\) as fixed at the beginning of this section.
For every \(\beta>0\), every \(\lambda\) with \(|\lambda|\le\lambda_{\mathrm{ng}}(\beta)\), \(\Sigma\subset\Gamma\), \(Y\Subset\Gamma\) and \(A\in\A_Y\),
\[
 \sup_{\substack{\Psi\in\mathfrak J_1\\\Sigma(\Psi)=\Sigma}}
 \sup_{\rho\in\KMS_\beta(\tau^{\Phi_\lambda+\Psi})}
 |\rho(A)-\omega_{\beta,\lambda}(A)|
 \le C(\beta)\|A\|\,\cSbeta(Y,\Sigma).
\]
\end{enumerate}
\end{theorem}

\paragraph{Decay of correlations.}

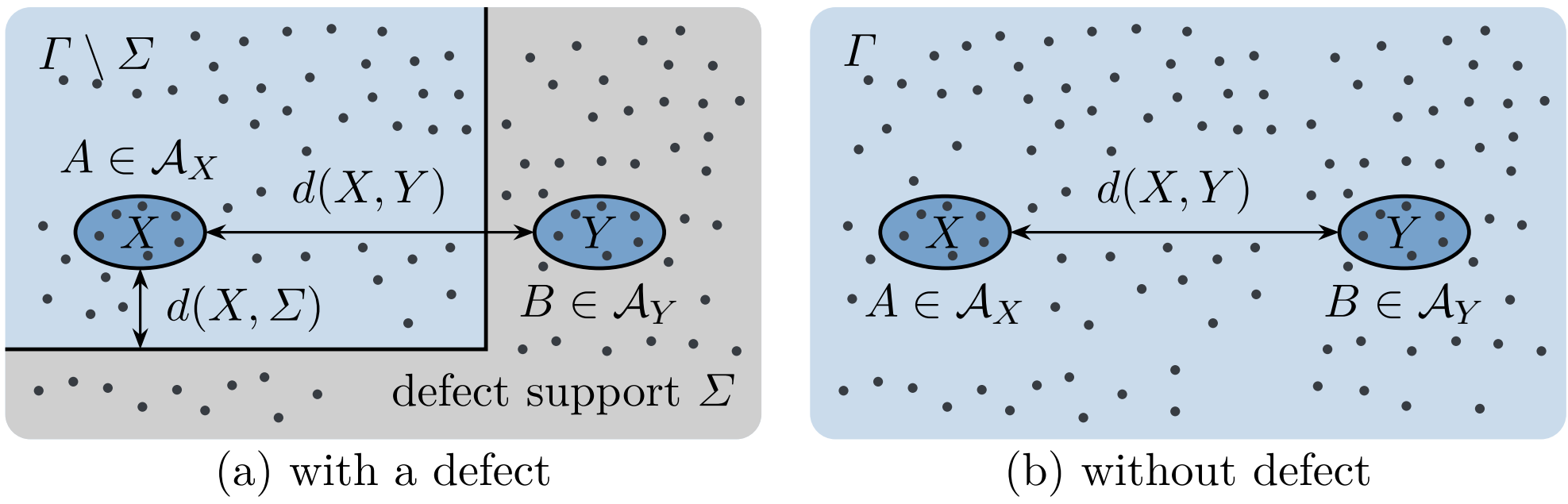
\begin{figure}[H]
\centering
\normalsize
\definecolor{lpplGamma}{RGB}{202,219,235}
\definecolor{lpplDefect}{RGB}{207,207,207}
\definecolor{lpplObs}{RGB}{118,161,203}
\definecolor{lpplSite}{RGB}{55,60,66}
\begin{tikzpicture}[>={Stealth[length=5pt]},x=0.9cm,y=0.75cm,font=\normalsize]
  \def\lpplCorrelationSites{%
    3.02/4.56,6.25/4.54,2.61/4.53,3.49/4.53,1.76/4.48,5.90/4.43,2.21/4.42,5.29/4.37,%
    4.11/4.26,4.86/4.24,3.79/4.20,3.34/4.17,6.14/4.16,6.55/4.15,1.93/4.14,2.82/4.02,%
    0.54/3.99,5.54/3.99,0.85/3.95,5.07/3.94,2.37/3.90,1.55/3.88,1.22/3.84,3.87/3.84,%
    4.20/3.83,3.40/3.80,2.02/3.78,6.80/3.76,6.10/3.75,6.46/3.73,2.61/3.65,5.77/3.65,%
    5.44/3.58,3.13/3.57,2.31/3.50,4.64/3.50,3.63/3.48,3.96/3.44,4.27/3.44,6.51/3.36,%
    6.20/3.24,2.79/3.20,5.52/3.09,5.13/3.07,4.81/3.06,5.83/3.06,3.91/3.04,6.49/2.98,%
    2.37/2.75,4.98/2.73,4.64/2.71,1.27/2.59,5.52/2.59,2.06/2.57,1.03/2.50,5.28/2.50,%
    1.58/2.48,5.83/2.48,6.29/2.45,6.59/2.38,0.35/2.37,0.87/2.26,5.12/2.26,1.61/2.19,%
    5.86/2.19,3.31/2.13,4.12/2.13,1.32/2.04,5.57/2.04,2.78/2.03,2.33/2.01,0.56/2.00,%
    3.77/2.00,6.14/1.97,4.98/1.92,3.45/1.77,4.13/1.61,0.39/1.56,6.48/1.55,3.73/1.29,%
    5.10/1.09,5.98/1.09,6.37/1.06,4.79/1.00,5.57/0.98,6.77/0.98,2.40/0.69,0.63/0.64,%
    2.10/0.60,0.95/0.57,1.59/0.55,0.31/0.54,2.89/0.50,1.27/0.36,1.85/0.32,2.53/0.24%
  }
  \begin{scope}
  \fill[lpplGamma, rounded corners=6pt] (0.00,0) rectangle (7.00,4.8);
  \begin{scope}
    \clip[rounded corners=6pt] (0.00,0) rectangle (7.00,4.8);
    \fill[lpplDefect] (0.00,0) -- (7.00,0) -- (7.00,4.8)
      -- (4.45,4.8) -- (4.45,1.00) -- (0.00,1.00) -- cycle;
  \end{scope}
  \draw[line width=0.9pt] (0.00,1.00) -- (4.45,1.00) -- (4.45,4.8);
  \filldraw[fill=lpplObs, line width=0.9pt] (1.25,2.30) ellipse [x radius=0.6, y radius=0.4];
  \filldraw[fill=lpplObs, line width=0.9pt] (5.50,2.30) ellipse [x radius=0.6, y radius=0.4];
  \draw[<->, line width=0.6pt] (1.25,1.00) -- (1.25,1.90);
  \draw[<->, line width=0.6pt] (1.85,2.30) -- (4.90,2.30);
  \foreach \x/\y in \lpplCorrelationSites {
    \fill[lpplSite] (\x,\y) circle (1.15pt);
  }
  \foreach \x/\y in {0.93/1.80,1.09/1.47,0.79/1.39} {
    \fill[lpplSite] (\x,\y) circle (1.15pt);
  }
  \node (aX) at (1.25,2.30) {$X$};
  \node[fill=lpplGamma, inner sep=1pt] (aA) at (1.25,3.10) {$A\in\A_X$};
  \node (aY) at (5.50,2.30) {$Y$};
  \node (aB) at (5.50,1.50) {$B\in\A_Y$};
  \node[anchor=south, fill=lpplGamma, inner sep=2pt] (aD) at (3.375,2.40) {$d(X,Y)$};
  \node[anchor=north west] (aG) at (0.15,4.68) {$\Gamma\setminus\Sigma$};
  \node (aT) at (3.50,-0.38) {(a) with a defect};
  \node[anchor=west, fill=lpplGamma, inner sep=2pt] (aE) at (1.40,1.45) {$d(X,\Sigma)$};
  \node[align=center] (aS) at (5.17,0.50) {defect support $\Sigma$};
  \end{scope}

  \begin{scope}[shift={(7.45,0)}]
  \fill[lpplGamma, rounded corners=6pt] (0.00,0) rectangle (7.00,4.8);
  \filldraw[fill=lpplObs, line width=0.9pt] (1.25,2.30) ellipse [x radius=0.6, y radius=0.4];
  \filldraw[fill=lpplObs, line width=0.9pt] (5.50,2.30) ellipse [x radius=0.6, y radius=0.4];
  \draw[<->, line width=0.6pt] (1.85,2.30) -- (4.90,2.30);
  \foreach \x/\y in \lpplCorrelationSites {
    \fill[lpplSite] (\x,\y) circle (1.15pt);
  }
  \foreach \x/\y in {1.36/3.26,0.71/3.45,0.45/3.22,2.05/3.21,0.96/2.87,3.38/0.77,6.20/0.34,5.52/0.37,5.13/0.54,4.70/0.59,3.38/0.31,3.07/1.67,2.58/1.27,2.27/1.47,1.45/4.37,1.15/4.26} {
    \fill[lpplSite] (\x,\y) circle (1.15pt);
  }
  \node (bX) at (1.25,2.30) {$X$};
  \node (bA) at (1.25,1.50) {$A\in\A_X$};
  \node (bY) at (5.50,2.30) {$Y$};
  \node (bB) at (5.50,1.50) {$B\in\A_Y$};
  \node[anchor=south, fill=lpplGamma, inner sep=2pt] (bD) at (3.375,2.40) {$d(X,Y)$};
  \node[anchor=north west] (bG) at (0.15,4.68) {$\Gamma$};
  \node (bT) at (3.50,-0.38) {(b) without defect};
  \end{scope}
\end{tikzpicture}
\caption[Decay of correlations with and without defects.]{Geometry of the correlation bounds. (a) A defect supported in the gray region \(\Sigma\) adds an error controlled by the distance of $X$ to \(\Sigma\) (Theorem~\ref{thm:correlations}). (b) Without defects, correlations are controlled by the separation of the observable supports \(X\) and \(Y\) (Corollary~\ref{cor:background-correlations}).}
\label{fig:correlations}
\end{figure}

We next prove decay of correlations in the presence of admissible defects, uniformly in their strength and geometry. For a state \(\omega\), define \(\Cov_\omega(A,B):=\omega(AB)-\omega(A)\omega(B)\). The following estimate is one-sided: the defect may overlap one observable, while the bound depends on the separation of the other observable from both that observable and the defect. Figure~\ref{fig:correlations} illustrates the geometry.

\begin{theorem}[One-sided decay of correlations]\label{thm:correlations}\leavevmode
\begin{enumerate}[label=(\roman*),leftmargin=*,beginpenalty=10000]
\item \emph{With a free spectral gap.}
Assume the hypotheses of Theorem~\ref{thm:uniqueness}(i), with \(\lambda_{\mathrm g}\) as chosen there.
In regime~\ref{ass:exponential}, fix \(0<\theta<\theta_{\rm in}\).
There are constants \(C>0\) and, in regime \ref{ass:exponential}, \(m>0\), depending only on the model data and, in regime \ref{ass:exponential}, on \(\theta\), with the following property. For every \(\lambda\) with \(|\lambda|\le\lambda_{\mathrm g}\), every \(\beta>0\), \(\Sigma\subset\Gamma\), \(X,Y\Subset\Gamma\), \(A\in\A_X\) and \(B\in\A_Y\),
\[
 \sup_{\substack{\Psi\in\mathfrak J_1\\\Sigma(\Psi)=\Sigma}}
 \sup_{\rho\in\KMS_\beta(\tau^{\Phi_\lambda+\Psi})}
 \lvert\Cov_\rho(A,B)\rvert
 \le C\|A\|\|B\|\bigl[\cS(X,Y)+\cS(X,\Sigma)\bigr].
\]
\item \emph{Without a gap assumption.}
Assume the hypotheses of Theorem~\ref{thm:uniqueness}(ii), with \(\lambda_{\mathrm{ng}}(\beta)\) as chosen there.
In regime \ref{ass:exponential}, take \(\theta=\theta_{\rm in}\). Use \(C(\beta)\) and \(G_\beta\) as fixed at the beginning of this section.
For every \(\beta>0\), every \(\lambda\) with \(|\lambda|\le\lambda_{\mathrm{ng}}(\beta)\), \(\Sigma\subset\Gamma\), \(X,Y\Subset\Gamma\), \(A\in\A_X\) and \(B\in\A_Y\),
\[
 \sup_{\substack{\Psi\in\mathfrak J_1\\\Sigma(\Psi)=\Sigma}}
 \sup_{\rho\in\KMS_\beta(\tau^{\Phi_\lambda+\Psi})}
 \lvert\Cov_\rho(A,B)\rvert
 \le C(\beta)\|A\|\|B\| \bigl[\cSbeta(X,Y)+\cSbeta(X,\Sigma)\bigr] .
\]
\end{enumerate}
The same bounds hold with \(X\) and \(Y\) exchanged.
\end{theorem}

Setting \(\Psi=0\) in Theorem~\ref{thm:correlations} gives the usual decay of correlations without defects. In the gapped case, the constants are uniform in temperature, and the convergence established in Corollary~\ref{cor:low-temperature} extends the estimate to the unique ground state.

\begin{corollary}[Decay of correlations]\label{cor:background-correlations}
Assume the hypotheses of Theorem~\ref{thm:correlations} and use the same constants and decay profiles.
\begin{enumerate}[label=(\roman*),leftmargin=*]
\item \emph{With a free spectral gap.}
For every \(\lambda\) with \(|\lambda|\le\lambda_{\mathrm g}\), every \(\beta\in(0,\infty]\), \(X,Y\Subset\Gamma\), \(A\in\A_X\) and \(B\in\A_Y\),
\[
 \lvert\Cov_{\omega_{\beta,\lambda}}(A,B)\rvert
 \le C\|A\|\|B\|\min\{\cS(X,Y),\cS(Y,X)\}.
\]
\item \emph{Without a gap assumption.}
For every \(\beta>0\), every \(\lambda\) with \(|\lambda|\le\lambda_{\mathrm{ng}}(\beta)\), \(X,Y\Subset\Gamma\), \(A\in\A_X\) and \(B\in\A_Y\),
\[
 \lvert\Cov_{\omega_{\beta,\lambda}}(A,B)\rvert
 \le C(\beta)\|A\|\|B\|\min\{\cSbeta(X,Y),\cSbeta(Y,X)\}.
\]
\end{enumerate}
\end{corollary}

Clustering at every positive temperature is known for one-dimensional spin systems with translation-invariant finite-range interactions~\cite{Araki69} and with sufficiently decaying long-range interactions~\cite{KK25}, with temperature-dependent bounds.
For finite-range weakly interacting fermions, Tong and Zhan~\cite{TZ25} prove exponential clustering at each fixed temperature, with a temperature-dependent coupling threshold and one observable supported on a bounded number of adjacent sites.
For correlations between odd observables, more is known.
Hastings~\cite{Hastings04} showed that, for every lattice fermion Hamiltonian with finite-range interactions, the correlation \(\langle AB\rangle_\beta\) of two odd observables in the Gibbs state of a finite system decays exponentially in the distance between their supports at every positive temperature, with a decay length of order \(\beta\) at low temperature; neither weak coupling nor a gap is required.
Hern\'andez-Santana, Gogolin, Cirac and Ac\'in~\cite[Theorem~3]{HGCA17} extended this to two-site interactions that decay like \(r^{-\alpha}\) with \(\alpha>2D\) on a \(D\)-dimensional hypercubic lattice, where they obtain power-law decay with an exponent arbitrarily close to \(\alpha\).
As pointed out in~\cite{Hastings04}, such results cannot hold for even observables without further assumptions, since these may exhibit long-range order at positive temperature.
Corollary~\ref{cor:background-correlations} covers observables of both parities, long-range interactions and arbitrary finite observation supports; under the free gap assumption, both the coupling interval and the correlation bounds are independent of temperature.
Without a gap, in regime~\ref{ass:exponential} with \(\theta_{\rm in}=1\), the decay length of \(G_\beta\) is of order \(1+\beta\), the same temperature dependence as in~\cite{Hastings04}, which is optimal in general according to the examples of free fermions cited there.
High-temperature results for broader interaction classes, together with a comparison of spatial decay rates and support dependence, are discussed in Section~\ref{sec:results-high-temperature}.

\paragraph{Finite-volume Gibbs states.}
We next establish a form of local indistinguishability~\cite[Definition~3]{CMTW25}, also known as locality of temperature~\cite[Section~II.C]{KGKRE14}. Here, we compare local expectations in the unique infinite-volume KMS state with those in finite-volume Gibbs states, also allowing arbitrary even defects in the finite system. For \(\Lambda\Subset\Gamma\), \(B\subset\Lambda\) and \(K=K^*\in\A_B^+\), we set 
\[
 H^{(\lambda)}_{\Lambda,K}
   :=\sum_{Z\subset\Lambda}\Phi_\lambda(Z)+K.
\]
Denote by $\Tr$ the trace of the Fock representation of $\A_\Lambda$ and define the finite-volume Gibbs state by
\begin{equation}\label{eq:finite-gibbs-definition}
 \rho^{(\lambda)}_{\beta,\Lambda,K}(A)
   :=\frac{\Tr(\E^{-\beta H^{(\lambda)}_{\Lambda,K}}A)}
                {\Tr(\E^{-\beta H^{(\lambda)}_{\Lambda,K}})},
\end{equation}
for all $A \in \A_\Lambda$. For each fixed local observable, this expectation converges to its expectation in the unique KMS state as the distance of its support to \(\Lambda^c\cup B\) tends to infinity. The constants in these bounds are independent of the volume \(\Lambda\), the region \(B\) and the defect term \(K\). Figure~\ref{fig:lppl-and-gibbs}(b) illustrates the geometry.

\begin{theorem}[Finite-volume Gibbs states]\label{thm:gibbs}\leavevmode
\begin{enumerate}[label=(\roman*),leftmargin=*]
\item \emph{With a free spectral gap.}
Assume the hypotheses of Theorem~\ref{thm:uniqueness}(i), with \(\lambda_{\mathrm g}\) as chosen there.
In regime~\ref{ass:exponential}, fix \(0<\theta<\theta_{\rm in}\).
There are constants \(C>0\) and, in regime \ref{ass:exponential}, \(m>0\), depending only on the model data and, in regime \ref{ass:exponential}, on \(\theta\), with the following property. For every \(\lambda\) with \(|\lambda|\le\lambda_{\mathrm g}\), every \(\beta>0\), \(\Lambda\Subset\Gamma\), \(B\subset\Lambda\), \(Y\subset\Lambda\) and \(A\in\A_Y\),
\[
 \sup_{K=K^*\in\A_B^+}
 |\rho^{(\lambda)}_{\beta,\Lambda,K}(A)-\omega_{\beta,\lambda}(A)|
 \le C\|A\|\,\cS(Y,\Lambda^c\cup B).
\]
\item \emph{Without a gap assumption.}
Assume the hypotheses of Theorem~\ref{thm:uniqueness}(ii), with \(\lambda_{\mathrm{ng}}(\beta)\) as chosen there.
In regime \ref{ass:exponential}, take \(\theta=\theta_{\rm in}\). Use \(C(\beta)\) and \(G_\beta\) as fixed at the beginning of this section.
For every \(\beta>0\), every \(\lambda\) with \(|\lambda|\le\lambda_{\mathrm{ng}}(\beta)\), \(\Lambda\Subset\Gamma\), \(B\subset\Lambda\), \(Y\subset\Lambda\) and \(A\in\A_Y\),
\[
 \sup_{K=K^*\in\A_B^+}
 |\rho^{(\lambda)}_{\beta,\Lambda,K}(A)-\omega_{\beta,\lambda}(A)|
 \le C(\beta)\|A\|\,\cSbeta(Y,\Lambda^c\cup B).
\]
\end{enumerate}
\end{theorem}

\begin{remark}[Relation to strong LPPL]\label{rem:gibbs-condition}
Let
\[
 W_\Lambda:=
 \sum_{\substack{Z\cap\Lambda\ne\varnothing\\
                 Z\cap\Lambda^c\ne\varnothing}}
 \Phi_\lambda(Z)
\]
be the sum of the interaction terms crossing the boundary of \(\Lambda\).
Removing \(W_\Lambda\) and adding \(K\) decouples \(\Lambda\) from its exterior and gives the Hamiltonian \(H^{(\lambda)}_{\Lambda,K}\) inside \(\Lambda\).
Every \(\beta\)-KMS state of the modified dynamics therefore restricts to the Gibbs state \(\rho^{(\lambda)}_{\beta,\Lambda,K}\) defined in~\eqref{eq:finite-gibbs-definition}.
The Araki--Ion Gibbs condition expresses the corresponding property of equilibrium states: for even states, the Araki perturbation removing the boundary coupling factorizes into the finite-volume Gibbs state and an exterior state~\cite{ArakiIon74,ArakiMoriyaGibbs03}.

For finite-range \(\Phi_\lambda\), the modification \(K-W_\Lambda\) is supported in \(B\) together with a boundary layer.
Theorem~\ref{thm:kms-lppl} then gives a finite-volume comparison away from this region.
For long-range \(\Phi_\lambda\), terms in \(W_\Lambda\) may reach arbitrarily far into the sample.
The proof of Theorem~\ref{thm:gibbs} estimates their influence using the decay of \(\Phi_\lambda\), giving the stated bounds in terms of \(\cS(Y,\Lambda^c\cup B)\) and \(\cSbeta(Y,\Lambda^c\cup B)\).

In part~(i), the gap condition \eqref{eq:free-band} concerns the infinite-volume operator \(h\) only; its restriction to \(\Lambda\) may have eigenvalues in the gap, as for the edge modes of topological insulators and superconductors.
\end{remark}

\subsection{High-temperature results for general fermionic interactions}
\label{sec:results-high-temperature}

We next derive high-temperature uniqueness, decay of correlations and strong LPPL for arbitrary fermionic interactions satisfying Hypothesis~\ref{hyp:locality}. These conclusions follow directly from Theorems~\ref{thm:uniqueness}, \ref{thm:kms-lppl} and~\ref{thm:correlations} by setting \(h=0\) and choosing $\beta$ sufficiently small.

\begin{theorem}[High-temperature uniqueness, decay of correlations and strong LPPL]
\label{thm:high-temperature}
Assume Hypothesis~\ref{hyp:locality} with \(h=0\) and \(\cV \not = 0\). Choose \(J>0\) as in~\eqref{eq:interaction-strength-bound}.
In regime~\ref{ass:polynomial-weight}, take \(p_\Phi=p_{\cV}\).
There exist model constants \(c_{\rm HT}>0\), \(C>0\) and, in regime~\ref{ass:exponential}, \(m>0\), all independent of \(J\).
Use the site sum \(\cS\) from Definition~\ref{def:decay-profile} with the profile
\[
 G(D):=\begin{cases}
 \E^{-mD^{\theta_{\rm in}}},&\text{in \ref{ass:exponential}},\\
 (1+D)^{-p_{\cV}},&\text{in \ref{ass:polynomial-weight}},
 \end{cases}
 \qquad D\ge0,
 \qquad G(\infty):=0.
\]
For every \(\beta\) with \(0<\beta\le\beta_{\rm HT}:=c_{\rm HT}/J\), the following hold.
\begin{enumerate}[label=(\roman*),leftmargin=*]
\item \emph{Uniqueness.} We have
\[
 \KMS_\beta(\tau^{\cV})=\{\omega_\beta\}.
\]
\item \emph{Decay of correlations.} For every \(\Sigma\subset\Gamma\), \(X,Y\Subset\Gamma\), \(A\in\A_X\) and \(B\in\A_Y\),
\[
 \sup_{\substack{\Psi\in\mathfrak J_1\\\Sigma(\Psi)=\Sigma}}
 \sup_{\rho\in\KMS_\beta(\tau^{\cV+\Psi})}
 \lvert\Cov_\rho(A,B)\rvert
 \le C\|A\|\|B\|\bigl[\cS(X,Y)+\cS(X,\Sigma)\bigr].
\]
The same bound holds with \(X\) and \(Y\) exchanged.
\item \emph{Strong LPPL.} For every \(\Sigma\subset\Gamma\), \(Y\Subset\Gamma\) and \(A\in\A_Y\),
\[
 \sup_{\substack{\Psi\in\mathfrak J_1\\\Sigma(\Psi)=\Sigma}}
 \sup_{\rho\in\KMS_\beta(\tau^{\cV+\Psi})}
 |\rho(A)-\omega_\beta(A)|
 \le C\|A\|\,\cS(Y,\Sigma).
\]
\end{enumerate}
\end{theorem}

\begin{proof}
Take \(h=0\).
Set \(\cV_0:=\cV/J\), whose interaction norm is at most \(1\).
Choose the constants \(c_0,N\) in Theorem~\ref{thm:uniqueness}(ii) for interaction-norm bound \(1\), and set
\[
 c_{\rm HT}:=\min\{1,c_0/2^{N-1}\}.
\]
This choice is independent of \(J\). Let \(0<\beta\le\beta_{\rm HT}\). Then \(\beta J\le c_{\rm HT}\le\lambda_{\mathrm{ng}}(1)\). For every \(\Psi\in\mathfrak J_1\), rescaling time gives
\[
 \KMS_\beta(\tau^{\cV+\Psi})
 =\KMS_1\bigl(\tau^{(\beta J)\cV_0+\beta\Psi}\bigr).
\]
Apply Theorems~\ref{thm:uniqueness}(ii), \ref{thm:correlations}(ii) and~\ref{thm:kms-lppl}(ii) to \(\cV_0\) at inverse temperature \(1\) and coupling \(\lambda=\beta J\). The map \(\Psi\mapsto\beta\Psi\) is a bijection on \(\mathfrak J_1\) preserving defect supports, so their conclusions give all three assertions. Take the prefactor \(C:=C(1)\) and, in \ref{ass:exponential}, the rate \(m:=m_0/2\), for this normalized interaction. These constants and the spatial profile are fixed throughout \(0<\beta\le\beta_{\rm HT}\).
\end{proof}

We now compare Theorem~\ref{thm:high-temperature} with existing high-temperature results. Uniqueness of KMS states at high temperature is a classical result for spin systems: see, for example, \cite{BR97} and \cite[Theorem~2.1]{FU15}.

High-temperature decay of correlations has been studied extensively for short- and long-range systems~\cite{FU15,KGKRE14,KKS25}. For finite-range interactions, Kliesch et al.\ prove decay of correlations, also in the fermionic case \cite[Theorem~4]{KGKRE14}. Their bound decays exponentially with the distance between the observable supports, with a prefactor linear in the boundary size of one of the supports.
Theorem~\ref{thm:high-temperature}(ii) extends their high-temperature clustering result for the ordinary covariance to interactions with stretched-exponential or polynomial decay.

For \(\Lambda\Subset\Gamma\), define \(\cV_\Lambda(Z):=\cV(Z)\) when \(Z\subset\Lambda\), and \(\cV_\Lambda(Z):=0\) otherwise.
Deleting interaction terms does not increase either norm in \eqref{eq:interaction-strength-bound}, so the same bound \(J\) applies to \(\cV_\Lambda\).
The finite-volume Gibbs state of \(\sum_{Z\subset\Lambda}\cV(Z)\), extended by the tracial state on \(\A_{\Lambda^c}\), is a \(\beta\)-KMS state of \(\tau^{\cV_\Lambda}\).
Applying Theorem~\ref{thm:high-temperature}(ii) to \(\cV_\Lambda\) with \(\Psi=0\) gives the same clustering estimate for these Gibbs states, with constants independent of \(\Lambda\).

For exponentially decaying interactions, Theorem~\ref{thm:high-temperature}(ii) with \(\Psi=0\) and Lemma~\ref{lem:site-sum-bounds} give exponential decay of correlations. The prefactor is linear in the boundary size of the support of either observable. This establishes the fermionic analogue of \cite[Conjecture~12]{CMTW25}, which asks for exponential decay with polynomial dependence on the observable support sizes.\footnote{This assertion was already claimed in \cite[Theorem~3.2]{FU15}. As explained in \cite[Remark~14]{CMTW25}, the proof actually yields an exponential support prefactor rather than the stated polynomial one.}

For long-range \(k\)-body interactions in spin and fermionic systems, Kim, Kuwahara and Saito \cite[Theorem~2]{KKS25} prove decay of correlations with a prefactor that grows exponentially in the support sizes of the observables. 
However, their theorem allows slower polynomial decay of the interaction and gives a larger clustering exponent for pointwise two-site power laws.

The high-temperature condition \(\beta J\le c_{\rm HT}\) constrains only the background interaction \(\cV\).
Earlier LPPL estimates derived from clustering depend on the perturbation strength~\cite{CMTW25,MSAC26,Becker26}.
For quantum spin systems with weak finite-range interactions around gapped on-site Hamiltonians, Adhikari, Henheik, Lemm and Wessel prove clustering and local indistinguishability uniformly in temperature~\cite{AHLW26}.
Their LPPL estimate for an arbitrary bounded defect \(W\) contains the factor \(\E^{2\beta\|W\|}\); a strength-independent comparison follows when both Hamiltonians satisfy their weak-interaction assumptions~\cite[Theorem~4.1 and the subsequent discussion]{AHLW26}.
To the best of the authors' knowledge, part~(iii) gives the first strong LPPL theorem at positive temperature for a general class of noncommuting quantum lattice systems that covers arbitrary admissible defects, applies to every perturbed KMS state, and has a spatial bound uniform in both defect strength and extent.

\subsection{Stability under uniformly weak extensive perturbations}
\label{sec:results-continuity}

We finally show that local expectations are stable under perturbations that may act throughout the lattice. Their effect is controlled by the perturbation strength per site: for \(\Psi\in\mathfrak J_1\), we define
\begin{equation}\label{eq:perturbation-intensity}
 J_\Psi:=\sup_{x\in\Gamma}\sum_{Z\ni x}\|\Psi(Z)\|.
\end{equation}
The following theorem bounds the change in local expectations linearly in \(J_\Psi\).

\begin{theorem}[Local stability under uniformly weak extensive perturbations]\label{thm:continuity}\leavevmode
\begin{enumerate}[label=(\roman*),leftmargin=*,beginpenalty=10000]
\item \emph{With a free spectral gap.}
Assume the hypotheses of Theorem~\ref{thm:uniqueness}(i), with \(\lambda_{\mathrm g}\) as chosen there.
There is a constant \(C>0\), depending only on the model data, with the following property. For every \(\lambda\) with \(|\lambda|\le\lambda_{\mathrm g}\), every \(\beta>0\), \(\Psi\in\mathfrak J_1\), \(Y\Subset\Gamma\) and \(A\in\A_Y\),
\[
 \sup_{\rho\in\KMS_\beta(\tau^{\Phi_\lambda+\Psi})}
 |\rho(A)-\omega_{\beta,\lambda}(A)|
 \le C|Y|\|A\|J_\Psi.
\]
\item \emph{Without a gap assumption.}
Assume the hypotheses of Theorem~\ref{thm:uniqueness}(ii), with \(\lambda_{\mathrm{ng}}(\beta)\) as chosen there.
Use \(C(\beta)\) as fixed at the beginning of this section.
For every \(\beta>0\), every \(\lambda\) with \(|\lambda|\le\lambda_{\mathrm{ng}}(\beta)\), \(\Psi\in\mathfrak J_1\), \(Y\Subset\Gamma\) and \(A\in\A_Y\),
\[
 \sup_{\rho\in\KMS_\beta(\tau^{\Phi_\lambda+\Psi})}
 |\rho(A)-\omega_{\beta,\lambda}(A)|
 \le \frac{\beta}{1+\beta}C(\beta)|Y|\|A\|J_\Psi.
\]
\end{enumerate}
\end{theorem}

Araki proved analytic dependence of local equilibrium expectations on finite-range interactions in one dimension at every fixed positive temperature~\cite{Araki69}.
For weakly interacting gapped fermions, De Roeck and Salmhofer obtain temperature- and volume-uniform analytic control in norms of normal-ordered coefficients~\cite[Section~2.4 and Theorem~2]{DRS19}.
Their observable norms can grow exponentially with the support size even when the operator norm is one.

Volume-independent comparison bounds for finite-volume Gibbs states were established in~\cite[Theorem~34]{CMTW25} and~\cite[Theorem~III.1]{MSAC26}.
Becker proves analogous bounds and differentiability for selected paths of infinite-volume fermionic KMS states~\cite[Theorem~4.16]{Becker26}.
These results assume clustering uniformly along the interpolation; high-temperature clustering theorems supply this assumption under suitable uniform interaction bounds.

Theorem~\ref{thm:continuity} controls every perturbed KMS state using the unweighted intensity \(J_\Psi\).
The spatial decay and weak-coupling assumptions concern only the background interaction \(\Phi_\lambda\), so the perturbation may lie outside the decay classes required in those comparison theorems, and the perturbed dynamics may have several KMS states.
The estimate uses the operator norm \(\|A\|\) with a factor linear in \(|Y|\).
The bounds in~\cite[Theorem~34]{CMTW25} and~\cite[Theorem~4.16]{Becker26} contain an additional support factor inherited from the assumed clustering estimate, while~\cite[Theorem~III.1]{MSAC26} has explicit exponential dependence on \(|Y|\).
Under our free gap assumption, the constant is uniform over all positive temperatures, whereas the general comparison bounds derived from quantum belief propagation have temperature-dependent prefactors.

\section{A quantum Dobrushin comparison theorem}
\label{sec:comparison}

The central new tool of this paper is Theorem~\ref{thm:comparison}, a quantitative quantum analogue of Dobrushin's comparison theorem \cite{Dobrushin70}.
It provides a common basis for the uniqueness, local stability and correlation bounds of Section~\ref{sec:results}.
We formulate it for odd quasi-local operator families \((F_u)_{u\in\Gamma\times\{1,\ldots,n\}}\) on the CAR algebra \(\cC=\operatorname{CAR}(\ell^2(\Gamma,\C^n))\).
The operators \(F_u\) are small local perturbations of a linear Majorana family \((F_u^0)\) satisfying a normalized completeness condition (the Parseval condition \eqref{eq:parseval-condition}).
The main assumptions are weighted Schur bounds on the coefficients of \((F_u^0)\) and sufficiently small local corrections.
The precise bounds are stated in Hypothesis~\ref{hyp:perturbed-family}.

The key result is an exact decomposition of every local observable \(A\), with spatial bounds on its coefficients (Theorem~\ref{thm:comparison}(i)):
\[
 A=\omega_F(A)\1+\sum_{u\in\Gamma\times\{1,\ldots,n\}}
       (F_u^*X_u+Y_uF_u),
 \qquad X_u,Y_u\in\cC.
\]
The state \(\omega_F\) is determined by \((F_u)\): it is the unique invariant state of the auxiliary evolution of Section~\ref{sec:roadmap-comparison}, and the only state that can be annihilated by the family. The coefficient norms are summable in \(u\), and their decay away from \(\supp A\) is controlled by the locality weight in Hypothesis~\ref{hyp:perturbed-family}.

For a difference of states \(\varphi=\omega'-\omega\), the scalar term cancels.
Bounds on \(\varphi(F_u^*E)\) and \(\varphi(GF_u)\), for \(E,G\in\cC\), therefore give bounds on \(\varphi(A)\).
Theorem~\ref{thm:comparison}(ii) makes this implication quantitative, with coefficients that decay with the distance between the site indexed by \(u\) and \(\supp A\).
If two states agree on all products \(F_u^*E\) and \(GF_u\), they coincide.
If their differences on these products decay away from a defect, the theorem gives the same spatial decay for differences of local expectations.
The consequences parallel those of the classical Dobrushin comparison theorem: uniqueness, decay of correlations, and quantitative bounds on the influence of distant perturbations and boundary conditions.
We derive the corresponding abstract estimates below and explain their relation to the classical results in Subsection~\ref{sec:comparison-remarks}.

To apply the abstract comparison theorem to KMS states and ground states of weakly interacting fermion systems, we extend them to a doubled CAR algebra in Section~\ref{sec:thermofield}.
For free systems, \(F_u^0\) annihilates the GNS vector of the corresponding extension; weak interactions give perturbed families \(F_u\) that annihilate the interacting extensions' GNS vectors.
Defects change \(F_u\), and the comparison theorem turns bounds on these changes into bounds on local expectations.
Restriction to the physical algebra then yields the results of Section~\ref{sec:results}.

\subsection{Setting}
\label{sec:comparison-setting}

We use the setting of Section~\ref{sec:setting} with \(n\in\mathbb N\) fermionic modes per site on the lattice $\Gamma$ and write
\[
 \cC:=\operatorname{CAR}(\ell^2(\Gamma,\C^n)).
\]
The local algebras \(\cC_X\) and \(\cC_\loc\), parity automorphism \(\Theta\), and the even and odd subspaces \(\cC_X^\pm\) are defined as before. We denote the canonical annihilation operators by \(c_{x,j}\), where $x \in \Gamma$ and $j \in \{1, \ldots, n\}$, and the corresponding Majoranas by \(\gamma_{x,k}\), where $k \in \{1, \ldots, 2 n\}$. 

We use the index sets
\[
 \cI:=\Gamma\times\{1,\ldots,n\}
 \qquad\mbox{and}\qquad 
 \cJ:=\Gamma\times\{1,\ldots,2n\}
\]
for the annihilation families $(F_u)_{u\in\cI}$ and the Majorana operators $(\gamma_v)_{v\in\cJ}$.
For \(u=(x,j)\in\cI\) and \(v=(y,k)\in\cJ\), we write \(x_u=x\) and \(x_v=y\) for the associated sites.

For an odd operator \(F\in\cC^-\), we define the bounded linear \emph{graded derivative} \(\partial_F:\cC\to\cC\) by
\begin{equation*}
 \partial_F A:=FA-\Theta(A)F.
\end{equation*}
For even \(A\), this is the commutator \([F,A]\), while for odd \(A\) it is the anticommutator \(\{F,A\}\). Thus \(\|\partial_FA\|\) measures the failure of graded commutation between \(F\) and \(A\).

For Majoranas $\gamma_v$ we also write \(\partial_v:=\partial_{\gamma_v}\) and call these maps the Majorana derivatives. In the expansion of a local operator into products of pairwise distinct Majoranas, \(\partial_v\) annihilates the terms without \(\gamma_v\) and removes \(\gamma_v\) from each remaining term, up to a sign and a factor of two. Thus \(\|\partial_vA\|\) measures how strongly \(A\) depends on \(\gamma_v\). Lemma~\ref{lem:majorana-averaging} makes this precise.

The graded derivative satisfies the graded Leibniz rule \(\partial_F(AB)=(\partial_FA)B+\Theta(A)\partial_FB\). For \(F \in \cC_X^-\) and \(A\in\cC_Y\) with \(X \cap Y =\varnothing
\), the anticommutation relations give \(\partial_FA=0\).

The unperturbed family \((F_u^0)_{u\in\cI}\) consists of linear combinations of Majoranas and is defined as follows.

\begin{definition}[Parseval condition]\label{def:parseval-family}
We say that complex coefficients \((\sa_{uv})_{u\in\cI,v\in\cJ}\) satisfy the \emph{Parseval condition} if the following sum converges absolutely and satisfies
\begin{equation}\label{eq:parseval-condition}
 \RealPart\sum_{u\in\cI}\overline{\sa_{uv}}\sa_{uw}
 =\delta_{vw},
\end{equation}
for every \(v,w\in\cJ\). We call the family $(F_u^0)_{u\in\cI}$ with
\[
 F_u^0:=\tfrac12\sum_{v\in\cJ}\sa_{uv}\gamma_v,
\]
the corresponding \emph{linear reference family}.
\end{definition}

Equivalently, the real row vectors \((\RealPart\sa_{uv})_{v\in\cJ}\) and \((\operatorname{Im}\sa_{uv})_{v\in\cJ}\) taken together form a Parseval frame of \(\ell^2(\cJ,\R)\cong\ell^2(\Gamma,\R^{2n})\). The next lemma establishes square summability of each \((\sa_{uv})_{v\in\cJ}\), norm convergence of the operator series and a reconstruction formula for the Majoranas. All lemmas in this subsection are proved in Section~\ref{sec:comparison-lemma-proofs}.

\begin{lemma}\label{lem:parseval-reconstruction}
Let \((\sa_{uv})_{u\in\cI,v\in\cJ}\) satisfy the Parseval condition of Definition~\ref{def:parseval-family}. Then \(\sum_{v\in\cJ}|\sa_{uv}|^2\le2\) for every \(u\in\cI\). The series defining each \(F_u^0\) converges in norm with \(\|F_u^0\|\le1\). Moreover,
\begin{equation}\label{eq:reference-reconstruction}
 \gamma_v=\sum_{u\in\cI}\bigl(\overline{\sa_{uv}}F_u^0+\sa_{uv}(F_u^0)^*\bigr),
\end{equation}
for all $v\in\cJ$, where the sum converges in norm. In particular, \((F_u^0)_{u\in\cI}\) generates \(\cC\) as a unital \(C^*\)-algebra.
\end{lemma}

\begin{remark}\label{rem:Bogoliubov}
Bogoliubov transformations of the annihilation operators on $\cC$ provide an important class of examples satisfying the Parseval condition. Indeed, let \(\sigma\) be any such automorphism and define the family \(F_u^0:=\sigma(c_{x,j})\) for \(u=(x,j)\in\cI\). The action of \(\sigma\) on the Majoranas is given by a real orthogonal matrix \(O\). For all \(w\in\cJ\),
\[
 \sigma(\gamma_w)=\sum_{v\in\cJ}O_{wv}\gamma_v.
\]
Thus we can write \(F_u^0=\frac12\sum_{v\in\cJ}\sa_{uv}\gamma_v\) with complex coefficients \(\sa_{uv}=O_{(x,2j-1),v}+\I O_{(x,2j),v}\). These coefficients satisfy \(\sa_{uv}\1=\{\gamma_v,F_u^0\}\).
For each \(v\in\cJ\), we have 
\[
 \sum_{u\in\cI}|\sa_{uv}|^2=\sum_{w\in\cJ}O_{wv}^2=1.
\]
The sum below therefore converges absolutely by the Cauchy--Schwarz inequality. A direct computation shows that
\[
 \RealPart\sum_{u\in\cI}\overline{\sa_{uv}}\sa_{uw}\1
 =\tfrac12\{\sigma^{-1}(\gamma_v),\sigma^{-1}(\gamma_w)\}
 =\delta_{vw}\1,
\]
where we have used the anticommutation relations of the Majoranas in the second equality. Thus the coefficients of $(F_u^0)_{u\in\cI}$ satisfy the Parseval condition of Definition~\ref{def:parseval-family}.
\end{remark}

In addition to the Parseval condition, we require spatial locality of \((F_u^0)_{u\in\cI}\). We quantify this through the decay of its coefficients \((\sa_{uv})_{u\in\cI,v\in\cJ}\) in terms of the distance \(d(x_u,x_v)\), thereby controlling contributions to \(F_u^0\) from Majoranas far from \(x_u\). We express these bounds using weights and their associated Schur norms.

\begin{definition}\label{def:weight}
A \emph{weight} is a nondecreasing function \(W:[0,\infty]\to[1,\infty]\) with \(W(0)=1\), \(W(\infty)=\infty\), and
\[
 W(R+R')\le W(R)W(R'),\qquad R,R'\ge0.
\]
We set   \(1/W(\infty):=0\).
For a nonnegative kernel \(\mathsf H=(\mathsf H_{ij})_{i\in\cI_1,j\in\cI_2}\) with \(\cI_1,\cI_2\in\{\cI,\cJ\}\), we write
\[
 \|\mathsf H\|^{\rm Sch}_W:=\max\Bigl\{
  \sup_{i\in\cI_1}\sum_{j\in\cI_2}W(d(x_i,x_j))\mathsf H_{ij},\
  \sup_{j\in\cI_2}\sum_{i\in\cI_1}W(d(x_i,x_j))\mathsf H_{ij}\Bigr\}. \qedhere
\]
\end{definition}

In the following, we use the stretched-exponential weight \(W(R)=\E^{\mu R^\theta}\) with \(\mu>0\) and \(0<\theta\le1\), and the polynomial weight \(W(R)=(1+R)^p\) with \(p\ge0\). For comparison with a merely summable interaction, we use the trivial weight \(W=1\). The next lemma shows that the weighted Schur norm is submultiplicative and that applying a kernel with finite weighted Schur norm preserves decay of the form \(W(d(\cdot,S))^{-1}\) away from a set \(S\). 

\begin{lemma}\label{lem:weighted-Schur-properties}
Let \(\cI_1,\cI_2,\cI_3\in\{\cI,\cJ\}\), and let \(\mathsf H_1\) and \(\mathsf H_2\) be nonnegative kernels on \(\cI_1\times\cI_2\) and \(\cI_2\times\cI_3\), respectively. Let $W$ be a weight. Then
\[
 \|\mathsf H_1\mathsf H_2\|^{\rm Sch}_W
 \le\|\mathsf H_1\|^{\rm Sch}_W\|\mathsf H_2\|^{\rm Sch}_W.
\]
For a nonnegative kernel \(\mathsf H\) on \(\cI_1\times\cI_2\) and \(S\subset\Gamma\),
\begin{equation*}
 \sum_{j\in\cI_2}\frac{\mathsf H_{ij}}{W(d(x_j,S))}
 \le\frac{\|\mathsf H\|^{\rm Sch}_W}{W(d(x_i,S))},
 \qquad i\in\cI_1.
\end{equation*}
\end{lemma}

We can now state the assumptions on $(F_u)_{u\in\cI}$. This family consists of a linear reference family satisfying the Parseval condition and a small perturbation, with locality bounds on both parts.

\begin{hypothesis}\label{hyp:perturbed-family}
Let \(W\) be a weight, \(C_0\ge1\) and \(\varepsilon\ge0\).
\begin{enumerate}[label=(\roman*),leftmargin=*]
\item \emph{Reference family.} Let the family \(F_u^0=\frac12\sum_{v\in\cJ}\sa_{uv}\gamma_v\), $u\in\cI$, have complex coefficients $(\sa_{uv})_{u\in\cI,v\in\cJ}$ satisfying the Parseval condition of Definition~\ref{def:parseval-family}, with
\[
 \|(|\sa_{uv}|)\|^{\rm Sch}_W\le C_0.
\]
\item \emph{Corrections.} For each \(u\in\cI\) and each finite \(Z\ni x_u\), there is an odd   \(\Delta F_{u,Z}\in\cC_Z^-\) so that
\begin{equation*}
 \sup_{z\in\Gamma}\sum_{u\in\cI}\sum_{Z\ni z,x_u}
       |Z|W(\diam Z)\|\Delta F_{u,Z}\|\le\varepsilon.\qedhere
\end{equation*}
\end{enumerate}
\end{hypothesis}

If the families $F_u^0$ and $\Delta F_{u,Z}$ satisfy Hypothesis~\ref{hyp:perturbed-family}, the \emph{perturbed family}
\begin{equation}\label{eq:F_def}
 F_u:=F_u^0+\sum_{Z\ni x_u}\Delta F_{u,Z} \qquad (u\in\cI)
\end{equation}
is well defined, since
\[
 \frac12\sum_{v\in\cJ}\|\sa_{uv}\gamma_v\|\le\frac12C_0,
 \qquad
 \sum_{Z\ni x_u}\|\Delta F_{u,Z}\|\le\varepsilon,
\]
for all $u \in \cI$. Hence, both series defining \(F_u\) converge absolutely in norm.

We now define three kernels that we will use repeatedly below.

\begin{definition}\label{def:comparison-kernels}
For the family of Hypothesis~\ref{hyp:perturbed-family}, define, for \(u\in\cI\) and \(v,w\in\cJ\), the kernels
\begin{equation*}
 \begin{gathered}
 \mathsf L_{uv}:=\tfrac12|\sa_{uv}|+
       \sum_{Z:\,x_u,x_v\in Z}\|\Delta F_{u,Z}\|,\\
 \mathsf K_{vw}:=2\sum_{u\in\cI}\sum_{Z\ni x_u}\|\Delta F_{u,Z}\|
 \Bigl[|\sa_{uv}|\mathbf1_Z(x_w)
 +2\mathbf1_Z(x_v)\bigl(\mathsf L_{uw}+\|F_u\|\mathbf1_Z(x_w)\bigr)\Bigr].
 \end{gathered}
\end{equation*}
Here \(\mathbf1_Z\) denotes the indicator of \(Z\).
If \(\|\mathsf K^m\|^{\rm Sch}_W<1\) for some integer \(m\ge1\), we also define the kernel
\[
 \mathsf T:=\sum_{j\ge0}\mathsf L\mathsf K^j
          =\mathsf L(\1-\mathsf K)^{-1}. \qedhere
\]
\end{definition}

For \(m\) as in the definition, grouping the powers modulo \(m\) gives
\[
 \sum_{j\ge0}\|\mathsf L\mathsf K^j\|^{\rm Sch}_W
 \le\|\mathsf L\|^{\rm Sch}_W
      \frac{\sum_{b=0}^{m-1}\|\mathsf K^b\|^{\rm Sch}_W}{1-\|\mathsf K^m\|^{\rm Sch}_W}
 <\infty.
\]
Here the norms of \(\mathsf L\) and \(\mathsf K\) are finite by Lemma~\ref{lem:comparison-kernel-bounds}.

\begin{lemma}\label{lem:comparison-kernel-bounds}
Let the family $(F_u)_{u \in \cI}$ satisfy Hypothesis~\ref{hyp:perturbed-family}. It follows that for every \(u\in\cI\),
\[
 \|F_u^0\|\le1,
 \qquad \|F_u\|\le1+\varepsilon.
\]
The kernels of Definition~\ref{def:comparison-kernels} satisfy
\[
 \|\mathsf L\|^{\rm Sch}_W\le\tfrac12 C_0+2n\varepsilon,\qquad
 \|\mathsf K\|^{\rm Sch}_W\le4nC_0\varepsilon+8n\varepsilon\|\mathsf L\|^{\rm Sch}_W+8n(1+\varepsilon)\varepsilon.
\]
If \(\varepsilon\le[32n(1+C_0)]^{-1}\), then
\begin{equation*}
 \begin{aligned}
  \|\mathsf L\|^{\rm Sch}_W&\le\tfrac12(1+C_0), \quad
  \|\mathsf K\|^{\rm Sch}_W&\le\tfrac12, \quad
  \|\mathsf T\|^{\rm Sch}_W&\le 1+C_0.
 \end{aligned}
\end{equation*}
Moreover, for any \(u \in \cI\) and \(A\in\cC\), 
\begin{equation*}
 \max\bigl\{\|\partial_{F_u}A\|,
              \|\partial_{F_u^*}A\|,
              \|\partial_{F_u}(A^*)\|\bigr\}
 \le\sum_{v\in\cJ}\mathsf L_{uv}\|\partial_vA\|.
\end{equation*}
\end{lemma}

\subsection{The comparison theorem}
\label{sec:comparison-theorem}

We now state the comparison theorem, which connects the operator families constructed above with the main results of this paper. It shows that controlling the difference of two states on products containing \(F_u\) on the right or \(F_u^*\) on the left suffices to control their expectations of every local observable. The resulting comparison preserves spatial decay in these bounds. This allows us to obtain uniqueness, locality and stability estimates from the same principle. Part~(i) provides the decomposition of observables underlying this argument. Part (ii) turns it into a quantitative comparison of states.

\begin{theorem}[Comparison theorem]\label{thm:comparison}
Let $(F_u)_{u \in \cI}$ be a family of operators that satisfies Hypothesis~\ref{hyp:perturbed-family}, and assume that the kernel \(\mathsf K\) of Definition~\ref{def:comparison-kernels} satisfies \(\|\mathsf K^m\|^{\rm Sch}_W<1\) for some \(m\in\mathbb N\); by Lemma~\ref{lem:comparison-kernel-bounds}, this holds with \(m=1\) if \(\varepsilon\le[32n(1+C_0)]^{-1}\).
Let \(\mathsf T\) be the kernel of Definition~\ref{def:comparison-kernels}. Then the following statements hold.
\begin{enumerate}[label=(\roman*),leftmargin=*]
\item There is a state \(\omega_F\) on \(\cC\) with the following property. For every \(A\in\cC_\loc\), there are two families \((X_u)_{u\in\cI}\), \((Y_u)_{u\in\cI}\) of operators in \(\cC\), with \(\sum_{u\in\cI}(\|X_u\|+\|Y_u\|)<\infty\), such that
\begin{equation*}
 A=\omega_F(A)\1+\sum_{u\in\cI}(F_u^*X_u+Y_uF_u),
\end{equation*}
where the series converges absolutely in norm. For every \(u \in \cI\), these operators satisfy
\[
 \max\{\|X_u\|,\|Y_u\|\}
       \le\sum_{v\in\cJ}\mathsf T_{uv}\|\partial_vA\|.
\]
Every state \(\omega\) on \(\cC\) with \(\omega(F_u^*F_u)=0\) for all \(u\in\cI\) coincides with \(\omega_F\).
\item Let \(\varphi\) be a bounded linear functional on \(\cC\) with \(\varphi(\1)=0\). Assume that there is a bounded family of nonnegative numbers \((\mathfrak s_u)_{u \in \cI}\) such that, for every \(E,G\in\cC\) and every \(u \in \cI\),
\begin{equation}\label{eq:annihilator-linear-source}
 |\varphi(F_u^*E)|+|\varphi(GF_u)|
 \le\mathfrak s_u\max\{\|E\|,\|G\|\}.
\end{equation}
Then
\begin{equation*}
 |\varphi(A)|\le\sum_{\substack{u\in\cI\\v\in\cJ}}\mathfrak s_u\mathsf T_{uv}\|\partial_vA\|,
 \qquad A\in\cC_\loc.
\end{equation*}
\end{enumerate}
\end{theorem}

\begin{proof}[Proof of Theorem~\ref{thm:comparison}(ii) from Theorem~\ref{thm:comparison}(i)]
Let \(\varphi\) be as in the statement and use the decomposition from (i). Since the series converges in norm and \(\varphi\) is bounded, we may apply \(\varphi\) term by term. The scalar term vanishes because \(\varphi(\1)=0\), giving
\[
 \begin{aligned}
 |\varphi(A)|
 &\le\sum_{u\in\cI}
       \bigl(|\varphi(F_u^*X_u)|+|\varphi(Y_uF_u)|\bigr)\le\sum_{u\in\cI}\mathfrak s_u\max\{\|X_u\|,\|Y_u\|\}\le\sum_{\substack{u\in\cI\\v\in\cJ}}
       \mathfrak s_u\mathsf T_{uv}\|\partial_vA\|. 
 \end{aligned} \qedhere
\]
\end{proof}

The strategy for proving part (i) is explained in Section~\ref{sec:roadmap-comparison}, and its proof is given in Section~\ref{sec:comparison-proof}.

\begin{remark}[Relation to the ground-state comparison]\label{rem:ground-state-comparison}
The ground-state comparison of~\cite{LPPLV3} and Theorem~\ref{thm:comparison} both control local expectations through errors in annihilation conditions.
The ground-state argument uses a quasi-local automorphism to identify the bulk ground state with a product vacuum.
In these coordinates, let \(c_{x,j}\) be the on-site annihilation operators and \(Q_{x,j}\) their filtered counterparts, which also annihilate the vacuum.
The analogue of our Hypothesis~\ref{hyp:perturbed-family} on the family \((F_u)\) in \eqref{eq:F_def} is an operator inequality named \emph{local excitation detectability}.
Writing \(n_x:=\sum_jc_{x,j}^*c_{x,j}\) and \(n_{Q,x}:=\sum_jQ_{x,j}^*Q_{x,j}\),  it reads
\[
 n_x\le 2n_{Q,x}+\sum_z\mathscr K(x,z)n_z,
\]
where \(\mathscr K\) is nonnegative and \(\sup_x\sum_z\mathscr K(x,z)<1\).
Taking expectations in a state \(\rho\) and iterating bounds \(\rho(n_x)\) by \(2\sum_z(\mathbf1-\mathscr K)^{-1}(x,z)\rho(n_{Q,z})\).
Comparison with the local vacuum then bounds deviations of local expectations from those of the bulk ground state.
In~\cite{LPPLV3} it is shown that local excitation detectability  holds for sufficiently weak coupling.

In the present paper, the expectations \(\omega(F_u^*F_u)\) play the role of the filtered excitation densities \(\rho(n_{Q,z})\) in~\cite{LPPLV3}.
Just as the ground-state argument compares \(\rho\) with the product vacuum annihilated by the \(Q_{x,j}\), we compare \(\omega\) with the state \(\omega_F\), which in our applications satisfies \(\omega_F(F_u^*F_u)=0\) for every \(u\).
In the ground-state comparison, summing the powers of \(\mathscr K\) bounds the local excitation densities \(\rho(n_x)\) used to compare \(\rho\) with the vacuum.
Here the corresponding sum \(\mathsf T=\sum_{j\ge0}\mathsf L\mathsf K^j\) bounds the coefficients in the decomposition of a local observable used to compare \(\omega\) with \(\omega_F\) (Theorem~\ref{thm:comparison}).
\end{remark}

\subsection{Consequences}
\label{sec:comparison-corollaries}

We now derive the consequences used to prove the main results of Section~\ref{sec:results}. We first consider errors in the annihilation identities that decay with the distance from a fixed set \(S \subset \Gamma\). The comparison theorem transfers the decay to expectations of local operators. We use this estimate to prove strong LPPL, low-temperature comparison, finite-volume Gibbs comparison, and stability under uniformly weak perturbations.

\begin{corollary}\label{cor:localized}
Assume the hypotheses of Theorem~\ref{thm:comparison}. Let \(S\subset\Gamma\) and let \((\mathfrak s_u)_{u\in\cI}\) be a bounded family of nonnegative numbers such that, for some constants \(\mathfrak s_0,C_S\ge0\) and all $u \in \cI$, it holds that
\[
 \mathfrak s_u\le\mathfrak s_0+\frac{C_S}{W(d(x_u,S))}.
\]
Let \(\varphi\) be a bounded linear functional on \(\cC\) such that \(\varphi(\1)=0\) and
\[
 |\varphi(F_u^*E)|+|\varphi(GF_u)|
 \le\mathfrak s_u\max\{\|E\|,\|G\|\},
\]
for every \(E,G\in\cC\) and \(u\in\cI\).
Then, for every \(Y\Subset\Gamma\) and \(A\in\cC_Y\) it holds that
\begin{equation*}
 |\varphi(A)|\le4n\|\mathsf T\|^{\rm Sch}_W\|A\|\sum_{x\in Y}
       \Bigl[\mathfrak s_0+\frac{C_S}{W(d(x,S))}\Bigr].
\end{equation*}
\end{corollary}
\begin{proof}
Fix \(v\in\cJ\). Lemma~\ref{lem:weighted-Schur-properties} gives
\[
 \sum_{u\in\cI}\mathfrak s_u\mathsf T_{uv}
 \le\mathfrak s_0\sum_{u\in\cI}\mathsf T_{uv}
   +C_S\sum_{u\in\cI}\frac{\mathsf T_{uv}}{W(d(x_u,S))}
 \le\|\mathsf T\|^{\rm Sch}_W\Bigl[\mathfrak s_0+\frac{C_S}{W(d(x_v,S))}\Bigr].
\]
The Majorana derivatives of \(A\in\cC_Y\) vanish when \(x_v\notin Y\) and they satisfy \(\|\partial_vA\|\le2\|A\|\). The second assertion of Theorem~\ref{thm:comparison} therefore gives
\[
 |\varphi(A)|
 \le2\|\mathsf T\|^{\rm Sch}_W\|A\|\sum_{\substack{v\in\cJ\\x_v\in Y}}
       \left[\mathfrak s_0+\frac{C_S}{W(d(x_v,S))}\right]=4n\|\mathsf T\|^{\rm Sch}_W\|A\|\sum_{x\in Y}
       \left[\mathfrak s_0+\frac{C_S}{W(d(x,S))}\right],
\]
where we have used that each site has \(2n\) Majorana labels.
\end{proof}

To apply this estimate to the difference \(\varphi=\omega'-\omega\) of two states, we need to bound \(\varphi(F_u^*E)\) and \(\varphi(GF_u)\). The next lemma expresses these bounds in terms of \(\omega'(F_u^*F_u)^{1/2}\) when \(\omega(F_u^*F_u)=0\). If \(\omega'\) satisfies the same condition for another family \((F'_u)_{u \in \cI}\), they can be estimated using \(\|F_u-F'_u\|\).

\begin{lemma}\label{lem:stability}
Let \((F_u)_{u\in\cI}\subset\cC\), and let \(\omega\) be a state on \(\cC\) such that $\omega(F_u^*F_u)=0$ for every \(u\in\cI\). For every state \(\omega'\) on \(\cC\), the functional \(\varphi:=\omega'-\omega\) satisfies \(\varphi(\1)=0\) and
\[
 |\varphi(F_u^*E)|+|\varphi(GF_u)|
 \le\mathfrak s_u\max\{\|E\|,\|G\|\},
\]
for every \(E,G\in\cC\) and \(u\in\cI\), with \(\mathfrak s_u=2\omega'(F_u^*F_u)^{1/2}\).

If \((F'_u)_{u\in\cI}\) is a family of bounded operators in \(\cC\) such that \(\omega'((F'_u)^*F'_u)=0\) for every \(u\in\cI\), then
\[
 \omega'(F_u^*F_u)^{1/2}\le\|F_u-F'_u\|,
 \qquad u\in\cI.
\]
\end{lemma}

\begin{proof}
We first notice that the Cauchy--Schwarz inequality gives
\(\omega(F_u^*E)=\omega(EF_u)=0\) for all $E \in \cC$. This gives, for any $E, G \in \cC$, 
\[
 \begin{aligned}
 |\varphi(F_u^*E)|+|\varphi(GF_u)|
 &=|\omega'(F_u^*E)|+|\omega'(GF_u)|\\ 
 &\le\omega'(F_u^*F_u)^{1/2}
       \bigl(\omega'(E^*E)^{1/2}+\omega'(GG^*)^{1/2}\bigr)\\
 &\le2\omega'(F_u^*F_u)^{1/2}\max\{\|E\|,\|G\|\},
 \end{aligned}
\]
where we have again used the Cauchy--Schwarz inequality. To prove the remaining assertion, let \((\pi_{\omega'},\mathcal H_{\omega'},\Omega_{\omega'})\) be the GNS triple of \(\omega'\). The assumption \(\omega'((F'_u)^*F'_u)=0\) gives \(\|\pi_{\omega'}(F'_u)\Omega_{\omega'}\|^2=0\). Hence
\[
 \omega'(F_u^*F_u)^{1/2}
 =\|\pi_{\omega'}(F_u)\Omega_{\omega'}\|
 =\|\pi_{\omega'}(F_u-F'_u)\Omega_{\omega'}\|
 \le\|F_u-F'_u\|.  \qedhere
\]
\end{proof}

The comparison theorem also bounds correlations in a state approximately annihilated by $(F_u)_{u\in\cI}$.  In the following corollary, we obtain a bound on the covariance that retains the decay profile \(W^{-1}\) specified by the locality bounds on the reference coefficients and local corrections in Hypothesis~\ref{hyp:perturbed-family}, with an additional term accounting for approximate annihilation.

\begin{corollary}\label{cor:covariance}
Assume the hypotheses of Theorem~\ref{thm:comparison}. Let \(S\subset\Gamma\), \(\eta\ge0\), and let \(\omega\) be a state on \(\cC\) that satisfies
\[
 \omega(F_u^*F_u)^{1/2}\le\frac{\eta}{W(d(x_u,S))}
\]
for every \(u\in\cI\). Then, for every \(X,Y\Subset\Gamma\), \(A\in\cC_X\) and \(B\in\cC_Y\) it holds that
\begin{equation*}
 \begin{aligned}
 \lvert\Cov_\omega(A,B)\rvert
 &\le8n\|A\|\|B\|\Biggl[\|\mathsf T^{\top}\mathsf L\|^{\rm Sch}_W
 \min\Bigl\{\sum_{x\in X}\frac1{W(d(x,Y))},\
            \sum_{y\in Y}\frac1{W(d(y,X))}\Bigr\}\\
 &\hspace{2.6cm}+2\eta\|\mathsf T\|^{\rm Sch}_W\min\Bigl\{\sum_{x\in X}\frac1{W(d(x,S))},\
            \sum_{y\in Y}\frac1{W(d(y,S))}\Bigr\}\Biggr].
 \end{aligned}
\end{equation*}
\end{corollary}
\begin{proof}
Fix \(B\in\cC_Y\) and define the linear functional
\begin{equation*}
 \varphi_B(E):=\omega(EB)-\omega(E)\omega(B).
\end{equation*}
This functional is bounded, namely \(\|\varphi_B\|\le2\|B\|\), vanishes on the identity \(\varphi_B(\1)=0\), and implements the covariance \(\varphi_B(A)=\Cov_\omega(A,B)\). Using the assumption, the Cauchy--Schwarz inequality gives
\[
 |\omega(F_u^*C)|, \, |\omega(CF_u)| \le \omega(F_u^*F_u)^{1/2}\|C\|,
\]
for every \(C\in\cC\). It follows that
\begin{equation*}
 \begin{gathered}
  |\varphi_B(F_u^*C)|\le2\omega(F_u^*F_u)^{1/2}\|B\|\|C\|,\\
  |\varphi_B(CF_u)-\omega(C\,\partial_{F_u}B)|
       \le2\omega(F_u^*F_u)^{1/2}\|B\|\|C\|,
 \end{gathered}
\end{equation*}
for all $C\in \cC$. Here, we have used that \(F_uB=\partial_{F_u}B+\Theta(B)F_u\). For every \(E,G\in\cC\), these bounds and Lemma~\ref{lem:comparison-kernel-bounds} give
\[
 \begin{aligned}
  |\varphi_B(F_u^*E)|+|\varphi_B(GF_u)|
 &\le\|G\|\,\|\partial_{F_u}B\|
       +2\omega(F_u^*F_u)^{1/2}\|B\|(\|E\|+\|G\|)\\
 &\le\max\{\|E\|,\|G\|\}\Bigl[
       \sum_{w\in\cJ}\mathsf L_{uw}\|\partial_wB\|
       +4\omega(F_u^*F_u)^{1/2}\|B\|\Bigr].
 \end{aligned}
\]
The second assertion of Theorem~\ref{thm:comparison} therefore gives
\begin{equation}\label{eq:covariance-inequality}
 \begin{aligned}
 \lvert\Cov_\omega(A,B)\rvert &= |\varphi_B (A)| \\
 &\le\sum_{\substack{v,w\in\cJ}}\|\partial_vA\|
       (\mathsf T^{\top}\mathsf L)_{vw}\|\partial_wB\|
 +4\|B\|\sum_{\substack{u\in\cI\\v\in\cJ}}
       \omega(F_u^*F_u)^{1/2}\mathsf T_{uv}\|\partial_vA\|.
 \end{aligned}
\end{equation}
Lemmas~\ref{lem:comparison-kernel-bounds} and~\ref{lem:weighted-Schur-properties} give
\[
 \|\mathsf T^{\top}\mathsf L\|^{\rm Sch}_W
 \le\|\mathsf T\|^{\rm Sch}_W\|\mathsf L\|^{\rm Sch}_W<\infty.
\]
For \(x_v\in X\), the definition of the Schur norm and the preceding estimate give
\[
 \begin{aligned}
 \sum_{\substack{w\in\cJ\\x_w\in Y}}(\mathsf T^{\top}\mathsf L)_{vw}
 &\le\frac1{W(d(x_v,Y))}
       \sum_{w\in\cJ}W(d(x_v,x_w))(\mathsf T^{\top}\mathsf L)_{vw}\le\frac{\|\mathsf T^{\top}\mathsf L\|^{\rm Sch}_W}{W(d(x_v,Y))}.
 \end{aligned}
\]
For $x_w \in Y$ we find in the same way
\[
 \sum_{\substack{v\in\cJ\\x_v\in X}}(\mathsf T^{\top}\mathsf L)_{vw}
 \le\frac{\|\mathsf T^{\top}\mathsf L\|^{\rm Sch}_W}{W(d(x_w,X))}.
\]
The Majorana derivatives vanish outside the respective supports and satisfy \(\|\partial_vA\|\le2\|A\|\) and \(\|\partial_wB\|\le2\|B\|\). Summing over the \(2n\) Majoranas at each site therefore gives
\[
 \begin{aligned}
 \sum_{v,w\in\cJ}\|\partial_vA\|
       (\mathsf T^{\top}\mathsf L)_{vw}\|\partial_wB\|
 &\le4\|A\|\|B\|\sum_{v\in\cJ:\,x_v\in X}\sum_{w\in\cJ:\,x_w\in Y}
       (\mathsf T^{\top}\mathsf L)_{vw}\\
 &\le8n\|\mathsf T^{\top}\mathsf L\|^{\rm Sch}_W\|A\|\|B\|
       \min\Bigl\{\sum_{x\in X}\frac1{W(d(x,Y))},\
            \sum_{y\in Y}\frac1{W(d(y,X))}\Bigr\}.
 \end{aligned}
\]
For the second term in Equation \eqref{eq:covariance-inequality}, the assumption on \(\omega\) and Lemma~\ref{lem:weighted-Schur-properties} give
\[
 \sum_{u\in\cI}\omega(F_u^*F_u)^{1/2}\mathsf T_{uv}
 \le\frac{\eta\|\mathsf T\|^{\rm Sch}_W}{W(d(x_v,S))}.
\]
Substituting this bound into Equation \eqref{eq:covariance-inequality} gives
\[
 \begin{aligned}
 4\|B\|\sum_{\substack{u\in\cI\\v\in\cJ}}
       \omega(F_u^*F_u)^{1/2}\mathsf T_{uv}\|\partial_vA\|
 &\le16n\|\mathsf T\|^{\rm Sch}_W\eta\|A\|\|B\|
       \sum_{x\in X}\frac1{W(d(x,S))}.
 \end{aligned}
\]
Applying this estimate to \(B^*\) and \(A^*\) with \(\Cov_\omega(A,B)=\overline{\Cov_\omega(B^*,A^*)}\) gives the sum over \(Y\). Taking the minimum proves the assertion.
\end{proof}

\subsection{Classical analogues}
\label{sec:comparison-remarks}

Theorem~\ref{thm:comparison} is a quantum analogue of Dobrushin's comparison theorem \cite{Dobrushin70}. We recall its classical form in the notation of \cite[Theorem~8.20]{Georgii88}. For simplicity, let \(\Omega=E^S\), with \(S\) countable and \(E\) finite. Let \(\gamma,\widetilde\gamma\) be specifications, and write \(\gamma_i^0(\cdot\mid\omega)\) for the conditional distribution of the spin at \(i\) given the exterior configuration. For a real local function \(f\), its oscillation at site \(i\) and Dobrushin's interdependence matrix are defined by
\[
 \begin{aligned}
 \delta_i(f)&:=\sup_{\omega_{S\setminus\{i\}}=\eta_{S\setminus\{i\}}}
                       |f(\omega)-f(\eta)|,\\
 C_{ij}(\gamma)&:=\sup_{\omega_{S\setminus\{j\}}=\eta_{S\setminus\{j\}}}
       \|\gamma_i^0(\cdot\mid\omega)-\gamma_i^0(\cdot\mid\eta)\|,
 \end{aligned}
\]
where \(\|p-q\|:=\sup_{B\subset E}|p(B)-q(B)|=\tfrac12\sum_{a\in E}|p(a)-q(a)|\) is the uniform distance between probability measures. Thus \(C_{ij}(\gamma)\) measures the dependence of the conditional distribution at \(i\) on the spin at \(j\). 
Let \(\mu\in\mathcal G(\gamma)\) and \(\widetilde\mu\in\mathcal G(\widetilde\gamma)\) be Gibbs measures for the two specifications.
For each \(j\in S\), let \(b_j:\Omega\to[0,\infty)\) be a measurable function bounding the difference between their conditional distributions at \(j\) for the same exterior configuration:
\[
 \|\gamma_j^0(\cdot\mid\omega)
       -\widetilde\gamma_j^0(\cdot\mid\omega)\|
 \le b_j(\omega)
 \qquad(\omega\in\Omega).
\]
Assume that \(\gamma\) is quasi-local and satisfies Dobrushin's condition
\[
 c(\gamma):=\sup_{i\in S}\sum_{j\in S}C_{ij}(\gamma)<1,
\]
or, more generally, the conditions
\[
 c(\gamma)<\infty,
 \qquad
 \sup_{i\in S}\sum_{j\in S}(C(\gamma)^m)_{ij}<1,
\]
for some $m \in \mathbb N$. Define
\[
 D(\gamma):=\sum_{k\ge0}C(\gamma)^k
            =(\1-C(\gamma))^{-1}.
\]
Then Dobrushin's comparison theorem in the single-site version of Rebeschini and van Handel~\cite[Corollary~2.4 and Corollary~2.6, condition~4]{RebeschiniVanHandel14} gives, for every real local function \(f\),
\begin{equation}\label{eq:classical_Dobrushin}
 |\mu(f)-\widetilde\mu(f)|
 \le\sum_{i,j\in S}
       \delta_i(f)D_{ij}(\gamma)\widetilde\mu(b_j).
\end{equation}

In our quantum version of the bound \eqref{eq:classical_Dobrushin} in Theorem~\ref{thm:comparison}(ii), the norms \(\|\partial_vA\|\) of the Majorana derivatives play the role of the single-site oscillations \(\delta_i(f)\).
The classical error vector \((\widetilde\mu(b_j))_j\) corresponds to \(\mathsf L^{\top}\mathfrak s\), where \(\mathfrak s=(\mathfrak s_u)_{u\in\cI}\) consists of the error bounds in \eqref{eq:annihilator-linear-source}.
Here \(\mathsf L^{\top}\) converts bounds indexed by the operators \(F_u\) into bounds indexed by the Majorana labels \(v\).
The kernels \(\mathsf K^{\top}\) and \((\1-\mathsf K^{\top})^{-1}\) play the roles of \(C(\gamma)\) and \(D(\gamma)\), respectively.
In our applications, Lemma~\ref{lem:comparison-kernel-bounds} gives the contraction bound \(\|\mathsf K\|^{\rm Sch}_W\le\tfrac12\), corresponding to Dobrushin's condition in the classical theorem.
The condition \(\|\mathsf K^m\|^{\rm Sch}_W<1\) of Theorem~\ref{thm:comparison} corresponds to the more general condition above.

Corollary~\ref{cor:covariance} is a quantum analogue of the classical covariance estimates of Föllmer and Künsch \cite{Foellmer82,Kuensch82}. For \(\mu\in\mathcal G(\gamma)\) and real local functions \(f,g\), Dobrushin's comparison theorem yields
\[
 |\mu(fg)-\mu(f)\mu(g)|
 \le\frac14\sum_{i,j\in S}\delta_i(f)D_{ij}(\gamma)\delta_j(g),
\]
see \cite[Proposition~8.34]{Georgii88}.

The quantum covariance bound \eqref{eq:covariance-inequality}, for a state satisfying \(\omega(F_u^*F_u)=0\) for every \(u\), contains \(\mathsf T^{\top}\mathsf L\) in place of the classical covariance kernel \(\tfrac14D(\gamma)\), with
\[
 \mathsf T^{\top}\mathsf L
 =(\1-\mathsf K^{\top})^{-1}\mathsf L^{\top}\mathsf L.
\]
The two kernels differ in how they describe correlations in the absence of conditional dependence or perturbative corrections, respectively.
For \(C(\gamma)=0\), the classical kernel is diagonal.
When all corrections \(\Delta F_{u,Z}\) vanish, the quantum kernel reduces to \(\mathsf L^{\top}\mathsf L\), which can still connect different sites because the linear operators \(F_u^0\) can involve Majoranas at different sites.

\paragraph{Classical restrictions.}
One might try to apply the classical comparison theorem to a quantum ground state by measuring a fixed observable at each site and applying the theorem to the joint distribution of the outcomes.
Such a distribution determines only the expectations of functions of the measured observables, and it need not have a quasi-local specification.
Consider the transverse-field Ising chain with formal Hamiltonian
\[
 H=-J\sum_{j\in\Z}\sigma_j^x\sigma_{j+1}^x+h\sum_{j\in\Z}\sigma_j^z,
 \qquad h>|J|>0,
\]
and let \(\mu_z\) be the joint distribution of the outcomes of measuring \(\sigma_i^z\) at all sites in its ground state.
For \(|J|/h\) small enough, De Roeck, Maes, Neto\v{c}n\'y and Sch\"utz show that \(\mu_z\) is not quasi-local \cite[Theorem~4.3]{DMNS15}.
With \(\Gamma_L:=\{-L^2,\ldots,L^2\}\), the conditional probability of the outcome \(+1\) at the origin, given the record \(-1\) at all other sites of \(\Gamma_L\), tends to zero as \(L\to\infty\), whereas it tends to one if the record is changed to \(+1\) at the site \(L\) \cite[Proposition~8.1]{DMNS15}.
To apply the classical comparison theorem with \(\mu=\mu_z\), one would need a quasi-local specification \(\gamma\) for \(\mu_z\).
Quasi-locality requires the conditional expectation of each local observable to be continuous in the exterior configuration \(\eta\) for the product topology, in particular the conditional probability \(\gamma_0^0(+1\mid\eta)\) obtained by taking the indicator of a \(+1\) spin at the origin.
Continuity at the all-minus exterior configuration would force the two conditional probabilities to have the same limit, since both records agree with that configuration on intervals of increasing length around the origin.
The classical comparison theorem therefore cannot be applied with \(\mu=\mu_z\).

We now apply the quantum comparison theorem to the ground state of the same transverse-field Ising chain.
With the Jordan--Wigner convention \(\sigma_j^z=2c_j^*c_j-\mathbf1\), the Hamiltonian on the infinite chain has the formal expression
\[
 H=-J\sum_{j\in\mathbb Z}(c_j^*-c_j)(c_{j+1}^*+c_{j+1})
    \mathbin{+}h\sum_{j\in\mathbb Z}(2c_j^*c_j-\mathbf1).
\]
Thus \(\mu_z\) is the joint distribution of the fermion occupations \(n_j=c_j^*c_j\), with outcomes relabelled by \(n_j\mapsto 2n_j-1\).
The fermion Hamiltonian is quadratic and has a one-particle gap for \(h>|J|\), which includes the regime of the example.
Its ground state therefore has the exponentially localized linear annihilation family \(F_u^0=\mathsf F_u^{\infty,0}\) of Lemma~\ref{lem:ground-free-family} on the doubled algebra of Section~\ref{sec:thermofield}, and its product extension \(\omega_{\st}\) from \eqref{eq:stacked-states} satisfies \(\omega_{\st}((F_u^0)^*F_u^0)=0\) for every \(u\).
Since \(F_u=F_u^0\) here, the quantum influence matrix \(\mathsf K\) vanishes, and Theorem~\ref{thm:comparison} determines all local ground-state expectations from these annihilation identities, even though the occupation law \(\mu_z\) admits no quasi-local specification.

\subsection{Proof strategy for the comparison theorem}
\label{sec:roadmap-comparison}

We explain the main steps of the proof of Theorem~\ref{thm:comparison}(i); the details are given in Section~\ref{sec:comparison-proof}.
Let \(A\in\cC_Y\) with \(Y\Subset\Gamma\).
We look for a state \(\omega_F\) on \(\cC\) and operators \(X_u,Y_u\in\cC\) such that
\[
 A=\omega_F(A)\1+\sum_{u\in\cI}(F_u^*X_u+Y_uF_u).
\]
To this end we construct an auxiliary evolution \((T_t)_{t\ge0}\) from the operators \(F_u\), with \(T_tA\to\omega_F(A)\1\) in norm.
Its generator \(\cL\) is a sum of terms with \(F_u^*\) on the left or \(F_u\) on the right.
Integrating \(-\frac{\dd}{\dd t}T_tA=-\cL(T_tA)\) over \(t\ge0\) therefore gives the desired decomposition of \(A-\omega_F(A)\1\).
Bounds on the Majorana derivatives \(\partial_vT_tA\) give the convergence of \(T_tA\) and the bounds on \(X_u\) and \(Y_u\).

The limit has a simple interpretation. If \(\omega\) is a state with \(\omega(F_u^*F_u)=0\) for every \(u\in\cI\), then \(\omega\circ\cL=0\) by the Cauchy--Schwarz inequality, hence \(\omega(T_tA)=\omega(A)\) for all \(t\ge0\) and \(\omega_F(A)=\omega(A)\). Thus the auxiliary evolution relaxes every local observable to its expectation in the state annihilated by the family \((F_u)\), whenever such a state exists. The construction of \(\omega_F\) does not require such a state to exist.

\paragraph{The auxiliary evolution.}
Let \(\cC^1\) be the space of all \(E\in\cC\) with \(\sum_{v\in\cJ}\|\partial_vE\|<\infty\); it contains \(\cC_\loc\).
For \(E\in\cC^1\) we define
\begin{equation}\label{eq:roadmap-generator}
 \cL(E):=-\sum_{u\in\cI}
 \bigl(F_u^*\partial_{F_u}E+(\partial_{F_u}(E^*))^*F_u\bigr)
 =\sum_{u\in\cI}
 \bigl(2F_u^*\Theta(E)F_u-F_u^*F_uE-EF_u^*F_u\bigr).
\end{equation}
The series converges absolutely by Lemma~\ref{lem:comparison-kernel-bounds}, and the second expression has Lindblad form~\cite{Lindblad76}, because \(\Theta\) is a \(*\)-automorphism.
In Lemma~\ref{lem:auxiliary-evolution} we use the Lumer--Phillips theorem~\cite[Theorem~II.3.15]{EngelNagel00} to show that \(\cL\) is closable and that its closure generates a strongly continuous semigroup \((T_t)_{t\ge0}\) of unital completely positive contractions on \(\cC\).
This semigroup maps \(\cC^1\) into itself and satisfies \(\frac{\dd}{\dd t}T_tE=\cL(T_tE)\) for every \(E\in\cC^1\).
The construction follows that of the quantum Feller semigroups of Bertini, De Sole, Posta and Presilla \cite[Theorems~2.2 and~5.2]{BDSP25}, which adapts the classical construction of interacting particle systems \cite[Chapter~I]{Liggett85}.

\paragraph{Decay of the Majorana derivatives.}
By Lemma~\ref{lem:majorana-averaging}, in the limit \(Z\nearrow\Gamma\), every \(E\in\cC\) satisfies
\[
 \|E-\omega^{\rm tr}(E)\1\|
 \le\tfrac12\sum_{v\in\cJ}\|\partial_vE\|,
\]
where \(\omega^{\rm tr}\) is the tracial state on \(\cC\).
Decay of the Majorana derivatives of \(T_tA\) therefore controls the distance of \(T_tA\) from the scalars.
To prove this decay, we split \(\cL=\cL_0+\cL_{\rm res}\), where \(\cL_0\) is given by \eqref{eq:roadmap-generator} with the reference family \((F_u^0)\) in place of \((F_u)\).
For \(v\in\cJ\), the anticommutation relations give \(\partial_vF_u^0=\sa_{uv}\1\).
The graded Leibniz rule and the reconstruction formula \eqref{eq:reference-reconstruction} then yield
\[
 \partial_v\cL_0(E)-\cL_0(\partial_vE)
 =-\sum_{u\in\cI}
      \bigl(\overline{\sa_{uv}}\partial_{F_u^0}E
             +\sa_{uv}\partial_{(F_u^0)^*}E\bigr)=-\partial_vE.
\]
This is where the Parseval condition enters.
For the reference family alone, that is, for \(\varepsilon=0\), the identity gives \(\partial_vT_t^0A=\E^{-t}T_t^0(\partial_vA)\) for the corresponding semigroup \((T_t^0)_{t\ge0}\): every Majorana derivative is damped at unit rate.

Every term of \(\cL_{\rm res}\) contains at least one correction \(\Delta F_{u,Z}\).
We expand
\[
 \mathcal R_v(E):=\partial_v\cL_{\rm res}(E)-\cL_{\rm res}(\partial_vE)
\]
with the graded Leibniz rule and estimate the resulting terms with Lemmas~\ref{lem:comparison-kernel-bounds} and~\ref{lem:local-graded-commutator}.
This gives
\[
 \|\mathcal R_v(E)\|
 \le\sum_{w\in\cJ}\mathsf K_{vw}\|\partial_wE\|,
 \qquad E\in\cC^1,
\]
which is the origin of the kernel \(\mathsf K\) of Definition~\ref{def:comparison-kernels}.
Adding the two commutation relations, we obtain
\[
 \partial_v\cL(E)
 =\cL(\partial_vE)-\partial_vE+\mathcal R_v(E),
 \qquad E\in\cC^1,
\]
and variation of constants gives
\[
 \partial_vT_tA
 =\E^{-t}T_t(\partial_vA)
   +\int_0^t\E^{-(t-s)}T_{t-s}\mathcal R_v(T_sA)\,\dd s.
\]
Since \(T_t\) is a contraction, taking norms and iterating the resulting integral inequality yields
\begin{equation}\label{eq:roadmap-majorana-derivative-bound}
 \|\partial_vT_tA\|
 \le\sum_{w\in\cJ}\bigl(\E^{t(\mathsf K-\1)}\bigr)_{vw}\|\partial_wA\|.
\end{equation}
The argument leading to \eqref{eq:roadmap-majorana-derivative-bound}, damping of the derivatives at unit rate by the reference generator and a bound on their commutators with the remainder by a nonnegative matrix, is that of \cite[Proposition~2]{Matsui94}, where the reference generator is the one at infinite temperature; see also \cite[Remark~3.6]{BDSP25}.
Matsui sums over the sites, which gives the analogue of~\eqref{eq:roadmap-total-derivative-bound} and with it uniqueness; the comparison theorem rests on keeping \eqref{eq:roadmap-majorana-derivative-bound} as a matrix inequality and integrating it in time.
Choose \(\delta\in(0,1)\) with \(\|\mathsf K^m\|^{\rm Sch}_W\le(1-\delta)^m\). For \(\ell=j\bmod m\), we have
\[
 \|\mathsf K^j\|^{\rm Sch}_W
 \le(1-\delta)^{j-\ell}\|\mathsf K^\ell\|^{\rm Sch}_W
 \le(1-\delta)^j\max_{0\le k<m}(1-\delta)^{-k}\|\mathsf K^k\|^{\rm Sch}_W
 =:C(1-\delta)^j.
\]
Since \(W\ge1\), the column sums of \(\mathsf K^j\) are at most \(\|\mathsf K^j\|^{\rm Sch}_W\), and summing \eqref{eq:roadmap-majorana-derivative-bound} over \(v\) after expanding the exponential gives
\begin{equation}\label{eq:roadmap-total-derivative-bound}
 \begin{aligned}
 \sum_{v\in\cJ}\|\partial_vT_tA\|
 &\le\E^{-t}\sum_{j\ge0}\frac{t^j}{j!}\|\mathsf K^j\|^{\rm Sch}_W
         \sum_{w\in\cJ}\|\partial_wA\|\\
 &\le C\E^{-t}\sum_{j\ge0}\frac{[t(1-\delta)]^j}{j!}
          \sum_{w\in\cJ}\|\partial_wA\|
          =C\E^{-\delta t}\sum_{w\in\cJ}\|\partial_wA\|,
 \end{aligned}
\end{equation}
where the right-hand side is finite because \(\partial_wA=0\) for \(x_w\notin Y\).
Since \(T_s\) is a unital contraction, it follows that, for \(s,t\ge0\),
\[
 \|T_{t+s}A-T_tA\|
 \le2\|T_tA-\omega^{\rm tr}(T_tA)\1\|
 \le\sum_{v\in\cJ}\|\partial_vT_tA\|
 \le C\E^{-\delta t}\sum_{v\in\cJ}\|\partial_vA\|.
\]
Hence \(T_tA\) is Cauchy in norm and its distance from the scalars tends to zero, so it converges to a multiple of the identity, which we write as \(\omega_F(A)\1\).
The same holds for every \(A\in\cC\) by density, and
\[
 \omega_F(A)=\lim_{t\to\infty}\omega^{\rm tr}(T_tA).
\]
Since every \(T_t\) is positive and unital, \(\omega_F\) is a state on \(\cC\).

\paragraph{Construction of the coefficients.}
Since \(T_s\) is positive, \((T_sA)^*=T_s(A^*)\).
Integrating \(\frac{\dd}{\dd s}T_sA=\cL(T_sA)\) and inserting \eqref{eq:roadmap-generator} therefore gives, for \(t\ge0\),
\begin{equation}\label{eq:roadmap-finite-time-decomposition}
 A=T_tA+\sum_{u\in\cI}\left[
 F_u^*\int_0^t\partial_{F_u}T_sA\,\dd s
 +\left(\int_0^t\partial_{F_u}T_s(A^*)\,\dd s\right)^*F_u\right].
\end{equation}
The last assertion of Lemma~\ref{lem:comparison-kernel-bounds} and \eqref{eq:roadmap-majorana-derivative-bound} give, for every \(u\in\cI\) and \(t\ge0\),
\[
 \max\{\|\partial_{F_u}T_tA\|,\|\partial_{F_u}T_t(A^*)\|\}
 \le\sum_{v\in\cJ}\mathsf L_{uv}\|\partial_vT_tA\|
 \le\sum_{w\in\cJ}
       \bigl(\mathsf L\E^{t(\mathsf K-\1)}\bigr)_{uw}\|\partial_wA\|,
\]
and expanding the exponential gives
\[
 \int_0^\infty\mathsf L\E^{t(\mathsf K-\1)}\,\dd t
 =\mathsf L\sum_{j\ge0}\mathsf K^j=\mathsf T.
\]
Hence the Bochner integrals
\[
 X_u:=\int_0^\infty\partial_{F_u}T_tA\,\dd t,
 \qquad
 Y_u:=\left(\int_0^\infty\partial_{F_u}T_t(A^*)\,\dd t\right)^*
\]
converge in norm and satisfy the bound of Theorem~\ref{thm:comparison}(i),
\[
 \max\{\|X_u\|,\|Y_u\|\}
 \le\sum_{w\in\cJ:\,x_w\in Y}\mathsf T_{uw}\|\partial_wA\|.
\]
By the bound following Definition~\ref{def:comparison-kernels}, \(\sum_{u\in\cI}\mathsf T_{uw}\le\|\mathsf T\|^{\rm Sch}_W<\infty\), and Lemma~\ref{lem:comparison-kernel-bounds} gives \(\|F_u\|\le1+\varepsilon\). The sum over \(u\) in \eqref{eq:roadmap-finite-time-decomposition} therefore converges absolutely, uniformly in \(t\), and \(\sum_{u\in\cI}(\|X_u\|+\|Y_u\|)<\infty\). Letting \(t\to\infty\) in \eqref{eq:roadmap-finite-time-decomposition} and using \(T_tA\to\omega_F(A)\1\), we obtain the decomposition of Theorem~\ref{thm:comparison}(i).

The kernel \(\mathsf T\) also carries the spatial information.
Since \(W(d(x_u,x_w))\ge W(d(x_u,Y))\) for \(x_w\in Y\) and \(\|\partial_wA\|\le2\|A\|\), the bound on the coefficients implies
\[
 \max\{\|X_u\|,\|Y_u\|\}
 \le\frac{2\|\mathsf T\|^{\rm Sch}_W\,\|A\|}{W(d(x_u,Y))},
 \qquad u\in\cI.
\]

\section{State extensions and annihilation families on the doubled CAR algebra}
\label{sec:thermofield}

To apply the comparison theorem of Section~\ref{sec:comparison}, we extend physical states to the fixed doubled CAR algebra
\[
 \A^{\st}=\A\widehat\otimes\A_{\aux},
 \qquad \A_{\aux}:=\operatorname{CAR}(\Hsp_{\aux}),
\]
where \(\Hsp_{\aux}\) is a second copy of \(\Hsp\).
The algebra and the embeddings \(A^{(1)}:=A\widehat\otimes\1\), \(B^{(2)}:=\1\widehat\otimes B\) are independent of the state, interaction and temperature.
We use the complex-linear Majorana reversal \(\ft\), characterized by \(\ft(\I\gamma_u)=\I\gamma_u\) and \(\ft(AB)=\ft(B)\ft(A)\), whose properties are collected in Appendix~\ref{sec:doubled-notation}.

For an even \(\beta\)-KMS state \(\rho\) of \(\tau^\Xi\), \(\widehat\rho\) denotes its \emph{diagonal thermal extension}, or simply its \emph{thermal extension}.
It combines the modular right action of Haag--Hugenholtz--Winnink \cite{HHW67} with the fermionic diagonal construction of Crismale, Duvenhage and Fidaleo \cite{CDF21}.
For quantum spin systems, the corresponding extension is the purification of KMS states introduced by Matsui \cite[Definition~1]{Matsui94}, and the annihilation identities \eqref{eq:annihilation-identities} of Proposition~\ref{prop:lift} are the fermionic counterpart of \cite[equations~(39) and~(40)]{Matsui94}.
Proposition~\ref{prop:lift} constructs this extension on the GNS space of \(\rho\) and gives the marginals
\[
 \widehat\rho(A^{(1)})=\rho(A),\qquad
 \widehat\rho(B^{(2)})=\rho(\ft(B)).
\]
The two copies are generally correlated; in finite dimensions, the extension is the fermionic form of the purification described in Section~\ref{sec:tfd-finite}, where the Gibbs state is represented by the square root of its density matrix.

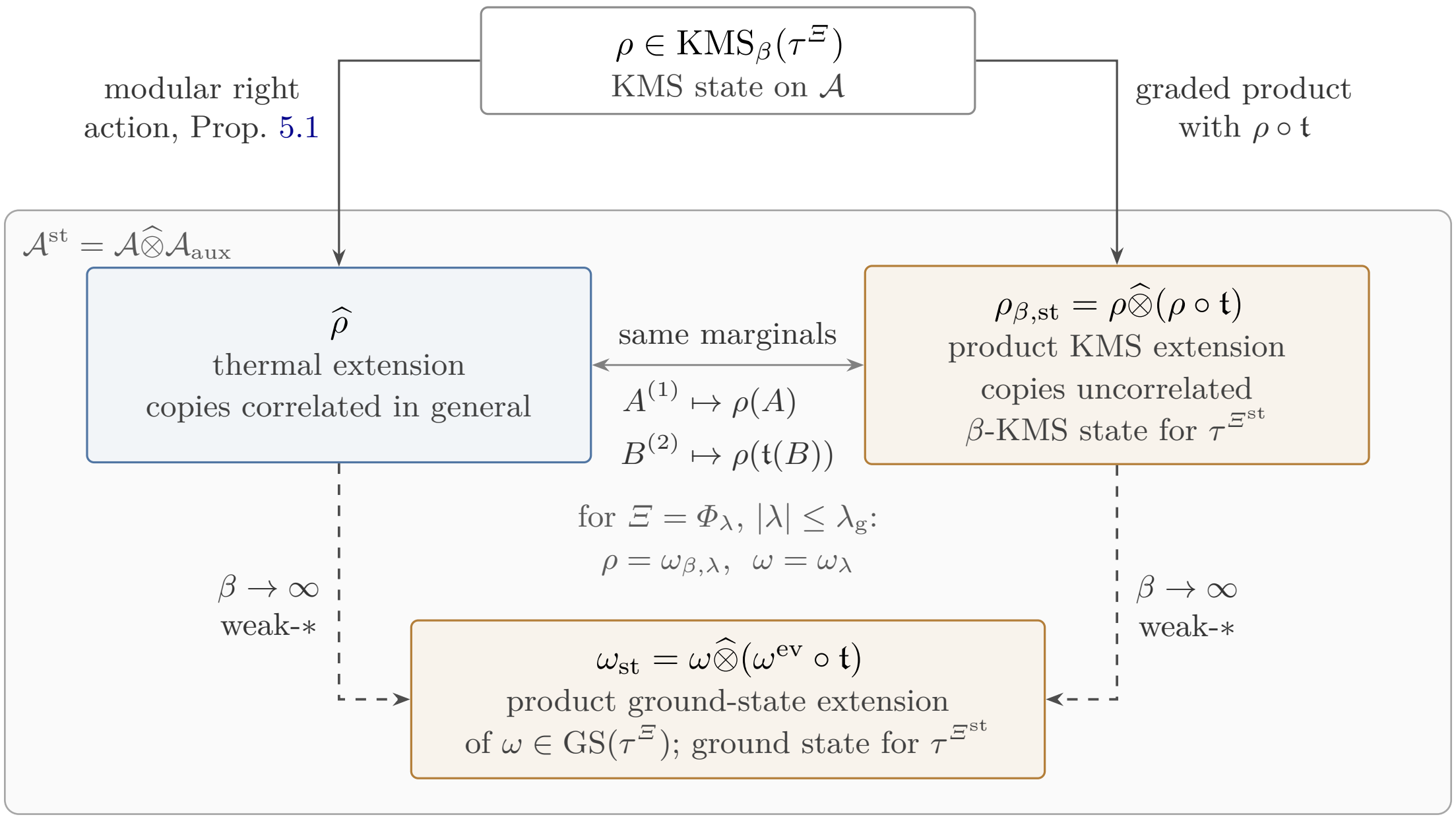
\begin{figure}[!ht]
\centering
\begin{tikzpicture}[lppl]
  \draw[draw=black!35, fill=black!2, rounded corners=4pt, line width=0.5pt]
    (-7.25,-4.5) rectangle (7.25,1.55);
  \node[lbl, anchor=north west, text=black!65] at (-7.15,1.47)
    {$\A^{\st}=\A\widehat\otimes\A_{\aux}$};

  \node[neutralbox, text width=4.6cm] (phys) at (0,3.05)
    {$\rho\in\KMS_\beta(\tau^\Xi)$\\
     \lpplsub{KMS state on $\A$}};

  \node[thermalbox, text width=4.7cm, minimum height=1.95cm] (therm) at (-3.9,0)
    {$\widehat\rho$\\
     \lpplsub{thermal extension}\\
     \lpplsub{copies correlated in general}};
  \node[productbox, text width=4.7cm, minimum height=1.95cm] (prod) at (3.9,0)
    {$\rho_{\beta,\st}=\rho\widehat\otimes(\rho\circ\ft)$\\
     \lpplsub{product KMS extension}\\
     \lpplsub{copies uncorrelated}\\
     \lpplsub{$\beta$-KMS state for $\tau^{\Xi^{\st}}$}};

  \node[productbox, text width=6.0cm, minimum height=1.45cm] (gs) at (0,-3.35)
    {$\omega_\st=\omega\widehat\otimes(\omega^{\rm ev}\circ\ft)$\\
     \lpplsub{product ground-state extension}\\
     \lpplsub{of $\omega\in\GS(\tau^\Xi)$; ground state for $\tau^{\Xi^{\st}}$}};

  \draw[construct] (phys.west) -| (therm.north)
    node[lbl, pos=0.62, left=3pt] {modular right\\action, Prop.~\ref{prop:lift}};
  \draw[construct] (phys.east) -| (prod.north)
    node[lbl, pos=0.62, right=3pt] {graded product\\with $\rho\circ\ft$};

  \draw[<->, line width=0.5pt, draw=black!50] (therm.east) -- (prod.west)
    node[lbl, midway, above=2pt] {same marginals}
    node[lbl, midway, below=2pt]
      {$\begin{aligned}A^{(1)}&\mapsto\rho(A)\\[-1pt]
        B^{(2)}&\mapsto\rho(\ft(B))\end{aligned}$};

  \draw[limit] (therm.south) |- (gs.west)
    node[lbl, pos=0.3, left=3pt] {$\beta\to\infty$\\weak-$*$};
  \draw[limit] (prod.south) |- (gs.east)
    node[lbl, pos=0.3, right=3pt] {$\beta\to\infty$\\weak-$*$};
  \node[lbl, text=black!65] at (0,-1.75)
    {for $\Xi=\Phi_\lambda$, $|\lambda|\le\lambda_{\mathrm g}$:\\[1pt]
     $\rho=\omega_{\beta,\lambda}$, \ $\omega=\omega_\lambda$};
\end{tikzpicture}
\caption[Thermal and product extensions of KMS states.]{%
Thermal and product extensions of \(\rho\in\KMS_\beta(\tau^\Xi)\), \(\Xi\in\mathfrak J_1\), to the fixed algebra \(\A^{\st}\) (shaded).
The solid arrows from \(\rho\) denote the constructions of Proposition~\ref{prop:lift} and \eqref{eq:stacked-states}; the horizontal double arrow records their common marginals \eqref{eq:diagonal-marginals}.
Dashed arrows are weak-\(*\) limits for \(\Xi=\Phi_\lambda\), \(|\lambda|\le\lambda_{\mathrm g}\), under the hypotheses of Theorem~\ref{thm:uniqueness}(i); see Remark~\ref{rem:zero-temperature-extensions}.
Colors match the annihilation families of Figure~\ref{fig:doubling-annihilation-families}.}
\label{fig:doubling-state-extensions}
\end{figure}

For ground-state and low-temperature comparison, we also use product extensions with a reversed copy of the full physical system.
For \(\Xi\in\mathfrak J_1\), define \((\ft\Xi)(Z):=\ft(\Xi(Z))\) and
\[
 \Xi^{\st}(Z):=\Xi(Z)^{(1)}+(\ft\Xi)(Z)^{(2)}.
\]
The auxiliary dynamics thus includes the reversed interaction and any reversed defect.
For \(\rho\in\KMS_\beta(\tau^\Xi)\) and \(\omega\in\GS(\tau^\Xi)\), the product extensions used in Section~\ref{sec:ground-construction} are
\[
 \rho_{\beta,\st}:=\rho\widehat\otimes(\rho\circ\ft),
 \qquad
 \omega_\st:=\omega\widehat\otimes(\omega^{\rm ev}\circ\ft),
 \qquad
 \omega^{\rm ev}:=\tfrac12(\omega+\omega\circ\Theta).
\]
We call \(\omega_\st\) the \emph{product ground-state extension} of \(\omega\), and \(\rho_{\beta,\st}\) the \emph{product KMS extension} of \(\rho\).
The parity average makes the auxiliary factor even, so the product exists for every ground state \(\omega\), including those with \(\omega\circ\Theta\ne\omega\).
Reversal and graded products preserve the corresponding equilibrium conditions: \(\rho_{\beta,\st}\) is a KMS state and \(\omega_\st\) is a ground state for the dynamics generated by \(\Xi^{\st}\).
For a KMS state \(\rho\), the thermal and product KMS extensions have the same marginals, but the product extension satisfies \(\rho_{\beta,\st}(A^{(1)}B^{(2)})=\rho(A)\rho(\ft(B))\).
This factorization distinguishes it from the correlated thermal extension \(\widehat\rho\) at positive temperature.
Figure~\ref{fig:doubling-state-extensions} illustrates the two extensions and their common zero-temperature limit for the background system.

We construct two thermal families and one ground-state family of annihilation operators in the doubled algebra \(\A^{\st}\).
We write the interaction generating the dynamics in square brackets.
The filters and normalizers are chosen using the free Hamiltonian \(h\) and, for the thermal families, the inverse temperature \(\beta\); we use the same choices when replacing the background dynamics by defect dynamics.
A defect \(\Psi\) is included by taking the interaction to be \(\Xi=\Phi_\lambda+\Psi\).

When the free Hamiltonian has a spectral gap, we construct the \emph{cutoff thermal family} \(\mathsf F_u^\beta[\Xi]\) using a filter supported at strictly negative energy transfers.
Without assuming a spectral gap for the free Hamiltonian, we construct the \emph{Fermi-weighted thermal family} \(\mathsf G_u^\beta[\Xi]\) using a Gaussian frequency filter multiplied by a Fermi weight.
Both families annihilate the GNS vector of every thermal extension of a KMS state of \(\tau^\Xi\); see Section~\ref{sec:thermal-family-statement}.
The cutoff construction gives estimates uniform in \(\beta\) and a temperature-independent coupling interval; the Fermi-weighted estimates and coupling interval depend on \(\beta\).

Under the free gap assumption, Section~\ref{sec:ground-construction} constructs the temperature-independent \emph{ground-state family} \(\mathsf F_u^\infty[\Xi]\) by negative-frequency filtering of the doubled dynamics, following the energy-filtering strategy of \cite{LPPLV3}.
This family annihilates the GNS vector of every product ground-state extension for \(\Xi\).
For product KMS extensions, the ground-state annihilation identities hold up to errors that decay exponentially as \(\beta\to\infty\); see Proposition~\ref{prop:ground-family}.
Applying the comparison theorem to these approximate identities gives the low-temperature comparison with the ground state.

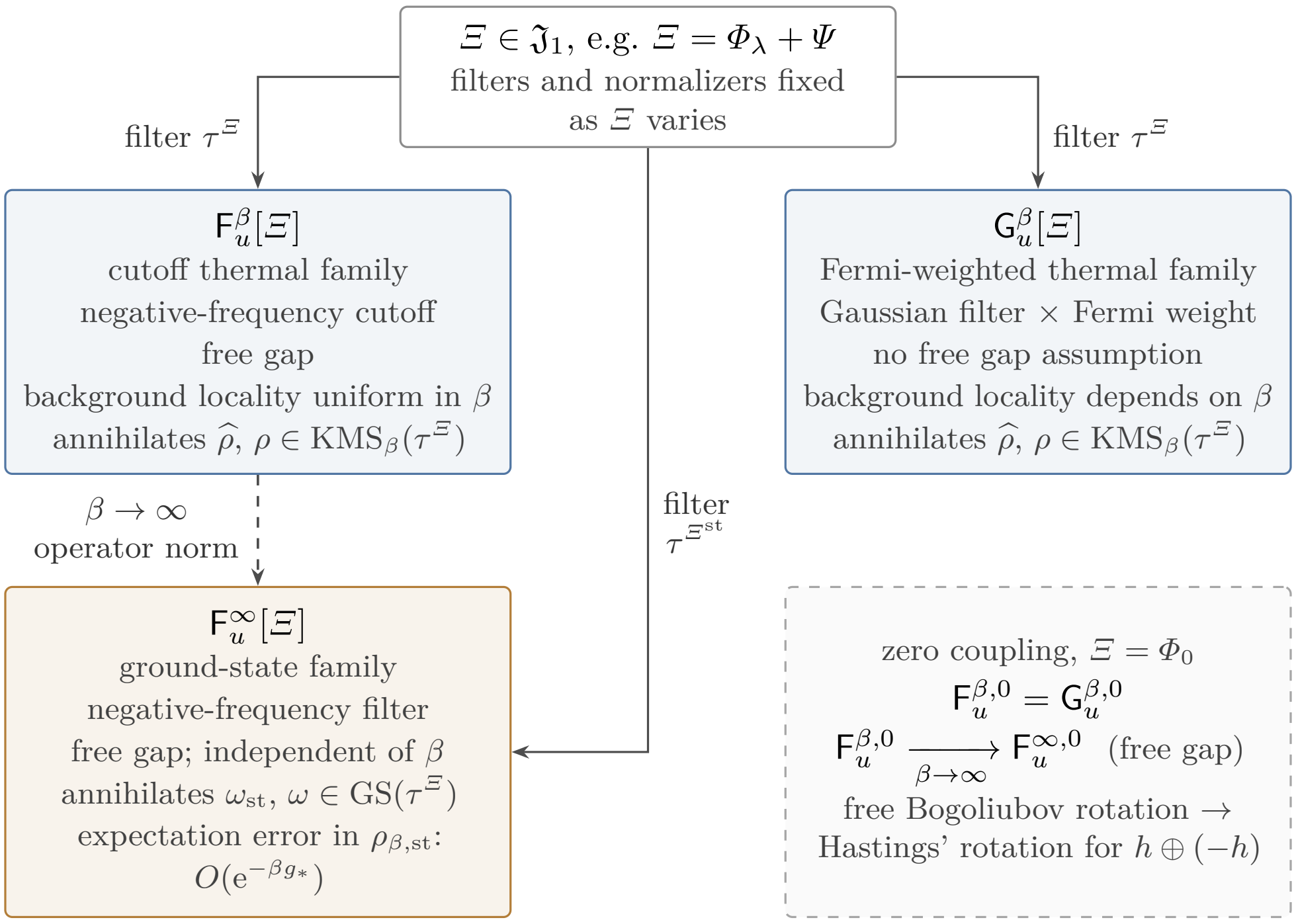
\begin{figure}[!ht]
\centering
\begin{tikzpicture}[lppl]
  \node[neutralbox, text width=4.6cm] (xi) at (0,2.55)
    {$\Xi\in\mathfrak J_1$, e.g.\ $\Xi=\Phi_\lambda+\Psi$\\
     \lpplsub{filters and normalizers fixed}\\[-1pt]
     \lpplsub{as $\Xi$ varies}};

  \node[thermalbox, text width=4.7cm, minimum height=2.4cm] (cut) at (-3.9,0)
    {$\mathsf F_u^\beta[\Xi]$\\
     \lpplsub{cutoff thermal family}\\
     \lpplsub{negative-frequency cutoff}\\
     \lpplsub{free gap}\\
     \lpplsub{background locality uniform in $\beta$}\\
     \lpplsub{annihilates $\widehat\rho$, $\rho\in\KMS_\beta(\tau^\Xi)$}};
  \node[thermalbox, text width=4.7cm, minimum height=2.4cm] (fw) at (3.9,0)
    {$\mathsf G_u^\beta[\Xi]$\\
     \lpplsub{Fermi-weighted thermal family}\\
     \lpplsub{Gaussian filter $\times$ Fermi weight}\\
     \lpplsub{no free gap assumption}\\
     \lpplsub{background locality depends on $\beta$}\\
     \lpplsub{annihilates $\widehat\rho$, $\rho\in\KMS_\beta(\tau^\Xi)$}};

  \node[productbox, text width=4.7cm, minimum height=3.3cm] (gsf) at (-3.9,-4.2)
    {$\mathsf F_u^\infty[\Xi]$\\
     \lpplsub{ground-state family}\\
     \lpplsub{negative-frequency filter}\\
     \lpplsub{free gap; independent of $\beta$}\\
     \lpplsub{annihilates $\omega_\st$, $\omega\in\GS(\tau^\Xi)$}\\
     \lpplsub{expectation error in $\rho_{\beta,\st}$:}\\
     \lpplsub{$O(\E^{-\beta g_*})$}};

  \node[notebox, text width=4.7cm, minimum height=3.3cm] (free) at (3.9,-4.2)
    {\lpplsub{zero coupling, $\Xi=\Phi_0$}\\[2pt]
     $\mathsf F_u^{\beta,0}=\mathsf G_u^{\beta,0}$\\[2pt]
     $\mathsf F_u^{\beta,0}\xrightarrow[\beta\to\infty]{}\mathsf F_u^{\infty,0}$
     \ \lpplsub{(free gap)}\\[1pt]
     \lpplsub{free Bogoliubov rotation $\to$}\\[-1pt]
     \lpplsub{Hastings' rotation for $h\oplus(-h)$}};

  \draw[construct] (xi.west) -| (cut.north)
    node[lbl, pos=0.75, left=2pt] {filter $\tau^{\Xi}$};
  \draw[construct] (xi.east) -| (fw.north)
    node[lbl, pos=0.75, right=2pt] {filter $\tau^{\Xi}$};
  \draw[construct] (xi.south) -- (0,-4.2)
    node[lbl, pos=0.62, right=2pt] {filter\\$\tau^{\Xi^{\st}}$}
    -- (gsf.east);

  \draw[limit] (cut.south) -- (gsf.north)
    node[lbl, midway, left=3pt] {$\beta\to\infty$\\operator norm};
\end{tikzpicture}
\caption[State-independent annihilation families.]{%
State-independent annihilation families in \(\A^{\st}\), obtained by filtering the dynamics indicated on the arrows.
Blue: the thermal families annihilate the GNS vector of the thermal extension \(\widehat\rho\) of every \(\rho\in\KMS_\beta(\tau^\Xi)\) (Proposition~\ref{prop:thermal-family}).
The locality statements concern the background families \(\Xi=\Phi_\lambda\) on the respective weak-coupling intervals.
Amber: the ground-state family annihilates the GNS vector of the product ground-state extension \(\omega_\st\) of every \(\omega\in\GS(\tau^\Xi)\); expectation errors in the product KMS extension \(\rho_{\beta,\st}\) obey \eqref{eq:ground-family-kms}.
The dashed arrow is the operator-norm limit under the free gap condition~\eqref{eq:free-band}, for each fixed \(\Xi\), including defect dynamics; see Remark~\ref{rem:zero-temperature-extensions}.
The dashed box collects the zero-coupling identities; its zero-temperature limits require \eqref{eq:free-band}, as in Remark~\ref{rem:zero-temperature-extensions}.}
\label{fig:doubling-annihilation-families}
\end{figure}

For the background interaction, we write \(\mathsf F_u^{\beta,\lambda}:=\mathsf F_u^\beta[\Phi_\lambda]\), \(\mathsf G_u^{\beta,\lambda}:=\mathsf G_u^\beta[\Phi_\lambda]\) and \(\mathsf F_u^{\infty,\lambda}:=\mathsf F_u^\infty[\Phi_\lambda]\), with \(\lambda=0\) denoting the corresponding free families.
All these operators are fixed by the dynamics, filters and normalizers, independently of the equilibrium state to be compared.
On the respective weak-coupling intervals, the background families are small, spatially localized perturbations of complete free annihilation families.
For either thermal family, Theorem~\ref{thm:comparison}(i) identifies the limiting state \(\omega_F\) with the diagonal thermal extension \(\widehat\omega_{\beta,\lambda}\).
Analogously, for the ground-state family, it identifies \(\omega_F\) with the product ground-state extension \(\omega_{\lambda,\st}\).
It also bounds the difference between local expectations in the background and defect states in terms of the difference between their annihilation families.
Restricting the extended states to the first copy gives uniqueness and local comparison for the physical states.
Figure~\ref{fig:doubling-annihilation-families} summarizes the annihilation families and the zero-temperature limit of the cutoff thermal family.

At zero coupling, the cutoff and Fermi-weighted thermal families give the same free annihilation family \(\mathsf F_u^{\beta,0}\) from \eqref{eq:smooth-free-family}.
For number-conserving dynamics, its vacuum is the Araki--Wyss purification after a particle--hole change in the right copy \cite{ArakiWyss64}.
Under the free gap condition~\eqref{eq:free-band}, this family converges in operator norm to the free ground-state family \(\mathsf F_u^{\infty,0}\) as \(\beta\to\infty\).
The corresponding free Bogoliubov rotation converges to Hastings' rotation for \(h\oplus(-h)\) \cite{Hastings19}; see Remark~\ref{rem:hastings}.

Under the same free gap assumption, the cutoff thermal family \(\mathsf F_u^{\beta,\lambda}\) converges in operator norm to the ground-state family \(\mathsf F_u^{\infty,\lambda}\) as \(\beta\to\infty\), including at nonzero coupling.
The same convergence holds for each fixed defect dynamics; see Remark~\ref{rem:zero-temperature-extensions}.

For each fixed \(|\lambda|\le\lambda_{\mathrm g}\) under the hypotheses of Theorem~\ref{thm:uniqueness}(i), both the thermal and product KMS extensions of \(\omega_{\beta,\lambda}\) converge weak-\(*\) to the product ground-state extension of the unique physical ground state \(\omega_\lambda\) as \(\beta\to\infty\); see Remark~\ref{rem:zero-temperature-extensions}.
For systems with several ground states, a zero-temperature limit of the thermal extensions can remain correlated between the copies.

To illustrate the idea behind the thermal extension, we first present its finite-dimensional construction before giving its formulation for the CAR algebra.

\subsection{The finite-dimensional construction}
\label{sec:tfd-finite}

Let \(H=H^*\) on \(\C^M\), \(M\ge2\), with orthonormal eigenbasis \(H|n\rangle=E_n|n\rangle\), and let \(\beta>0\). The Gibbs density matrix is
\[
 r_\beta=\frac{\E^{-\beta H}}{\Tr(\E^{-\beta H})}
        =\sum_n p_n|n\rangle\langle n|,
 \qquad
 p_n:=\frac{\E^{-\beta E_n}}{\sum_j\E^{-\beta E_j}}.
\]
Since \(r_\beta>0\), an operator \(B\) satisfying \(\Tr(r_\beta B^*B)=0\) must vanish. We therefore represent the Gibbs state on a larger Hilbert space, where it admits nonzero annihilation operators.

On the Hilbert--Schmidt space \(\mathcal B(\C^M)\), with inner product \(\langle X,Y\rangle_{\rm HS}:=\Tr(X^*Y)\), the unit vector \(r_\beta^{1/2}\) represents the Gibbs state:
\[
 \langle r_\beta^{1/2},Ar_\beta^{1/2}\rangle_{\rm HS}
 =\Tr(r_\beta A).
\]
The coefficients of \(r_\beta^{1/2}\) in the diagonal matrix units are \(\sqrt{p_n}\). Furthermore, denoting the matrix units by \(A_{mn}:=|m\rangle\langle n|\), we define
\[
 \mathsf D_{mn}(X)
 :=A_{mn}X-\E^{\beta(E_m-E_n)/2}XA_{mn}.
\]
Hence, $r_\beta^{1/2}A_{mn}=\sqrt{p_m}A_{mn}$ and $A_{mn}r_\beta^{1/2}=\sqrt{p_n}A_{mn}$. 
Using \(\E^{\beta(E_m-E_n)/2}=\sqrt{p_n/p_m}\), we obtain
\[
 \mathsf D_{mn}(r_\beta^{1/2})
 =\bigl(\sqrt{p_n}-\sqrt{p_n/p_m}\sqrt{p_m}\bigr)A_{mn}=0.
\]
Thus \(\mathsf D_{mn}\) annihilates the vector representing the Gibbs state.

We define the \emph{thermofield Hamiltonian} by summing the corresponding positive operators:
\[
 \mathsf H_{\rm TFD}(X)
 :=\sum_{m,n}p_m\mathsf D_{mn}^*\mathsf D_{mn}(X)
 =2X-2r_\beta^{1/2}\langle r_\beta^{1/2},X\rangle_{\rm HS},
\]
where \(*\) denotes the Hilbert--Schmidt adjoint.
Its ground-state space is \(\C r_\beta^{1/2}\), and its gap is \(2\).
Hence the Gibbs state may be represented by the unique normalized ground-state vector of \(\mathsf H_{\rm TFD}\), up to a phase.

This construction uses the full spectral data of \(H\) and does not provide spatial locality estimates. For fermionic lattice systems, instead, we will construct annihilation operators whose locality follows from bounds on the dynamics.

To express this representation on a doubled Hilbert space, we may use the identification
\(\mathcal B(\C^M)\simeq\C^M\otimes\overline{\C^M}\):
the vector \(r_\beta^{1/2}\) corresponds then to
\[
\Omega_\beta:=\sum_n\sqrt{p_n}|n\rangle\otimes\overline{|n\rangle},
\]
left multiplication by \(A\) acts as \(A\otimes\1\), and right multiplication by \(B\) as \(\1\otimes\overline{B^*}\); moreover,
the restriction of the vector state to the first factor yields thermal expectations, namely
\[
\langle\Omega_\beta, (A\otimes\1) \Omega_\beta\rangle = \Tr(r_\beta A).
\]
Finally, for \(\tau_t(A):=\E^{\I tH}A\E^{-\I tH}\), evaluation at imaginary time $-\I\beta/2$ gives the factor in the definition of \(\mathsf D_{mn}\):
\[
 \tau_{-\I\beta/2}(A_{mn})
       =\E^{\beta(E_m-E_n)/2}A_{mn}.
\]
By linearity, the annihilation identities for the matrix units give
\begin{equation}\label{eq:finite-thermal-identity}
 A r_\beta^{1/2}
   =r_\beta^{1/2}\tau_{-\I\beta/2}(A)
 \qquad(A\in\mathcal B(\C^M)).
\end{equation}
We note that this identity does not depend on the choice of eigenbasis. In analogy with \eqref{eq:finite-thermal-identity}, in infinite volume we shall obtain the corresponding relation between the left and right actions from the KMS condition. For the doubled CAR algebra, these actions must also satisfy graded commutation relations.

\subsection{The diagonal thermal extension}
\label{sec:canonical-extension}

Let \(\tau\) be a strongly continuous even dynamics, let \(\beta>0\), and let \(\rho\in\KMS_\beta(\tau)\) be even, with GNS triple \((\pi,\mathcal H,\Omega)\).

We now construct the analogue of right multiplication on the GNS space of \(\rho\).

For analytic \(B,C\), we define
\[
 R(B)\pi(C)\Omega:=\pi(C\tau_{\I\beta/2}(B))\Omega.
\]
The KMS condition gives the bound
\begin{equation}\label{eq:KMS-right-action-bound}
 \|R(B)\pi(C)\Omega\|\le\|B\|\,\|\pi(C)\Omega\|.
\end{equation}
In the finite-dimensional identification discussed in Section~\ref{sec:tfd-finite}, this is right multiplication:
\[
 R(B)\pi(C)\Omega
 \longleftrightarrow C\tau_{\I\beta/2}(B)r_\beta^{1/2}
 =Cr_\beta^{1/2}B.
\]
The bound \eqref{eq:KMS-right-action-bound} allows us to extend \(R\) to all of \(\A\). One can then check that the resulting right action satisfies
\[
 R(AB)=R(B)R(A),\qquad R(B^*)=R(B)^*,\qquad [R(B),\pi(A)]=0.
\]
For analytic \(A\), the definition gives the analogue of the finite-dimensional identity~\eqref{eq:finite-thermal-identity}:
\[
 R(\tau_{-\I\beta/2}(A))\Omega
 =\pi(\tau_{\I\beta/2}(\tau_{-\I\beta/2}(A)))\Omega
 =\pi(A)\Omega.
\]

To compare different KMS states, we need to express the relation between the left and right actions using elements of a fixed CAR algebra, independent of the state.
We use the doubled CAR algebra and the embeddings introduced at the beginning of this section. We write
\[
 \A^\st_X:=\A_X\widehat\otimes\A_{\aux,X},
 \qquad
 \gamma_{u,1}:=\gamma_u^{(1)},\qquad
 \gamma_{u,2}:=\gamma_u^{(2)}.
\]
For a homogeneous element \(B\), we write \(\Theta(B)=(-1)^{|B|}B\), where \(|B|\in\{0,1\}\).
The graded tensor product satisfies, for $A$ and $B$ homogeneous,
\[
 B^{(2)}A^{(1)}=(-1)^{|A||B|}A^{(1)}B^{(2)}.
\]

The action \(R\) reverses products and commutes with the physical representation. To obtain a representation satisfying the graded commutation relations above, we combine it with the reversal \(\ft\) introduced above, and parity.
Let \(U\) be the parity unitary. Then
\[
 \begin{aligned}
 U\pi(A)\Omega&=\pi(\Theta(A))\Omega,
       &UR(B)U&=R(\Theta(B)).
 \end{aligned}
\]
For homogeneous \(B\), we define
\[
 r(B):=\I^{|B|}R(\ft(B))U^{|B|},
\]
and extend it linearly to \(\A\). For homogeneous \(A,B\), \(r\) satisfies
\[
 \begin{aligned}
 r(AB)&=r(A)r(B),\qquad r(B^*)=r(B)^*,\\
 r(B)\pi(A)&=(-1)^{|A||B|}\pi(A)r(B).
 \end{aligned}
\]
Hence \(\pi\) and \(r\) define a representation of \(\A^\st\):
\[
\widehat\pi(A^{(1)}B^{(2)}):=\pi(A)r(B).
\]
For odd analytic \(A\), we use \(U\Omega=\Omega\) to obtain
\[
 \begin{aligned}
 \widehat\pi\bigl(A^{(1)}
       +\I(\ft(\tau_{-\I\beta/2}(A)))^{(2)}\bigr)\Omega
 &=\pi(A)\Omega+\I r(\ft(\tau_{-\I\beta/2}(A)))\Omega\\
 &=\pi(A)\Omega-R(\tau_{-\I\beta/2}(A))\Omega=0.
 \end{aligned}
\]

We prove boundedness of the right action and the representation properties in Section~\ref{sec:lift-proof}. The marginal identities are proved there as well.
The following proposition summarizes the construction.

\begin{proposition}[Diagonal thermal extension]\label{prop:lift}
Let \(\tau\) be a strongly continuous even dynamics on \(\A\), let \(\beta>0\), and let \(\rho\in\KMS_\beta(\tau)\) be even, with GNS triple \((\pi,\mathcal H,\Omega)\).
There is a unique representation \(\widehat\pi\) of \(\A^{\st}\) on \(\mathcal H\) satisfying \(\widehat\pi(A^{(1)})=\pi(A)\) for all \(A\in\A\) and, for every odd \(A\in\A_\an(\tau)\),
\begin{equation}\label{eq:annihilation-identities}
 \begin{aligned}
 \widehat\pi\bigl(A^{(1)}+
       \I(\ft(\tau_{-\I\beta/2}(A)))^{(2)}\bigr)\Omega&=0,\\
 \widehat\pi\bigl((\ft(A^*))^{(2)}-
       \I((\tau_{-\I\beta/2}(A))^*)^{(1)}\bigr)\Omega&=0.
 \end{aligned}
\end{equation}
The vector state \(\widehat\rho(\widehat C):=\langle\Omega,\widehat\pi(\widehat C)\Omega\rangle\) on \(\A^\st\) is even, does not depend on the choice of the GNS triple, and has marginals
\begin{equation}\label{eq:diagonal-marginals}
 \widehat\rho(A^{(1)})=\rho(A),\qquad
 \widehat\rho(B^{(2)})=\rho(\ft(B))
 \qquad(A,B\in\A).
\end{equation}
We call it the \emph{diagonal thermal extension} of \(\rho\).
Under the reversal \(\ft\), which identifies the right CAR copy with the opposite algebra, it is the fermionic diagonal state with our parity-twist convention.
\end{proposition}

Notice that the evenness assumption holds for all KMS states of the interacting dynamics considered in this work, by Proposition~\ref{prop:kms-nonempty}.

Moreover, the extension $\widehat\rho$ depends on \(\rho\), whereas the two elements of the doubled CAR algebra appearing in \eqref{eq:annihilation-identities} depend only on \(\tau,\beta\) and \(A\). Notice also that, crucially, the diagonal extension is uniquely determined by the identities \eqref{eq:annihilation-identities}.

We will use these identities to construct the free annihilation family in Section~\ref{sec:thermal-free-coordinates} and its interacting counterpart in Section~\ref{sec:thermal-family-statement}.

\subsection{The free annihilation family}
\label{sec:thermal-free-coordinates}

We now apply Proposition~\ref{prop:lift} to the free dynamics \(\tau^{\Phi_h}\). This gives a concrete family of operators satisfying the annihilation identities of Section~\ref{sec:canonical-extension}. We choose their coefficients so that they also satisfy the CAR and, together with their adjoints, generate the doubled algebra. This family will serve as the reference for the interacting construction in Section~\ref{sec:thermal-family-statement}.

Since \(h\) is bounded, the Majoranas are entire analytic for the free dynamics. By \eqref{eq:Majorana-free-dynamics}, for \(\beta>0\),
\[
 \tau_{-\I\beta/2}^{\Phi_h}(\gamma_u)
 =\sum_v(\E^{-\beta h/2})_{uv}\gamma_v.
\]
The reversal \(\ft\) leaves the expression above unchanged, since it is a linear combination of Majoranas. Thus the first identity in \eqref{eq:annihilation-identities} has coefficient matrix \(\1\) in the first copy and \(\I\E^{-\beta h/2}\) in the second.
To obtain coefficient functions bounded uniformly in \(\beta\ge0\), we normalize the corresponding scalar coefficients \(1\) and \(\E^{-\beta \varepsilon/2}\) by defining, for real \(\varepsilon\),
\begin{equation}\label{eq:free_normalization}
    a_\beta(\varepsilon):=(1+\E^{-\beta \varepsilon})^{-1/2}.
\end{equation}
Notice that their squares \(a_\beta(\varepsilon)^2\) and \(a_\beta(-\varepsilon)^2\) are complementary Fermi weights. Also, for a positive energy \(\varepsilon\), the ratio of the coefficients is \(a_\beta(-\varepsilon)/a_\beta(\varepsilon)=\E^{-\beta\varepsilon/2}\), as in the finite-dimensional annihilation identities of Section~\ref{sec:tfd-finite}.

Using such normalization factors, given \(u\in\Gamma\times\{1,\ldots,2r\}\) we define the corresponding free annihilation operator by
\begin{equation}\label{eq:smooth-free-family}
 \mathsf F_u^{\beta,0}
 :=\frac1{\sqrt2}\sum_v
       \bigl[a_\beta(h)_{uv}\gamma_{v,1}
              +\I a_\beta(-h)_{uv}\gamma_{v,2}\bigr].
\end{equation}
For \(\beta>0\), the identity \(a_\beta(h)\E^{-\beta h/2}=a_\beta(-h)\) gives
\[
 \mathsf F_u^{\beta,0}
 =\frac1{\sqrt2}\sum_v a_\beta(h)_{uv}
 \Bigl[\gamma_{v,1}
       +\I\bigl(\ft(\tau_{-\I\beta/2}^{\Phi_h}(\gamma_v))\bigr)^{(2)}\Bigr].
\]
Applying the first identity of Proposition~\ref{prop:lift} to each Majorana \(\gamma_v\) therefore gives
\[
 \widehat\pi(\mathsf F_u^{\beta,0})\Omega=0.
\]
Consequently, the thermal extension \(\widehat\omega_{\beta,0}\) of the free KMS state given by Proposition~\ref{prop:lift} satisfies
\begin{equation}\label{eq:free-annihilation}
 \widehat\omega_{\beta,0}\bigl((\mathsf F_u^{\beta,0})^*\mathsf F_u^{\beta,0}\bigr)
 =\|\widehat\pi(\mathsf F_u^{\beta,0})\Omega\|^2=0,
\end{equation}
for any \(u\in\Gamma\times\{1,\ldots,2r\}\).
We next verify the CAR and show that this family determines the state uniquely.

We construct a Bogoliubov automorphism sending this family to the on-site annihilation operators \(c_u:=\frac12(\gamma_{u,1}+\I\gamma_{u,2})\). Let \(\omega_\at\) be the atomic vacuum, characterized by \(\omega_\at(c_u^*c_u)=0\) for all \(u\).
We define the rotation, written in two blocks corresponding to the two copies,
\begin{equation*}
 O^\beta_\st:=\frac1{\sqrt2}
 \begin{pmatrix}
 a_\beta(h)+a_\beta(-h)&\I\bigl(a_\beta(h)-a_\beta(-h)\bigr)\\
 \I\bigl(a_\beta(h)-a_\beta(-h)\bigr)&a_\beta(h)+a_\beta(-h)
 \end{pmatrix},
\end{equation*}
and use the identities
\[
 a_\beta(h)^\top=a_\beta(-h),\qquad
 \overline{a_\beta(h)}=a_\beta(-h),\qquad
 a_\beta(h)^2+a_\beta(-h)^2=\1
\]
to see that it is real orthogonal.
The corresponding parity-preserving Bogoliubov automorphism acts by
\begin{equation}\label{eq:thermal-Bogoliubov-automorphism}
 \sigma^\beta_\st(\gamma_{u,\iota})
 =\sum_{\substack{v\in\Gamma\times\{1,\ldots,2r\}\\\kappa\in\{1,2\}}}
 (O^\beta_\st)_{(u,\iota),(v,\kappa)}\gamma_{v,\kappa},
 \qquad\iota\in\{1,2\}.
\end{equation}
Applying the inverse $(\sigma^{\beta}_\st)^{-1}$, obtained by replacing \(O^\beta_\st\) by \((O^\beta_\st)^\top\) in the definition of \(\sigma^\beta_\st\), to \(c_u=(\gamma_{u,1}+\I\gamma_{u,2})/2\), we obtain
\[
 \begin{aligned}
 (\sigma^\beta_\st)^{-1}(c_u)
 &=\frac{1}{2}\bigl[(\sigma^\beta_\st)^{-1}(\gamma_{u,1})
       +\I(\sigma^\beta_\st)^{-1}(\gamma_{u,2})\bigr]\\
 &=\frac1{\sqrt2}\sum_v
       \bigl[a_\beta(h)_{uv}\gamma_{v,1}
              +\I a_\beta(-h)_{uv}\gamma_{v,2}\bigr]=\mathsf F_u^{\beta,0}.
 \end{aligned}
\]
In particular, it follows that
\[
 \{\mathsf F_u^{\beta,0},\mathsf F_v^{\beta,0}\}=0,\qquad
 \{\mathsf F_u^{\beta,0},(\mathsf F_v^{\beta,0})^*\}=\delta_{uv}\1.
\]
Since the operators \(c_u\) and their adjoints generate \(\A^\st\), the same holds for the free family.

Consequently, \(\omega_\at\circ\sigma^\beta_\st\) is a pure quasifree state annihilated by every \(\mathsf F_u^{\beta,0}\). Indeed, by the defining property of the atomic vacuum,
\[
 (\omega_\at\circ\sigma^\beta_\st)
       \bigl((\mathsf F_u^{\beta,0})^*\mathsf F_u^{\beta,0}\bigr)
 =\omega_\at(c_u^*c_u)=0.
\]
By repeated use of the CAR, every polynomial in these operators and their adjoints can be written as a scalar plus terms beginning with an adjoint or ending with an annihilator.
These latter terms have zero expectation in every state satisfying these annihilation identities, by Cauchy--Schwarz. All such states therefore agree on these polynomials and, by norm density, on all of \(\A^\st\). This proves uniqueness of the state annihilated by the free family, and for $\beta>0$, by \eqref{eq:free-annihilation} we get
\[
 \widehat\omega_{\beta,0}=\omega_\at\circ\sigma^\beta_\st.
\]

\begin{remark}\label{rem:hastings}
Under the free spectral-gap assumption \eqref{eq:free-band}, functional calculus gives \mbox{\(a_\beta(\pm h)\to P_\pm\)} in operator norm as \(\beta\to\infty\). It follows that
\[
 O^\beta_\st\xrightarrow[\beta\to\infty]{}O_\st:=
 \frac1{\sqrt2}\begin{pmatrix}\1&\I\sgnh\\\I\sgnh&\1\end{pmatrix},
\]
in operator norm. $O_\st$ is Hastings' rotation for \(h\oplus(-h)\) \cite[Section~III]{Hastings19}, used in \cite[Section~4]{StabilityBdG26} and \cite[Section~5]{LPPLV3} to obtain the atomic vacuum \(\omega_\at\).
\end{remark}

The two-point matrix of a state \(\omega\) has entries \(\omega(\gamma_{u,j}\gamma_{v,k})\), grouped into blocks according to the copy indices \(j,k\in\{1,2\}\). For the atomic vacuum, this matrix is
\[
\begin{pmatrix}\1&\I\1\\-\I\1&\1\end{pmatrix}.
\]
Using \(a_\beta(h)^2-a_\beta(-h)^2=\tanh(\beta h/2)\) and \(2a_\beta(h)a_\beta(-h)=\operatorname{sech}(\beta h/2)\)\footnote{For real $x$, we write \(\operatorname{sech}(x):=1/\cosh(x)\).}, we find that the two-point matrix of $\widehat\omega_{\beta,0}$ is
\[
 O^\beta_\st
 \begin{pmatrix}\1&\I\1\\-\I\1&\1\end{pmatrix}
 (O^\beta_\st)^\top
 =
 \begin{pmatrix}
 \1+\tanh(\beta h/2)&\I\operatorname{sech}(\beta h/2)\\
 -\I\operatorname{sech}(\beta h/2)&\1-\tanh(\beta h/2)
 \end{pmatrix}.
\]
In particular, the two-point functions within the physical copy and between the two copies are
\[
 \begin{aligned}
  (\omega_\at\circ\sigma^\beta_\st)(\gamma_{u,1}\gamma_{v,1})
    &=\bigl[\1+\tanh(\beta h/2)\bigr]_{uv},\\
  (\omega_\at\circ\sigma^\beta_\st)(\gamma_{u,1}\gamma_{v,2})
    &=\I\bigl[\operatorname{sech}(\beta h/2)\bigr]_{uv}.
 \end{aligned}
\]
For any \(\beta\), the first identity gives the two-point matrix of the free KMS state \cite{Araki70}.

For nonnegative scalar matrices \(\mathsf H\), we use the weighted Schur norm of Definition~\ref{def:weight}. 
In particular, we denote by \(\|\mathsf H\|^{\rm Sch}_{\mu,\theta}\) and \(\|\mathsf H\|^{\rm Sch,pol}_p\) the norms corresponding to the weights \(W(R)=\exp(\mu R^\theta)\) and \(W(R)=(1+R)^p\), respectively.

Cauchy's integral formula expresses \(a_\beta(\pm h)\) in terms of resolvents of \(h\). Their spatial decay is controlled by Combes--Thomas estimates. In the gapped case, we can keep the contours fixed around the positive and negative spectral bands and obtain bounds uniform in \(\beta\). Without a gap, the complex singularities of \(a_\beta\) can approach the spectrum as the temperature decreases. To keep Cauchy's formula valid, the contours must then move closer to the spectrum, which weakens the resolvent bounds. We obtain the following proposition.

\begin{proposition}[The free annihilation family]\label{prop:free-thermal-family}
Assuming Hypothesis~\ref{hyp:locality}, for every \(\beta\ge0\) the series in \eqref{eq:smooth-free-family} converge absolutely in norm. Their Majorana coefficients
\[
 \mathsf a^\beta_{uv}\1:=\{\gamma_v,\mathsf F_u^{\beta,0}\},
 \qquad u\in\Gamma\times\{1,\ldots,2r\},\quad
 v\in\Gamma\times\{1,\ldots,2r\}\times\{1,2\},
\]
satisfy the Parseval condition of Definition~\ref{def:parseval-family}, namely
\[
 \RealPart\sum_u\overline{\mathsf a^\beta_{uv}}\mathsf a^\beta_{uw}
 =\delta_{vw},
\]
for all \(v, w\). Furthermore, in regimes \ref{ass:exponential} and \ref{ass:polynomial-weight} their scalar Schur norms satisfy the following.
\begin{enumerate}[label=(\roman*),leftmargin=*]
\item \emph{With a free spectral gap.} Assuming \eqref{eq:free-band}, there are model constants \(\mu>0\) and \(C_0\ge1\) such that, uniformly for \(\beta\ge0\),
\[
 \ref{ass:exponential}:\quad
 \bigl\|(|\mathsf a^\beta_{uv}|)\bigr\|^{\rm Sch}_{\mu,\theta_{\rm in}}\le C_0,
 \qquad
 \ref{ass:polynomial-weight}:\quad
 \bigl\|(|\mathsf a^\beta_{uv}|)\bigr\|^{\rm Sch,pol}_{p_h}\le C_0.
\]
\item \emph{Without a gap assumption.} There are model constants \(K,N_0,\bar \mu>0\) such that, for every \(\beta\ge0\),
\begin{equation}\label{eq:thermal-free-coefficient-bound-ng}
 \ref{ass:exponential}:\quad
 \bigl\|(|\mathsf a^\beta_{uv}|)\bigr\|^{\rm Sch}_{\bar\mu/(1+\beta),\theta_{\rm in}}\le K(1+\beta)^{N_0},
 \qquad
 \ref{ass:polynomial-weight}:\quad
 \bigl\|(|\mathsf a^\beta_{uv}|)\bigr\|^{\rm Sch,pol}_{p_h}\le K(1+\beta)^{N_0}.
\end{equation}
\end{enumerate}
\end{proposition}

\subsection{Interacting thermal annihilation families}
\label{sec:thermal-family-statement}

In this section, we introduce interacting annihilation operators that reduce to \eqref{eq:smooth-free-family} at zero coupling. In order to construct the analytic, odd elements $A$ needed in Proposition~\ref{prop:lift}, we average the dynamics of single-site Majorana operators over a spectral filter. More precisely, for \(q\in L^1(\R)\) and \(\Xi\in\mathfrak J_1\), our candidates will be of the form
\begin{equation}\label{eq:filtered-Majorana-definition}
 \gamma_u[q;\Xi]:=\int_\mathbb R q(t) \tau^\Xi_t(\gamma_{u})\,\dd t,
\end{equation}
where, for a function $\hat q\in L^1(\mathbb R)$, we use the Fourier convention
\begin{equation}\label{eq:thermal-filter-definition}
 q(t):=\frac1{2\pi}\int_\R \hat q(\omega)\E^{-\I t\omega}\,\dd\omega.
\end{equation}
Their spatial locality properties will be determined by combining the decay of appropriately chosen time kernels $q$ with propagation bounds for the interacting dynamics $\tau^{\Phi_\lambda}$.

For the analytic filters used below, Lemma~\ref{lem:analytic-filtering} gives
\begin{equation}\label{eq:filter-imaginary-time-shift}
\tau^\Xi_{-\I\beta/2}(\gamma_u[q;\Xi]) = \int_{\R} q(t+\I\beta/2)\tau^\Xi_t(\gamma_{u})\,\dd t.
\end{equation}
In particular, the modified filter has Fourier transform $e^{\beta\omega/2}\hat q(\omega)$. This allows us to construct the identities in Proposition~\ref{prop:lift} by simply choosing $q$ appropriately for each case at hand.

\paragraph{The cutoff thermal family under a free gap.}
Let us assume the free gap condition \eqref{eq:free-band} and fix \(\sG>1\). We choose a real smooth cutoff function \(0\le\widehat f\le1\), of Gevrey order $\sG$.
To this end, we take \(\widehat f\) to be the function given by Lemma~\ref{lem:gevrey-cutoff} with \(I=[-C_h,-g]\) and \(\delta=g/8\). In particular,
\[
 \widehat f=1\text{ on }[-C_h,-g],\qquad
 \supp\widehat f\subset[-2C_h,-7g/8].
\]
Employing the Fourier convention \eqref{eq:thermal-filter-definition}, we set the filter function to be $f$.
The imaginary-time shift of $f$ then has Fourier representation
\[
 f(t+\I\beta/2)=\frac1{2\pi}\int_\R
 \E^{\beta\omega/2}\widehat f(\omega)\E^{-\I t\omega}\,\dd\omega,
\]
and we have that
\[
\E^{\beta\omega/2}\widehat f(\omega)\le\E^{-\beta g/4}\qquad \text{on}\ \supp\hat f.
\]

Together with the derivative bounds for \(\widehat f\), this implies fast decay of both \(f\) and \(f(\,\cdot\,+\I\beta/2)\), uniformly in \(\beta\), as proved in Lemma~\ref{lem:Gevrey}.

We combine the two kernels by setting, for \(\beta\ge0\) and \(  t\in\R\),
\begin{equation}\label{eq:filtered-Majoranas}
 f_\beta(t):=f(t)+\overline{f(t+\I\beta/2)}.
\end{equation} The Fourier transform of $f_\beta$ is real-valued, belongs to \(C_c^\infty(\R)\), and satisfies
\[
 \begin{gathered}
 \widehat f_\beta(\omega)
 =\widehat f(\omega)+\E^{-\beta\omega/2}\widehat f(-\omega),\\
 \E^{\beta\omega/2}\widehat f_\beta(\omega)=\widehat f_\beta(-\omega).
 \end{gathered}
\]
For \(\Xi\in\mathfrak J_1\) and \(u\in\Gamma\times\{1,\ldots,2r\}\), the filtered operator \(\gamma_u[f_\beta;\Xi]\) is entire analytic by Lemma~\ref{lem:analytic-filtering} and odd because the dynamics is even. The preceding identity and the real-valuedness of \(\widehat f_\beta\) imply \(f_\beta(t+\I\beta/2)=\overline{f_\beta(t)}\), so~\eqref{eq:filter-imaginary-time-shift} gives
\begin{equation}\label{eq:CD}
 \tau_{-\I\beta/2}^\Xi
 \bigl(\gamma_u[f_\beta;\Xi]\bigr)
 =\gamma_u[f_\beta;\Xi]^*.
\end{equation}
Analogously to \eqref{eq:free_normalization}, we set \(E_\beta:=\E^{-\beta|h|/2}\) and we define the normalization operator
\begin{equation}\label{eq:gapped-thermal-normalizer}
 \Rb:=(\1+E_\beta^2)^{-1/2}.
\end{equation}
Motivated by the first identity in \eqref{eq:annihilation-identities}, we define the \emph{cutoff thermal family} by
\begin{equation}\label{eq:thermal-coordinates}
 \mathsf F_u^{\beta,\lambda}
 :=\frac1{\sqrt2}\sum_{v\in\Gamma\times\{1,\ldots,2r\}}(\Rb)_{uv}\Bigl[
       \gamma_v[f_\beta;\Phi_\lambda]^{(1)}
       +\I\bigl(\ft(\gamma_v[f_\beta;\Phi_\lambda]^*)\bigr)^{(2)}\Bigr].
\end{equation}
The normalization is chosen so that at zero coupling \eqref{eq:thermal-coordinates} agrees with the free family \eqref{eq:smooth-free-family}, as asserted in Proposition~\ref{prop:thermal-family}.

\paragraph{The Fermi-weighted thermal family at fixed temperature.}
\leavevmode We now work without assuming a spectral gap for \(h\). We set \(a:=(1+C_h)^{-2}\) and, for finite $\beta>0$, we define the \emph{Fermi-weighted Gaussian} filter
\begin{equation}\label{eq:FG-kernel}
\hat q_\beta(\omega):=\frac{\E^{-a(\beta\omega/(1+\beta))^2}}{\sqrt{1+\E^{\beta\omega}}},
\end{equation}
which then satisfies the condition\footnote{This is sometimes referred to as the \emph{detailed-balance} condition.}
\[
 \hat q_\beta(-\omega)=\E^{\beta\omega/2}\hat  q_\beta(\omega).
\]
For fixed \(\beta>0\), \(\widehat q_\beta\) and its first two derivatives have Gaussian decay. For \(\Xi\in\mathfrak J_1\) and \(u\in\Gamma\times\{1,\ldots,2r\}\), the filtered operator \(\gamma_u[q_\beta;\Xi]\) is entire analytic by Lemma~\ref{lem:analytic-filtering} and odd because the dynamics is even. Since \(\widehat q_\beta\) is real-valued, the preceding identity implies \(q_\beta(t+\I\beta/2)=\overline{q_\beta(t)}\), so~\eqref{eq:filter-imaginary-time-shift} gives
\begin{equation}\label{eq:FG-fields}
 \tau_{-\I\beta/2}^\Xi
 \bigl(\gamma_u[q_\beta;\Xi]\bigr)
 =\gamma_u[q_\beta;\Xi]^*.
\end{equation}
We define the normalization operator
\begin{equation}\label{eq:FG-normalizer}
 \mathsf M:=M_\beta(h),\qquad
 M_\beta(z):=\frac1{\sqrt2}\E^{a(\beta z/(1+\beta))^2}.
\end{equation}
Again, motivated by the first identity in \eqref{eq:annihilation-identities}, we define the \emph{Fermi-weighted thermal family} by
\begin{equation}\label{eq:FG-coordinates}
 \mathsf G_u^{\beta,\lambda}
 :=\sum_{v\in\Gamma\times\{1,\ldots,2r\}}\mathsf M_{uv}\Bigl[
       \gamma_v[q_\beta;\Phi_\lambda]^{(1)}
       +\I\bigl(\ft(\gamma_v[q_\beta;\Phi_\lambda]^*)\bigr)^{(2)}\Bigr].
\end{equation}
The normalization is chosen so that at zero coupling \eqref{eq:FG-coordinates} agrees with the free family \eqref{eq:smooth-free-family}, as asserted in Proposition~\ref{prop:thermal-family}.

\begin{remark}
The filter \(q_\beta\) decays exponentially in time with rate \(1/(2\beta)\to0\) as \(\beta\to\infty\). The filtered operator
\[
 \gamma_u[q_\beta;\Phi_\lambda]
 =\int_\R q_\beta(t)\tau_t^{\Phi_\lambda}(\gamma_u)\,\dd t
\]
therefore has contributions from longer evolution times as \(\beta\to\infty\). During these times, the initially local Majorana \(\gamma_u\) can spread farther. Combining this time decay with Lieb--Robinson bounds gives the temperature-dependent locality estimates for the Fermi-weighted family, contained in Proposition~\ref{prop:thermal-family} below.
A related mechanism underlies Hastings' proof that correlations between odd observables decay exponentially at positive temperature with a decay length of order \(\beta\) \cite{Hastings04}: there the Fermi function enters through an integral representation of \(\langle AB\rangle_\beta\) whose time kernel decays at rate \(\pi/\beta\), and this is combined with a Lieb--Robinson bound for anticommutators.
\end{remark}

\paragraph{Families for general dynamics.}
For any interaction \(\Xi\in\mathfrak J_1\), we may employ the same filters to define families \((\mathsf F_u^\beta[\Xi])_{u\in\Gamma\times\{1,\ldots,2r\}}\) and \((\mathsf G_u^\beta[\Xi])_{u\in\Gamma\times\{1,\ldots,2r\}}\) associated to \(\tau^\Xi\).
The modified families are then given by
\begin{equation}\label{eq:modified-thermal-family}
 \begin{aligned}
 \mathsf F_u^\beta[\Xi]
 &=\frac1{\sqrt2}\sum_v(\Rb)_{uv}\Bigl[
       \gamma_v[f_\beta;\Xi]^{(1)}
       +\I\bigl(\ft(\gamma_v[f_\beta;\Xi]^*)\bigr)^{(2)}\Bigr],\\
 \mathsf G_u^\beta[\Xi]
 &=\sum_v\mathsf M_{uv}\Bigl[
       \gamma_v[q_\beta;\Xi]^{(1)}
       +\I\bigl(\ft(\gamma_v[q_\beta;\Xi]^*)\bigr)^{(2)}\Bigr].
 \end{aligned}
\end{equation}
We now summarize the properties of the annihilation families at nonzero coupling. The locality estimates in part~(ii) use the propagation bounds of~\cite{FermionicLR}, in particular the lossless bounds that allow us to retain the exponent \(p_\Phi\) in regime~\ref{ass:polynomial-weight}.

\begin{proposition}[The thermal annihilation families]\label{prop:thermal-family}
Let us assume Hypothesis~\ref{hyp:locality}. Consider the operator families $(\mathsf F_u^{\beta,\lambda}, \mathsf G_u^{\beta,\lambda})_{u\in\Gamma\times\{1,\dots,2r\}}$ defined in \eqref{eq:thermal-coordinates} and \eqref{eq:FG-coordinates}, respectively.

Then, their defining series converge absolutely in norm for every \(\lambda\in\R\), and for every \(\beta\) for which they were introduced. In particular, at zero coupling,
\begin{equation}\label{eq:normalized-constraints}
 \mathsf F_u^{\beta}[\Phi_0]=\mathsf F_u^{\beta,0},\qquad 
 \mathsf G_u^{\beta,0}=\mathsf F_u^{\beta,0},
\end{equation}
where \((\mathsf F_u^{\beta,0})_u\) is the free family defined in \eqref{eq:smooth-free-family}.

Finally, the following hold for any Majorana label $u\in\Gamma\times\{1,\dots,2r\}$.
\begin{enumerate}[label=(\roman*),leftmargin=*]
\item \emph{Annihilation.} For every \(\beta>0\), \(\Xi\in\mathfrak J_1\) and \(\rho\in\KMS_\beta(\tau^\Xi)\), the families from \eqref{eq:modified-thermal-family} satisfy
\[
 \widehat\rho((\mathsf F_u^\beta[\Xi])^*\mathsf F_u^\beta[\Xi])
 =\widehat\rho((\mathsf G_u^\beta[\Xi])^*\mathsf G_u^\beta[\Xi])=0.
\]
\item \emph{Local decompositions.}
\[
 \begin{aligned}
 \mathsf F_u^{\beta,\lambda}-\mathsf F_u^{\beta,0}
 &=\sum_{\Gamma\Supset Z\ni x_u}\Delta\mathsf F_{u,Z}^{\beta,\lambda},\\
 \mathsf G_u^{\beta,\lambda}-\mathsf F_u^{\beta,0}
 &=\sum_{\Gamma\Supset Z\ni x_u}\Delta\mathsf G_{u,Z}^{\beta,\lambda},
 \end{aligned}
 \qquad \Delta\mathsf F_{u,Z}^{\beta,\lambda},\Delta\mathsf G_{u,Z}^{\beta,\lambda}\in(\A_Z^\st)^-.
\]
Both series converge in norm.

\emph{Cutoff family.} There is a model constant \(C\ge1\) with the following property. Setting \(\theta:=\theta_{\rm in}/\sG\), there is a rate \(\mu>0\), depending only on the model data and on \(\theta\), such that, uniformly in \(\beta\ge0\) and \(|\lambda|\le1\), the local terms satisfy
\[
 \begin{aligned}
 \sup_z\sum_u\sum_{Z\ni z,x_u}|Z|\E^{\mu\diam(Z)^\theta}
       \|\Delta \mathsf F_{u,Z}^{\beta,\lambda}\|
 &\le C|\lambda|^{1/2}&&\text{in regime \ref{ass:exponential}},\\
 \sup_z\sum_u\sum_{Z\ni z,x_u}|Z|(1+\diam Z)^{p_\Phi}
       \|\Delta \mathsf F_{u,Z}^{\beta,\lambda}\|
 &\le C|\lambda|&&\text{in regime  \ref{ass:polynomial-weight}}.
 \end{aligned}
\]

\emph{Fermi-weighted family.} Put \(\eta:=\beta/(1+\beta)\). There are model constants \(C,M\ge1\) and a model rate \(\mu_{\rm FG}>0\), with the following property. For every \(\beta>0\) and for any $\lambda$ such that \(|\eta\lambda|\le1\), we have
\[
 \begin{aligned}
 \sup_z\sum_u\sum_{Z\ni z,x_u}|Z|\E^{\mu\diam(Z)^{\theta_{\rm in}}}
       \|\Delta \mathsf G_{u,Z}^{\beta,\lambda}\|
 &\le C(1+\beta)^M|\eta\lambda|^{1/2}&&\text{in regime \ref{ass:exponential}},\\
 \sup_z\sum_u\sum_{Z\ni z,x_u}|Z|(1+\diam Z)^{p_\Phi}
       \|\Delta \mathsf G_{u,Z}^{\beta,\lambda}\|
 &\le C(1+\beta)^M|\eta\lambda|&&\text{in regime \ref{ass:polynomial-weight}},
 \end{aligned}
\]
where \(\mu:=\mu_{\rm FG}/(1+\beta)\) in regime~\ref{ass:exponential}.
\end{enumerate}
\end{proposition}

Next, we trace the change in the annihilation families of Proposition~\ref{prop:thermal-family} when a (possibly infinite) defect is inserted.
For an admissible defect \(\Psi\) as in Definition~\ref{def:defect}, we put
\begin{equation*}
 \Xi:=\Phi_\lambda+\Psi.
\end{equation*}
Furthermore, for \(\Lambda\Subset\Gamma\), a subset \(B\subset\Lambda\) and an operator \(K=K^*\in\A_B^+\), we denote the associated finite-volume interaction by
\begin{equation}\label{eq:cut-interaction}
 \Xi_{\Lambda,K}(Z)
 :=\1_{\{Z\subset\Lambda\}}\Phi_\lambda(Z)
       +\1_{\{Z=B\}}K.
\end{equation}
 We then consider the perturbed operators \(\mathsf F_u^\beta[\Xi]\) and \(\mathsf G_u^\beta[\Xi]\) defined in \eqref{eq:modified-thermal-family} by replacing \(\tau^{\Phi_\lambda}\) with either \(\tau^{\Xi}= \tau^{\Phi_\lambda + \Psi}\) or \(\tau^{\Xi_{\Lambda,K}}\). In the next proposition, we bound their difference from \(\mathsf F_u^{\beta,\lambda}\) and \(\mathsf G_u^{\beta,\lambda}\), respectively, in terms of the perturbation strength.

\begin{proposition}[Thermal comparison under perturbations]\label{prop:thermal-comparison}
Assume the hypotheses of Proposition~\ref{prop:thermal-family}. In the cutoff case, we consider any \(\beta\ge0\) and any $\lambda\in\mathbb R$ such that \(|\lambda|\le1\). In the Fermi-weighted case, we take any \(\beta>0\); setting \(\eta:=\beta/(1+\beta)\), we consider any \(\lambda\) such that \(|\eta\lambda|\le1\). 
Then, for every Majorana label $u$ the following hold.
\begin{enumerate}[label=(\roman*),leftmargin=*]
\item \emph{Perturbation bound.} There exist model constants \(C,M\ge1\) such that, for every \(\Psi\in\mathfrak J_1\), the families defined for \(\Xi=\Phi_\lambda+\Psi\) satisfy
\begin{equation}\label{eq:thermal-family-perturbation}
 \begin{aligned}
 \sup_u\|\mathsf F_u^{\beta,\lambda}-\mathsf F_u^\beta[\Xi]\|
 &\le C J_\Psi,\\
 \sup_u\|\mathsf G_u^{\beta,\lambda}-\mathsf G_u^\beta[\Xi]\|
 &\le C(1+\beta)^M\eta J_\Psi.
 \end{aligned}
\end{equation}
\item \emph{Spatial comparison.} Let \(\Xi=\Phi_\lambda+\Psi\), with \(\Psi\in\mathfrak J_1\) and \(S=\Sigma(\Psi)\), or let \(\Xi=\Xi_{\Lambda,K}\), as defined in \eqref{eq:cut-interaction}, with \(S=\Lambda^c\cup B\).

\emph{Cutoff family.} In regime~\ref{ass:exponential}, we put \(\theta:=\theta_{\rm in}/\sG\). Then, there exist \(C,\mu>0\), depending only on the model data and on \(\theta\), such that the bound below holds.
\[
 \|\mathsf F_u^{\beta,\lambda}-\mathsf F_u^\beta[\Xi]\|
 \le C\begin{cases}
 \E^{-\mu d(x_u,S)^\theta}&\text{in regime \ref{ass:exponential}},\\
 (1+d(x_u,S))^{-p_\Phi}&\text{in regime  \ref{ass:polynomial-weight}}.
 \end{cases}
\]
In regime~\ref{ass:polynomial-weight}, \(C>0\) is a model constant.

\emph{Fermi-weighted family.} There exist model constants \(C,\bar\mu>0\), \(M\ge1\) such that
\[
 \|\mathsf G_u^{\beta,\lambda}-\mathsf G_u^\beta[\Xi]\|
 \le C\begin{cases}
 \log(3+\beta)\E^{-\mu d(x_u,S)^{\theta_{\rm in}}}&\text{in regime  \ref{ass:exponential}},\\
 (1+\beta)^M(1+d(x_u,S))^{-p_\Phi}&\text{in regime \ref{ass:polynomial-weight}},
 \end{cases}
\]
where \(\mu:=\bar\mu/(1+\beta)\) in regime~\ref{ass:exponential}.
\end{enumerate}
\end{proposition}

\subsection{Ground-state family and low-temperature estimates}
\label{sec:ground-construction}

Throughout this subsection we assume the free gap condition~\eqref{eq:free-band} with \(g>0\).

For \(\Xi\in\mathfrak J_1\), we use the reversed interaction \(\ft\Xi\) and the doubled interaction \(\Xi^\st\) introduced at the beginning of Section~\ref{sec:thermofield}. Since \(\|\Xi^\st\|_1\le2\|\Xi\|_1\), Proposition~\ref{prop:weak-dynamics} shows that $\Xi^\st$ generates a dynamics on \(\A^\st\). For any \(\Psi\in\mathfrak J_1\), we write \(\tau^{\st,\lambda,\Psi}:=\tau^{(\Phi_\lambda+\Psi)^\st}\) and \(\tau^{\st,\lambda}:=\tau^{\st,\lambda,0}\).

We use the filter \(f\) from Section~\ref{sec:thermal-family-statement}. For \(u\in\Gamma\times\{1,\ldots,2r\}\), the filtered operator \(\gamma_u[f;\Phi_\lambda]\) is odd and entire analytic. We define the \emph{ground-state family} by
\begin{equation}\label{eq:ground-filtered-family}
 \mathsf F_u^{\infty,\lambda}
 :=\frac1{\sqrt2}\Bigl[
       \gamma_u[f;\Phi_\lambda]^{(1)}
       +\I\bigl(\ft(\gamma_u[f;\Phi_\lambda]^*)\bigr)^{(2)}\Bigr].
\end{equation}
For \(\Xi\in\mathfrak J_1\), we define \(\mathsf F_u^\infty[\Xi]\) by replacing \(\Phi_\lambda\) with \(\Xi\) in \eqref{eq:ground-filtered-family}. Since \(\widehat f\) is real, Lemma~\ref{lem:reversed-interactions} gives the equivalent expression
\begin{equation}\label{eq:ground-doubled-family}
 \mathsf F_u^\infty[\Xi]
 =\frac1{\sqrt2}\int_\R f(t)\tau_t^{\Xi^\st}(\gamma_{u,1}+\I\gamma_{u,2})\,\dd t.
\end{equation}

At \(\lambda=0\), Equation~\eqref{eq:quadratic-Majorana-interaction} gives \(\ft\Phi_h=\Phi_{-h}\), so \(\Phi_0^\st=\Phi_h^{(1)}+\Phi_{-h}^{(2)}\). Inserting the free dynamics~\eqref{eq:Majorana-free-dynamics} into~the preceding expression, we obtain, for \(u\in\Gamma\times\{1,\ldots,2r\}\), 
\begin{equation}\label{eq:ground-free-family}
 \begin{aligned}
 \mathsf F_u^{\infty,0}
 &=\frac1{\sqrt2}\sum_{v\in\Gamma\times\{1,\ldots,2r\}}
 \bigl(\widehat f(-h)_{uv}\gamma_{v,1}+\I\widehat f(h)_{uv}\gamma_{v,2}\bigr)\\
 &=\frac1{\sqrt2}\sum_{v\in\Gamma\times\{1,\ldots,2r\}}
 \bigl((P_+)_{uv}\gamma_{v,1}+\I(P_-)_{uv}\gamma_{v,2}\bigr),
 \end{aligned}
\end{equation}
where we have used that \(\widehat f(-h)=P_+\) and \(\widehat f(h)=P_-\) by the choice of \(\widehat f\) in Section~\ref{sec:thermal-family-statement}. This is the free annihilation family used in \cite[Section~5]{LPPLV3}. In the following lemma we collect some properties of $\mathsf F_u^{\infty,0}$.

\begin{lemma}\label{lem:ground-free-family}
Assume Hypothesis~\ref{hyp:locality} and the free gap condition~\eqref{eq:free-band} with \(g>0\).
For every \(u\in\Gamma\times\{1,\ldots,2r\}\), the series in Equation~\eqref{eq:ground-free-family} converges absolutely in operator norm. The coefficients of the family \((\mathsf F_u^{\infty,0})_{u\in\Gamma\times\{1,\ldots,2r\}}\)
\[
 \sa_{u,(v,1)}=\sqrt2(P_+)_{uv},
 \qquad
 \sa_{u,(v,2)}=\sqrt2\I(P_-)_{uv}
\]
satisfy the Parseval condition of Definition~\ref{def:parseval-family}. There are model constants \(C_0\ge1\) and, in regime~\ref{ass:exponential}, \(\mu_1>0\), such that
\begin{equation}\label{eq:ground-free-coefficients}
 \ref{ass:exponential}:\quad
 \bigl\|(|\sa_{u,(v,j)}|)\bigr\|^{\rm Sch}_{\mu_1,\theta_{\rm in}}
 \le C_0,
 \qquad
 \ref{ass:polynomial-weight}:\quad
 \bigl\|(|\sa_{u,(v,j)}|)\bigr\|^{\rm Sch,pol}_{p_h}
 \le C_0,
\end{equation}
where \(u,v\in\Gamma\times\{1,\ldots,2r\}\) and \(j\in\{1,2\}\).
\end{lemma}

For \(\Upsilon,\Psi\in\mathfrak J_1\), \(\beta>0\), \(\omega\in\GS(\tau^{\Phi_\lambda+\Upsilon})\) and \(\rho\in\KMS_\beta(\tau^{\Phi_\lambda+\Psi})\) we use the product ground-state and product KMS extensions introduced at the beginning of Section~\ref{sec:thermofield}:
\begin{equation}\label{eq:stacked-states}
 \begin{aligned}
 \omega_\st=\omega\widehat\otimes(\omega^{\rm ev}\circ\ft),\qquad
 \rho_{\beta,\st}=\rho\widehat\otimes(\rho\circ\ft).
 \end{aligned}
\end{equation}
Parity invariance and convexity preserve the ground-state condition under parity averaging, while \(\rho\) is even by Proposition~\ref{prop:kms-nonempty}. Thus the auxiliary states are even, so the products exist by Lemma~\ref{lem:graded-product}. Lemmas~\ref{lem:reversed-interactions}, \ref{lem:graded-ground} and~\ref{lem:graded-KMS} give \(\omega_\st\in\GS(\tau^{\Phi^\st_\lambda+\Upsilon^\st})\) and \(\rho_{\beta,\st}\in\KMS_\beta(\tau^{\st,\lambda,\Psi})\).

\begin{proposition}[The ground-state and low-temperature annihilation family]\label{prop:ground-family}
Assume Hypothesis~\ref{hyp:locality} and the free gap condition~\eqref{eq:free-band} with \(g>0\). For every \(|\lambda|\le1\) and every \(u\in\Gamma\times\{1,\ldots,2r\}\), the following hold.
\begin{enumerate}[label=(\roman*),leftmargin=*]
\item \emph{Annihilation and low-temperature bound.} For every \(\Upsilon\in\mathfrak J_1\) and \(\omega\in\GS(\tau^{\Phi_\lambda+\Upsilon})\), let \((\pi,\mathcal H,\Omega)\) be the GNS representation of \(\omega_\st\), with GNS Hamiltonian \(L\) implementing \(\tau^{\st,\lambda,\Upsilon}\). Then
\[
 \pi(\mathsf F_u^{\infty}[\Phi_\lambda+\Upsilon])\1_{[0,g_*]}(L)=0.
\]
In particular, \(\omega_\st\bigl((\mathsf F_u^{\infty}[\Phi_\lambda+\Upsilon])^*\mathsf F_u^{\infty}[\Phi_\lambda+\Upsilon]\bigr)=0\).

There is a model constant \(C>0\) such that, for every \(\beta>0\), \(\Psi\in\mathfrak J_1\), \(\rho\in\KMS_\beta(\tau^{\Phi_\lambda+\Psi})\) and \(E,G\in\A^\st\),
\begin{equation}\label{eq:ground-family-kms}
 |\rho_{\beta,\st}((\mathsf F_u^{\infty}[\Phi_\lambda+\Psi])^*E)|
 +|\rho_{\beta,\st}(G\mathsf F_u^{\infty}[\Phi_\lambda+\Psi])|
 \le C\E^{-\beta g_*}\max\{\|E\|,\|G\|\}.
\end{equation}
\item \emph{Local decompositions.} In regime~\ref{ass:exponential}, set \(\theta:=\theta_{\rm in}/\sG\). In regime~\ref{ass:polynomial-weight}, choose \(\sG=2\). There are a model constant \(C_{\rm gs}\ge1\) and, in regime~\ref{ass:exponential}, a rate \(\mu>0\), depending only on the model data and \(\theta\), with the following property. For all \(u\in\Gamma\times\{1,\ldots,2r\}\) and \(Z\Subset\Gamma\) with \(x_u\in Z\), there exist operators \(\Delta \mathsf F_{u,Z}^{\infty,\lambda}\in(\A_Z^\st)^-\) such that
\[
 \mathsf F_u^{\infty,\lambda}-\mathsf F_u^{\infty,0}=\sum_{\substack{Z\Subset\Gamma\\x_u\in Z}}\Delta \mathsf F_{u,Z}^{\infty,\lambda},
\]
where the sum converges in norm. Moreover, it holds that
\begin{equation}\label{eq:ground-family-correction}
 \begin{aligned}
 \sup_{z\in\Gamma}\sum_{u\in\Gamma\times\{1,\ldots,2r\}}
 \sum_{\substack{Z\Subset\Gamma\\z,x_u\in Z}}|Z|\E^{\mu\diam(Z)^\theta}\|\Delta \mathsf F_{u,Z}^{\infty,\lambda}\|
 &\le C_{\rm gs}|\lambda|^{1/2}&&\text{in regime~\ref{ass:exponential}},\\
 \sup_{z\in\Gamma}\sum_{u\in\Gamma\times\{1,\ldots,2r\}}
 \sum_{\substack{Z\Subset\Gamma\\z,x_u\in Z}}|Z|(1+\diam Z)^{p_\Phi}\|\Delta \mathsf F_{u,Z}^{\infty,\lambda}\|
 &\le C_{\rm gs}|\lambda|&&\text{in regime~\ref{ass:polynomial-weight}}.
 \end{aligned}
\end{equation}

\end{enumerate}
\end{proposition}

\begin{proposition}[Ground-state comparison under defects]\label{prop:ground-comparison}
Assume Hypothesis~\ref{hyp:locality} and the free gap condition~\eqref{eq:free-band} with \(g>0\). In regime~\ref{ass:exponential}, set \(\theta:=\theta_{\rm in}/\sG\). In regime~\ref{ass:polynomial-weight}, choose \(\sG=2\).

There exist constants \(C>0\) and, in regime~\ref{ass:exponential}, \(\mu>0\), depending only on the model data and, in regime~\ref{ass:exponential}, on \(\theta\). Set \(W(R):=\exp(\mu R^\theta)\) in regime~\ref{ass:exponential} and \(W(R):=(1+R)^{p_\Phi}\) in regime~\ref{ass:polynomial-weight}. Then, for every \(|\lambda|\le1\), \(\Psi\in\mathfrak J_1\) and \(u\in\Gamma\times\{1,\ldots,2r\}\),
\begin{equation}\label{eq:ground-family-defect}
 \|\mathsf F_u^{\infty,\lambda}-\mathsf F_u^{\infty}[\Phi_\lambda+\Psi]\|
 \le C/W(d(x_u,\Sigma(\Psi))).
\end{equation}
\end{proposition}

\begin{remark}[Zero-temperature limits]\label{rem:zero-temperature-extensions}
Assume Hypothesis~\ref{hyp:locality} and the free gap condition~\eqref{eq:free-band}.
Fix \(\lambda\in\R\) and \(\Psi\in\mathfrak J_1\), put \(\Xi=\Phi_\lambda+\Psi\), and let \(\mathsf F_u^\beta[\Xi]\) be the cutoff family associated with \(\tau^\Xi\) in \eqref{eq:modified-thermal-family}.
For every \(u\in\Gamma\times\{1,\ldots,2r\}\),
\begin{equation}\label{eq:cutoff-zero-temperature-limit}
 \lim_{\beta\to\infty}
 \|\mathsf F_u^\beta[\Xi]-\mathsf F_u^\infty[\Xi]\|=0.
\end{equation}
The two estimates needed for this limit are
\[
 \sup_u\sum_v|(\Rb)_{uv}-\delta_{uv}|\longrightarrow0,
 \qquad
 \|f_\beta-f\|_{L^1(\R)}=\|f(\,\cdot\,+\I\beta/2)\|_{L^1(\R)}\longrightarrow0.
\]
For the first estimate, use the contours \(\mathscr C_\pm\) from the proof of Lemma~\ref{lem:uniform-free-calculus}: the scalar functions \((1+\E^{\mp\beta z})^{-1/2}-1\) tend uniformly to zero there, and Lemma~\ref{lem:weighted-contour-calculus} bounds the Schur norms of the resolvents on these contours.
For the second estimate, the Fourier representation \eqref{eq:thermal-filter-definition} and \(\supp\widehat f\subset[-2C_h,-7g/8]\) give \(f(t+\I\beta/2)\to0\) for each \(t\); the integrable bound \eqref{eq:Gevrey-tail} permits dominated convergence. Comparing the definition \eqref{eq:ground-filtered-family} with \eqref{eq:modified-thermal-family}, and using the uniform row bound \eqref{eq:thermal-absolute-row-sums}, gives a constant \(C\) independent of \(\beta,\lambda,\Psi\) such that
\[
 \sup_u\|\mathsf F_u^\beta[\Xi]-\mathsf F_u^\infty[\Xi]\|
 \le C\left(\sup_u\sum_v|(\Rb)_{uv}-\delta_{uv}|
       +\|f(\,\cdot\,+\I\beta/2)\|_{L^1(\R)}\right)
 \longrightarrow0.
\]

Under the hypotheses of Theorem~\ref{thm:uniqueness}(i), fix \(|\lambda|\le\lambda_{\mathrm g}\) and use the cutoff chosen in Section~\ref{sec:gapped-choices}.
Uniqueness of the physical ground state makes \(\omega_\lambda\) even, since \(\omega_\lambda\circ\Theta\) is also a ground state.
Both extensions then have the same zero-temperature limit:
\begin{equation}\label{eq:extensions-zero-temperature-limit}
 \begin{aligned}
 \widehat\omega_{\beta,\lambda}
 &\xrightarrow[\beta\to\infty]{\mathrm{weak}\text{-}*}
 \omega_{\lambda,\st}
 :=\omega_\lambda\widehat\otimes(\omega_\lambda\circ\ft),\\
 \omega_{\beta,\lambda}\widehat\otimes
       (\omega_{\beta,\lambda}\circ\ft)
 &\xrightarrow[\beta\to\infty]{\mathrm{weak}\text{-}*}
 \omega_{\lambda,\st}.
 \end{aligned}
\end{equation}
Theorem~\ref{thm:uniqueness}(i) gives \(\omega_{\beta,\lambda}\to\omega_\lambda\) weak-\(*\), and factorization gives the second limit on products \(A^{(1)}B^{(2)}\), whose linear span is norm dense in \(\A^\st\).
For any weak-\(*\) accumulation point \(\chi\) of \(\widehat\omega_{\beta,\lambda}\), Equation~\eqref{eq:cutoff-zero-temperature-limit} and Proposition~\ref{prop:thermal-family}(i) give
\[
 \chi\bigl((\mathsf F_u^{\infty,\lambda})^*\mathsf F_u^{\infty,\lambda}\bigr)=0
 \qquad\text{for every }u\in\Gamma\times\{1,\ldots,2r\}.
\]
Proposition~\ref{prop:ground-family}(i) gives the same identities for \(\omega_{\lambda,\st}\).
Lemma~\ref{lem:gapped-hypotheses} and Theorem~\ref{thm:comparison}(i) therefore imply \(\chi=\omega_{\lambda,\st}\), proving the first limit in \eqref{eq:extensions-zero-temperature-limit}.
\end{remark}

\subsection{Extension to weakly interacting spin systems}
\label{sec:spin-systems}

The strategy of this paper also applies to weakly interacting quantum spin systems.
Nevertheless, the adaptation is not merely notational: one first has to find the spin counterparts of the objects on which the fermionic proofs rest.
Site oscillations replace the Majorana derivatives, and the reset identity for an on-site reference family described below replaces the Parseval condition.
The direct analogue of the fermionic annihilation family does not suffice; the thermal annihilation operators are instead sums of products \(G\,\mathsf D\), where \(\mathsf D\) is an annihilator built from filtered on-site matrix units and \(G\) is an on-site coefficient.
Once this dictionary is available, the proofs follow the same strategy as in the fermionic case.
Since the fermionic case is the more demanding one, we do not write out the spin proofs.
We give the dictionary, verify the identities and estimates behind it, and describe the resulting statements at the end of the subsection.
For spin-\(\tfrac12\) systems near infinite temperature, the uniqueness part of this strategy is the method of Matsui \cite{Matsui94}: the extension used below is his purification \cite[Definition~1]{Matsui94}, and the identity \eqref{eq:spin-annihilation} is his equation~(39).
The dictionary adapts his construction to a reference system of independent sites at arbitrary temperature and to the comparison theorem.

\paragraph{Setting.}
Let \(\mathcal B\) be the quasi-local algebra over \(\Gamma\) with site algebras \(M_q(\C)\), \(q\ge2\), and local algebras \(\mathcal B_Z\).
Interactions, dynamics and KMS states on \(\mathcal B\) are defined as in Section~\ref{sec:setting}, without parity conditions.
Let \(\mathcal B^\st:=\mathcal B\otimes\mathcal B\), with \(\mathcal B^\st_Z:=\mathcal B_Z\otimes\mathcal B_Z\), \(A^{(1)}:=A\otimes\1\) and \(B^{(2)}:=\1\otimes B\).
Let \(h_x=h_x^*\in M_q(\C)\) act at the site \(x\), with \(\|h_x\|\le C_h\), orthonormal eigenbasis \((\psi_{x,j})_{j=0}^{q-1}\) and eigenvalues \(E_{x,j}\), indexed so that \(E_{x,0}=\min_j E_{x,j}\).
Set \(\Phi_0(\{x\}):=h_x\), \(\Phi_0(Z):=0\) for \(|Z|\ne1\), and \(\Phi_\lambda:=\Phi_0+\lambda\cV\), where \(\cV\) satisfies the interaction-norm bound of either regime in Hypothesis~\ref{hyp:locality}.
The analogue of the free gap condition \eqref{eq:free-band} is a simple on-site ground state with a uniform gap \(g>0\),
\begin{equation}\label{eq:spin-gap}
 E_{x,0}=0,\qquad E_{x,j}\ge g\quad(1\le j<q,\ x\in\Gamma).
\end{equation}

\paragraph{Thermal extension.}
Let \(B\mapsto B^\top\) and \(B\mapsto\overline B:=(B^*)^\top\) be the transposition and the complex conjugation on \(\mathcal B\) in the product of these eigenbases.
For an interaction \(\Xi\) on \(\mathcal B\) and \(\rho\in\KMS_\beta(\tau^\Xi)\) with GNS triple \((\pi,\mathcal H,\Omega)\), the construction of the right action \(R\) in Section~\ref{sec:canonical-extension} uses only the KMS condition.
Since the two tensor factors commute, no parity twist is needed: \(\widehat\pi(A^{(1)}B^{(2)}):=\pi(A)R(B^\top)\) is a representation of \(\mathcal B^\st\), its vector state \(\widehat\rho\) at \(\Omega\) is the thermal extension of \(\rho\), and \eqref{eq:annihilation-identities} becomes
\begin{equation}\label{eq:spin-annihilation}
 \widehat\pi\bigl(A^{(1)}-(\tau^\Xi_{-\I\beta/2}(A)^\top)^{(2)}\bigr)\Omega=0,
 \qquad A\in\mathcal B_\an(\tau^\Xi).
\end{equation}
At zero coupling, \(\widehat\rho\) is the product of the vector states of the on-site purifications \(v_x:=\sum_jp_{x,j}^{1/2}\,\psi_{x,j}\otimes\psi_{x,j}\) of Section~\ref{sec:tfd-finite}, where \(p_{x,j}\) are the Gibbs weights of \(h_x\); see also \cite{Matsui94}.

\paragraph{Reset family and site oscillations.}
Let \(P_x\) be the projection onto \(\C v_x\), complete \(v_{x,0}:=v_x\) to an orthonormal basis \((v_{x,a})_{a=0}^{q^2-1}\) of \(\C^q\otimes\C^q\), and define operators in \(\mathcal B^\st_{\{x\}}\) by
\[
 F_{x,a}^0:=\tfrac1{\sqrt2}|v_{x,0}\rangle\langle v_{x,a}|
 \quad(1\le a<q^2),
 \qquad
 F_{x,0}^0:=\tfrac1{\sqrt2}(\1-P_x).
\]
They annihilate \(v_x\).
For a family \((F_{y,a})\) we use the auxiliary generator of Section~\ref{sec:roadmap-comparison} with commutators in place of the graded derivatives, and group its terms by sites:
\[
 \cL:=\sum_{y\in\Gamma}\cL_y,
 \qquad
 \cL_y(A):=-\sum_{a=0}^{q^2-1}
 \bigl(F_{y,a}^*[F_{y,a},A]+[A,F_{y,a}^*]F_{y,a}\bigr).
\]
For the reference family \((F^0_{y,a})\), a direct computation gives the \emph{reset identity}
\[
 \cL_y^0=\mathbb E_y-\id=:-Q_y,
\]
where \(\mathbb E_y\) is the conditional expectation onto \(\mathcal B^\st_{\Gamma\setminus\{y\}}\) determined by \(\mathbb E_y(BC)=\langle v_y,Bv_y\rangle\,C\) for \(B\in\mathcal B^\st_{\{y\}}\) and \(C\in\mathcal B^\st_{\Gamma\setminus\{y\}}\).
The maps \(Q_y\) are commuting idempotents.
The reset identity takes the place of the Parseval condition in the proof of Theorem~\ref{thm:comparison}, and the site oscillations \(\operatorname{osc}_x(A):=\|Q_xA\|\) take the place of the norms of the Majorana derivatives.
For finite \(U\subset\Gamma\), telescoping the product \(\mathbb E_U:=\prod_{z\in U}\mathbb E_z\) and using \([B,\mathbb E_U(A)]=0\) for \(B\in\mathcal B^\st_U\) gives
\begin{equation}\label{eq:spin-osc}
 \|A-\mathbb E_U(A)\|\le\sum_{z\in U}\operatorname{osc}_z(A),
 \qquad
 \|[B,A]\|\le2\|B\|\sum_{z\in U}\operatorname{osc}_z(A)
 \quad(B\in\mathcal B^\st_U).
\end{equation}
This part uses only that the \(v_x\) are unit vectors.

\paragraph{Damping of oscillations.}
Consider operators \(F_{y,a}=F^0_{y,a}+\sum_{Z\ni y}\Delta F_{(y,a),Z}\) with corrections \(\Delta F_{(y,a),Z}\in\mathcal B^\st_Z\) that satisfy the weighted bound of Hypothesis~\ref{hyp:perturbed-family}(ii) with constant \(\varepsilon\).
Then the series defining \(\cL(A)\) converges absolutely for local \(A\).
To reconstruct local observables as in Section~\ref{sec:roadmap-comparison}, we need decay of \(\operatorname{osc}_x(T_tA)\) along the auxiliary evolution \(T_t\) generated by \(\cL\).
Let \(\Delta\cL_y:=\cL_y+Q_y\) be the deviation of \(\cL_y\) from its reference value, and set \(\cL^{[x]}:=\sum_{y\ne x}\cL_y\).
Since the \(Q_y\) are commuting idempotents,
\begin{equation}\label{eq:spin-damping}
 Q_x\cL=(\cL^{[x]}-\id)Q_x+\mathcal R_x,
 \qquad
 \mathcal R_x:=Q_x\,\Delta\cL_x+\sum_{y\ne x}[Q_x,\Delta\cL_y].
\end{equation}
This replaces the identity \(\partial_v\cL(E)=\cL(\partial_vE)-\partial_vE+\mathcal R_v(E)\) of Section~\ref{sec:roadmap-comparison}, and \(\mathcal R_x=0\) for the reference family.
Expanding \(\Delta\cL_y\) gives terms that are linear or quadratic in the corrections.
A term containing corrections with supports \(Z\), or \(Z\) and \(Z'\), involves only operators in \(\mathcal B^\st_U\), where \(U=Z\), respectively \(U=Z\cup Z'\).
It commutes with \(Q_x\) if \(x\notin U\), and it vanishes on \(\mathcal B^\st_{\Gamma\setminus U}\), so that \(A\) may be replaced by \(A-\mathbb E_U(A)\) and \eqref{eq:spin-osc} applies.
Since \(|U|\le|Z|+|Z'|\) and \(W(\diam U)\le W(\diam Z)W(\diam Z')\), summing these contributions with Hypothesis~\ref{hyp:perturbed-family}(ii) gives a nonnegative kernel \((\mathsf K_{xz})_{x,z\in\Gamma}\) with
\[
 \|\mathcal R_x(A)\|\le\sum_{z\in\Gamma}\mathsf K_{xz}\operatorname{osc}_z(A),
 \qquad
 \|\mathsf K\|^{\rm Sch}_W\le C(\varepsilon+\varepsilon^2),
\]
where \(C\) is an absolute constant.
The generator \(\cL^{[x]}\) has the same form as \(\cL\) and generates a contraction semigroup, as in Lemma~\ref{lem:auxiliary-evolution}.
If \(\|\mathsf K^m\|^{\rm Sch}_W<1\) for some \(m\ge1\), variation of constants in \eqref{eq:spin-damping} then gives \eqref{eq:roadmap-majorana-derivative-bound} with site oscillations in place of Majorana derivatives, and the remaining steps of Section~\ref{sec:roadmap-comparison} yield the spin analogue of Theorem~\ref{thm:comparison}.

For spin-\(\tfrac12\) systems and \(\beta=0\), where \(v_x\) is the maximally entangled vector, \(\cL_y^0\) is the free motion of \cite[equation~(47)]{Matsui94}, and the scheme above is that of \cite[Proposition~2]{Matsui94}: his hypothesis~(54) is a bound of the same form as the bound on \(\mathcal R_x\) above, with his differentials in place of site oscillations and a nonnegative matrix in place of \(\mathsf K\).
He sums the resulting inequalities over the sites and obtains exponential decay of the total oscillation when the column sums of this matrix are smaller than one; here the matrix inequality is kept and integrated in time, which gives the spatial kernel of the comparison inequality.
Bertini, De Sole, Posta and Presilla use the same scheme for perturbations of independent on-site dissipative evolutions that are self-adjoint with respect to a full-rank  stationary state, and obtain a unique stationary state with exponential relaxation of local observables~\cite[Theorems~2.2 and~2.3]{BDSP25}.
The stationary state \(P_x\) of our on-site reset has rank one, so their result does not apply directly.

\paragraph{Thermal annihilation family.}
To apply this comparison to KMS states, we need operators that annihilate the GNS vectors of their thermal extensions and reduce to \(F^0_{x,a}\) at zero coupling.
By \eqref{eq:spin-annihilation}, every operator of the form \(D=A^{(1)}-(\tau^\Xi_{-\I\beta/2}(A)^\top)^{(2)}\) satisfies \(\widehat\pi(D)\Omega=0\).
Hence so does every left multiple \(GD\) with \(G\in\mathcal B^\st\), because \(\widehat\pi(GD)\Omega=\widehat\pi(G)\widehat\pi(D)\Omega\), and so does every finite sum of such products.
Unlike in the fermionic case, the operators \(D\) alone do not suffice.
At \(\beta=0\) and on a single site, they act in the Hilbert--Schmidt representation of Section~\ref{sec:tfd-finite} as commutators \(X\mapsto AX-XA\).
These vanish on \(v_x=\1/\sqrt q\) and have traceless range, so the auxiliary evolution they generate fixes \(P_x\), whereas the reset evolution sends \(P_x\) to \(\1\) as \(t\to\infty\).
We therefore look for the thermal family in the form \(\sum_\ell G_\ell D_\ell\), with operators \(D_\ell\) of this form and coefficients \(G_\ell\in\mathcal B^\st_{\{x\}}\) chosen so that the sum equals \(F^0_{x,a}\) at zero coupling.

Fix \(\beta>0\), let \(A_{x,jk}:=|\psi_{x,j}\rangle\langle\psi_{x,k}|\), so that \(\tau_t^{\Phi_0}(A_{x,jk})=\E^{\I t\nu_{x,jk}}A_{x,jk}\) with \(\nu_{x,jk}:=E_{x,j}-E_{x,k}\), and let \(\mathcal P:=\{(0,j),(j,0):1\le j<q\}\).
For arbitrary on-site spectra we use the Fermi-weighted filter \(\varphi_\beta=q_\beta\) of \eqref{eq:FG-kernel}.
Under \eqref{eq:spin-gap}, the transitions in \(\mathcal P\) satisfy \(g\le|\nu_{x,jk}|\le C_h\), and we may instead use the cutoff filter \(\varphi_\beta=f_\beta\) of \eqref{eq:filtered-Majoranas}, whose time-decay bounds are uniform in \(\beta\).
For \(A\in\mathcal B\) let \(A[\varphi_\beta;\Xi]:=\int_\R\varphi_\beta(t)\tau_t^\Xi(A)\,\dd t\), as in \eqref{eq:filtered-Majorana-definition}.
As in \eqref{eq:CD} and \eqref{eq:FG-fields}, \(\tau^\Xi_{-\I\beta/2}(A[\varphi_\beta;\Xi])=(A^*[\varphi_\beta;\Xi])^*\), so by \eqref{eq:spin-annihilation} the operators
\[
 \mathsf D^\beta_{x,jk}[\Xi]
 :=\frac{a_\beta(-\nu_{x,jk})}{\widehat\varphi_\beta(\nu_{x,jk})}
   \Bigl[A_{x,jk}[\varphi_\beta;\Xi]^{(1)}
       -\bigl(\overline{A_{x,kj}[\varphi_\beta;\Xi]}\bigr)^{(2)}\Bigr],
 \qquad(j,k)\in\mathcal P,
\]
with \(a_\beta\) from \eqref{eq:free_normalization}, annihilate the GNS vector of the thermal extension of every \(\rho\in\KMS_\beta(\tau^\Xi)\).
The prefactor equals \(\sqrt2\,M_\beta(\nu_{x,jk})\), with \(M_\beta\) from \eqref{eq:FG-normalizer}, for \(\varphi_\beta=q_\beta\), and \(a_\beta(|\nu_{x,jk}|)\) for \(\varphi_\beta=f_\beta\); it is bounded uniformly in \(\beta\) and \(x\), since \(|\nu_{x,jk}|\le2C_h\).
For the free dynamics, \(A_{x,jk}[\varphi_\beta;\Phi_0]=\widehat\varphi_\beta(\nu_{x,jk})A_{x,jk}\), so both filters give the same on-site operator
\[
 \mathsf D^\beta_{x,jk}[\Phi_0]=\mathsf D_{x,jk}
 :=a_\beta(-\nu_{x,jk})A_{x,jk}^{(1)}-a_\beta(\nu_{x,jk})A_{x,kj}^{(2)},
\]
a scalar multiple of the operator \(\mathsf D_{jk}\) of Section~\ref{sec:tfd-finite} for \(H=h_x\).

It remains to write each reset operator as a combination of these on-site annihilators with coefficients on the left,
\begin{equation}\label{eq:spin-left-multiples}
 F^0_{x,a}=\sum_{(j,k)\in\mathcal P}G_{x,a,jk}\,\mathsf D_{x,jk},
 \qquad G_{x,a,jk}\in\mathcal B^\st_{\{x\}},
\end{equation}
with norms \(\|G_{x,a,jk}\|\) bounded uniformly in \(\beta\) and \(x\).
Such coefficients exist because \(M_x:=\sum_{(j,k)\in\mathcal P}\mathsf D_{x,jk}^*\mathsf D_{x,jk}\) satisfies
\begin{equation}\label{eq:spin-parent}
 M_x\ge\1-P_x
\end{equation}
for every \(\beta\ge0\), where the on-site formulas for \(v_x\), \(\mathsf D_{x,jk}\) and \(M_x\) are extended to \(\beta=0\).
To see this, fix \(x\), write \(p_j=p_{x,j}\) and \(|mn\rangle=\psi_{x,m}\otimes\psi_{x,n}\), and note that \(p_0\ge p_j\) and \(a_\beta(-\nu_{x,jk})^2=p_j/(p_j+p_k)\).
The operator \(M_x\) leaves the span of the vectors \(|jj\rangle\) invariant, and for \(z=\sum_jz_j|jj\rangle\) the two transitions between \(0\) and each \(j>0\) give
\[
 \langle z,M_x z\rangle
 =\sum_{j>0}\frac{2p_0}{p_0+p_j}
        \left|z_j-\sqrt{p_j/p_0}\,z_0\right|^2
 \ge \left\|z-\frac{z_0}{\sqrt{p_0}}v_x\right\|^2
 \ge \|(\1-P_x)z\|^2 .
\]
For \(m\ne n\), \(M_x|mn\rangle=(w_m+w_n)|mn\rangle\) with \(w_j=p_0/(p_0+p_j)\) for \(j>0\) and \(w_0=\sum_{j>0}p_j/(p_0+p_j)\), and \(w_m+w_n\ge1\) because \(w_j\ge\tfrac12\) and \(w_0+w_j\ge1\) for every \(j>0\).

Since \(M_xv_x=0\), \eqref{eq:spin-parent} shows that the inverse \(M_x^+\) of \(M_x\) on \(v_x^\perp\), extended by zero, satisfies \(\|M_x^+\|\le1\) and \(M_x^+M_x=\1-P_x\).
As \(F^0_{x,a}P_x=0\), we get \(F^0_{x,a}=F^0_{x,a}M_x^+M_x\), which is \eqref{eq:spin-left-multiples} with
\[
 G_{x,a,jk}:=F^0_{x,a}M_x^+\mathsf D_{x,jk}^*,
 \qquad
 \|G_{x,a,jk}\|\le1.
\]
The norm bound holds since \(\mathsf D_{x,jk}^*\mathsf D_{x,jk}\le M_x\) implies \(\|M_x^+\mathsf D_{x,jk}^*\|^2\le\|M_x^+\|\).
Replacing each on-site annihilator in \eqref{eq:spin-left-multiples} by its interacting counterpart defines
\[
 \mathsf F^\beta_{x,a}[\Xi]
 :=\sum_{(j,k)\in\mathcal P}G_{x,a,jk}\,\mathsf D^\beta_{x,jk}[\Xi].
\]
Each \(\mathsf F^\beta_{x,a}[\Xi]\) is a sum of left multiples of annihilators and therefore annihilates the GNS vector of every thermal extension for \(\tau^\Xi\), and \(\mathsf F^\beta_{x,a}[\Phi_0]=F^0_{x,a}\) by \eqref{eq:spin-left-multiples}.

\paragraph{Resulting statements and previous work.}
Two locality estimates are needed to apply the spin comparison to KMS states: the weighted bound of Hypothesis~\ref{hyp:perturbed-family}(ii) for the local terms of \(\mathsf F^\beta_{x,a}[\Phi_\lambda]-F^0_{x,a}\), and a bound on \(\|\mathsf F^\beta_{x,a}[\Xi]-\mathsf F^\beta_{x,a}[\Phi_\lambda]\|\) in terms of the distance from \(x\) to a defect or a finite-volume boundary.
They follow by the arguments of Lemmas~\ref{lem:normalized-field-differences} and~\ref{lem:defect-source}.
Indeed, \(\Xi\) enters \(\mathsf F^\beta_{x,a}[\Xi]\) only through the filtered matrix units \(A_{x,jk}[\varphi_\beta;\Xi]\), the propagation bounds used in both proofs apply to spin observables~\cite[Remark~6.6]{FermionicLR}, and left multiplication by the on-site contractions \(G_{x,a,jk}\) neither enlarges a support \(Z\ni x\) nor increases a norm.
Since \(\Phi_0\) has only on-site terms, the polynomial weight has exponent \(p_{\cV}\).
As in Lemma~\ref{lem:defect-source}, only bulk interaction terms enter the second estimate, which is therefore independent of the strength of the defect or boundary term.

With these estimates, the arguments of Section~\ref{sec:main-proofs} give the spin analogues of our positive-temperature results: uniqueness of the \(\beta\)-KMS state, clustering with a prefactor linear in the smaller support size, strong LPPL for every defect KMS state, and insensitivity of bulk expectations to boundary terms in finite systems.
Under \eqref{eq:spin-gap} and with the filter \(f_\beta\), the coupling interval and all bounds are uniform in \(\beta>0\).
For arbitrary on-site spectra they depend on \(\beta\), as in parts~(ii) of Theorems~\ref{thm:uniqueness}, \ref{thm:kms-lppl} and~\ref{thm:correlations}.

For weak finite-range perturbations of on-site Hamiltonians satisfying \eqref{eq:spin-gap}, Yarotsky proved that the influence of boundary conditions on ground-state expectations decays exponentially with the distance to the boundary, and deduced uniqueness of the infinite-volume ground state~\cite{Yarotsky05}.
Henheik, Teufel and Wessel used this estimate to prove strong LPPL for ground states, with a bound independent of the strength of the local perturbation even when it closes the spectral gap~\cite{HTW22}.
At positive temperature, Adhikari, Henheik, Lemm and Wessel proved exponential clustering and local indistinguishability uniformly in \(\beta\) for finite-range perturbations~\cite[Theorems~2.2--2.3]{AHLW26}.
Their clustering prefactor grows exponentially in the observable support sizes, and their LPPL bound for a local perturbation \(W\) contains the factor \(\E^{2\beta\|W\|}\)~\cite[Theorem~4.1]{AHLW26}.
The statements above are free of both dependences and include uniqueness among all infinite-volume \(\beta\)-KMS states, as conjectured in~\cite[after Corollary~2.5]{AHLW26}.

\section{Proofs of the results in Section~\ref{sec:results}}
\label{sec:main-proofs}

We now prove the results of Section~\ref{sec:results} using the comparison theorem of Section~\ref{sec:comparison} and the families on the doubled algebra constructed in Section~\ref{sec:thermofield}. We first choose constants and spatial weights and verify the hypotheses of the comparison theorem; then we prove the thermal, ground-state and low-temperature assertions.

\subsection{Choice of constants and weights}
\label{sec:thermal-hypotheses}

Assume Hypothesis~\ref{hyp:locality}, and let \(J>0\) be given
by \eqref{eq:interaction-strength-bound}.

\paragraph{Normalization.}
We set \(\cV_0:=\cV/J\) and \(s:=J\lambda\), so that
\begin{equation*}
 \Phi_\lambda=\Phi_h+s\cV_0.
\end{equation*}
The interaction norm of \(\cV_0\) in the respective regime equals \(1\). Throughout this section, we apply the bounds from Section~\ref{sec:thermofield} to \(\cV_0\) with coupling \(s\), while retaining the index \(\lambda\) on the dynamics, states and annihilation families. The constants \(c_{\mathrm g},c_0,N\) chosen below therefore depend only on the model data other than \(J\).

We apply the comparison theorem on \(\A^\st\), with \(n=2r\). For each spatial weight \(W\) below, put \(G:=1/W\) in the notation of Definition~\ref{def:decay-profile}. We write \(G,\cS\) for the corresponding decay profile and support sum. Without a free gap, these functions depend on the inverse temperature and we write \(G_\beta,\cSbeta\).

\subsubsection{Under a free gap}
\label{sec:gapped-choices}

\paragraph{Bounds on the coefficients of the free annihilation operators.}
In regime~\ref{ass:exponential}, we choose a model rate \(\mu_1>0\) less than or equal to the rates in Proposition~\ref{prop:free-thermal-family}(i) and in Lemma~\ref{lem:ground-free-family}. We use the weights
\[
 w(R):=\begin{cases}
 \exp(\mu_1R^{\theta_{\rm in}})&\ref{ass:exponential},\\
 (1+R)^{p_h}&\ref{ass:polynomial-weight}.
 \end{cases}
\]
We choose a model constant \(C_0\ge1\) such that Proposition~\ref{prop:free-thermal-family}(i) and Lemma~\ref{lem:ground-free-family} give the bound
\[
 \max\left\{
 \sup_{\beta\ge0}\bigl\|(|\mathsf a^\beta_{uv}|)\bigr\|^{\rm Sch}_w,
 \bigl\|(|\mathsf a_{u,(v,j)}|)\bigr\|^{\rm Sch}_w
 \right\}
 \le C_0\begin{cases}
 \exp(-\mu_1)&\ref{ass:exponential},\\
 1&\ref{ass:polynomial-weight}.
 \end{cases}
\]

\paragraph{Coupling thresholds.}
We set $\varepsilon:=1/[64r(1+C_0)]$. Using the model constants \(C\) and \(C_{\rm gs}\) from Propositions~\ref{prop:thermal-family}(ii) and~\ref{prop:ground-family}(ii), respectively, we define the following model constant
\begin{equation}\label{eq:gap-coupling-choice}
 c_{\mathrm g}:=\begin{cases}
 \min\{1,(\varepsilon/\max\{C,C_{\rm gs}\})^2\}&\ref{ass:exponential},\\
 \min\{1,\varepsilon/\max\{C,C_{\rm gs}\}\}&\ref{ass:polynomial-weight}.
 \end{cases}
\end{equation}
Recall from Theorem~\ref{thm:uniqueness}(i) that \(\lambda_{\mathrm g}=c_{\mathrm g}/J\), so \(|\lambda|\le\lambda_{\mathrm g}\) is equivalent to \(|s|\le c_{\mathrm g}\).

\paragraph{Decay profiles in the main results.}
\par\smallskip\noindent
\emph{Regime~\ref{ass:exponential}.} Fix \(0<\theta<\theta_{\rm in}\) and choose the cutoff function \(\widehat f\) from Section~\ref{sec:thermal-family-statement} with Gevrey order \(\sG=\theta_{\rm in}/\theta\). We choose a constant $m> 0$, depending only on the model data and $\theta$, that is less than or equal to the rates given by Propositions~\ref{prop:thermal-family}(ii), \ref{prop:thermal-comparison}(ii), \ref{prop:ground-comparison} and~\ref{prop:ground-family}(ii), and to \(\mu_1\). We set
\[
 W(R):=\exp(mR^\theta).
\]

\par\smallskip\noindent
\emph{Regime~\ref{ass:polynomial-weight}.} We choose the cutoff function \(\widehat f\) of Gevrey order \(\sG=2\), and set
\[
 W(R):=(1+R)^{p_\Phi}.
\]

\paragraph{Prefactors in the main results.}
The prefactors \(C\) in the bounds proved below under the free gap assumption depend only on the model data and, for spatial estimates in regime~\ref{ass:exponential}, on \(\theta\). All these prefactors are independent of the inverse temperature \(\beta\).

\subsubsection{Without a free gap}
\label{sec:ng-choices}

\paragraph{Bounds on the coefficients of the free annihilation operators.}
Fix \(\beta>0\). In regime~\ref{ass:exponential}, let \(\bar\mu_1:=\bar\mu\) be the model rate from Proposition~\ref{prop:free-thermal-family}(ii). We use the weights
\[
 w(R):=\begin{cases}
 \exp (\bar\mu_1R^{\theta_{\rm in}}/(1+\beta))&\ref{ass:exponential},\\
 (1+R)^{p_h}&\ref{ass:polynomial-weight}.
 \end{cases}
\]
For the coefficients \(\mathsf a^\beta_{uv}\) of the free family \(\mathsf F_u^{\beta,0}\), define
\[
 C_0(\beta):=\max\left\{1,
 \bigl\|(|\mathsf a^\beta_{uv}|)\bigr\|^{\rm Sch}_w\right\}.
\]
Proposition~\ref{prop:free-thermal-family}(ii) gives model constants \(K\ge1\) and \(N_0>0\) such that
\[
 1+C_0(\beta)\le K(1+\beta)^{N_0}.
\]

\paragraph{Coupling thresholds.}
Writing \(C_{\rm FG},M_{\rm FG}\ge1\) for the constants \(C,M\) in the Fermi-weighted part of Proposition~\ref{prop:thermal-family}(ii), we define the model constants
\begin{equation}\label{eq:ng-coupling-choice}
 (N,c_0):=
 \begin{cases}
 \bigl(\lceil2(M_{\rm FG}+N_0)\rceil,(32nKC_{\rm FG})^{-2}\bigr)&\ref{ass:exponential},\\
 \bigl(\lceil M_{\rm FG}+N_0\rceil,(32nKC_{\rm FG})^{-1}\bigr)&\ref{ass:polynomial-weight}.
 \end{cases}
\end{equation}
Recall from Theorem~\ref{thm:uniqueness}(ii) that \(\lambda_{\mathrm{ng}}(\beta)=c_0/[J\beta(1+\beta)^{N-1}]\), so \(|\lambda|\le\lambda_{\mathrm{ng}}(\beta)\) is equivalent to \(|s|\le c_0/[\beta(1+\beta)^{N-1}]\).

\paragraph{Decay profiles in the main results.}
\par\smallskip\noindent
\emph{Regime~\ref{ass:exponential}.} Let \(\mu_{\rm FG}>0\) be the model rate from Proposition~\ref{prop:thermal-family}(ii), and write \(\mu_{\rm def,FG}:=\bar\mu>0\) for the model rate in the Fermi-weighted part of Proposition~\ref{prop:thermal-comparison}(ii). Set $m_0:=\min\{\bar\mu_1,\mu_{\rm FG},\mu_{\rm def,FG}\}/2$. We use the weight
\[
 W(R):=\exp \bigl(m_0R^{\theta_{\rm in}}/(1+\beta)\bigr).
\]
\par\smallskip\noindent
\emph{Regime~\ref{ass:polynomial-weight}.} We use the weight
\[
 W(R):=(1+R)^{p_\Phi}.
\]

\paragraph{Prefactors in the main results.}
Choose model constants \(C_{\rm cmp},M_{\rm cmp}\ge1\) such that \(C_{\rm cmp}(1+\beta)^{M_{\rm cmp}}\) bounds the prefactors in the Fermi-weighted estimates of Proposition~\ref{prop:thermal-comparison}(i) and~(ii). Set \(C_*:=8nK^2C_{\rm cmp}\) and \(M:=2N_0+M_{\rm cmp}\), and define
\[
 C(\beta):=C_*(1+\beta)^M.
\]
In particular,
\begin{equation}\label{eq:ng-prefactor-bound}
 8n(1+C_0(\beta))^2C_{\rm cmp}(1+\beta)^{M_{\rm cmp}}
 \le C(\beta).
\end{equation}

\subsection{Verification of the comparison hypotheses}

We verify Hypothesis~\ref{hyp:perturbed-family} for the families once under a free gap and once without. We use the constants, filters and weights of Section~\ref{sec:thermal-hypotheses}.

\subsubsection{Under a free gap}

\begin{lemma}\label{lem:gapped-hypotheses}
Assume Hypothesis~\ref{hyp:locality} and the free gap condition \eqref{eq:free-band}  with $g >0$. Use the constants and cutoff chosen in Section~\ref{sec:thermal-hypotheses}. For every \(\lambda\) with \(|\lambda|\le\lambda_{\mathrm g}\) and every \(\beta>0\), both families
\[
 (\mathsf F_u^{\beta,\lambda})_{u\in\Gamma\times\{1,\ldots,2r\}}
 \qquad\text{and}\qquad
 (\mathsf F_u^{\infty,\lambda})_{u\in\Gamma\times\{1,\ldots,2r\}}
\]
satisfy Hypothesis~\ref{hyp:perturbed-family} with \(C_0\), \(\varepsilon=1/[64r(1+C_0)]\) and either \(W=1\) or
\[
 W(R)=\begin{cases}
 \E^{mR^\theta}&\text{in regime~\ref{ass:exponential}},\\
 (1+R)^{p_\Phi}&\text{in regime~\ref{ass:polynomial-weight}}.
 \end{cases}
\]
In regime~\ref{ass:exponential}, \(m>0\) and \(0<\theta<\theta_{\rm in}\) are the parameters fixed in Section~\ref{sec:gapped-choices}.
\end{lemma}
\begin{proof}
Proposition~\ref{prop:free-thermal-family}(i) and Lemma~\ref{lem:ground-free-family} show that the coefficients \(\mathsf a^\beta_{uv}\) of \(\mathsf F_u^{\beta,0}\) and \(\mathsf a_{u,(v,j)}\) of \(\mathsf F_u^{\infty,0}\), respectively, satisfy the Parseval condition~\eqref{eq:parseval-condition}. With \(w\), \(\mu_1\) and \(C_0\) chosen in Section~\ref{sec:gapped-choices}, we have
\[
 W(R)\le\begin{cases}
 \E^{\mu_1}w(R)&\text{in regime~\ref{ass:exponential}},\\
 w(R)&\text{in regime~\ref{ass:polynomial-weight}},
 \end{cases}
\]
for all $R\ge0$. By our choice of \(C_0\) we therefore find
\[
 \max\left\{
 \sup_{\beta\ge0}\bigl\|(|\mathsf a^\beta_{uv}|)\bigr\|^{\rm Sch}_W,
 \bigl\|(|\mathsf a_{u,(v,j)}|)\bigr\|^{\rm Sch}_W
 \right\}\le C_0.
\]
This verifies Hypothesis~\ref{hyp:perturbed-family}(i) for both families.

We now verify Hypothesis~\ref{hyp:perturbed-family}(ii). Propositions~\ref{prop:thermal-family}(ii) and~\ref{prop:ground-family}(ii) give, for every \(u\in\Gamma\times\{1,\ldots,2r\}\), the norm-convergent decompositions
\[
 \begin{aligned}
 \mathsf F_u^{\beta,\lambda}-\mathsf F_u^{\beta,0}
 &=\sum_{\substack{Z\Subset\Gamma\\Z\ni x_u}}\Delta \mathsf F_{u,Z}^{\beta,\lambda},\\
 \mathsf F_u^{\infty,\lambda}-\mathsf F_u^{\infty,0}
 &=\sum_{\substack{Z\Subset\Gamma\\Z\ni x_u}}\Delta \mathsf F_{u,Z}^{\infty,\lambda},
 \end{aligned}
\]
where both \(\Delta \mathsf F_{u,Z}^{\beta,\lambda}\) and \(\Delta \mathsf F_{u,Z}^{\infty,\lambda} \in (\A_Z^\st)^-\). With the spatial weight \(W\) fixed in Section~\ref{sec:gapped-choices}, the propositions give
\begin{equation*}
 \max\left\{\begin{aligned}
 &\sup_{z\in\Gamma}\sum_{u\in\Gamma\times\{1,\ldots,2r\}}\sum_{\substack{Z\Subset\Gamma\\Z\ni z,x_u}}
 |Z|W(\diam Z)\|\Delta \mathsf F_{u,Z}^{\beta,\lambda}\|,\\
 &\sup_{z\in\Gamma}\sum_{u\in\Gamma\times\{1,\ldots,2r\}}\sum_{\substack{Z\Subset\Gamma\\Z\ni z,x_u}}
 |Z|W(\diam Z)\|\Delta \mathsf F_{u,Z}^{\infty,\lambda}\|
 \end{aligned}\right\}
 \le\max\{C,C_{\rm gs}\}\begin{cases}
 |s|^{1/2}&\ref{ass:exponential},\\
 |s|&\ref{ass:polynomial-weight}.
 \end{cases}
\end{equation*}
The right-hand side is at most \(\varepsilon\) in both regimes by \(|s|\le c_{\mathrm g}\) and the choice~\eqref{eq:gap-coupling-choice}.
Thus both decompositions satisfy Hypothesis~\ref{hyp:perturbed-family}(ii).
Since \(W(R)\ge1\), the same bounds hold with \(W=1\).
\end{proof}

\subsubsection{Without a free gap}

\begin{lemma}\label{lem:polynomial-temperature-thresholds}
Assume Hypothesis~\ref{hyp:locality} and use the choices of Section~\ref{sec:ng-choices}. For every \(\beta>0\) and every \(\lambda\) with \(|\lambda|\le\lambda_{\mathrm{ng}}(\beta)\), the family
\[
 (\mathsf G_u^{\beta,\lambda})_{u\in\Gamma\times\{1,\ldots,2r\}}
\]
satisfies Hypothesis~\ref{hyp:perturbed-family} with \(C_0=C_0(\beta)\), \(\varepsilon=1/[64r(1+C_0(\beta))]\) and either \(W=1\) or
\[
 W(R)=\begin{cases}
 \E^{m_0R^{\theta_{\rm in}}/(1+\beta)}&\text{in regime~\ref{ass:exponential}},\\
 (1+R)^{p_\Phi}&\text{in regime~\ref{ass:polynomial-weight}}.
 \end{cases}
\]
In regime~\ref{ass:exponential}, \(m_0>0\) is the parameter fixed in Section~\ref{sec:ng-choices}.
\end{lemma}
\begin{proof}
Proposition~\ref{prop:free-thermal-family}(ii) shows that the coefficients \(\mathsf a^\beta_{uv}\) of \(\mathsf F_u^{\beta,0}\) satisfy the Parseval condition~\eqref{eq:parseval-condition}. With \(w\) and \(C_0(\beta)\) chosen in Section~\ref{sec:ng-choices}, we have \(W(R)\le w(R)\) for all \(R\ge0\), and hence
\[
 \bigl\|(|\mathsf a^\beta_{uv}|)\bigr\|^{\rm Sch}_W
 \le\bigl\|(|\mathsf a^\beta_{uv}|)\bigr\|^{\rm Sch}_w
 \le C_0(\beta).
\]
This verifies Hypothesis~\ref{hyp:perturbed-family}(i).

Since \(\beta|s|/(1+\beta)\le c_0/(1+\beta)^N\le1\), Proposition~\ref{prop:thermal-family}(ii) gives, for every \(u\in\Gamma\times\{1,\ldots,2r\}\), the norm-convergent decomposition
\[
 \mathsf G_u^{\beta,\lambda}-\mathsf F_u^{\beta,0}
 =\sum_{\substack{Z\Subset\Gamma\\Z\ni x_u}}\Delta \mathsf G_{u,Z}^{\beta,\lambda},
\]
where \(\Delta \mathsf G_{u,Z}^{\beta,\lambda} \in (\A_Z^\st)^-\). With the spatial weight \(W\) fixed in Section~\ref{sec:ng-choices}, the bound in this proposition and the choices~\eqref{eq:ng-coupling-choice} give
\begin{equation*}
 \begin{aligned}
 \sup_{z\in\Gamma}\sum_{u\in\Gamma\times\{1,\ldots,2r\}}
 \sum_{\substack{Z\Subset\Gamma\\Z\ni z,x_u}}|Z|W(\diam Z)\|\Delta \mathsf G_{u,Z}^{\beta,\lambda}\|&\le C_{\rm FG}(1+\beta)^{M_{\rm FG}}
 \begin{cases}
 \bigl(\beta|s|/(1+\beta)\bigr)^{1/2}&\ref{ass:exponential},\\
 \beta|s|/(1+\beta)&\ref{ass:polynomial-weight}
 \end{cases}\\
 &\le1/[32nK(1+\beta)^{N_0}]\le\varepsilon.
 \end{aligned}
\end{equation*}
The second inequality follows from \(|s|\le c_0/[\beta(1+\beta)^{N-1}]\) and the choices in Equation~\eqref{eq:ng-coupling-choice}. The last inequality uses \(1+C_0(\beta)\le K(1+\beta)^{N_0}\) and \(n=2r\). This proves Hypothesis~\ref{hyp:perturbed-family}(ii).
Since \(W(R)\ge1\), the same bounds hold with \(W=1\).
\end{proof}

\subsection{Thermal results}

Throughout this subsection, assume Hypothesis~\ref{hyp:locality} and fix \(\beta>0\). Under the free spectral-gap condition~\eqref{eq:free-band}, we take \(|\lambda|\le\lambda_{\mathrm g}\) and use the choices of Section~\ref{sec:gapped-choices}. Without a free gap, we take \(|\lambda|\le\lambda_{\mathrm{ng}}(\beta)\) and use the choices of Section~\ref{sec:ng-choices}. For \(\Xi\in\mathfrak J_1\) and \(u\in\Gamma\times\{1,\ldots,2r\}\), we write
\[
 (F_u,\widetilde F_u):=\begin{cases}
 (\mathsf F_u^{\beta,\lambda},\mathsf F_u^\beta[\Xi])&\text{under a free gap},\\
 (\mathsf G_u^{\beta,\lambda},\mathsf G_u^\beta[\Xi])&\text{without a free gap}.
 \end{cases}
\]
 Lemmas~\ref{lem:gapped-hypotheses} and~\ref{lem:polynomial-temperature-thresholds} verify Hypothesis~\ref{hyp:perturbed-family} with the smallness required by Lemma~\ref{lem:comparison-kernel-bounds}. Theorem~\ref{thm:comparison} and its corollaries therefore apply with \(m=1\), and \(\|\mathsf T\|^{\rm Sch}_W\le1+C_0\) and \(\|\mathsf T^{\top}\mathsf L\|^{\rm Sch}_W\le\tfrac12(1+C_0)^2\). For spatial estimates, we use the weight \(W\) from the respective part of Section~\ref{sec:thermal-hypotheses}.

By Proposition~\ref{prop:kms-nonempty}, every KMS state \(\rho\) considered below is even. Proposition~\ref{prop:lift} therefore gives it a diagonal thermal extension, which we denote by \(\widehat\rho\). For any \(A\in\A_Y\), we have $A^{(1)}\in\A_Y^\st$, $\widehat\rho(A^{(1)})=\rho(A)$ and $\|A^{(1)}\|=\|A\|$.

\begin{proof}[KMS assertions of Theorem~\ref{thm:uniqueness}]
In the gapped case, we take \(\sG=2\) in the cutoff construction.
The set \(\KMS_\beta(\tau^{\Phi_\lambda})\) is nonempty by Proposition~\ref{prop:kms-nonempty}. For any \({\rho}\in\KMS_\beta(\tau^{\Phi_\lambda})\), Proposition~\ref{prop:thermal-family}(i) gives
\[
 \widehat\rho\bigl((F_u)^*F_u\bigr)=0,
\]
for every $u \in \Gamma \times \{1, \ldots, 2r\}$. Theorem~\ref{thm:comparison}(i), with \(W=1\), therefore gives \(\widehat\rho=\omega_F\). For every \(A\in\A\) we obtain
\[
 \rho(A)=\widehat\rho(A^{(1)})=\omega_F(A^{(1)}).
\]
Thus there is a unique KMS state, which we denote by \(\omega_{\beta,\lambda}\).
\end{proof}

\begin{proof}[Proof of Theorem~\ref{thm:kms-lppl}]
Take an arbitrary perturbation \(\Psi\in\mathfrak J_1\), and let \(\rho\in\KMS_\beta(\tau^{\Phi_\lambda+\Psi})\). We take \(\Xi=\Phi_\lambda+\Psi\). Proposition~\ref{prop:thermal-family}(i) gives
\[
 \widehat\rho(\widetilde F_u^*\widetilde F_u)=0,
 \qquad
 \widehat\omega_{\beta,\lambda}
 \bigl((F_u)^*F_u\bigr)=0,
\]
for all $u \in \Gamma \times \{1, \ldots, 2r\}$. For \(\varphi:=\widehat\rho-\widehat\omega_{\beta,\lambda}\), we have \(\varphi(\1)=0\). Lemma~\ref{lem:stability} and Proposition~\ref{prop:thermal-comparison}(ii) give, for every \(E,G\in\A^\st\) and \(u\in\Gamma\times\{1,\ldots,2r\}\),
\[
 \begin{aligned}
 |\varphi((F_u)^*E)|+|\varphi(GF_u)|&\le2\|F_u-\widetilde F_u\|
       \max\{\|E\|,\|G\|\}\\
 &\le\frac{\max\{\|E\|,\|G\|\}}{W(d(x_u,\Sigma(\Psi)))}
 \begin{cases}
 C&\text{with a free gap},\\
 2C_{\rm cmp}(1+\beta)^{M_{\rm cmp}}&\text{without a free gap}.
 \end{cases}
 \end{aligned}
\]
Corollary~\ref{cor:localized}, with \(S=\Sigma(\Psi)\) and \(\mathfrak s_0=0\), and the bound~\eqref{eq:ng-prefactor-bound} give, for \(Y\Subset\Gamma\) and \(A\in\A_Y\),
\[
 \begin{aligned}
 &|\rho(A)-\omega_{\beta,\lambda}(A)|=|\varphi(A^{(1)})|\le\|A\|\begin{cases}
 C\cS(Y,\Sigma(\Psi))&\text{with a free gap},\\
 C(\beta)\cSbeta(Y,\Sigma(\Psi))&\text{without a free gap}.
 \end{cases}
 \end{aligned}
\]
Since \(\rho\) and \(\Psi\) were arbitrary, taking the suprema in the theorem proves both assertions.
\end{proof}

\begin{proof}[Proof of Theorem~\ref{thm:correlations}]
Take an arbitrary perturbation \(\Psi\in\mathfrak J_1\), and let \(\rho\in\KMS_\beta(\tau^{\Phi_\lambda+\Psi})\). We take \(\Xi=\Phi_\lambda+\Psi\). Proposition~\ref{prop:thermal-family}(i) gives, for every \(u\in\Gamma\times\{1,\ldots,2r\}\),
\[
 \widehat\rho(\widetilde F_u^*\widetilde F_u)=0.
\]
Lemma~\ref{lem:stability} and Proposition~\ref{prop:thermal-comparison}(ii) give, for every \(u\in\Gamma\times\{1,\ldots,2r\}\),
\[
 \begin{aligned}
 \widehat\rho\bigl((F_u)^*F_u\bigr)^{1/2}
 &\le\|F_u-\widetilde F_u\|\\
 &\le\frac1{W(d(x_u,\Sigma(\Psi)))}\begin{cases}
 C&\text{with a free gap},\\
 C_{\rm cmp}(1+\beta)^{M_{\rm cmp}}&\text{without a free gap}.
 \end{cases}
 \end{aligned}
\]
Corollary~\ref{cor:covariance} and the bound~\eqref{eq:ng-prefactor-bound} give, for \(X,Y\Subset\Gamma\), \(A\in\A_X\) and \(B\in\A_Y\),
\[
 \begin{aligned}
 \lvert\Cov_\rho(A,B)\rvert&=\lvert\Cov_{\widehat\rho}(A^{(1)},B^{(1)})\rvert\\
 &\le\|A\|\|B\|\begin{cases}
 C\bigl[\cS(X,Y)+\cS(X,\Sigma(\Psi))\bigr]&\text{with a free gap},\\
 C(\beta)\bigl[\cSbeta(X,Y)+\cSbeta(X,\Sigma(\Psi))\bigr]&\text{without a free gap}.
 \end{cases}
 \end{aligned}
\]
Applying the preceding estimate to \(B^*,A^*\) and using
\[
 \Cov_\rho(A,B)=\overline{\Cov_\rho(B^*,A^*)}
\]
gives the bound with \(X\) and \(Y\) exchanged. Since \(\rho\) and \(\Psi\) are arbitrary, taking the suprema in the theorem proves both assertions.
\end{proof}

\begin{proof}[Proof of Theorem~\ref{thm:gibbs}]
Let \(\Lambda\Subset\Gamma\), \(B\subset\Lambda\), and let \(K\in\A_B^+\) be self-adjoint. We extend the finite-volume Gibbs state \(\rho^{(\lambda)}_{\beta,\Lambda,K}\), defined in Equation~\eqref{eq:finite-gibbs-definition}, to the full algebra by setting
\[
 \rho:=\rho^{(\lambda)}_{\beta,\Lambda,K}\widehat\otimes\tr_{\Lambda^c}.
\]
For the interaction \(\Xi_{\Lambda,K}\) defined in Equation~\eqref{eq:cut-interaction}, Proposition~\ref{prop:weak-dynamics} gives, for \(A\in\A_\Lambda\) and \(D\in\A_{\Lambda^c}\)
\[
 \tau_t^{\Xi_{\Lambda,K}}(AD)
 =\E^{\I tH^{(\lambda)}_{\Lambda,K}}A
  \E^{-\I tH^{(\lambda)}_{\Lambda,K}}D,
\]
for all $t \in \R$. The Gibbs state and \(\tr_{\Lambda^c}\) are even \(\beta\)-KMS states for $\tau^{\Xi_{\Lambda,K}}$ on \(\A_\Lambda\) and the identity on \(\A_{\Lambda^c}\), respectively. Lemma~\ref{lem:graded-KMS} therefore gives \(\rho\in\KMS_\beta(\tau^{\Xi_{\Lambda,K}})\).

We take \(\Xi=\Xi_{\Lambda,K}\). Proposition~\ref{prop:thermal-family}(i) gives, for every \(u\in\Gamma\times\{1,\ldots,2r\}\),
\[
 \widehat\rho(\widetilde F_u^*\widetilde F_u)=0,
 \qquad
 \widehat\omega_{\beta,\lambda}
 \bigl((F_u)^*F_u\bigr)=0.
\]
For \(\varphi:=\widehat\rho-\widehat\omega_{\beta,\lambda}\), we have \(\varphi(\1)=0\). Lemma~\ref{lem:stability} and Proposition~\ref{prop:thermal-comparison}(ii), with \(S=\Lambda^c\cup B\), give, for every \(E,G\in\A^\st\) and \(u\in\Gamma\times\{1,\ldots,2r\}\),
\[
 \begin{aligned}
 |\varphi((F_u)^*E)|+|\varphi(GF_u)|&\le2\|F_u-\widetilde F_u\|
       \max\{\|E\|,\|G\|\}\\
 &\le\frac{\max\{\|E\|,\|G\|\}}{W(d(x_u,\Lambda^c\cup B))}
 \begin{cases}
 C&\text{with a free gap},\\
 2C_{\rm cmp}(1+\beta)^{M_{\rm cmp}}&\text{without a free gap}.
 \end{cases}
 \end{aligned}
\]
Corollary~\ref{cor:localized}, with \(S=\Lambda^c\cup B\) and \(\mathfrak s_0=0\), and the bound~\eqref{eq:ng-prefactor-bound} give, for \(Y\subset\Lambda\) and \(A\in\A_Y\),
\[
 \begin{aligned}
 &|\rho^{(\lambda)}_{\beta,\Lambda,K}(A)-\omega_{\beta,\lambda}(A)|
 =|\varphi(A^{(1)})|\le\|A\|\begin{cases}
 C\cS(Y,\Lambda^c\cup B)&\text{with a free gap},\\
 C(\beta)\cSbeta(Y,\Lambda^c\cup B)&\text{without a free gap}.
 \end{cases}
 \end{aligned}
\]
Since \(K\) is arbitrary, taking the supremum over all such perturbations proves both assertions.
\end{proof}

\begin{proof}[Proof of Theorem~\ref{thm:continuity}]
In the gapped case, we take \(\sG=2\) in the cutoff construction. Take \(\Psi\in\mathfrak J_1\) and let \(\rho\in\KMS_\beta(\tau^{\Phi_\lambda+\Psi})\) be arbitrary. We take \(\Xi=\Phi_\lambda+\Psi\). We set $\varphi:=\widehat\rho-\widehat\omega_{\beta,\lambda}$. By Proposition~\ref{prop:thermal-family}(i) we find that
\[
 \widehat\omega_{\beta,\lambda}\bigl((F_u)^*F_u\bigr)=0,
 \qquad
 \widehat\rho(\widetilde F_u^*\widetilde F_u)=0,
\]
for every \(u\in\Gamma\times\{1,\ldots,2r\}\). We have \(\varphi(\1)=0\). Lemma~\ref{lem:stability} and Proposition~\ref{prop:thermal-comparison}(i) give, for every \(E,G\in\A^\st\) and \(u\in\Gamma\times\{1,\ldots,2r\}\),
\[
 \begin{aligned}
 |\varphi((F_u)^*E)|+|\varphi(GF_u)|&\le2\|F_u-\widetilde F_u\|
       \max\{\|E\|,\|G\|\}\\
 &\le J_\Psi\max\{\|E\|,\|G\|\}\begin{cases}
 C&\text{with a free gap},\\
 2C_{\rm cmp}(1+\beta)^{M_{\rm cmp}}\dfrac{\beta}{1+\beta}&\text{without a free gap}.
 \end{cases}
 \end{aligned}
\]
Corollary~\ref{cor:localized}, with \(W=1\) and \(C_S=0\), and~\eqref{eq:ng-prefactor-bound} give, for \(Y\Subset\Gamma\) and \(A\in\A_Y\),
\[
 \begin{aligned}
 &|\rho(A)-\omega_{\beta,\lambda}(A)|=|\varphi(A^{(1)})|\le |Y|\|A\|J_\Psi\begin{cases}
 C&\text{with a free gap},\\
 \dfrac{\beta}{1+\beta}C(\beta)&\text{without a free gap}.
 \end{cases}
 \end{aligned}
\]
Taking the supremum over all such KMS states proves the assertions.
\end{proof}

\subsection{Ground-state and low-temperature results}

Throughout this subsection, assume Hypothesis~\ref{hyp:locality} and the free gap condition~\eqref{eq:free-band} with $g >0$, and let \(|\lambda|\le\lambda_{\mathrm g}\). We use the choices of Section~\ref{sec:gapped-choices} and the product ground-state and KMS extensions defined in Equation~\eqref{eq:stacked-states}.
Lemma~\ref{lem:gapped-hypotheses} verifies Hypothesis~\ref{hyp:perturbed-family} for the ground-state family with the smallness required by Lemma~\ref{lem:comparison-kernel-bounds}.
Hence Theorem~\ref{thm:comparison} and its corollaries apply with \(m=1\) and \(\|\mathsf T\|^{\rm Sch}_W\le1+C_0\).

\begin{proof}[Ground-state assertion of Theorem~\ref{thm:uniqueness}(i)]
We take \(\sG=2\). Lemma~\ref{lem:ground-state-existence} shows that \(\GS(\tau^{\Phi_\lambda})\) is not empty. Let \(\omega\in\GS(\tau^{\Phi_\lambda})\) be arbitrary. By Proposition~\ref{prop:ground-family}(i) we find that
\[
 \omega_\st\bigl((\mathsf F_u^{\infty,\lambda})^*\mathsf F_u^{\infty,\lambda}\bigr)=0,
\]
for every \(u\in\Gamma\times\{1,\ldots,2r\}\). Choosing \(W=1\) in Theorem~\ref{thm:comparison}(i), we find that \(\omega_\st=\omega_F\). Hence, we obtain for every \(A\in\A\)
\[
 \omega(A)=\omega_\st(A^{(1)})=\omega_F(A^{(1)}),
\]
which shows the assertion.
\end{proof}

\begin{proof}[Proof of Proposition~\ref{prop:ground-gap}]
We take \(\sG=2\). Let \((\pi,\mathcal H,\Omega)\) be the GNS triple of \(\omega_{\lambda,\st}\). Lemmas~\ref{lem:graded-ground} and~\ref{lem:ground-state-equivalence} give a Hamiltonian \(L\ge 0\) on $\mathcal H$ such that $L\Omega=0$ and
\[
 \pi(\tau_t^{\st,\lambda}(A))=\E^{\I tL}\pi(A)\E^{-\I tL},
\]
for all \(A\in\A^\st\) and \(t\in\R\). Proposition~\ref{prop:ground-family}(i) gives, for every \(u\in\Gamma\times\{1,\ldots,2r\}\),
\[
 \pi(\mathsf F_u^{\infty,\lambda})\1_{[0,g_*]}(L)=0.
\]
Take \(\psi\in\1_{[0,g_*]}(L)\mathcal H\) such that \(\psi \perp \Omega\), and define the bounded linear functional \(\varphi:\A^\st\to\C\) by \(\varphi(A)=\langle\psi,\pi(A)\Omega\rangle\). It follows that $\varphi(\1)=0$. Since \(\pi(\mathsf F_u^{\infty,\lambda})\Omega = \pi(\mathsf F_u^{\infty,\lambda})\psi = 0\), it follows that
\[
 \varphi((\mathsf F_u^{\infty,\lambda})^*E)=\varphi(B\mathsf F_u^{\infty,\lambda})=0
\]
for all \(E,B\in\A^\st\) and every \(u\in\Gamma\times\{1,\ldots,2r\}\). Theorem~\ref{thm:comparison}(ii), with \(W=1\), gives
\[
 \langle\psi,\pi(A)\Omega\rangle=\varphi(A)=0,
\]
for all $A\in\A_\loc^\st$. Since \(\pi(\A_\loc^\st)\Omega\) is dense in \(\mathcal H\) it follows that \(\psi=0\). We conclude that \(\1_{[0,g_*]}(L)=|\Omega\rangle\langle\Omega|\), and hence
\begin{equation}\label{eq:stacked-gns-gap-bound}
 L\ge g_*(\1-|\Omega\rangle\langle\Omega|).
\end{equation}
Finally, we restrict this estimate to the left algebra copy. For \(A\in\A_\loc\), we use that
\[
 \delta_{\Phi_\lambda^\st}(A^{(1)})=\bigl(\delta_{\Phi_\lambda}(A)\bigr)^{(1)},
 \qquad
 \omega_{\lambda,\st}(A^{(1)})=\omega_\lambda(A).
\] 
Since \(A^{(1)}\) is local, \(t\mapsto\tau_t^{\st,\lambda}(A^{(1)})\) is norm differentiable, so \(\pi(A^{(1)})\Omega\in\operatorname{Dom}(L)\). Differentiating \(\E^{\I tL}\pi(A^{(1)})\Omega=\pi(\tau_t^{\st,\lambda}(A^{(1)}))\Omega\) at \(t=0\) gives
\[
 L\pi(A^{(1)})\Omega
 =\pi\!\left(\delta_{\Phi_\lambda^\st}(A^{(1)})\right)\Omega
 =\pi\!\left((\delta_{\Phi_\lambda}(A))^{(1)}\right)\Omega.
\]
Applying the bound~\eqref{eq:stacked-gns-gap-bound} to \(\pi(A^{(1)})\Omega\), we obtain
\[
 \begin{aligned}
 \RealPart\omega_\lambda(A^*\delta_{\Phi_\lambda}(A))
 &=\langle\pi(A^{(1)})\Omega,L\pi(A^{(1)})\Omega\rangle\\
 &\ge g_*\left(\|\pi(A^{(1)})\Omega\|^2
 -|\langle\Omega,\pi(A^{(1)})\Omega\rangle|^2\right)\\
 &=g_*\bigl(\omega_\lambda(A^*A)-|\omega_\lambda(A)|^2\bigr).
 \end{aligned}
\]
This proves the proposition.
\end{proof}

\begin{proof}[Proof of Theorem~\ref{thm:ground}]
Take \(\Psi,\Upsilon\in\mathfrak J_1\) and arbitrary states \(\rho\in\KMS_\beta(\tau^{\Phi_\lambda+\Psi})\) and \(\omega\in\GS(\tau^{\Phi_\lambda+\Upsilon})\), which exist by Proposition~\ref{prop:kms-nonempty} and Lemma~\ref{lem:ground-state-existence}. Set \(\varphi:=\rho_{\beta,\st}-\omega_\st\).

Proposition~\ref{prop:ground-family}(i) and the Cauchy--Schwarz inequality give, for all \(E,B\in\A^\st\) and \(u\in\Gamma\times\{1,\ldots,2r\}\),
\begin{equation}\label{eq:low-temperature-ground-annihilation}
 \omega_\st\bigl((\mathsf F_u^{\infty}[\Phi_\lambda+\Upsilon])^*E\bigr)
 =\omega_\st\bigl(B\mathsf F_u^{\infty}[\Phi_\lambda+\Upsilon]\bigr)=0.
\end{equation}
We use the weight \(W\) fixed in Section~\ref{sec:gapped-choices}. We choose a constant \(C\ge1\) depending only on the model data, and in regime \ref{ass:exponential} on $\theta$, that is larger than or equal to the constants obtained in Propositions~\ref{prop:ground-family}(i) and~\ref{prop:ground-comparison}.
Inserting the definition of $\varphi$ we find
\[
 \begin{gathered}
 \varphi((\mathsf F_u^{\infty,\lambda})^*E)\\
 =\rho_{\beta,\st}((\mathsf F_u^{\infty}[\Phi_\lambda+\Psi])^*E) 
 +\rho_{\beta,\st}((\mathsf F_u^{\infty,\lambda}-\mathsf F_u^{\infty}[\Phi_\lambda+\Psi])^*E)
 -\omega_\st((\mathsf F_u^{\infty,\lambda}-\mathsf F_u^{\infty}[\Phi_\lambda+\Upsilon])^*E),
 \end{gathered}
\]
where the term \(\omega_\st((\mathsf F_u^{\infty}[\Phi_\lambda+\Upsilon])^*E)\) vanishes by Equation~\eqref{eq:low-temperature-ground-annihilation}. Applying the triangle inequality to this expression and the corresponding one for \(\varphi(B\mathsf F_u^{\infty,\lambda})\), and using \(\|\rho_{\beta,\st}\|=\|\omega_\st\|=1\), gives
\[
 \begin{aligned}
 &|\varphi((\mathsf F_u^{\infty,\lambda})^*E)|+|\varphi(B\mathsf F_u^{\infty,\lambda})|\\
 &\quad\le
 |\rho_{\beta,\st}((\mathsf F_u^{\infty}[\Phi_\lambda+\Psi])^*E)|
 +|\rho_{\beta,\st}(B\mathsf F_u^{\infty}[\Phi_\lambda+\Psi])|\\
 &\qquad+
 \bigl[\|\mathsf F_u^{\infty,\lambda}-\mathsf F_u^{\infty}[\Phi_\lambda+\Psi]\|
 +\|\mathsf F_u^{\infty,\lambda}-\mathsf F_u^{\infty}[\Phi_\lambda+\Upsilon]\|\bigr]
 (\|E\|+\|B\|)\\
 &\quad\le\bigl[C\E^{-\beta g_*}
 +2\|\mathsf F_u^{\infty,\lambda}-\mathsf F_u^{\infty}[\Phi_\lambda+\Psi]\|
 +2\|\mathsf F_u^{\infty,\lambda}-\mathsf F_u^{\infty}[\Phi_\lambda+\Upsilon]\|\bigr]
 \max\{\|E\|,\|B\|\}.
 \end{aligned}
\]
The last inequality uses Proposition~\ref{prop:ground-family}(i) and \(\|E\|+\|B\|\le2\max\{\|E\|,\|B\|\}\). We now apply Proposition~\ref{prop:ground-comparison} to \(\Psi\) and \(\Upsilon\) and find
\[
 \begin{aligned}
 2\|\mathsf F_u^{\infty,\lambda}-\mathsf F_u^{\infty}[\Phi_\lambda+\Psi]\|
  +2\|\mathsf F_u^{\infty,\lambda}-\mathsf F_u^{\infty}[\Phi_\lambda+\Upsilon]\|&\le2C\left[
 \frac1{W(d(x_u,\Sigma(\Psi)))}
 +\frac1{W(d(x_u,\Sigma(\Upsilon)))}\right]\\
 &\le\frac{4C}{W(d(x_u,\Sigma(\Psi)\cup\Sigma(\Upsilon)))},
 \end{aligned}
\]
where we have used that $W$ is nondecreasing and
\[
 d(x_u,\Sigma(\Psi)\cup\Sigma(\Upsilon))
 =\min\{d(x_u,\Sigma(\Psi)),d(x_u,\Sigma(\Upsilon))\}.
\]
Since both states are normalized, we have \(\varphi(\1)=0\). We can therefore apply Corollary~\ref{cor:localized} with \(S=\Sigma(\Psi)\cup\Sigma(\Upsilon)\), \(\mathfrak s_0=C\E^{-\beta g_*}\) and \(C_S=4C\). Using \(\|\mathsf T\|^{\rm Sch}_W\le1+C_0\) from Lemma~\ref{lem:comparison-kernel-bounds}, we find that for all \(Y\Subset\Gamma\) and \(A\in\A_Y\),
\[
 \begin{aligned}
 &|\rho(A)-\omega(A)|=|\varphi(A^{(1)})|\le16n(1+C_0)C\|A\|\bigl[|Y|\E^{-\beta g_*}
 +\cS(Y,\Sigma(\Psi)\cup\Sigma(\Upsilon))\bigr].
 \end{aligned}
\]
The prefactor \(16n(1+C_0)C\) depends only on the model data and, in regime~\ref{ass:exponential}, on \(\theta\).
Since \(\rho\), \(\omega\), \(\Psi\) and \(\Upsilon\) were arbitrary, taking the suprema over them proves the assertion.
\end{proof}

\section{Proof of the comparison theorem}
\label{sec:comparison-proof}

We prove Theorem~\ref{thm:comparison}(i), following the strategy outlined in Section~\ref{sec:roadmap-comparison}. Throughout, \((F_u)_{u\in\cI}\) is the family given in that theorem. We begin with the deferred proofs of the lemmas from Section~\ref{sec:comparison}.

\subsection{Proofs of the lemmas from Section~\ref{sec:comparison}}
\label{sec:comparison-lemma-proofs}

\begin{proof}[Proof of Lemma~\ref{lem:parseval-reconstruction}]

We prove that, for each fixed \(u\in\cI\), \(\sum_{v\in\cJ}|\sa_{uv}|^2\le2\). To this end, we first notice that for every finitely supported family of real coefficients \((b_v)_{v \in \cJ}\) the Parseval condition gives
\[
 \sum_{u'\in\cI}\left|\sum_{v\in\cJ}\sa_{u'v}b_v\right|^2
 =\sum_{v,w\in\cJ}b_vb_w\RealPart
       \sum_{u'\in\cI}\overline{\sa_{u'v}}\sa_{u'w}
 =\sum_{v\in\cJ}|b_v|^2.
\]
For complex \(b_v\), use \(|z_1+\I z_2|^2\le2(|z_1|^2+|z_2|^2)\) and apply this identity to the real and imaginary parts of $b_v$ to obtain
\[
 \sum_{u'\in\cI}\left|\sum_{v\in\cJ}\sa_{u'v}b_v\right|^2
 \le2\sum_{v\in\cJ}|b_v|^2.
\]
Fix \(u\in\cI\) and a finite \(J_0\subset\cJ\). Taking \(b_v=\overline{\sa_{uv}}\mathbf1_{J_0}(v)\) yields
\[
 \begin{aligned}
 \left(\sum_{v\in J_0}|\sa_{uv}|^2\right)^2
 &=\left|\sum_{v\in\cJ}\sa_{uv}b_v\right|^2\le\sum_{u'\in\cI}\left|\sum_{v\in\cJ}\sa_{u'v}b_v\right|^2
 \le2\sum_{v\in J_0}|\sa_{uv}|^2.
 \end{aligned}
\]
Thus \(\sum_{v\in J_0}|\sa_{uv}|^2\le2\), and increasing \(J_0\) proves \(\sum_{v\in\cJ}|\sa_{uv}|^2\le2\).

To prove norm convergence of the series defining $(F_u^0)_{u\in\cI}$, first take a finitely supported family $(z_v)_{v\in\cJ}$ of complex coefficients. For positive operators \(0\le P\le Q\), the operator norm satisfies \(\|P\|\le\|Q\|\). Using the \(C^*\)-identity, we obtain
\[
 \begin{aligned}
 \left\|\sum_{v\in\cJ}z_v\gamma_v\right\|^2
 &=\left\|\left(\sum_{v\in\cJ}z_v\gamma_v\right)^*
              \left(\sum_{v\in\cJ}z_v\gamma_v\right)\right\|\\
 &\le\left\|\left(\sum_{v\in\cJ}z_v\gamma_v\right)^*
              \left(\sum_{v\in\cJ}z_v\gamma_v\right)
       +\left(\sum_{v\in\cJ}z_v\gamma_v\right)
              \left(\sum_{v\in\cJ}z_v\gamma_v\right)^*\right\|=2\|z\|_2^2,
 \end{aligned}
\]
where we have used the anticommutation relations of the Majoranas. Fix \(u\in\cI\). For finite \(J_0\subset J_1\subset\cJ\), substituting \(z_v=\tfrac12\sa_{uv}\mathbf1_{J_1\setminus J_0}(v)\) into the bound above gives
\[
 \left\|\frac12\sum_{v\in J_1}\sa_{uv}\gamma_v
          -\frac12\sum_{v\in J_0}\sa_{uv}\gamma_v\right\|^2
 \le\frac12\sum_{v\in J_1\setminus J_0}|\sa_{uv}|^2.
\]
Since \(\sum_{v\in\cJ}|\sa_{uv}|^2\le2\), the right-hand side tends to zero as \(J_0\uparrow\cJ\), uniformly over finite \(J_1\supset J_0\). Thus the operators \(\tfrac12\sum_{v\in J_0}\sa_{uv}\gamma_v\), indexed by finite subsets \(J_0\subset\cJ\), are Cauchy in norm as \(J_0\uparrow\cJ\), proving norm convergence of the series defining \(F_u^0\). Substituting instead \(z_v=\tfrac12\sa_{uv}\mathbf1_{J_0}(v)\) into \(\|\sum_{v\in\cJ}z_v\gamma_v\|^2\le2\|z\|_2^2\) and passing to the norm limit yields
\[
 \|F_u^0\|^2
 =\lim_{J_0\uparrow\cJ}
    \left\|\frac12\sum_{v\in J_0}\sa_{uv}\gamma_v\right\|^2
 \le\frac12\sum_{v\in\cJ}|\sa_{uv}|^2\le1.
\]

To prove Equation \eqref{eq:reference-reconstruction}, we first fix \(v\in\cJ\) and a finite region \(\Lambda\Subset\Gamma\). By the definition of $F_u^0$, we find that
\[
 \sum_{\substack{u\in\cI\\x_u\in\Lambda}}
       \bigl(\overline{\sa_{uv}}F_u^0+\sa_{uv}(F_u^0)^*\bigr)
 =\sum_{w\in\cJ}
       \left(\RealPart\sum_{\substack{u\in\cI\\x_u\in\Lambda}}
                    \overline{\sa_{uv}}\sa_{uw}\right)\gamma_w.
\]
For the fixed label \(v\) and region \(\Lambda\), define the coefficients \(a^{v,\Lambda}=(a_w^{v,\Lambda})_{w\in\cJ}\). The Parseval condition gives
\[
 a_w^{v,\Lambda}:=\delta_{vw}-\RealPart
       \sum_{\substack{u\in\cI\\x_u\in\Lambda}}
             \overline{\sa_{uv}}\sa_{uw}
     =\RealPart\sum_{\substack{u\in\cI\\x_u\notin\Lambda}}
             \overline{\sa_{uv}}\sa_{uw}.
\]
For each fixed \(u\in\cI\), we have already proved that the coefficient family \((\sa_{uw})_{w\in\cJ}\) satisfies \(\sum_{w\in\cJ}|\sa_{uw}|^2\le2\). Since the sum over \(x_u\in\Lambda\) is finite, it follows that \(a^{v,\Lambda}\in\ell^2(\cJ,\R)\). The preceding norm estimate therefore shows that $\sum_{w\in\cJ}a_w^{v,\Lambda}\gamma_w$ converges in norm. Applying the anticommutation relations first to finite sums and then passing to the limit gives
\[
 \left(\sum_{w\in\cJ}a_w^{v,\Lambda}\gamma_w\right)^2=\|a^{v,\Lambda}\|_2^2\1.
\]
Putting these estimates together gives
\[
 \begin{aligned}
 &\left\|\gamma_v-\sum_{\substack{u\in\cI\\x_u\in\Lambda}}
       \bigl(\overline{\sa_{uv}}F_u^0+\sa_{uv}(F_u^0)^*\bigr)\right\|^2=\left\|\left(\sum_{w\in\cJ}a_w^{v,\Lambda}\gamma_w\right)^2\right\|
 =\|a^{v,\Lambda}\|_2^2.
 \end{aligned}
\]
It remains to estimate the norm \(\|a^{v,\Lambda}\|_2\). To this end, let $b\in \ell^2 (\cJ, \mathbb R)$ be finitely supported. We can interchange its finite sum over \(w\) with the absolutely convergent coefficient sums over \(u\). Inserting the definition of $a_w^{v, \Lambda}$ gives
\[
 \sum_{w\in\cJ}a_w^{v,\Lambda}b_w
 =\RealPart\sum_{\substack{u\in\cI\\x_u\notin\Lambda}}
       \overline{\sa_{uv}}\sum_{w\in\cJ}\sa_{uw}b_w.
\]
Using the Cauchy--Schwarz inequality and the Parseval condition gives
\[
 \begin{aligned}
 \left|\sum_{w\in\cJ}a_w^{v,\Lambda}b_w\right|
 &\le\left(\sum_{\substack{u\in\cI\\x_u\notin\Lambda}}
                  |\sa_{uv}|^2\right)^{1/2}
       \left(\sum_{u\in\cI}
          \left|\sum_{w\in\cJ}\sa_{uw}b_w\right|^2\right)^{1/2}=\left(\sum_{\substack{u\in\cI\\x_u\notin\Lambda}}
                  |\sa_{uv}|^2\right)^{1/2}\|b\|_2.
 \end{aligned}
\]
Taking the supremum over finitely supported real \(b\) with \(\|b\|_2\le1\) gives
\[
 \|a^{v,\Lambda}\|_2\le
 \left(\sum_{\substack{u\in\cI\\x_u\notin\Lambda}}|\sa_{uv}|^2\right)^{1/2}
 \longrightarrow0\qquad\text{as }\Lambda\uparrow\Gamma.
\]
This proves norm convergence in Equation \eqref{eq:reference-reconstruction}. Since the Majoranas generate \(\cC\), the family \((F_u^0)_{u\in\cI}\) does as well.
\end{proof}

\begin{proof}[Proof of Lemma~\ref{lem:weighted-Schur-properties}]
Submultiplicativity of weights and Tonelli's theorem give, for each \(i\),
\[
 \begin{aligned}
 \sum_{j\in\cI_3} W(d(x_i,x_j))(\mathsf H_1\mathsf H_2)_{ij}
 &\le\sum_{k\in\cI_2} W(d(x_i,x_k))(\mathsf H_1)_{ik}
          \sum_{j\in\cI_3} W(d(x_k,x_j))(\mathsf H_2)_{kj}\\
 &\le\|\mathsf H_1\|^{\rm Sch}_W\|\mathsf H_2\|^{\rm Sch}_W.
 \end{aligned}
\]
By the same calculation, we find for any $j$,
\[
 \begin{aligned}
 \sum_{i\in\cI_1}W(d(x_i,x_j))(\mathsf H_1\mathsf H_2)_{ij}
 &\le\|\mathsf H_1\|^{\rm Sch}_W\|\mathsf H_2\|^{\rm Sch}_W.
 \end{aligned}
\]
Together, these estimates prove submultiplicativity of the weighted Schur norm. Now consider a nonempty set \(S\subset\Gamma\). Monotonicity and submultiplicativity of the weight give
\[
 W(d(x_i,S))\le W(d(x_i,x_j))W(d(x_j,S)).
\]
Hence, we find that
\[
 \sum_{j\in\cI_2}\frac{\mathsf H_{ij}}{W(d(x_j,S))}
 \le\frac{\sum_{j\in\cI_2}W(d(x_i,x_j))\mathsf H_{ij}}{W(d(x_i,S))}
 \le\frac{\|\mathsf H\|^{\rm Sch}_W}{W(d(x_i,S))}.
\]
For \(S=\emptyset\), both sides vanish, completing the proof.
\end{proof}

\begin{proof}[Proof of Lemma~\ref{lem:comparison-kernel-bounds}]
Lemma~\ref{lem:parseval-reconstruction} gives \(\|F_u^0\|\le1\). Since \(W\ge1\), Hypothesis~\ref{hyp:perturbed-family} gives
\[
 \|F_u\|\le\|F_u^0\|+\sum_{Z\ni x_u}\|\Delta F_{u,Z}\|
 \le1+\varepsilon,
\]
for all $u \in \cI$. To bound the Schur norms, we use Tonelli's theorem to obtain
\begin{equation}\label{eq:correction-row-bound}
 \begin{aligned}
 &\sup_{u\in\cI}\sum_{v\in\cJ} W(d(x_u,x_v))\Bigl(\sum_{Z:\,x_u,x_v\in Z}\|\Delta F_{u,Z}\|\Bigr)\\
 &\qquad\le\sup_{u\in\cI}\sum_{Z\ni x_u}\|\Delta F_{u,Z}\|
       \sum_{v\in\cJ:\,x_v\in Z}W(\diam Z)\\
 &\qquad=2n\sup_{u\in\cI}\sum_{Z\ni x_u}|Z|W(\diam Z)\|\Delta F_{u,Z}\|\\
 &\qquad\le2n\sup_{z\in\Gamma}\sum_{u\in\cI}\sum_{Z\ni z,x_u}|Z|W(\diam Z)\|\Delta F_{u,Z}\|
 \le2n\varepsilon,
 \end{aligned}
\end{equation}
where we have used that each site in \(Z\) has \(2n\) Majorana labels. Likewise, we find that
\begin{equation}\label{eq:correction-column-bound}
 \sup_{v\in\cJ}\sum_{u\in\cI} W(d(x_u,x_v))\Bigl(\sum_{Z:\,x_u,x_v\in Z}\|\Delta F_{u,Z}\|\Bigr)
 \le\sup_{z\in\Gamma}\sum_{u\in\cI}\sum_{Z\ni z,x_u}
       W(\diam Z)\|\Delta F_{u,Z}\|
 \le\varepsilon.
\end{equation}
Combining these bounds with the assumptions on the reference family gives \(\|\mathsf L\|^{\rm Sch}_W\le\tfrac12 C_0+2n\varepsilon\).

We now proceed to estimate the Schur norm of \(\mathsf K\). It consists of three different contributions and we estimate them separately. By Lemma~\ref{lem:weighted-Schur-properties} the Schur norm is submultiplicative. Hence, the assumption and Equations \eqref{eq:correction-row-bound} and \eqref{eq:correction-column-bound} give for the first term
\[
 \begin{aligned}
 \left\|\left(2\sum_{u\in\cI}|\sa_{uv}|
          \sum_{Z\ni x_u,x_w}\|\Delta F_{u,Z}\|\right)_{v,w\in\cJ}\right\|^{\rm Sch}_W
 &\le2C_0(2n\varepsilon).
 \end{aligned}
\]
The same argument bounds the second contribution:
\[
 \begin{aligned}
 \left\|\left(4\sum_{u\in\cI}
          \sum_{Z\ni x_u,x_v}\|\Delta F_{u,Z}\|\mathsf L_{uw}\right)_{v,w\in\cJ}\right\|^{\rm Sch}_W
 &\le4(2n\varepsilon)\|\mathsf L\|^{\rm Sch}_W.
 \end{aligned}
\]
For the last contribution, Tonelli's theorem gives
\[
 \begin{aligned}
 &\sup_{v\in\cJ}\sum_{w\in\cJ}W(d(x_v,x_w))
   4\sum_{u\in\cI}\|F_u\|
       \sum_{Z\ni x_u,x_v,x_w}\|\Delta F_{u,Z}\|\\
 &\quad\le8n(1+\varepsilon)
       \sup_{v\in\cJ}\sum_{u\in\cI}\sum_{Z\ni x_u,x_v}
       |Z|W(\diam Z)\|\Delta F_{u,Z}\|\\
 &\quad\le8n(1+\varepsilon)\varepsilon,
 \end{aligned}
\]
where we have used $\lVert F_u \rVert \le 1 + \varepsilon$. This last contribution is symmetric in \(v\) and \(w\), so the same bound holds for its column sums.
Adding the three Schur norm bounds proves
\[
 \|\mathsf K\|^{\rm Sch}_W\le4nC_0\varepsilon+8n\varepsilon\|\mathsf L\|^{\rm Sch}_W+8n(1+\varepsilon)\varepsilon.
\]
For \(\varepsilon\le[32n(1+C_0)]^{-1}\), we have \(\|\mathsf L\|^{\rm Sch}_W\le\tfrac12(1+C_0)\) and \(\varepsilon\le C_0\). Substituting into the bound above gives
\[
 \|\mathsf K\|^{\rm Sch}_W\le\tfrac18+\tfrac18+\tfrac14=\tfrac12.
\]
For $\mathsf T$, submultiplicativity from Lemma~\ref{lem:weighted-Schur-properties} gives
\[
 \sum_{j\ge0}\|\mathsf L\mathsf K^j\|^{\rm Sch}_W
 \le\|\mathsf L\|^{\rm Sch}_W\sum_{j\ge0}
       \bigl(\|\mathsf K\|^{\rm Sch}_W\bigr)^j
 =\frac{\|\mathsf L\|^{\rm Sch}_W}{1-\|\mathsf K\|^{\rm Sch}_W},
\]
which proves convergence of the series defining $\mathsf T$. Since \(\|\mathsf K\|^{\rm Sch}_W\le\tfrac12\) we find
\[
 \|\mathsf T\|^{\rm Sch}_W
 \le2\|\mathsf L\|^{\rm Sch}_W\le 1+C_0.
\]

Finally, we show how \(\mathsf L\) controls the graded derivatives.
By Lemma~\ref{lem:local-graded-commutator} we have for any odd \(B\in\cC_Z\) the bound
\(\|\partial_BA\|\le\|B\|\sum_{v\in\cJ:\,x_v\in Z}\|\partial_vA\|\).
Linearity of the graded derivatives and \(F_u^0=\frac12\sum_{v\in\cJ} \sa_{uv}\gamma_v\) give
\[
 \|\partial_{F_u}A\|
 \le\frac12\sum_{v\in\cJ}|\sa_{uv}|\|\partial_vA\|
   +\sum_{Z\ni x_u}\|\Delta F_{u,Z}\|
                    \sum_{v\in\cJ:\,x_v\in Z}\|\partial_vA\|
 =\sum_{v\in\cJ}\mathsf L_{uv}\|\partial_vA\|.
\]
The same argument bounds \(\|\partial_{F_u^*}A\|\). We also notice that
\[
 \partial_v(A^*)= \gamma_v A^* - \Theta(A^*) \gamma_v = \gamma_v(\partial_vA)^*\gamma_v,
\] 
where we have used that $\gamma_v^2 = 1$ and that $\Theta$ is a $*$-automorphism. Since $\gamma_v$ is a unitary it follows that \(\|\partial_v(A^*)\|=\|\partial_vA\|\), and hence also \(\|\partial_{F_u}(A^*)\|\le\sum_{v\in\cJ}\mathsf L_{uv}\|\partial_vA\|\).
\end{proof}

\subsection{The auxiliary evolution}
\label{sec:kernels}

We start with the construction of a semigroup of contractive operators \((T_t)_{t\ge0}\). For this construction, we use the Banach space
\[
 \cC^1:=\left\{E\in\cC:\sum_{v\in\cJ}\|\partial_vE\|<\infty\right\},
\]
equipped with the norm \(\|E\|_{\cC^1}:=\|E\|+\sum_{v\in\cJ}\|\partial_vE\|\).

\begin{lemma}\label{lem:derivative-space-complete}
The space \(\cC^1\), equipped with \(\|\cdot\|_{\cC^1}\), is a Banach space. For any \(v\in\cJ\) the Majorana derivative \(\partial_v\) maps \(\cC^1\) into itself and satisfies
\[
 \|\partial_vE\|_{\cC^1}\le2\|E\|_{\cC^1},
 \qquad E\in\cC^1.
\]
\end{lemma}
\begin{proof}
Let \((E_j)_{j\in \mathbb N}\) be Cauchy in \(\cC^1\). Completeness of \(\cC\) and \(\ell^1(\cJ,\cC)\) gives an operator \(E\in\cC\) and a family \((B_v)_{v\in\cJ}\in\ell^1(\cJ,\cC)\) so that
\[
 \|E_j-E\|+\sum_{v\in\cJ}\|\partial_vE_j-B_v\|\longrightarrow0.
\]
For each \(v\in\cJ\), the bound
\(\|\partial_v(E_j-E)\|\le2\|E_j-E\|\) implies that \(B_v=\partial_vE\). Hence \(E\in\cC^1\), and the displayed convergence is precisely \(\|E_j-E\|_{\cC^1}\to0\).
The anticommutation relations for the Majoranas imply that \(\partial_w \partial_v = - \partial_v \partial_w \). Together with \(\|\partial_v\|_{\cC \to \cC}\le 2\), this gives
\[
 \|\partial_vE\|_{\cC^1}
 =\|\partial_vE\|+\sum_{w\in\cJ}\|\partial_v\partial_wE\|\le2\|E\|_{\cC^1}. \qedhere
\]
\end{proof}

Since the Majorana derivatives \(\partial_v E\) vanish when \(x_v\) lies outside the support of \(E\), it follows that \(\cC_\loc\subset\cC^1\). The local operators are dense in \(\cC\) with respect to the $C^*$-norm and so is $\cC^1$. For \(E\in\cC^1\), we define the linear operator
\begin{equation*}
 \cL(E):=-\sum_{u\in\cI}
 \bigl(F_u^*\partial_{F_u}E+(\partial_{F_u}(E^*))^*F_u\bigr).
\end{equation*}
Using Lemma~\ref{lem:comparison-kernel-bounds} we find the bound
\[
 \sum_{u\in\cI}
 \|F_u^*\partial_{F_u}E+(\partial_{F_u}(E^*))^*F_u\|
 \le2(1+\varepsilon)\|\mathsf L\|^{\rm Sch}_W
       \sum_{v\in\cJ}\|\partial_vE\|.
\]
Thus the series converges absolutely in norm. It follows that $\cL$ defines a bounded linear map \(\cL:\cC^1\to\cC\).

The auxiliary evolution \((T_t)_{t\ge0}\) is the semigroup generated by the closure of \(\cL\).
The next lemma shows that it exists and that, for every \(E\in\cC^1\), it solves
\[
 \frac{\dd}{\dd t}T_tE=\cL(T_tE),\qquad T_0E=E,
\]
with the derivative taken in the norm of \(\cC\).

\begin{lemma}\label{lem:auxiliary-evolution}
Assume the hypotheses of Theorem~\ref{thm:comparison}. The operator \(\cL\) with domain \(\cC^1\) is closable, and its closure generates a strongly continuous semigroup \((T_t)_{t\ge0}\) of unital completely positive contractions on \(\cC\). For every \(E\in\cC^1\) and \(t\ge0\), we have \(T_tE\in\cC^1\). For each such \(E\), the map \(t\mapsto T_tE\) is differentiable in the norm of \(\cC\) and satisfies \(\frac{\dd}{\dd t}T_tE=\cL(T_tE)\).
\end{lemma}
\begin{proof}
We follow the strategy of \cite[Theorems~2.2 and~5.2]{BDSP25}, which goes back to the construction of classical interacting particle systems \cite[Chapter~I]{Liggett85}.
We use the Lumer--Phillips theorem \cite[Theorem~II.3.15]{EngelNagel00} to show that the closure of \(\cL\) generates a strongly continuous contraction semigroup \((T_t)_{t\ge0}\).
To apply this theorem, we verify that \(\cL\) is densely defined and dissipative, and that \(\eta-\cL\) has dense range for some \(\eta>0\).
These conditions also imply that \(\cL\) is closable.
We then prove that \(T_t\) is unital and completely positive and maps \(\cC^1\) into itself.

\smallskip
\noindent\emph{Dissipativity of \(\cL\).}
This follows by an approximation argument. To this end, for any finite \(J\subset\cI\) and $E \in \cC$, we define
\[
 \cL_J(E):=-\sum_{u\in J}
 \bigl(F_u^*\partial_{F_u}E+(\partial_{F_u}(E^*))^*F_u\bigr).
\]
Since the sum is over a finite set, it follows that $\cL_J$ is a well-defined bounded linear operator. Plugging the definition of the graded derivative into this expression we find that 
\[
 \cL_J(E)=\sum_{u\in J}
 \bigl(2F_u^*\Theta(E)F_u-F_u^*F_uE-EF_u^*F_u\bigr).
\]
This expression has the bounded-generator form of \cite[Theorem~3]{Lindblad76}, because \(\Theta\) is a \(*\)-automorphism. Thus \(\E^{t\cL_J}\) is completely positive. Since \(\cL_J(\1)=0\), it is unital. Complete positivity means that every matrix amplification of $\E^{t \cL_J}$ is therefore unital and completely positive, so the Kadison--Schwarz inequality gives, for \(k\ge1\) and \(E\in M_k(\cC)\),
\[
 \begin{aligned}
 \bigl((\id_{M_k}\otimes\E^{t\cL_J})(E)\bigr)^*
       (\id_{M_k}\otimes\E^{t\cL_J})(E)
 &\le(\id_{M_k}\otimes\E^{t\cL_J})(E^*E)\le\|E\|^2\1.
 \end{aligned}
\]
Thus \(\E^{t\cL_J}\) is completely contractive. It follows that for any \(\eta>0\) we have that
\[
 \|(\eta-\cL)E\|
 =\lim_{J\uparrow\cI}\|(\eta-\cL_J)E\|
 \ge\eta\|E\|,
\]
for any $E \in \cC^1$. Hence \(\cL\) is dissipative. Since its domain is dense, \cite[Proposition~II.3.14(iv)]{EngelNagel00} implies that \(\cL\) is closable and that its closure \(\overline{\cL}\) is again dissipative.

\smallskip
\noindent\emph{Majorana derivative bounds for \(\E^{t\cL_J}\).}
We first bound the Majorana derivatives of \(\E^{t\cL_J}E\), uniformly in \(J\). For any $v, w \in \cJ$ we define
\[
 \mathsf H_{vw}:=\sum_{u\in\cI}|\sa_{uv}|\,|\sa_{uw}|+\mathsf K_{vw},
\]
which satisfies the bound
\[
 \|\mathsf H\|^{\rm Sch}_W\le C_0^2+\|\mathsf K\|^{\rm Sch}_W.
\]
Using the graded Leibniz rule and the definition $\cL_J$ we find for any $E \in \cC^1$
\begin{equation}\label{eq:partial-generator-commutator}
\begin{aligned}
 \partial_v\cL_J(E)-\cL_J(\partial_vE)
 =\sum_{u\in J}\bigl[
 &-(\partial_vF_u)^*\partial_{F_u}E
 -(\partial_{F_u^*}E)\partial_vF_u\\
 &+[(\partial_vF_u)^*,\Theta(E)]F_u
 +F_u^*[\partial_vF_u,E]\bigr].
\end{aligned}
\end{equation}
To estimate this equation we use the identity
\[
 \partial_vF_u=\sa_{uv}\1
       +\sum_{Z:\,x_u,x_v\in Z}\partial_v\Delta F_{u,Z}.
\]
We first estimate the scalar contributions to the first two terms in Equation \eqref{eq:partial-generator-commutator}. Lemma~\ref{lem:comparison-kernel-bounds} and the definition of \(\mathsf L\) give
\[
 \begin{aligned}
 \|\overline{\sa_{uv}}\partial_{F_u}E
       +\sa_{uv}\partial_{F_u^*}E\|
 &\le2|\sa_{uv}|\sum_{w\in\cJ}\mathsf L_{uw}\|\partial_wE\|\\
 &=\sum_{w\in\cJ}\Bigl[
 |\sa_{uv}|\,|\sa_{uw}|
 +2|\sa_{uv}|\sum_{Z:\,x_u,x_w\in Z}\|\Delta F_{u,Z}\|
 \Bigr]\|\partial_wE\|.
 \end{aligned}
\]
The remaining contributions to the first two terms in Equation \eqref{eq:partial-generator-commutator} contain \(\partial_v(F_u-F_u^0)\). Since \(\|\partial_v\Delta F_{u,Z}\|\le2\|\Delta F_{u,Z}\|\), Lemma~\ref{lem:comparison-kernel-bounds} gives
\[
 \begin{aligned}
 &\|(\partial_v(F_u-F_u^0))^*\partial_{F_u}E\|
 +\|(\partial_{F_u^*}E)\partial_v(F_u-F_u^0)\|\\
 &\quad\le2\|\partial_v(F_u-F_u^0)\|
                 \sum_{w\in\cJ}\mathsf L_{uw}\|\partial_wE\|\\
 &\quad\le4\sum_{w\in\cJ}\mathsf L_{uw}
             \sum_{Z:\,x_u,x_v\in Z}\|\Delta F_{u,Z}\|\|\partial_wE\|.
 \end{aligned}
\]
In the last two terms of Equation \eqref{eq:partial-generator-commutator}, the scalar part of \(\partial_vF_u\) commutes with \(E\) and \(\Theta(E)\). We apply Lemma~\ref{lem:local-graded-commutator} to get
\[
 \|[\partial_v\Delta F_{u,Z},E]\|
 \le2\|\Delta F_{u,Z}\|
       \sum_{w\in\cJ:\,x_w\in Z}\|\partial_wE\|.
\]
The same estimate holds with \((\partial_v\Delta F_{u,Z})^*\) and \(\Theta(E)\). Hence, combining these two estimates gives the bound
\[
 \begin{aligned}
 &\|[(\partial_vF_u)^*,\Theta(E)]F_u\|
       +\|F_u^*[\partial_vF_u,E]\|\\
 &\quad\le4\|F_u\|\sum_{Z:\,x_u,x_v\in Z}\|\Delta F_{u,Z}\|
                 \sum_{w\in\cJ:\,x_w\in Z}\|\partial_wE\|\\
 &\quad=4\|F_u\|\sum_{w\in\cJ}
       \sum_{Z:\,x_u,x_v,x_w\in Z}\|\Delta F_{u,Z}\|\|\partial_wE\|.
 \end{aligned}
\]
Inserting these three bounds into Equation  \eqref{eq:partial-generator-commutator} and summing over \(u\in J\) yields
\begin{equation}\label{eq:partial-generator-commutator-bound}
 \begin{aligned}
 &\|\partial_v\cL_J(E)-\cL_J(\partial_vE)\|\\
 &\quad\le\sum_{w\in\cJ}\sum_{u\in J}\Bigl[
 |\sa_{uv}|\,|\sa_{uw}|
 +2|\sa_{uv}|\sum_{Z:\,x_u,x_w\in Z}\|\Delta F_{u,Z}\|\\
 &\qquad\qquad+4\mathsf L_{uw}\sum_{Z:\,x_u,x_v\in Z}\|\Delta F_{u,Z}\|
 +4\|F_u\|\sum_{Z:\,x_u,x_v,x_w\in Z}\|\Delta F_{u,Z}\|\Bigr]\|\partial_wE\|\\
 &\quad\le\sum_{w\in\cJ}\mathsf H_{vw}\|\partial_wE\|.
 \end{aligned}
\end{equation}
Summing over all \(v \in \cJ\) then shows that
\[
 \sum_{v\in\cJ}\|\partial_v\cL_J(E)\|
 \le\bigl(\|\cL_J\|_{\cC\to\cC}
              +\|\mathsf H\|_{\ell^1\to\ell^1}\bigr)
       \sum_{w\in\cJ}\|\partial_wE\|.
\]
It follows that $\lVert \cL_J \rVert_{\cC^1 \to \cC^1} \le 1 + \|\cL_J\|_{\cC\to \cC} + \| \mathsf H \|_{\ell^1 \to \ell^1}$. In particular, \(\E^{s\cL_J}\) preserves \(\cC^1\). For fixed \(t\ge0\) and \(E\in\cC^1\), the product rule gives
\[
 \frac{\dd}{\dd s}
 \bigl(\E^{(t-s)\cL_J}\partial_v\E^{s\cL_J}E\bigr)
 =\E^{(t-s)\cL_J}
   (\partial_v\cL_J-\cL_J\partial_v)
   \E^{s\cL_J}E.
\]
Integrating from \(0\) to \(t\), we obtain
\begin{equation*}
 \partial_v\E^{t\cL_J}E
 =\E^{t\cL_J}\partial_vE
 +\int_0^t\E^{(t-s)\cL_J}
    (\partial_v\cL_J-\cL_J\partial_v)
    \E^{s\cL_J}E\,\dd s.
\end{equation*}
The contractivity of $\E^{t \cL_J}$ together with the bound from Equation \eqref{eq:partial-generator-commutator-bound} gives the bound
\begin{equation}\label{eq:partial-evolution-majorana-bound}
 \begin{aligned}
 \|\partial_v\E^{t\cL_J}E\|
 &\le\|\partial_vE\|+
 \int_0^t\sum_{w\in\cJ}\mathsf H_{vw}
       \|\partial_w\E^{s\cL_J}E\|\,\dd s\le\sum_{w\in\cJ}(\E^{t\mathsf H})_{vw}\|\partial_wE\|,
 \end{aligned}
\end{equation}
where the second inequality follows by iteration of the first one. Summing over \(v\in\cJ\) and using contractivity on \(\cC\), we obtain
\[
 \begin{aligned}
 \|\E^{t\cL_J}E\|_{\cC^1}
 &\le\|E\|+\E^{t\|\mathsf H\|_{\ell^1\to\ell^1}}
                \sum_{v\in\cJ}\|\partial_vE\|\le\E^{t\|\mathsf H\|_{\ell^1\to\ell^1}}\|E\|_{\cC^1}.
 \end{aligned}
\]
This bound is independent of \(J\).

\smallskip
\noindent\emph{Majorana derivative bounds for \((\eta-\cL_J)^{-1}\).}
Fix \(\eta>C_0^2+\|\mathsf K\|^{\rm Sch}_W\). Since
\(\|\mathsf H\|_{\ell^1\to\ell^1}\le C_0^2+\|\mathsf K\|^{\rm Sch}_W<\eta\), it follows that
\[
 \int_0^\infty\|\E^{-\eta t}\E^{t\cL_J}E\|_{\cC^1}\,\dd t
 \le\frac{\|E\|_{\cC^1}}
 {\eta-\|\mathsf H\|_{\ell^1\to\ell^1}}<\infty.
\]
Thus the Laplace representation of the resolvent converges in \(\cC^1\) and gives
\[
 (\eta-\cL_J)^{-1}E
 =\int_0^\infty\E^{-\eta t}\E^{t\cL_J}E\,\dd t\in\cC^1,
\]
for any $E \in \cC^1$. Since \(\partial_v\) is bounded on \(\cC\), it can be passed through the integral. Equation~\eqref{eq:partial-evolution-majorana-bound} therefore gives
\[
 \begin{aligned}
 \|\partial_v(\eta-\cL_J)^{-1}E\|
 &\le\int_0^\infty\E^{-\eta t}
          \|\partial_v\E^{t\cL_J}E\|\,\dd t\\
 &\le\sum_{w\in\cJ}
       \left(\int_0^\infty\E^{-\eta t}(\E^{t\mathsf H})_{vw}\,\dd t\right)
       \|\partial_wE\|\\
 &=\sum_{w\in\cJ}((\eta-\mathsf H)^{-1})_{vw}\|\partial_wE\|,
 \end{aligned}
\]
where we have used Tonelli's theorem and that \(\eta>\|\mathsf H\|_{\ell^1\to\ell^1}\).

\smallskip
\noindent\emph{Dense range of \(\eta-\cL\) and generation of \(T_t\).}
The preceding estimates show that for $E \in \cC^1$ it holds that \((\eta-\cL_J)^{-1}E \in \cC^1\), hence we can apply $\cL$ to it.
We first note that
\[
 \sum_{\substack{u\in\cI\\v,w\in\cJ}}
 \mathsf L_{uv}((\eta-\mathsf H)^{-1})_{vw}\|\partial_wE\|
 \le\frac{\|\mathsf L\|^{\rm Sch}_W}
 {\eta-\|\mathsf H\|_{\ell^1\to\ell^1}}
 \sum_{w\in\cJ}\|\partial_wE\|<\infty.
\]
The last sum is finite because \(E\in\cC^1\). Lemma~\ref{lem:comparison-kernel-bounds} then gives
\[
 \begin{aligned}
 &\|(\eta-\cL)(\eta-\cL_J)^{-1}E-E\| =\|(\cL_J-\cL)(\eta-\cL_J)^{-1}E\|\\
 &\qquad\le2(1+\varepsilon)
 \sum_{\substack{u\in\cI\setminus J\\v,w\in\cJ}}
 \mathsf L_{uv}((\eta-\mathsf H)^{-1})_{vw}\|\partial_wE\|.
 \end{aligned}
\]
The summands are nonnegative and independent of \(J\), so their sum over \(u\notin J\) tends to zero as \(J\uparrow\cI\). Hence \((\eta-\cL)(\eta-\cL_J)^{-1}E\to E\) in norm. It follows that for any \(B\in\cC\) there is a sequence \(E_n\in\cC^1\) such that \((\eta-\cL)E_n\to B\). We next show that \(\eta-\overline{\cL}\) is surjective. To this end, we note that dissipativity of $\cL$ gives
\[
 \eta\|E_n-E_m\|
 \le\|(\eta-\cL)E_n-(\eta-\cL)E_m\|,
\]
so \(E_n\to E\) for some \(E\in\cC\), and \(\cL E_n\to\eta E-B\). By definition of the closure of a linear operator it follows that $E\in\operatorname{Dom}(\overline{\cL})$ and
\[
 (\eta-\overline{\cL})E=B.
\]
Thus \(\eta-\overline{\cL}\) is surjective. The Lumer--Phillips theorem now yields a strongly continuous contraction semigroup \((T_t)_{t\ge0}\) generated by \(\overline{\cL}\).

\smallskip
\noindent\emph{Unitality, complete positivity and preservation of \(\cC^1\).}
Since \(\cC^1\) is a core for \(\overline{\cL}\), \(\cL_JE\to\cL E\) on this core, and all semigroups are contractions, the first Trotter--Kato approximation theorem \cite[Theorem~III.4.8]{EngelNagel00} gives, for every \(E\in\cC\) and \(t_*>0\),
\begin{equation*}
 \sup_{0\le t\le t_*}\|\E^{t\cL_J}E-T_tE\|\longrightarrow0.
\end{equation*}
Unitality passes to the limit, as does complete positivity. Thus \(T_t\) is unital and completely positive.

For \(E\in\cC^1\), boundedness of \(\partial_v\) allows us to pass to the limit in Equation~\eqref{eq:partial-evolution-majorana-bound}. Summing the resulting estimate gives
\[
 \sum_{v\in\cJ}\|\partial_vT_tE\|
 \le\sum_{v,w\in\cJ}(\E^{t\mathsf H})_{vw}\|\partial_wE\|
 \le\E^{t\|\mathsf H\|_{\ell^1\to\ell^1}}
       \sum_{w\in\cJ}\|\partial_wE\|<\infty.
\]
Thus \(T_tE\in\cC^1\). Finally, \(\cC^1\subset\operatorname{Dom}(\overline{\cL})\) and its invariance give
\[
 \frac{\dd}{\dd t}T_tE
 =\overline{\cL}(T_tE)=\cL(T_tE),
\]
for all $E \in \cC^1$.
\end{proof}

We next estimate the Majorana derivatives of \(T_tE\). To this end, we separate the free part from the perturbation. For \(E\in\cC^1\), define the linear operator
\[
 \cL_0(E):=-\sum_{u\in\cI}
 \bigl((F_u^0)^*\partial_{F_u^0}E
       +(\partial_{F_u^0}(E^*))^*F_u^0\bigr),
\]
and set \(\cL_{\rm res}:=\cL-\cL_0\). By Lemma~\ref{lem:comparison-kernel-bounds} the defining series for \(\cL_0\) converges absolutely, since
\[
 \begin{aligned}
 \sum_{u\in\cI}
 \bigl\|(F_u^0)^*\partial_{F_u^0}E
       +(\partial_{F_u^0}(E^*))^*F_u^0\bigr\|&\le\sum_{u\in\cI}\sum_{v\in\cJ}
       |\sa_{uv}|\,\|\partial_vE\|\le C_0\sum_{v\in\cJ}\|\partial_vE\|<\infty.
 \end{aligned}
\]
We now calculate separately how \(\partial_v\) for any $v \in \cJ$ commutes with \(\cL_0\) and \(\cL_{\rm res}\). For \(E\in\cC^1\), the composition \(\partial_v\cL_0(E)\) is defined because \(\cL_0(E)\in\cC\) and \(\partial_v\) is bounded on \(\cC\). Conversely, \(\cL_0(\partial_vE)\) is defined because Lemma~\ref{lem:derivative-space-complete} gives \(\partial_vE\in\cC^1\). The same reasoning applies to \(\cL_{\rm res}\).

To calculate the commutation relation we first notice that by definition, we can also write the unperturbed operator $\cL_0$ as
\[
\cL_0(E)=\sum_{u\in\cI}(2(F_u^0)^*\Theta(E)F_u^0-(F_u^0)^*F_u^0E-E(F_u^0)^*F_u^0).
\]
Using the graded Leibniz rule together with the identities \( \partial_vF_u^0=\sa_{uv}\1\) and \(\partial_v\Theta = -\Theta \partial_v\), which directly follow from their respective definitions, we find
\[
 \begin{gathered}
 \partial_v\cL_0(E)-\cL_0(\partial_vE)\\
 =\sum_{u\in\cI}\Bigl[
2\overline{\sa_{uv}}\Theta(E)F_u^0
 -2\sa_{uv}(F_u^0)^*E
 -\bigl(\overline{\sa_{uv}}F_u^0-\sa_{uv}(F_u^0)^*\bigr)E
 -\Theta(E)\bigl(\overline{\sa_{uv}}F_u^0-\sa_{uv}(F_u^0)^*\bigr)\Bigr].
 \end{gathered}
\]
Regrouping and using the reconstruction identity of Lemma~\ref{lem:parseval-reconstruction}, we obtain
\begin{equation*}
 \begin{aligned}
 \partial_v\cL_0(E)-\cL_0(\partial_vE)
 &=-\left(\sum_{u\in\cI}
       \bigl(\overline{\sa_{uv}}F_u^0+\sa_{uv}(F_u^0)^*\bigr)\right)E
   +\Theta(E)\left(\sum_{u\in\cI}
       \bigl(\overline{\sa_{uv}}F_u^0+\sa_{uv}(F_u^0)^*\bigr)\right)\\
 &=-\gamma_vE+\Theta(E)\gamma_v=-\partial_vE.
 \end{aligned}
\end{equation*}
To see how the residual part $\cL_{\rm res}$ commutes with the Majorana derivative we define the linear operator
\[
\mathcal R_v (E) \coloneq \partial_v \cL_{\rm res} (E) - \cL_{\rm res} (\partial_v E).
\]
To compute this operator we first express the residual part by
\[
 \begin{aligned}
 \cL_{\rm res}(E)=\sum_{u\in\cI}\Bigl[
 &2F_u^*\Theta(E)F_u-2(F_u^0)^*\Theta(E)F_u^0-\bigl(F_u^*F_u-(F_u^0)^*F_u^0\bigr)E
 -E\bigl(F_u^*F_u-(F_u^0)^*F_u^0\bigr)\Bigr],
 \end{aligned}
\]
where we have used the definition of the graded derivative. We differentiate this expression using the graded Leibniz rule. The terms in which the derivative acts on \(E\) give \(\cL_{\rm res}(\partial_vE)\), since \(\partial_v\Theta=-\Theta\partial_v\) and \(F_u,F_u^0\) are odd. The terms in which it acts on \(F_u\) or \(F_u^0\) give
\[
 \begin{aligned}
\mathcal R_v (E) =\sum_{u\in\cI}\Bigl[
 &-(\partial_vF_u)^*\partial_{F_u}E
       +\overline{\sa_{uv}}\partial_{F_u^0}E\\
 &-(\partial_{F_u^*}E)\partial_vF_u
       +\sa_{uv}\partial_{(F_u^0)^*}E\\
 &+[(\partial_vF_u)^*,\Theta(E)]F_u
       +F_u^*[\partial_vF_u,E]\Bigr],
 \end{aligned}
\]
where we used \(\partial_vF_u^0=\sa_{uv}\1\) and regrouped the products containing \(E\) and \(\Theta(E)\) into graded derivatives and commutators.
We now insert \(\partial_vF_u=\sa_{uv}\1+\partial_v(F_u-F_u^0)\). The scalar terms in each of the first two lines combine by \(\partial_{F_u}-\partial_{F_u^0}=\partial_{F_u-F_u^0}\), while the scalar parts in the commutators vanish. Thus we find
\[
 \begin{aligned}
 \mathcal R_v(E)=\sum_{u\in\cI}\bigl[
 &-\overline{\sa_{uv}}\partial_{F_u-F_u^0}E
  -\sa_{uv}\partial_{(F_u-F_u^0)^*}E\\
 &-(\partial_v(F_u-F_u^0))^*\partial_{F_u}E
  -(\partial_{F_u^*}E)\partial_v(F_u-F_u^0)\\
 &+[(\partial_v(F_u-F_u^0))^*,\Theta(E)]F_u
  +F_u^*[\partial_v(F_u-F_u^0),E]\bigr].
 \end{aligned}
\]
To estimate these terms, we use that
\[
 \partial_v(F_u-F_u^0)
 =\sum_{Z:\,x_u,x_v\in Z}\partial_v\Delta F_{u,Z}.
\]
Each \(\partial_v\Delta F_{u,Z}\) is even, supported in \(Z\), and has norm at most \(2\|\Delta F_{u,Z}\|\). We estimate the three lines in the preceding expansion of \(\mathcal R_v(E)\) separately.

For the first line, Lemma~\ref{lem:local-graded-commutator} applied to the odd operators \(\Delta F_{u,Z}\) and \(\Delta F_{u,Z}^*\) gives
\[
 \max\{\|\partial_{\Delta F_{u,Z}}E\|,
          \|\partial_{\Delta F_{u,Z}^*}E\|\}
 \le\|\Delta F_{u,Z}\|
      \sum_{w:\,x_w\in Z}\|\partial_wE\|.
\]
Using \(F_u-F_u^0=\sum_{Z\ni x_u}\Delta F_{u,Z}\) and adding the two derivative bounds, we obtain
\[
 \begin{aligned}
 &|\sa_{uv}|\bigl(\|\partial_{F_u-F_u^0}E\|
                +\|\partial_{(F_u-F_u^0)^*}E\|\bigr)\le2|\sa_{uv}|\sum_{w\in\cJ}
       \sum_{Z:\,x_u,x_w\in Z}\|\Delta F_{u,Z}\|\,
       \|\partial_wE\|.
 \end{aligned}
\]

For the second line, the bound on \(\partial_v\Delta F_{u,Z}\) gives
\[
 \|\partial_v(F_u-F_u^0)\|
 \le2\sum_{Z:\,x_u,x_v\in Z}\|\Delta F_{u,Z}\|.
\]
Lemma~\ref{lem:comparison-kernel-bounds} bounds both \(\|\partial_{F_u}E\|\) and \(\|\partial_{F_u^*}E\|\) by \(\sum_w\mathsf L_{uw}\|\partial_wE\|\). Applying these bounds to the two products gives
\[
 \begin{aligned}
 \|(\partial_v(F_u-F_u^0))^*\partial_{F_u}E\|
  +\|(\partial_{F_u^*}E)\partial_v(F_u-F_u^0)\|&\le2\sum_{Z:\,x_u,x_v\in Z}\|\Delta F_{u,Z}\|
       \bigl(\|\partial_{F_u}E\|+\|\partial_{F_u^*}E\|\bigr)\\
 &\le4\sum_{w\in\cJ}\mathsf L_{uw}
       \sum_{Z:\,x_u,x_v\in Z}\|\Delta F_{u,Z}\|\,
       \|\partial_wE\|.
 \end{aligned}
\]

For the third line, we apply Lemma~\ref{lem:local-graded-commutator} to the even operators \(\partial_v\Delta F_{u,Z}\) and their adjoints. Since \(\partial_w\Theta=-\Theta\partial_w\) and \(\Theta\) is isometric, \(\|\partial_w\Theta(E)\|=\|\partial_wE\|\). We therefore find
\[
 \begin{aligned}
 \|[(\partial_v\Delta F_{u,Z})^*,\Theta(E)]\|
  +\|[\partial_v\Delta F_{u,Z},E]\| &\le2\|\partial_v\Delta F_{u,Z}\|
       \sum_{w:\,x_w\in Z}\|\partial_wE\|\\
 &\le4\|\Delta F_{u,Z}\|
       \sum_{w:\,x_w\in Z}\|\partial_wE\|.
 \end{aligned}
\]
Multiplying by \(\|F_u\|=\|F_u^*\|\) and summing over \(Z\) bounds the third line by
\[
 4\|F_u\|\sum_{w\in\cJ}
   \sum_{Z:\,x_u,x_v,x_w\in Z}\|\Delta F_{u,Z}\|\,
   \|\partial_wE\|.
\]
Adding these three bounds and summing over \(u\) gives
 \begin{equation}\label{eq:annihilator-gradient-remainder}
 \begin{aligned}
 \|\mathcal R_v(E)\|
 &\le\sum_{u\in\cI}\sum_{w\in\cJ}\Bigl[
 2|\sa_{uv}|\sum_{Z:\,x_u,x_w\in Z}\|\Delta F_{u,Z}\|+4\mathsf L_{uw}\sum_{Z:\,x_u,x_v\in Z}\|\Delta F_{u,Z}\|\\
 &\qquad+4\|F_u\|\sum_{Z:\,x_u,x_v,x_w\in Z}\|\Delta F_{u,Z}\|\Bigr]\|\partial_wE\|\\
&\le \sum_{w\in\cJ}\mathsf K_{vw}\|\partial_wE\|,
  \end{aligned}
 \end{equation}
where $\mathsf K$ is the kernel from Definition \ref{def:comparison-kernels}. Combining the free and residual parts yields
\begin{equation}\label{eq:annihilator-gradient-identity}
 \partial_v\cL(E)
 =\cL(\partial_vE)-\partial_vE+\mathcal R_v(E),
\end{equation}
for all $E \in \cC^1$. For \(E\in\cC^1\), Lemma~\ref{lem:auxiliary-evolution} ensures that \(T_tE\in\cC^1\) and that \(\frac{\dd}{\dd t}T_tE=\cL(T_tE)\) in the norm of \(\cC\). Applying Equation~\eqref{eq:annihilator-gradient-identity} to \(T_tE\) therefore gives
\[
 \begin{aligned}
 \frac{\dd}{\dd t}\partial_vT_tE
 &=\partial_v\!\left(\frac{\dd}{\dd t}T_tE\right)
  =\partial_v\cL(T_tE)=(\cL-\1)(\partial_vT_tE)+\mathcal R_v(T_tE),
 \end{aligned}
\]
where we have used that $\partial_v$ is bounded and \(\partial_vT_tE\in\cC^1\) by Lemma~\ref{lem:derivative-space-complete}. The map \(t\mapsto\mathcal R_v(T_tE)\) is norm continuous, since Equation~\eqref{eq:annihilator-gradient-remainder} and \(\|\partial_w\|\le2\) imply
\[
 \|\mathcal R_v(T_tE)-\mathcal R_v(T_sE)\|
 \le2\|\mathsf K\|^{\rm Sch}_W\|T_tE-T_sE\|\longrightarrow0,
\]
as $t \to s$. For fixed \(t>0\), we differentiate with respect to \(s\in(0,t)\). Since \(\partial_vT_sE\in\cC^1\), the product rule and the semigroup equation give
\[
 \begin{aligned}
 \frac{\dd}{\dd s}
 \bigl(\E^{-(t-s)}T_{t-s}(\partial_vT_sE)\bigr)
 &=\E^{-(t-s)}T_{t-s}
 \left(\frac{\dd}{\dd s}\partial_vT_sE
       -(\cL-\1)(\partial_vT_sE)\right)\\
 &=\E^{-(t-s)}T_{t-s}\mathcal R_v(T_sE).
 \end{aligned}
\]
Integrating from \(0\) to \(t\) yields
\[
 \partial_vT_tE
 =\E^{-t}T_t(\partial_vE)
  +\int_0^t\E^{-(t-s)}T_{t-s}\mathcal R_v(T_sE)\,\dd s.
\]
Contractivity of $T_t$ and Equation \eqref{eq:annihilator-gradient-remainder} now imply
\[
 \|\partial_vT_tE\|
 \le\E^{-t}\|\partial_vE\|
 +\int_0^t\E^{-(t-s)}\sum_{w\in\cJ}
       \mathsf K_{vw}\|\partial_wT_sE\|\,\dd s.
\]
Iterating this integral gives the estimate
\begin{equation}\label{eq:infinite-derivative-semigroup-bound}
 \|\partial_vT_tE\|
 \le\sum_{w\in\cJ}
       (\E^{t(\mathsf K-\1)})_{vw}\|\partial_wE\|.
\end{equation}
Choose \(\delta\in(0,1)\) with \(\|\mathsf K^m\|^{\rm Sch}_W\le(1-\delta)^m\). For \(\ell=j\bmod m\), we have
\[
 \|\mathsf K^j\|^{\rm Sch}_W
 \le(1-\delta)^{j-\ell}\|\mathsf K^\ell\|^{\rm Sch}_W
 \le(1-\delta)^j\max_{0\le k<m}(1-\delta)^{-k}\|\mathsf K^k\|^{\rm Sch}_W
 =:C(1-\delta)^j.
\]
Since \(W\ge1\), the column sums of \(\mathsf K^j\) are at most \(\|\mathsf K^j\|^{\rm Sch}_W\), and summing \eqref{eq:infinite-derivative-semigroup-bound} over \(v\) after expanding the exponential gives
\[
 \begin{aligned}
 \sum_{v\in\cJ}\|\partial_vT_tE\|
 &\le\E^{-t}\sum_{j\ge0}\frac{t^j}{j!}\|\mathsf K^j\|^{\rm Sch}_W
         \sum_{w\in\cJ}\|\partial_wE\|\\
 &\le C\E^{-t}\sum_{j\ge0}\frac{[t(1-\delta)]^j}{j!}
          \sum_{w\in\cJ}\|\partial_wE\|
          =C\E^{-\delta t}\sum_{w\in\cJ}\|\partial_wE\|.
 \end{aligned}
\]

We now show that this decay implies convergence of \(T_tE\) to a scalar. By Lemma~\ref{lem:majorana-averaging}, \(\mathbb E_Z(T_tE)\) converges in norm to \(\omega^{\rm tr}(T_tE)\1\) as \(Z\uparrow\Gamma\). This lemma gives
\[
 \begin{aligned}
 \|T_tE-\omega^{\rm tr}(T_tE)\1\|
 &\le\tfrac12\sum_{v\in\cJ}\|\partial_vT_tE\|\le\tfrac12C\E^{-\delta t}\sum_{v\in\cJ}\|\partial_vE\|.
 \end{aligned}
\]
Since \(T_s\) is unital and contractive, it acts trivially on \(\omega^{\rm tr}(T_tE)\1\), and therefore
\[
 \|T_{t+s}E-\omega^{\rm tr}(T_tE)\1\|
 =\|T_s(T_tE-\omega^{\rm tr}(T_tE)\1)\|
 \le\|T_tE-\omega^{\rm tr}(T_tE)\1\|.
\]
The triangle inequality now gives, for all \(s,t\ge0\),
\[
 \begin{aligned}
 \|T_{t+s}E-T_tE\|
 &\le2\|T_tE-\omega^{\rm tr}(T_tE)\1\|\\
 &\le C\E^{-\delta t}\sum_{v\in\cJ}\|\partial_vE\|.
 \end{aligned}
\]
This bound is uniform in \(s\), so \(T_tE\) is Cauchy in norm, and its distance from the closed subspace \(\C\1\) tends to zero.
Hence \(T_tE\) converges in norm to a multiple of the identity for every \(E\in\cC^1\).
Since \(\cC^1\) is dense in \(\cC\) and every \(T_t\) is a contraction, the same holds for every \(E\in\cC\).
We write the limit as \(\omega_F(E)\1\), so that
\begin{equation}\label{eq:auxiliary-scalar-limit}
 \lim_{t\to\infty}\|T_tE-\omega_F(E)\1\|=0,
 \qquad
 \omega_F(E)=\lim_{t\to\infty}\omega^{\rm tr}(T_tE),
 \qquad E\in\cC.
\end{equation}
The map \(\omega_F\) is linear, and it is positive and unital because every \(T_t\) is.
Thus \(\omega_F\) is a state on \(\cC\).
The semigroup property gives \(\omega_F(T_sE)=\omega_F(E)\) for all \(E\in\cC\) and \(s\ge0\), so \(\omega_F\) is invariant under the auxiliary evolution.
It is the only invariant state, since \(\nu\circ T_t=\nu\) for all \(t\ge0\) implies \(\nu(E)=\lim_{t\to\infty}\nu(T_tE)=\omega_F(E)\).

\subsection{Decomposition of local observables}
\label{sec:master-proof}

\begin{proof}[Proof of Theorem~\ref{thm:comparison}(i)]
Let \(\omega_F\) be the state defined by \eqref{eq:auxiliary-scalar-limit}, and let \(A\in\cC_Y\) be arbitrary. We obtain the asserted decomposition of \(A\) by integrating \(-\cL(T_tA)\) over time. The two terms in the generator determine the coefficients \(X_u\) and \(Y_u\). We first show that their defining integrals converge and satisfy the required bounds. Equation~\eqref{eq:infinite-derivative-semigroup-bound} and Lemma~\ref{lem:comparison-kernel-bounds} give
\[
 \max\{\|\partial_{F_u}T_tA\|,\|\partial_{F_u}T_t(A^*)\|\}
 \le\sum_{w\in\cJ}
      \bigl(\mathsf L\E^{t(\mathsf K-\1)}\bigr)_{uw}\|\partial_wA\|.
\]
The bound \(\|\mathsf K^m\|^{\rm Sch}_W<1\) for some \(m\in\mathbb N\) ensures convergence of the Neumann series. Expanding the exponential and using Tonelli's theorem therefore gives
\[
 \int_0^\infty\mathsf L\E^{t(\mathsf K-\1)}\,\dd t
 =\sum_{j\ge0}\mathsf L\mathsf K^j
       \int_0^\infty\E^{-t}\frac{t^j}{j!}\,\dd t
 =\mathsf T.
\]
Integrating the above estimate yields the bound
\begin{equation}\label{eq:integrated-coefficient-bound}
 \int_0^\infty\max\{\|\partial_{F_u}T_tA\|,\|\partial_{F_u}T_t(A^*)\|\}\,\dd t
 \le\sum_{w\in\cJ}\mathsf T_{uw}\|\partial_wA\|<\infty.
\end{equation}
This estimate allows us to define the coefficients by the norm-convergent Bochner integrals
\[
 \begin{aligned}
 X_u&:=\int_0^\infty\partial_{F_u}T_tA\,\dd t,\\
 Y_u&:=\left(\int_0^\infty\partial_{F_u}T_t(A^*)\,\dd t\right)^*.
 \end{aligned}
\]
The triangle inequality ensures the claimed bound
\[
 \max\{\|X_u\|,\|Y_u\|\}
 \le\sum_{w\in\cJ}\mathsf T_{uw}\|\partial_wA\|.
\]
To integrate the generator and exchange the sum and integral, we use the definition of \(\cL\), the equality \((T_tA)^*=T_t(A^*)\), and the estimate \(\|F_u\|\le1+\varepsilon\) from Lemma~\ref{lem:comparison-kernel-bounds}. They give
\[
 \begin{aligned}
 \int_0^\infty\|\cL(T_tA)\|\,\dd t
 &\le\sum_{u\in\cI}\|F_u\|
       \int_0^\infty\bigl(\|\partial_{F_u}T_tA\|
                        +\|\partial_{F_u}T_t(A^*)\|\bigr)\,\dd t\\
 &\le2(1+\varepsilon)
       \sum_{u\in\cI}\sum_{w\in\cJ}
           \mathsf T_{uw}\|\partial_wA\|\\
 &\le2(1+\varepsilon)\|\mathsf T\|^{\rm Sch}_W
       \sum_{w\in\cJ}\|\partial_wA\|<\infty.
 \end{aligned}
\]
Here Tonelli's theorem justifies exchanging the sum and integral, Equation~\eqref{eq:integrated-coefficient-bound} gives the second inequality, and the last sum is finite because \(A\) is local. By Equation~\eqref{eq:auxiliary-scalar-limit}, \(T_tA\to\omega_F(A)\1\). We can now integrate from \(0\) to \(\infty\) and insert the definitions of \(X_u\) and \(Y_u\). Absolute convergence and the definition of $\cL$ give
\[
 \begin{aligned}
 A-\omega_F(A)\1
 &=-\int_0^\infty\cL(T_tA)\,\dd t\\
 &=\sum_{u\in\cI}\left[
 F_u^*\int_0^\infty\partial_{F_u}T_tA\,\dd t
 +\left(\int_0^\infty\partial_{F_u}T_t(A^*)\,\dd t\right)^*F_u\right]\\
 &=\sum_{u\in\cI}(F_u^*X_u+Y_uF_u),
 \end{aligned}
\]
where we have used \(T_t(A^*)=(T_tA)^*\).
Finally, let \(\omega\) be a state with \(\omega(F_u^*F_u)=0\) for all \(u\in\cI\). The Cauchy--Schwarz inequality gives \(\omega(F_u^*X_u)=\omega(Y_uF_u)=0\) for every \(u\). Applying \(\omega\) term by term to the decomposition therefore yields \(\omega(A)=\omega_F(A)\) for every \(A\in\cC_\loc\), and hence \(\omega=\omega_F\) by density. This finishes the proof of (i).
\end{proof}

\section{Technical details: locality and filtering}
\label{sec:thermal-family}

We prove the results stated in Section~\ref{sec:thermofield}.

\subsection{The diagonal thermal extension}
\label{sec:lift-proof}

We prove Proposition~\ref{prop:lift} using the modular right action and parity twist described in Section~\ref{sec:canonical-extension}.

\begin{proof}[Proof of Proposition~\ref{prop:lift}]
We keep the given representation of the first copy and use the KMS identity of Definition~\ref{def:equilibrium}(i) to construct the action of the second copy. Together, these will define \(\widehat\pi\).
Fix \(B\in\A_\an(\tau)\).
For \(A\in\A_\an(\tau)\), we prescribe
\begin{equation}\label{eq:KMS-right-action}
 R(B)\pi(A)\Omega:=\pi(A\tau_{\I\beta/2}(B))\Omega.
\end{equation}
We first show that this depends only on the vector \(\pi(A)\Omega\) and defines a bounded operator.
For \(X\in\A_\an(\tau)\), the \(\tau\)-invariance of \(\rho\), extended to complex times by analyticity, and the KMS identity give
\begin{equation}\label{eq:KMS-half-shift-norm}
\begin{aligned}
\rho\!\left(\tau_{-\I\beta/2}(X)\tau_{-\I\beta/2}(X)^*\right)
&=\rho\!\left(\tau_{-\I\beta/2}(X)\tau_{\I\beta/2}(X^*)\right)\\
&=\rho\!\left(\tau_{-\I\beta/2}
       \bigl(X\tau_{\I\beta}(X^*)\bigr)\right)\\
&=\rho\!\left(X\tau_{\I\beta}(X^*)\right)=\rho(X^*X).
\end{aligned}
\end{equation}
Applying \eqref{eq:KMS-half-shift-norm} first to \(X=A\tau_{\I\beta/2}(B)\) and then to \(X=A\), and using \(\tau_{-\I\beta/2}(A\tau_{\I\beta/2}(B))=\tau_{-\I\beta/2}(A)B\), gives
\[
 \begin{aligned}
 \|R(B)\pi(A)\Omega\|^2
 &=\rho\bigl((A\tau_{\I\beta/2}(B))^*A\tau_{\I\beta/2}(B)\bigr)\\
 &=\rho\bigl(\tau_{-\I\beta/2}(A)BB^*
                   \tau_{-\I\beta/2}(A)^*\bigr)\\
 &\le\|B\|^2\rho\bigl(\tau_{-\I\beta/2}(A)
                   \tau_{-\I\beta/2}(A)^*\bigr)
 =\|B\|^2\|\pi(A)\Omega\|^2.
 \end{aligned}
\]
This bound shows that the map \(R(B)\) is well defined on \(\pi(\A_\an(\tau))\Omega\) and bounded.
Since \(\A_\an(\tau)\) is norm dense and \(\Omega\) is cyclic, \(\pi(\A_\an(\tau))\Omega\) is dense in \(\mathcal H\), and \(R(B)\) extends uniquely to a bounded operator with \(\|R(B)\|\le\|B\|\).
The map \(B\mapsto R(B)\) is linear on \(\A_\an(\tau)\) and contractive, so it extends to a bounded linear map \(R:\A\to\mathcal B(\mathcal H)\) with \(\|R(B)\|\le\|B\|\) and \(R(\1)=\1\).

The map \(R\) satisfies for all \(A,B\in \A\)
\begin{equation}\label{eq:KMS-right-properties}
  R(AB)=R(B)R(A),\quad
  R(B^*)=R(B)^*,\quad
  [R(B),\pi(A)]=0\,.
\end{equation}
For \(A,B,C\in\A_\an(\tau)\), the identities in \eqref{eq:KMS-right-properties} follow from \eqref{eq:KMS-right-action} together with the KMS identity: 
\[
R(B)R(A)\pi(C)\Omega=\pi(C\tau_{\I\beta/2}(A)\tau_{\I\beta/2}(B))\Omega=R(AB)\pi(C)\Omega\,,
\] 
\[
 \langle R(B)\pi(A)\Omega,\pi(C)\Omega\rangle
 =\rho(\tau_{-\I\beta/2}(B^*)A^*C)
 =\rho(A^*C\tau_{\I\beta/2}(B^*))
 =\langle\pi(A)\Omega,R(B^*)\pi(C)\Omega\rangle\,,
\]
\[
\begin{aligned}
 [R(B),\pi(A)]\pi(C)\Omega
 &=R(B)\pi(AC)\Omega-\pi(A)R(B)\pi(C)\Omega\\
 &=\pi(AC\tau_{\I\beta/2}(B))\Omega
   -\pi(A)\pi(C\tau_{\I\beta/2}(B))\Omega=0\,.
\end{aligned}
\]
Density, boundedness and norm continuity extend the three identities to \(\A\).

Next we construct the parity unitary.
On the dense subspace \(\pi(\A)\Omega\), we prescribe \(U\pi(A)\Omega:=\pi(\Theta(A))\Omega\).
Evenness of \(\rho\) gives \(\|\pi(\Theta(A))\Omega\|^2=\rho(\Theta(A^*A))=\|\pi(A)\Omega\|^2\), so \(U\) extends to an isometry.
Since \(\Theta^2=\id\), \(U\) is a self-adjoint unitary satisfying
\begin{equation}\label{eq:KMS-parity-unitary}
 U\Omega=\Omega,\qquad U\pi(B)U=\pi(\Theta(B)),\qquad B\in\A.
\end{equation}
Since \(\tau\) is even, analytic continuation gives \(\Theta\tau_z(B)=\tau_z(\Theta(B))\) for \(B\in\A_\an(\tau)\), so that
\[
\begin{aligned}
 UR(B)U\pi(A)\Omega
 &=U\pi(\Theta(A)\tau_{\I\beta/2}(B))\Omega\\
 &=\pi(A\tau_{\I\beta/2}(\Theta(B)))\Omega
 =R(\Theta(B))\pi(A)\Omega\,.
\end{aligned}
\]
Boundedness and density give
\begin{equation}\label{eq:KMS-right-parity}
 UR(B)U=R(\Theta(B)),\qquad B\in\A.
\end{equation}

Recall that \(\ft\) is the complex-linear Majorana reversal: it fixes each \(\gamma_u\) and satisfies \(\ft(AB)=\ft(B)\ft(A)\) and \(\ft(A^*)=\ft(A)^*\) (Lemma~\ref{lem:majorana-reversal}).
Together with \eqref{eq:KMS-right-properties}, these identities show that \(R\circ\ft\) is a unital \(*\)-representation whose range commutes with \(\pi(\A)\).
To implement the graded commutation relations between the two copies, we define, for homogeneous \(B\in\A\),
\begin{equation}\label{eq:rightrep}
 r(B):=\I^{|B|}R(\ft(B))U^{|B|},
\end{equation}
and extend linearly.
Since \(\ft\) preserves parity, \eqref{eq:KMS-right-parity} gives \(UR(\ft(B))=(-1)^{|B|}R(\ft(B))U\).
Hence, by \eqref{eq:rightrep}, for homogeneous \(A,B\),
\[
 r(A)r(B)
 =\I^{|A|+|B|}(-1)^{|A||B|}
       R(\ft(A))R(\ft(B))U^{|A|+|B|}
 =r(AB),
\]
and
\[
 r(B)^*
 =(-\I)^{|B|}U^{|B|}R(\ft(B^*))
 =\I^{|B|}R(\ft(B^*))U^{|B|}
 =r(B^*),
\]
so \(r\) is a unital \(*\)-representation.
The two copies graded-commute: for homogeneous \(A,B\),
\begin{equation}\label{eq:KMS-graded-copies}
 \begin{aligned}
 \pi(A)r(B)
 &=\I^{|B|}\pi(A)R(\ft(B))U^{|B|}\\
 &=\I^{|B|}R(\ft(B))\pi(A)U^{|B|}\\
 &=(-1)^{|A||B|}\I^{|B|}R(\ft(B))U^{|B|}\pi(A)\\
 &=(-1)^{|A||B|}r(B)\pi(A),
 \end{aligned}
\end{equation}
by \eqref{eq:KMS-right-properties} and \eqref{eq:KMS-parity-unitary}.
In particular, \(\{r(\gamma_u),r(\gamma_v)\}=2\delta_{uv}\1\) and \(\{\pi(\gamma_u),r(\gamma_v)\}=0\).
Together with the CAR for \(\pi(\gamma_u)\), these are the defining relations of \(\A^{\st}\). Its universal property therefore gives a unital \(*\)-representation \(\widehat\pi\) with \(\widehat\pi(A^{(1)}B^{(2)})=\pi(A)r(B)\).
Define the vector state
\[
 \widehat\rho(\widehat C):=\langle\Omega,\widehat\pi(\widehat C)\Omega\rangle,
 \qquad \widehat C\in\A^{\st}.
\]
Since \(r(\1)=\1\), we have \(\widehat\pi(A^{(1)})=\pi(A)\) and therefore
\[
 \widehat\rho(A^{(1)})=\langle\Omega,\pi(A)\Omega\rangle=\rho(A),
 \qquad A\in\A.
\]
Thus \(\widehat\rho\) restricts to \(\rho\) on the left copy.
For analytic \(D\), invariance of \(\rho\) gives
\[
 \langle\Omega,R(D)\Omega\rangle
 =\rho(\tau_{\I\beta/2}(D))=\rho(D).
\]
By density and continuity, the identity extends to every \(D\in\A\). Since \(U\Omega=\Omega\) and \(\rho\) vanishes on odd elements, we obtain
\[
 \widehat\rho(B^{(2)})=\I^{|B|}\rho(\ft(B))
                    =\rho(\ft(B))
\]
for homogeneous \(B\), and then for all \(B\) by linearity.

To prove evenness, \eqref{eq:KMS-right-parity} gives \(UR(\ft(B))U=R(\Theta(\ft(B)))=(-1)^{|B|}R(\ft(B))\) for homogeneous \(B\), so that \(Ur(B)U=(-1)^{|B|}r(B)=r(\Theta(B))\). Together with \eqref{eq:rightrep} and \eqref{eq:KMS-parity-unitary}, this gives
\[
 \begin{aligned}
 \widehat\rho\bigl((\Theta(A))^{(1)}(\Theta(B))^{(2)}\bigr)
 &=\langle\Omega,\pi(\Theta(A))r(\Theta(B))\Omega\rangle\\
 &=\langle\Omega,U\pi(A)U\,Ur(B)U\Omega\rangle\\
 &=\langle U\Omega,\widehat\pi(A^{(1)}B^{(2)})U\Omega\rangle\\
 &=\widehat\rho(A^{(1)}B^{(2)}).
 \end{aligned}
\]
Ordered products \(A^{(1)}B^{(2)}\) have dense linear span in \(\A^{\st}\), so \(\widehat\rho\) is even.

Formula \eqref{eq:KMS-right-action}, applied to the cyclic vector, gives
\begin{equation}\label{eq:KMS-right}
 R(\tau_{-\I\beta/2}(A))\Omega=\pi(A)\Omega,
 \qquad A\in\A_\an(\tau).
\end{equation}
For odd \(A\in\A_\an(\tau)\), \eqref{eq:rightrep} and \eqref{eq:KMS-right}, together with \(\ft^2=\id\) and \((\tau_{-\I\beta/2}(A))^*=\tau_{\I\beta/2}(A^*)\), finally give \eqref{eq:annihilation-identities}:
\[
 \begin{aligned}
 \widehat\pi\bigl(A^{(1)}+
       \I(\ft(\tau_{-\I\beta/2}(A)))^{(2)}\bigr)\Omega
 &=\pi(A)\Omega-R(\tau_{-\I\beta/2}(A))\Omega=0,\\
 \widehat\pi\bigl((\ft(A^*))^{(2)}-
       \I((\tau_{-\I\beta/2}(A))^*)^{(1)}\bigr)\Omega
 &=\I R(A^*)\Omega-\I\pi(\tau_{\I\beta/2}(A^*))\Omega=0.
 \end{aligned}
\]

For uniqueness, let \(\widehat\pi'\) be another representation with the prescribed left action and the first identity in \eqref{eq:annihilation-identities}.
For odd \(B\in\A_\an(\tau)\), that identity with \(A=\tau_{\I\beta/2}(B)\) gives \(\widehat\pi'((\ft(B))^{(2)})\Omega=\I\pi(\tau_{\I\beta/2}(B))\Omega=\widehat\pi((\ft(B))^{(2)})\Omega\).
For homogeneous \(C\in\A\), the graded commutation relation \eqref{eq:KMS-graded-copies}, valid in both representations, then gives
\[
 \begin{aligned}
 \widehat\pi'((\ft(B))^{(2)})\pi(C)\Omega
 &=(-1)^{|C|}\pi(C)\widehat\pi'((\ft(B))^{(2)})\Omega\\
 &=(-1)^{|C|}\pi(C)\widehat\pi((\ft(B))^{(2)})\Omega
 =\widehat\pi((\ft(B))^{(2)})\pi(C)\Omega.
 \end{aligned}
\]
Since these vectors span a dense subspace, the two operators representing the right copy agree.
Odd analytic elements are dense in the odd part of \(\A\), and \(\ft\) maps that part onto itself. Norm continuity gives equality on every odd right-copy element.
Writing an even \(B\) as \(\gamma_u(\gamma_uB)\) gives equality on even right-copy elements.
Since the two copies generate \(\A^\st\), \(\widehat\pi'=\widehat\pi\).

Finally, let \((\pi',\mathcal H',\Omega')\) be another GNS triple of \(\rho\), and let \(V:\mathcal H\to\mathcal H'\) be the unitary with \(V\pi(A)\Omega=\pi'(A)\Omega'\).
Then \(V\widehat\pi(\cdot)V^*\) is a representation of \(\A^\st\) on \(\mathcal H'\) with left action \(\pi'\), and it satisfies the first identity in \eqref{eq:annihilation-identities} for the vector \(\Omega'=V\Omega\).
By the uniqueness just proved, applied to \((\pi',\mathcal H',\Omega')\), it is the representation constructed from that triple, and the identity \(\langle\Omega',V\widehat\pi(\widehat C)V^*\Omega'\rangle=\langle\Omega,\widehat\pi(\widehat C)\Omega\rangle\), valid for all \(\widehat C\in\A^{\st}\), shows that the vector states agree.
\end{proof}

\subsection{Filters and the free annihilation family}
\label{sec:free-thermal}

We first establish decay estimates for the filters \(f\) and \(q_\beta\) introduced in Section~\ref{sec:thermal-family-statement}.

\begin{lemma}\label{lem:Gevrey}
Fix \(\sG>1\), and use the filter \(f\) chosen in Section~\ref{sec:thermal-family-statement}.
There are \(C_f,a_f>0\), depending only on the model data and \(\sG\), such that, for every \(\beta\ge0\),
\begin{equation}\label{eq:Gevrey-tail}
 |f_\beta(t)|\le |f(t)|+|f(t+\I\beta/2)|\le C_f\E^{-a_f|t|^{1/\sG}}, \qquad t \in \R.
\end{equation}
There are also model constants \(C_*,a_*>0\) such that, for every \(\beta\ge0\),
\begin{equation}\label{eq:Gevrey-coarse-tail}
 |f_\beta(t)|\le |f(t)|+|f(t+\I\beta/2)|\le C_*\E^{-a_*|t|^{1/2}}, \qquad t \in \R.
\end{equation}
\end{lemma}
\begin{proof}
Since \(\omega\le-g/2\) on \(\supp\widehat f\), for every integer \(j\ge0\), every \(\beta\ge0\), and every \(\omega\in\supp\widehat f\),
\[
 \left|\partial_\omega^j\E^{\beta\omega/2}\right|
 =(\beta/2)^j\E^{\beta\omega/2}
 \le(\beta/2)^j\E^{-\beta g/4} =(2/g)^j(\beta g/4)^j\E^{-\beta g/4}
 \le(2/g)^j j!.
\]
The last inequality uses \(x^j\E^{-x}\le j!\) for \(x\ge0\). Fix \(\sigma\in\{\sG,2\}\). Since \(\sigma>1\), we find for integers \(0\le j\le n\),
\[
 \binom nj j!((n-j)!)^{\sigma}
 =n!((n-j)!)^{\sigma-1}
 \le(n!)^{\sigma}.
\]
Leibniz's rule and Lemma~\ref{lem:gevrey-cutoff} therefore give, uniformly in \(\beta\ge0\),
\[
 \|(\E^{\beta\cdot/2}\widehat f)^{(n)}\|_1
 \le\sum_{j=0}^n\binom nj(2/g)^j j!\|\widehat f^{(n-j)}\|_1
 \le C_\sigma^{n+1}(n!)^\sigma.
\]
Together with the bounds for \(\widehat f\) in Lemma~\ref{lem:gevrey-cutoff}, this yields a constant \(K_\sigma\ge1\) that depends only on the model data and $\sigma$ (i.e. $K_2$ is independent of $\sG$) such that
\begin{equation*}
 \|q^{(n)}\|_1\le K_\sigma^{n+1}(n!)^\sigma,
\end{equation*}
for all $q\in\{\widehat f,\E^{\beta\cdot/2}\widehat f\}$ and $n\in\mathbb N_{0}$.
Since \(q\in C_c^\infty(\R)\), all boundary terms vanish when integrating by parts. For every integer \(n\ge0\) and every \(t\ne0\),
\[
 \check q(t)
 =\frac1{2\pi}\int_\R q(\omega)\E^{-\I t\omega}\,\dd\omega
 =\frac1{2\pi(\I t)^n}\int_\R q^{(n)}(\omega)\E^{-\I t\omega}\,\dd\omega,
\]
and hence
\[
 |\check q(t)|\le\frac{\|q^{(n)}\|_1}{2\pi|t|^n}
 \le K_\sigma^{n+1}(n!)^\sigma|t|^{-n}.
\]
For \(|t|\ge2^{\sigma+1}K_\sigma\), choose \(n=\lfloor(|t|/(2K_\sigma))^{1/\sigma}\rfloor\).
Then
\[
 n\ge\tfrac12(|t|/(2K_\sigma))^{1/\sigma},
 \qquad \frac{K_\sigma n^\sigma}{|t|}\le\frac12.
\]
Using \(n!\le n^n\), we obtain
\[
 |\check q(t)|
 \le K_\sigma\left(\frac{K_\sigma n^\sigma}{|t|}\right)^n
 \le K_\sigma2^{-n}
 \le K_\sigma\E^{-a_\sigma|t|^{1/\sigma}},
\]
where $a_\sigma:=\log2 / (2(2K_\sigma)^{1/\sigma})>0$. The bound \(|\check q(t)|\le\|q\|_1/(2\pi)\le K_\sigma\) covers the case \(|t|\le2^{\sigma+1}K_\sigma\). After increasing the prefactor, this gives \(|\check q(t)|\le C_\sigma\E^{-a_\sigma|t|^{1/\sigma}}\) for all \(t\in\R\).
Adding the bounds for \(q=\widehat f,\E^{\beta\cdot/2}\widehat f\), with \(\sigma=\sG\) and \(\sigma=2\), proves \eqref{eq:Gevrey-tail} and \eqref{eq:Gevrey-coarse-tail}, respectively. The constants are uniform in \(\beta\), and for \(\sigma=2\) also in \(\sG\).
\end{proof}

Ding, Li and Lin \cite[Lemmas~29--30]{DLL25} use the same passage from Gevrey derivative bounds to stretched-exponential Fourier decay as in Lemma~\ref{lem:Gevrey} to obtain time kernels for Gibbs-sampler jump operators.

\begin{lemma}\label{lem:FG-filter}
Let \(q_\beta\) be defined by \eqref{eq:FG-kernel}. There is a model constant \(C>0\), and for every real \(j>0\) there is \(C_j>0\), depending only on the model data and \(j\), such that the following bounds hold for every \(\beta>0\) and \(t\in\R\):
\begin{equation}\label{eq:FG-moments}
 \begin{gathered}
 | q_\beta(t)|\le\frac{C}{\beta/(1+\beta)+|t|}\E^{-|t|/(2\beta)},
 \qquad \| q_\beta\|_1\le C\log(3+\beta),\\
 \int_\R|t|^j| q_\beta(t)|\,\dd t\le C_j\beta^j.
 \end{gathered}
\end{equation}
\end{lemma}
\begin{proof}
Fix \(\beta>0\) and put \(\eta:=\beta/(1+\beta)\). For \(z=x+\I y\) with \(|y|\le(2\beta)^{-1}\), the bound \(\beta|y|\le1/2\) gives
\[
 \RealPart(1+\E^{\beta z})
 =1+\E^{\beta x}\cos(\beta y)\ge1+\E^{\beta x}\cos(1/2)\ge\cos(1/2)(1+\E^{\beta x})>0.
\]
Since \(1+\E^{\beta z}\) lies in the right half-plane, we use its holomorphic square root with positive real part. This makes \(\widehat q_\beta\) holomorphic in a neighborhood of the closed strip. Here and below, \(C_a>0\) denotes a constant depending only on \(a\), whose value may change from line to line. Moreover, \(|\widehat q_\beta(x+\I y)|\le C_a\E^{-a(\eta x)^2}\), since \(|\eta y|\le1/2\). Integrating this bound and substituting \(u=\eta x\) gives \(\eta\|\widehat q_\beta(\cdot+\I y)\|_1\le C_a\).
Differentiation and the substitution \(u=\E^{\beta x}\) give, uniformly for \(|y|\le(2\beta)^{-1}\),
\[
 \int_\R\left|\partial_x(1+\E^{\beta(x+\I y)})^{-1/2}\right|\,\dd x
 \le C\int_0^\infty(1+u)^{-3/2}\,\dd u=2C.
\]
The product rule and the integrability of the Gaussian and its derivative yield
\[
 \sup_{|y|\le(2\beta)^{-1}}
 \left(\eta\|\widehat q_\beta(\cdot+\I y)\|_1
       +\|(\widehat q_\beta)'(\cdot+\I y)\|_1\right)
 \le C_a.
\]
We can now apply the contour-shift identity from \cite[Chapter~4, proof of Theorem~2.1]{SteinShakarchi03} to \(\widehat q_\beta\) and \((\widehat q_\beta)'\), shifting by \(-\I\operatorname{sgn}(t)/(2\beta)\). Together with \(\check{(\widehat q_\beta)'}(t)=\I t q_\beta(t)\) and the preceding \(L^1\) bounds, it gives
\[
 | q_\beta(t)|
 \le C_a\E^{-|t|/(2\beta)}\min\{\eta^{-1},|t|^{-1}\}
 \le\frac{2C_a}{\eta+|t|}\E^{-|t|/(2\beta)},
 \qquad t\ne0.
\]
At \(t=0\), the Fourier representation \eqref{eq:thermal-filter-definition} and the preceding \(L^1\) bound with \(y=0\) give
\[
 |q_\beta(0)|
 =\frac1{2\pi}\left|\int_\R\widehat q_\beta(\omega)\,\dd\omega\right|
 \le\frac1{2\pi}\|\widehat q_\beta\|_1
 \le\frac{C_a}{2\pi\eta}.
\]
Thus the pointwise estimate in \eqref{eq:FG-moments} holds also at \(t=0\) after increasing the constant. Integrating the pointwise estimate in \eqref{eq:FG-moments} over \(\R\) gives \(\| q_\beta\|_1\le C_a\log(3+\beta)\) by splitting at \(|t|=\beta\). Since \(\eta+|t|\ge|t|\), substituting \(|t|=\beta u\) also gives
\[
 \int_\R|t|^j| q_\beta(t)|\,\dd t\le C_{a,j}\beta^j
 \qquad(j>0). \qedhere
\]
\end{proof}

\paragraph{The free annihilation family.}

We now prove Proposition~\ref{prop:free-thermal-family}.
We adapt the functional-calculus estimates of \cite[Lemmas~A.7--A.8]{StabilityBdG26} to general spectral contours, adding an explicit polynomial bound in the distance to the spectrum for use in Lemma~\ref{lem:bounded-temperature-calculus}.

\begin{lemma}\label{lem:weighted-contour-calculus}
Assume either of the Schur locality assumptions on \(h\) in Hypothesis~\ref{hyp:locality}. In the respective locality regimes, there are model constants \(C_{\rm exp},c_{\rm exp}>0\) and \(C_{\rm pol}>0\) such that, for every \(0<\Delta\le1\) and every \(z\in\C\) with \(\dist(z,\operatorname{spec}(h))\ge\Delta\), the following bounds hold.
\begin{enumerate}[label=(\roman*),leftmargin=*]
\item In regime~\ref{ass:exponential},
\begin{equation}\label{eq:exponential-resolvent-bound}
 \|(z-h)^{-1}\|^{\rm Sch}_{c_{\rm exp}\Delta,\theta_{\rm in}}
 \le C_{\rm exp}\Delta^{-1-d_{\rm vol}/\theta_{\rm in}}.
\end{equation}
\item In regime~\ref{ass:polynomial-weight}, put \(m_d:=\lfloor d_{\rm vol}/2\rfloor+1\),
\begin{equation}\label{eq:polynomial-resolvent-bound}
 \|(z-h)^{-1}\|_{p_h}^{\rm Sch,pol}
 \le C_{\rm pol}\Delta^{-(m_d+1)(\lceil p_h\rceil+1)}.
\end{equation}
\end{enumerate}
\end{lemma}
\begin{proof}
In regime~\ref{ass:exponential}, fix \(0<\Delta\le1\) and \(z\) with \(\dist(z,\operatorname{spec}(h))\ge\Delta\).
For \(y\in\Gamma\), \(L>0\) and \(0<\eta\le\mu_h/2\), define the weights \(W_{y,L}(x):=\exp(\eta\min\{d(x,y)^{\theta_{\rm in}},L\})\).
Since \(x\mapsto\min\{d(x,y)^{\theta_{\rm in}},L\}\) is \(1\)-Lipschitz in the metric \(d^{\theta_{\rm in}}\), for \(x,x'\in\Gamma\) we have
\[
 \|(W_{y,L}hW_{y,L}^{-1}-h)_{xx'}\|
 =\left|\frac{W_{y,L}(x)}{W_{y,L}(x')}-1\right|\|h_{xx'}\|\le\bigl(\E^{\eta d(x,x')^{\theta_{\rm in}}}-1\bigr)\|h_{xx'}\|.
\]
The weighted Schur bound on \(h\) and the Schur test therefore give
\begin{equation*}
 \|W_{y,L}hW_{y,L}^{-1}-h\| \le C_h\sup_{s\ge0}(\E^{\eta s}-1)\E^{-\mu_hs}
 \le\frac{2C_h}{\E\mu_h}\eta\le C\eta,
\end{equation*}
uniformly in \(y,L\).
Choose \(\eta=c\Delta\), with a model constant \(c>0\) small enough that \(\eta\le\min\{1,\mu_h/2\}\) and \(C\eta\le\Delta/2\).
Then \(\|(W_{y,L}hW_{y,L}^{-1}-h)(z-h)^{-1}\|\le1/2\), so the Neumann series for the conjugated resolvent gives
\[
 \|W_{y,L}(z-h)^{-1}W_{y,L}^{-1}\|
 =\|(z-W_{y,L}hW_{y,L}^{-1})^{-1}\|
 \le\frac2\Delta.
\]
Since \(W_{y,L}(y)=1\) and taking a matrix block does not increase the operator norm, we have
\[
 \E^{\eta\min\{d(x,y)^{\theta_{\rm in}},L\}}\|((z-h)^{-1})_{xy}\|
 =\|(W_{y,L}(z-h)^{-1}W_{y,L}^{-1})_{xy}\|\le\|W_{y,L}(z-h)^{-1}W_{y,L}^{-1}\|\le\frac2\Delta.
\]
Dividing by the exponential and letting \(L\to\infty\), for fixed \(x,y\), yields
\[
 \|((z-h)^{-1})_{xy}\|
 \le\frac2\Delta\E^{-\eta d(x,y)^{\theta_{\rm in}}}.
\]
For \(x\in\Gamma\), Tonelli's theorem and the substitution \(t=\eta r^{\theta_{\rm in}}/2\) give
\[
 \begin{aligned}
 \sum_{y\in \Gamma}\E^{-\eta d(x,y)^{\theta_{\rm in}}/2}
 &=\sum_{y\in\Gamma}\int_{d(x,y)}^\infty
     \frac{\eta\theta_{\rm in}}2r^{\theta_{\rm in}-1}
     \E^{-\eta r^{\theta_{\rm in}}/2}\,\dd r =\int_0^\infty
     \bigl|\ball_{(2t/\eta)^{1/\theta_{\rm in}}}(x)\bigr|
     \E^{-t}\,\dd t\\
 &\le C_\Gamma\int_0^\infty
     \bigl(1+(2t/\eta)^{1/\theta_{\rm in}}\bigr)^{d_{\rm vol}}
     \E^{-t}\,\dd t \le C\eta^{-d_{\rm vol}/\theta_{\rm in}}.
 \end{aligned}
\]
The inequalities use \eqref{eq:growth}, \(0<\eta\le1\), and \(\int_0^\infty(1+t^{d_{\rm vol}/\theta_{\rm in}})\E^{-t}\,\dd t<\infty\).
By symmetry of \(d\), the same bound holds for \(\sup_y\sum_x\E^{-\eta d(x,y)^{\theta_{\rm in}}/2}\).
Consequently,
\[
 \|(z-h)^{-1}\|^{\rm Sch}_{\eta/2,\theta_{\rm in}}
 \le C\Delta^{-1}\eta^{-d_{\rm vol}/\theta_{\rm in}}
 \le C_{\rm exp}\Delta^{-1-d_{\rm vol}/\theta_{\rm in}}.
\]
This proves \eqref{eq:exponential-resolvent-bound} with \(c_{\rm exp}=c/2\).

In regime~\ref{ass:polynomial-weight}, fix \(0<\Delta\le1\) and \(z\) with \(\dist(z,\operatorname{spec}(h))\ge\Delta\).
For \(y\in\Gamma\), \(L>0\) and \(0<\eta\le1\), define the weights \(W_{y,L}(x):=(1+\eta\min\{d(x,y),L\})^{p_h}\).
Since \(x\mapsto\min\{d(x,y),L\}\) is \(1\)-Lipschitz, both \(W_{y,L}(x)/W_{y,L}(x')\) and its reciprocal are bounded by \((1+\eta d(x,x'))^{p_h}\).
The mean value theorem, \(p_h\ge1\) and \(\eta\le1\) give
\[
 \begin{aligned}
 \|(W_{y,L}hW_{y,L}^{-1}-h)_{xx'}\|
 &=\left|\frac{W_{y,L}(x)}{W_{y,L}(x')}-1\right|\|h_{xx'}\|\\
 &\le \bigl((1+\eta d(x,x'))^{p_h}-1\bigr)\|h_{xx'}\|\\
 &\le p_h\eta(1+d(x,x'))^{p_h}\|h_{xx'}\|.
 \end{aligned}
\]
Summing over \(x\) for fixed \(x'\) and using \eqref{eq:Schur-polynomial-norm}, we obtain
\[
 \sum_{x\in\Gamma}\|(W_{y,L}hW_{y,L}^{-1}-h)_{xx'}\|
 \le p_h\eta\sum_{x\in\Gamma}(1+d(x,x'))^{p_h}\|h_{xx'}\|
 \le p_h C_h\eta.
\]
The same bound holds when summing over \(x'\) for fixed \(x\). For a block matrix \(A\), the triangle inequality gives
\[
 \|A\|_{\ell^1\to\ell^1}
 \le\sup_{x'\in\Gamma}\sum_{x\in\Gamma}\|A_{xx'}\|.
\]
The displayed \(\ell^1\) operator norm bound and the Schur test on \(\ell^2\) give
\begin{equation}\label{eq:polynomial-conjugation-bound}
 \|W_{y,L}hW_{y,L}^{-1}-h\|_{\ell^j\to\ell^j}
 \le p_h C_h\eta,\qquad j\in\{1,2\}.
\end{equation}
Fix \(0<c\le1\) such that \(p_h C_h c\le1/2\).
For \(\eta=c\Delta\), \eqref{eq:polynomial-conjugation-bound} and \(\|(z-h)^{-1}\|_{\ell^2\to\ell^2}\le\Delta^{-1}\) give
\[
 \|(W_{y,L}hW_{y,L}^{-1}-h)(z-h)^{-1}\|_{\ell^2\to\ell^2}\le\frac12.
\]
The Neumann series for the conjugated resolvent on \(\ell^2\) gives
\begin{equation}\label{eq:polynomial-conjugated-resolvent}
 \|W_{y,L}(z-h)^{-1}W_{y,L}^{-1}\|_{\ell^2\to\ell^2}\le\frac2\Delta.
\end{equation}
For a block operator \(A\) bounded on \(\ell^j(\Gamma,\C^{2r})\), we have
\begin{equation}\label{eq:operator-column-sum}
 \sum_{x\in\Gamma}\|A_{xy}\|^j\le 2r\|A\|_{\ell^j\to\ell^j}^j,
 \qquad j\in\{1,2\}.
\end{equation}
Since \(d_{\rm vol}\ge1\), we have \(m_d\le d_{\rm vol}<p_h\).
Since \(W_{y,L}(y)=1\), \eqref{eq:operator-column-sum} with \(j=2\) and \eqref{eq:polynomial-conjugated-resolvent} give
\[
 \begin{aligned}
 \sum_{x\in\Gamma}W_{y,L}(x)^2\|((z-h)^{-1})_{xy}\|^2
 &=\sum_{x\in\Gamma}\|(W_{y,L}(z-h)^{-1}W_{y,L}^{-1})_{xy}\|^2\\
 &\le 2r\|W_{y,L}(z-h)^{-1}W_{y,L}^{-1}\|_{\ell^2\to\ell^2}^2
 \le 2r\left(\frac2\Delta\right)^2.
 \end{aligned}
\]
By monotone convergence, the limit \(L\to\infty\) replaces \(W_{y,L}(x)^2\) by \((1+\eta d(x,y))^{2p_h}\).
Using \(1+\eta d(x,y)\ge\eta(1+d(x,y))\) and \(m_d\le p_h\), we obtain
\[
 \sum_{x\in\Gamma}(1+d(x,y))^{2m_d}\|((z-h)^{-1})_{xy}\|^2
 \le 2r\eta^{-2m_d}\left(\frac2\Delta\right)^2
 \le C\Delta^{-2(m_d+1)}.
\]
Since \(2m_d>d_{\rm vol}\), volume growth and Cauchy--Schwarz give, for each \(y\in\Gamma\),
\[
 \sum_{x\in\Gamma}\|((z-h)^{-1})_{xy}\|
 \le C\Delta^{-(m_d+1)}
       \left(\sum_x(1+d(x,y))^{-2m_d}\right)^{1/2}\le C\Delta^{-(m_d+1)}.
\]
Applying the same argument to \(((z-h)^{-1})^*=(\bar z-h)^{-1}\) gives the bound \(C\Delta^{-(m_d+1)}\) for the sum over \(y\) at fixed \(x\). Thus
\[
 \|(z-h)^{-1}\|_0^{\rm Sch,pol}
 \le C\Delta^{-(m_d+1)}=:K.
\]
This gives \(\|(z-h)^{-1}\|_{\ell^1\to\ell^1}\le K\). We choose \(C\ge1\), so \(K\ge1\).

Now take \(\eta=c/K\). By \eqref{eq:polynomial-conjugation-bound}, we have
\[
 \|(W_{y,L}hW_{y,L}^{-1}-h)(z-h)^{-1}\|_{\ell^1\to\ell^1}\le\frac12.
\]
The Neumann series on \(\ell^1\) gives
\[
 \|W_{y,L}(z-h)^{-1}W_{y,L}^{-1}\|_{\ell^1\to\ell^1}\le2K.
\]
Since \(W_{y,L}(y)=1\), \eqref{eq:operator-column-sum} with \(j=1\) and the preceding estimate give
\[
 \begin{aligned}
 \sum_{x\in\Gamma} W_{y,L}(x)\|((z-h)^{-1})_{xy}\|
 &=\sum_{x\in\Gamma}\|(W_{y,L}(z-h)^{-1}W_{y,L}^{-1})_{xy}\|\\
 &\le 2r\|W_{y,L}(z-h)^{-1}W_{y,L}^{-1}\|_{\ell^1\to\ell^1}
 \le 4rK.
 \end{aligned}
\]
Letting \(L\to\infty\) by monotone convergence and using \(1+\eta d(x,y)\ge\eta(1+d(x,y))\), we obtain
\[
 \begin{aligned}
 \sum_{x\in\Gamma}(1+d(x,y))^{p_h}\|((z-h)^{-1})_{xy}\|
 &\le\eta^{-p_h}\sum_{x\in\Gamma}(1+\eta d(x,y))^{p_h}\|((z-h)^{-1})_{xy}\|\\
 &\le 4r\eta^{-p_h}K
 =4rc^{-p_h}K^{p_h+1}
 \le C K^{\lceil p_h\rceil+1}.
 \end{aligned}
\]
Applying the estimate to \(((z-h)^{-1})^*=(\bar z-h)^{-1}\) gives the same bound when summing over \(y\) at fixed \(x\). This proves \eqref{eq:polynomial-resolvent-bound}.
\end{proof}

\begin{lemma}\label{lem:uniform-free-calculus}
Assume either of the Schur locality assumptions on \(h\) in Hypothesis~\ref{hyp:locality}, and the spectral-gap condition
\eqref{eq:free-band} with \(g>0\).
There exist a model constant \(C_{\rm th}>0\) and, in regime~\ref{ass:exponential}, a model constant \(\mu_g\in(0,\mu_h)\) with the following property. With
\[
 w(R):=\begin{cases}
 \E^{\mu_gR^{\theta_{\rm in}}}&\text{in regime~\ref{ass:exponential}},\\
 (1+R)^{p_h}&\text{in regime~\ref{ass:polynomial-weight}},
 \end{cases}
\]
we have 
\begin{equation}\label{eq:thermal-functional-calculus}
 \sup_{\beta\ge0}\Bigl(
 \|P_\pm\|_w^{\rm Sch}
 +\|E_\beta P_\pm\|_w^{\rm Sch}
 +\|\Rb P_\pm\|_w^{\rm Sch}
 \Bigr)\le C_{\rm th}.
\end{equation}
\end{lemma}
\begin{proof}
Since \(\operatorname{spec}(h)\subset[-C_h,-g]\cup[g,C_h]\), let \(\mathscr C_\pm\) be the positively oriented boundaries of the rectangles
\[
 \pm[3g/4,C_h+g/4]+\I[-g/4,g/4].
\]
Both contours have length \(2C_h\) and distance at least \(g/4\) from \(\operatorname{spec}(h)\).
The contours enclose the respective spectral bands and satisfy
\(\pm\RealPart z>g/2\) on and inside each contour.
Apply Lemma~\ref{lem:weighted-contour-calculus} with \(\Delta=\min\{1,g/4\}\), taking \(\mu_g=\min\{c_{\rm exp}\Delta,\mu_h/2\}\) in regime~\ref{ass:exponential}. This gives
\begin{equation}\label{eq:uniform-thermal-resolvent}
 \sup_{z\in\mathscr C_+\cup\mathscr C_-}\|(z-h)^{-1}\|_w^{\rm Sch}
 \le C_{\rm res},
\end{equation}
for a model constant \(C_{\rm res}>0\). For each of the functions \(f(\zeta)=1,\ \zeta,\ (1+\zeta^2)^{-1/2}\), the identity \(|h|P_\pm=\pm hP_\pm\) gives
\begin{equation}\label{eq:thermal-contour-formula}
 f(E_\beta)P_\pm
 =f(\E^{\mp\beta h/2})P_\pm
 =\frac1{2\pi\I}\int_{\mathscr C_\pm}
       f(\E^{\mp\beta z/2})(z-h)^{-1}\,\dd z.
\end{equation}
For \(\beta\ge0\) and \(z\) on or inside \(\mathscr C_\pm\), we have
\[
 |\E^{\mp\beta z/2}|=\E^{\mp\beta\RealPart z/2}
 \le\E^{-\beta g/4}\le1.
\]
Using \(|\operatorname{Im}z|\le g/4\) and \(\cos t\ge-|t|\), we also have
\[
 \RealPart(1+\E^{\mp\beta z})
 =1+\E^{\mp\beta\RealPart z}\cos(\beta\operatorname{Im}z)\ge1-\frac{\beta g}{4}\E^{-\beta g/2}
 \ge\frac12.
\]
The last inequality uses \(u\E^{-u}\le1\) for \(u\ge0\).
We use the holomorphic square root of \(1+\E^{\mp\beta z}\) with positive real part. The resulting multipliers \(z\mapsto f(\E^{\mp\beta z/2})\) are holomorphic on a neighborhood of the closed region enclosed by \(\mathscr C_\pm\), and
\[
 \left|(1+\E^{\mp\beta z})^{-1/2}\right|
 \le\bigl(\RealPart(1+\E^{\mp\beta z})\bigr)^{-1/2}
 \le\sqrt2.
\]
Thus \(|f(\E^{\mp\beta z/2})|\le M:=\sqrt2\) for all three functions. Equations~\eqref{eq:thermal-contour-formula} and \eqref{eq:uniform-thermal-resolvent} give
\[
 \|f(E_\beta)P_\pm\|_w^{\rm Sch}
 \le\frac1{2\pi}\int_{\mathscr C_\pm}
       |f(\E^{\mp\beta z/2})|\,
       \|(z-h)^{-1}\|_w^{\rm Sch}\,|\dd z|\le\frac{2C_h}{2\pi}MC_{\rm res}.
\]
Summing these three bounds proves \eqref{eq:thermal-functional-calculus} with \(C_{\rm th}=3C_hMC_{\rm res}/\pi\).
\end{proof}

\begin{lemma}\label{lem:bounded-temperature-calculus}
Assume either of the Schur locality assumptions on \(h\) in Hypothesis~\ref{hyp:locality}.
There are model constants \(C\ge1\), \(N_0,C_{\mathsf M}>0\) and, in regime~\ref{ass:exponential}, \(\mu_0>0\), such that the following holds for every \(\beta\ge0\). With
\[
 w_\beta(R):=\begin{cases}
 \E^{\mu_0R^{\theta_{\rm in}}/(1+\beta)}&\text{in regime~\ref{ass:exponential}},\\
 (1+R)^{p_h}&\text{in regime~\ref{ass:polynomial-weight}},
 \end{cases}
\]
the functions \(a_\beta\) from Section~\ref{sec:thermal-free-coordinates} satisfy
\begin{equation}\label{eq:bounded-temperature-calculus}
 \|a_\beta(h)\|_{w_\beta}^{\rm Sch}+\|a_\beta(-h)\|_{w_\beta}^{\rm Sch}
 \le C(1+\beta)^{N_0}.
\end{equation}
The normalizer \(\mathsf M\) defined in \eqref{eq:FG-normalizer} satisfies
\(\|\mathsf M\|_{\mu_h,\theta_{\rm in}}^{\rm Sch}\le C_{\mathsf M}\) in regime~\ref{ass:exponential} and
\(\|\mathsf M\|_{p_h}^{\rm Sch,pol}\le C_{\mathsf M}\) in regime~\ref{ass:polynomial-weight}.
\end{lemma}
\begin{proof}
Fix \(\beta\ge0\). Let \(\mathscr C\) be the positively oriented boundary of
\[
 [-C_h-1,C_h+1]+\I[-(1+\beta)^{-1},(1+\beta)^{-1}].
\]
Its distance from \(\operatorname{spec}(h)\) is at least \((1+\beta)^{-1}\).
For \(z=x+\I y\) on or inside \(\mathscr C\), we have \(|\beta y|\le\beta/(1+\beta)\le1\), so
\[
 \begin{aligned}
 \RealPart\bigl(2\cosh(\beta z/2)\bigr)
 &=2\cosh(\beta x/2)\cos(\beta y/2)
 \ge \cos(1/2)\E^{|\beta x|/2}>0.
 \end{aligned}
\]
We use the holomorphic square root on the right half-plane with positive real part in
\[
 a_\beta(z)=\frac{\E^{\beta z/4}}{\sqrt{2\cosh(\beta z/2)}}.
\]
This gives a holomorphic extension of \(a_\beta\) to a neighborhood of the enclosed rectangle.
Thus \(|a_\beta(z)|\le\cos(1/2)^{-1/2}\), uniformly on the contour and in \(\beta\).
Since \(-\mathscr C=\mathscr C\), the same bound holds for \(|a_\beta(-z)|\) on \(\mathscr C\).
The holomorphic functional calculus gives
\begin{equation}\label{eq:bounded-temperature-contour-formula}
 a_\beta(\pm h)
 =\frac1{2\pi\I}\int_{\mathscr C} a_\beta(\pm z)(z-h)^{-1}\,\dd z.
\end{equation}

In regime~\ref{ass:exponential}, set \(\mu_0:=c_{\rm exp}\) from Lemma~\ref{lem:weighted-contour-calculus}(i). Applying that lemma with \(\Delta=(1+\beta)^{-1}\) gives
\[
 \sup_{z\in\mathscr C}\|(z-h)^{-1}\|_{w_\beta}^{\rm Sch}
 \le C_{\rm exp} (1+\beta)^{1+d_{\rm vol}/\theta_{\rm in}}.
\]
In regime~\ref{ass:polynomial-weight}, Lemma~\ref{lem:weighted-contour-calculus}(ii) instead gives
\[
 \sup_{z\in\mathscr C}\|(z-h)^{-1}\|_{p_h}^{\rm Sch,pol}
 \le C_{\rm pol}(1+\beta)^{(m_d+1)(\lceil p_h\rceil+1)}.
\]
Set
\[
 N_0:=
 \begin{cases}
  1+d_{\rm vol}/\theta_{\rm in},&\text{in regime~\ref{ass:exponential}},\\
  (m_d+1)(\lceil p_h\rceil+1),&\text{in regime~\ref{ass:polynomial-weight}}.
 \end{cases}
\]
Equation~\eqref{eq:bounded-temperature-contour-formula} gives
\[
 \|a_\beta(\pm h)\|_{w_\beta}^{\rm Sch}
 \le\frac1{2\pi}\int_{\mathscr C}|a_\beta(\pm z)|
                   \|(z-h)^{-1}\|_{w_\beta}^{\rm Sch}\,|\dd z|
 \le C(1+\beta)^{N_0}.
\]
This proves \eqref{eq:bounded-temperature-calculus}.
Finally, let \(w_{\rm in}(R)=\exp(\mu_hR^{\theta_{\rm in}})\) in regime~\ref{ass:exponential} and \(w_{\rm in}(R)=(1+R)^{p_h}\) in regime~\ref{ass:polynomial-weight}.
Submultiplicativity of the Schur norm and the definition \eqref{eq:FG-normalizer} give
\[
 \|\mathsf M\|_{w_{\rm in}}^{\rm Sch}
 \le\frac1{\sqrt2}\sum_{j\ge0}\frac{a^j(\|h\|_{w_{\rm in}}^{\rm Sch})^{2j}}{j!}
 \le\frac1{\sqrt2}\E^{aC_h^2}=:C_{\mathsf M}.\qedhere
\]
\end{proof}

\begin{proof}[Proof of Proposition~\ref{prop:free-thermal-family}]
As shown in Section~\ref{sec:thermal-free-coordinates}, for every \(\beta\ge0\), the operators \(\mathsf F_u^{\beta,0}\) defined in \eqref{eq:smooth-free-family} satisfy
\[
 \mathsf F_u^{\beta,0}=(\sigma^\beta_\st)^{-1}(c_u),
\]
for all $u \in \Gamma \times\{1, \ldots, 2r\}$. Here $c_u=\tfrac12(\gamma_{u,1}+\I\gamma_{u,2})$ and \(\sigma^\beta_\st\) is the Bogoliubov automorphism defined in Equation~\eqref{eq:thermal-Bogoliubov-automorphism}. Remark~\ref{rem:Bogoliubov}, applied to \((\sigma^\beta_\st)^{-1}\), shows that the coefficients \(\sa^\beta_{uv}\1=\{\gamma_v,\mathsf F_u^{\beta,0}\}\) satisfy the Parseval condition of Definition~\ref{def:parseval-family}. Lemma~\ref{lem:parseval-reconstruction} therefore gives norm convergence of the series in Equation~\eqref{eq:smooth-free-family} for every \(\beta\ge0\).

By definition \eqref{eq:smooth-free-family}, the coefficients \(\sa^\beta_{uv}\) of $\mathsf F_u^{\beta,0}$ are given by 
\[
 \sa^\beta_{u,(v,1)}=\sqrt2\,a_\beta(h)_{uv},\qquad
 \sa^\beta_{u,(v,2)}=\I\sqrt2\,a_\beta(-h)_{uv},
\]
where $u,v\in\Gamma\times\{1,\ldots,2r\}$. We notice that, for the \(2r\times2r\) blocks of \(a_\beta(\pm h)\), Cauchy--Schwarz gives, for all \(x,y\in\Gamma\),
\begin{equation}\label{eq:free-thermal-block-entry-bound}
 \begin{aligned}
 \max_{1\le i\le2r}\sum_{j=1}^{2r}
       |[a_\beta(\pm h)_{xy}]_{ij}|
 &\le\sqrt{2r}\,\|a_\beta(\pm h)_{xy}\|,\\
 \max_{1\le j\le2r}\sum_{i=1}^{2r}
       |[a_\beta(\pm h)_{xy}]_{ij}|
 &\le\sqrt{2r}\,\|a_\beta(\pm h)_{xy}\|.
 \end{aligned}
\end{equation}

\smallskip
\noindent\emph{(i) With a free spectral gap.}
Assume \eqref{eq:free-band}. Functional calculus and the definitions \eqref{eq:free_normalization}, \eqref{eq:PQE} and~\eqref{eq:gapped-thermal-normalizer} give
\begin{equation}\label{eq:free-band-coefficients}
  a_\beta(h)=\Rb(P_++E_\beta P_-),\qquad
  a_\beta(-h)=\Rb(E_\beta P_++P_-).
\end{equation}
Lemma~\ref{lem:uniform-free-calculus} gives model constants \(C_{\rm th}>0\) and, in regime~\ref{ass:exponential} \(\mu_g\in(0,\mu_h)\), together with the weight \(w\) defined there for each locality regime. Using \eqref{eq:free-band-coefficients} and Lemma~\ref{lem:weighted-Schur-properties}, we obtain, for every \(\beta\ge0\),
\[
 \begin{aligned}
 \|a_\beta(\pm h)\|_w^{\rm Sch}
 &\le\|\Rb P_\pm\|_w^{\rm Sch}
       +\|\Rb P_\mp\|_w^{\rm Sch}\|E_\beta P_\mp\|_w^{\rm Sch}\le C_{\rm th}+C_{\rm th}^2.
 \end{aligned}
\]
The definition \eqref{eq:smooth-free-family} and Equation~\eqref{eq:free-thermal-block-entry-bound} therefore give a model constant \(C_{0,\mathrm g}\ge1\) such that
\[
 \begin{aligned}
 \bigl\|(|\sa^\beta_{uv}|)\bigr\|_w^{\rm Sch}
 &\le2\sqrt r\bigl(\|a_\beta(h)\|_w^{\rm Sch}
                    +\|a_\beta(-h)\|_w^{\rm Sch}\bigr)\le C_{0,\mathrm g},
 \end{aligned}
\]
for all $\beta \ge 0$. Taking \(C_0=C_{0,\mathrm g}\) and, in regime~\ref{ass:exponential}, \(\mu=\mu_g\) proves (i).

\smallskip
\noindent\emph{(ii) Without a gap assumption.}
Lemma~\ref{lem:bounded-temperature-calculus} gives model constants \(C\ge1\), \(N_0>0\) and, in regime~\ref{ass:exponential}, \(\mu_0>0\), together with the weights \(w_\beta\) defined there for each locality regime. For every \(\beta\ge0\),
\[
 \|a_\beta(h)\|_{w_\beta}^{\rm Sch}
 +\|a_\beta(-h)\|_{w_\beta}^{\rm Sch}
 \le C(1+\beta)^{N_0}.
\]
The definition \eqref{eq:smooth-free-family} and Equation~\eqref{eq:free-thermal-block-entry-bound} therefore give a model constant \(C_{0,\mathrm{ng}}\ge1\) such that
\[
 \begin{aligned}
 \bigl\|(|\sa^\beta_{uv}|)\bigr\|_{w_\beta}^{\rm Sch}
 &\le2\sqrt r\bigl(\|a_\beta(h)\|_{w_\beta}^{\rm Sch}
                    +\|a_\beta(-h)\|_{w_\beta}^{\rm Sch}\bigr)\le C_{0,\mathrm{ng}}(1+\beta)^{N_0},
 \end{aligned}
\]
for all \(\beta\ge0\).
Taking \(K=C_{0,\mathrm{ng}}\) and, in regime~\ref{ass:exponential}, \(\bar\mu=\mu_0\) therefore gives \eqref{eq:thermal-free-coefficient-bound-ng} and proves (ii).
\end{proof}

\subsection{Locality and local decompositions}
\label{sec:local-decompositions}

Next we prove spatial approximation and perturbation bounds for the filtered Majoranas \(\gamma_u[k;\Phi_\lambda]\).
For a kernel \(k\in L^1(\R)\) we use the conditions
\begin{equation}\label{eq:kernel-conditions}
 \begin{aligned}
 \int_\R(1+|t|)^N|k(t)|\,\dd t
 &<\infty &&\text{for every }N\ge0,\\
 |k(t)|&\le C_k\E^{-a_k|t|^{1/\sG}}
 &&\text{in regime~\ref{ass:exponential}},
 \end{aligned}
\end{equation}
with \(a_k>0\) and \(\sG\ge1\).

\paragraph{Propagation input from \cite{FermionicLR}.}
We use the following propagation bounds for locality, local decompositions and defect comparison.
\begin{itemize}
\item \emph{Local approximation.}
In regime~\ref{ass:exponential}, we use \cite[Theorem~2.9 and Remark~2.10]{FermionicLR} with \(w(R)=\exp(\mu R^\theta)\). Let \(\Xi\) be a time-independent interaction with \(\|\Xi\|_{\mu,\theta}<\infty\). Then for \(X\Subset\Gamma\) and \(A\in\A_X\),
\[
 \|(\id-\Ftr{X^{(R)}})\tau^\Xi_t(A)\|
 \le 2|X|\|A\|\bigl(\E^{2|t|\|\Xi\|_{\mu,\theta}}-1\bigr)\E^{-\mu R^\theta}.
\]
\item \emph{Polynomial locality.}
In regime~\ref{ass:polynomial-weight}, Lemma~\ref{lem:rooted-parameter-derivative} uses the local representations of evolved Majoranas from \cite[Definition~2.5, Theorem~2.6(i) and Corollary~4.4]{FermionicLR}. The embedding of \cite[Lemma~4.3]{FermionicLR} lets us apply the coupling derivative estimate of \cite[Theorem~5.1(i)]{FermionicLR} to the Majoranas. This gives the bound \eqref{eq:rooted-parameter-derivative} with the full exponent \(p_\Phi\):
\[
 \sup_{|\lambda|\le1}\sup_z\sum_{X\ni z}|X|(1+\diam X)^{p_\Phi}
   \sum_{v:\,x_v\in X}
   \Bigl\|\partial_\lambda
       \Bigl[\mathcal T^{\Phi_\lambda}_{0,t,\{x_v\}}(\gamma_v)\Bigr](X)\Bigr\|
 \le C(1+|t|)^{m_B},
\]
where \(m_B\) is the model exponent from Lemma~\ref{lem:rooted-parameter-derivative}.
For the defect comparison in Lemma~\ref{lem:defect-source}, we use \cite[Proposition~5.3(ii)]{FermionicLR} to bound the sum of the commutator norms across the boundary in \eqref{eq:defect-commutator-B}, also with exponent \(p_\Phi\).
\item \emph{Shell estimates.}
For the Duhamel estimate in either regime, we use the dyadic shell maps and radii of Definition~\ref{def:tracial-shells}. For \(|\lambda|\le1\), \cite[Proposition~5.4 and Corollary~5.5]{FermionicLR} give
\[
 \bigl\|\Delta^{\rm tr}_{\{x_u\},i}\tau_t^{\Phi_\lambda}(\gamma_u)\bigr\|
 \le C(1+|t|)^m(1+R_i)^{-q},
 \qquad i\ge0,
\]
with model constants \(C,m>0\) and \(q>d_{\rm vol}\). The inequality \(q>d_{\rm vol}\) lets us sum the shell norms with the support sizes as weights.
\end{itemize}

\begin{lemma}\label{lem:filtered-field-locality}
Assume Hypothesis~\ref{hyp:locality}.
\begin{enumerate}[label=(\roman*),leftmargin=*]
\item \emph{Spatial bounds for general kernels.}
In regime~\ref{ass:exponential}, let \(k\) satisfy the stretched-exponential bound in \eqref{eq:kernel-conditions}. There are \(C_{\rm loc},c_{\rm loc}>0\), depending only on the model data and on \(C_k,a_k,\sG\), such that, for every \(u\in\Gamma\times\{1,\ldots,2r\}\), \(R\ge0\) and \(\lambda\) with \(|\lambda|\le1\),
\[
 \Bigl\|(\id-\Ftr{\ball_R(x_u)})
       \gamma_u[k;\Phi_\lambda]\Bigr\|
 \le C_{\rm loc}\E^{-c_{\rm loc}R^{\theta_{\rm in}/\sG}}.
\]
\item \emph{Temperature-dependent spatial bound for \(q_\beta\).}
In regime~\ref{ass:exponential}, there are model constants \(C_{\rm FG},c_{\rm FG}>0\) such that, for every \(u\in\Gamma\times\{1,\ldots,2r\}\), \(R\ge0\), \(\beta>0\) and \(\lambda\) with \(\beta|\lambda|/(1+\beta)\le1\),
\begin{equation}\label{eq:FG-field-locality}
 \Bigl\|(\id-\Ftr{\ball_R(x_u)})\gamma_u[q_\beta;\Phi_\lambda]\Bigr\|
 \le C_{\rm FG}\log(3+\beta)\E^{-c_{\rm FG}R^{\theta_{\rm in}}/(1+\beta)}.
\end{equation}
\item \emph{Dependence on the interaction.}
There are model constants \(C_D>0\) and \(m_D\ge0\) such that, for every \(u\in\Gamma\times\{1,\ldots,2r\}\), \(k\in L^1(\R)\), \(\Psi\in\mathfrak J_1\), \(0<b\le1\) and \(\lambda\in\R\) with \(b|\lambda|\le1\),
\begin{equation}\label{eq:filtered-field-perturbation}
 \Bigl\|\gamma_u[k;\Phi_\lambda]-\gamma_u[k;\Phi_\lambda+\Psi]\Bigr\|
 \le C_DJ_\Psi
       \int_\R|k(t)|\,|t|(1+|t|/b)^{m_D}\,\dd t.
\end{equation}
Here \(J_\Psi\) is defined in Equation~\eqref{eq:perturbation-intensity}.

\end{enumerate}
\end{lemma}
\begin{proof}
\emph{(i) Spatial bounds for general kernels.}
In regime~\ref{ass:exponential}, the norm \(\|\Phi_\lambda\|_{\min\{\mu_h,\mu_{\cV}\},\theta_{\rm in}}\) is bounded by a model constant for \(|\lambda|\le1\). By \cite[Theorem~2.9 and Remark~2.10]{FermionicLR}, there are model constants \(C,c,v>0\) such that
\[
 \|(\id-\Ftr{\ball_R(x_u)})\tau_t^{\Phi_\lambda}(\gamma_u)\|
 \le\min\{2,C\E^{v|t|-cR^{\theta_{\rm in}}}\}.
\]
With \(T=cR^{\theta_{\rm in}}/(2v)\), it follows that
\begin{equation}\label{eq:filtered-time-split}
 \begin{aligned}
 \Bigl\|(\id-\Ftr{\ball_R(x_u)})
       \gamma_u[k;\Phi_\lambda]\Bigr\|
 &\le\int_\R|k(t)|\min\{2,C\E^{v|t|-cR^{\theta_{\rm in}}}\}\,\dd t\\
 &\le C\E^{-cR^{\theta_{\rm in}}/2}\|k\|_1
       +2\int_{|t|>T}|k(t)|\,\dd t
 \le C\E^{-c'R^{\theta_{\rm in}/\sG}}.
 \end{aligned}
\end{equation}
For the last step, \eqref{eq:kernel-conditions} gives \(\int_{|t|>T}|k(t)|\,\dd t\le C\E^{-a_kT^{1/\sG}/2}\).
With \(T=cR^{\theta_{\rm in}}/(2v)\), both terms are bounded by a constant times \(\E^{-c'R^{\theta_{\rm in}/\sG}}\) for \(R\ge1\), where
\[
 c'=c_{\rm loc}:=\min\left\{\frac c2,
       \frac{a_k}{2}\left(\frac{c}{2v}\right)^{1/\sG}\right\}.
\]
Increasing the prefactor covers \(0\le R\le1\).

\emph{(ii) Temperature-dependent spatial bound for \(q_\beta\).}
For \(k= q_\beta\), put \(\eta:=\beta/(1+\beta)\). Put \(\mu:=\min\{\mu_h,\mu_{\cV}\}\). For \(\eta|\lambda|\le1\), \eqref{eq:quadratic-interaction-norm-bound} gives
\[
 \|\Phi_\lambda\|_{\mu,\theta_{\rm in}}
 \le 2rC_h+|\lambda|\|\cV\|_{\mu_{\cV},\theta_{\rm in}}
 \le\eta^{-1}\bigl(2rC_h+\|\cV\|_{\mu_{\cV},\theta_{\rm in}}\bigr).
\]
Applying \cite[Theorem~2.9 and Remark~2.10]{FermionicLR} again, we find
\[
 \|(\id-\Ftr{\ball_R(x_u)})\tau_t^{\Phi_\lambda}(\gamma_u)\|
 \le\min\{2,C\E^{v|t|/\eta-cR^{\theta_{\rm in}}}\}.
\]
By \eqref{eq:FG-moments}, for \(\beta>0\) and \(T\ge0\),
\begin{equation}\label{eq:FG-time-tail}
 \int_{|t|>T}| q_\beta(t)|\,\dd t
 \le C\E^{-T/(4\beta)}
       \int_\R\frac{\E^{-|t|/(4\beta)}}{\eta+|t|}\,\dd t\le C'\log(3+\beta)\E^{-T/(4\beta)}.
\end{equation}
The integral is bounded by splitting at \(|t|=\beta\): the inner part is at most \(2\log(1+\beta/\eta)\), and the substitution \(t=\beta s\) bounds the outer part by a constant.
Using \eqref{eq:FG-time-tail} with \(T=\eta cR^{\theta_{\rm in}}/(2v)\) and the \(L^1\) bound from Lemma~\ref{lem:FG-filter}, the first two inequalities of \eqref{eq:filtered-time-split}, applied with \(v|t|/\eta\) in place of \(v|t|\), give
\[
 \Bigl\|(\id-\Ftr{\ball_R(x_u)})\gamma_u[q_\beta;\Phi_\lambda]\Bigr\|
 \le C\log(3+\beta)\bigl(\E^{-cR^{\theta_{\rm in}}/2}
                  +\E^{-cR^{\theta_{\rm in}}/[8v(1+\beta)]}\bigr).
\]
Since \(1+\beta\ge1\), this gives \eqref{eq:FG-field-locality} with \(c_{\rm FG}:=\min\{c/2,c/(8v)\}\).

\emph{(iii) Dependence on the interaction.}
We first prove the estimate for \(b=1\). We use the interaction norm
\[
 \|\Xi\|^{\pol}_{a_{\rm LR}}:=\sup_x\sum_{Z\ni x}(1+\diam Z)^{a_{\rm LR}}\|\Xi(Z)\|,
 \qquad a_{\rm LR}\ge0,
\]
which is \(\|\Xi\|^{\pol}_{a_{\rm LR};0}\) in \cite{FermionicLR}.
Choose \(a_{\rm LR}>d_{\rm vol}\) with \(\sup_{|s|\le1}\|\Phi_s\|^{\pol}_{a_{\rm LR}}<\infty\): take \(a_{\rm LR}=d_{\rm vol}+1\) in \ref{ass:exponential}, where the stretched-exponential norm dominates every polynomial norm, and \(a_{\rm LR}=p_\Phi\) in \ref{ass:polynomial-weight}.
Fix a finite \(\Lambda\ni x_u\) and write \(\tau^{s,\Lambda}\) for the finite-volume dynamics generated by \(\sum_{Z\subset\Lambda}\Phi_s(Z)\). Let \(\widetilde\tau^\Lambda\) denote the corresponding finite-volume dynamics of \(\Phi_\lambda+\Psi\). Differentiating \(r\mapsto\widetilde\tau^\Lambda_r(\tau^{\lambda,\Lambda}_{t-r}(\gamma_u))\) gives
\begin{equation}\label{eq:perturbation-duhamel}
 \widetilde\tau^\Lambda_t(\gamma_u)-\tau^{\lambda,\Lambda}_t(\gamma_u)
 =\I\int_0^t\widetilde\tau^\Lambda_r
    \Bigl(\Bigl[\sum_{Z\subset\Lambda}\Psi(Z),\tau^{\lambda,\Lambda}_{t-r}(\gamma_u)\Bigr]\Bigr)\dd r.
\end{equation}
Using the tracial shell maps and dyadic radii of Definition~\ref{def:tracial-shells}, we obtain the telescoping sum
\begin{equation}\label{eq:finite-volume-shell-expansion}
 \sum_{i=0}^{\infty}\Delta^{\rm tr}_{\{x_u\},i}\tau_s^{\lambda,\Lambda}(\gamma_u)
 = \tau_s^{\lambda,\Lambda}(\gamma_u)\in\A_\Lambda.
\end{equation}
By Lemma~\ref{lem:conditional-expectations},
\[
 \Delta^{\rm tr}_{\{x_u\},i}\tau_s^{\lambda,\Lambda}(\gamma_u)
 \in\A_{\Lambda\cap\ball_{R_i}(x_u)},\qquad i\ge0.
\]
The shell term vanishes if \(i\ge1\) and \(\Lambda\subset\ball_{R_{i-1}}(x_u)\). See also \cite[Lemma~5.1 and Example~5.7]{NSY19} for decompositions by differences of conditional expectations. Set \(q=(a_{\rm LR}+d_{\rm vol})/2\), so that \(d_{\rm vol}<q<a_{\rm LR}\). Apply \cite[Proposition~5.4]{FermionicLR} to \(\Phi_\lambda\) restricted to \(\Lambda\), with decay exponent \(q\). It gives
\[
 \bigl\|\Delta^{\rm tr}_{\{x_u\},i}\tau_s^{\lambda,\Lambda}(\gamma_u)\bigr\|
 \le C(1+|s|)^m(1+R_i)^{-q},\qquad i\ge1,
\]
for an integer \(m\ge1\), uniformly in \(\Lambda\). For \(i=0\), contractivity gives the same bound after increasing \(C\).
Volume growth and \(q>d_{\rm vol}\) therefore give
\begin{equation}\label{eq:evolved-field-support-sum}
 \sum_{i\ge0}|\Lambda\cap\ball_{R_i}(x_u)|\,\bigl\|\Delta^{\rm tr}_{\{x_u\},i}\tau_s^{\lambda,\Lambda}(\gamma_u)\bigr\|
 \le C(1+|s|)^m
       \sum_{i\ge0}(1+R_i)^{d_{\rm vol}-q}
 \le C'(1+|s|)^m.
\end{equation}
For \(D\in\A_W\), evenness of the perturbation terms implies
\[
 \sum_{Z\subset\Lambda}\|[\Psi(Z),D]\|
 \le2\|D\|\sum_{x\in W}\sum_{Z\ni x}\|\Psi(Z)\|
 \le2J_\Psi|W|\|D\|.
\]
Apply this bound to the \(i\)th summand of \eqref{eq:finite-volume-shell-expansion} with \(W=\Lambda\cap\ball_{R_i}(x_u)\). Summing over \(i\) and using \eqref{eq:evolved-field-support-sum} gives
\[
 \Bigl\|\Bigl[\sum_{Z\subset\Lambda}\Psi(Z),\tau_s^{\lambda,\Lambda}(\gamma_u)\Bigr]\Bigr\|
 \le C J_\Psi(1+|s|)^m.
\]
Substituting this bound into \eqref{eq:perturbation-duhamel} gives
\begin{equation}\label{eq:perturbation-dynamics-bound}
 \|\widetilde\tau^\Lambda_t(\gamma_u)-\tau^{\lambda,\Lambda}_t(\gamma_u)\|
 \le C_D J_\Psi\,|t|(1+|t|)^{m_D},
\end{equation}
with model constants \(C_D\) and \(m_D:=m\), uniformly in \(\Lambda\). By Proposition~\ref{prop:weak-dynamics}, the finite-volume dynamics converge in norm on every observable, uniformly for \(t\) in compact sets. Passing to the limit gives the same bound for \(\tau^{\Phi_\lambda+\Psi}_t-\tau^{\Phi_{\lambda}}_t\). Integrating against \(|k(t)|\) proves \eqref{eq:filtered-field-perturbation} for \(b=1\).

Let \(0<b\le1\) with \(|b\lambda|\le1\). Since \(b\Phi_\lambda=\Phi_{bh}+(b\lambda)\cV\), we have
\[
 \|b\Phi_\lambda\|^{\pol}_{a_{\rm LR}}
 \le \|\Phi_h\|^{\pol}_{a_{\rm LR}}+\|\cV\|^{\pol}_{a_{\rm LR}}.
\]
We can therefore apply \eqref{eq:perturbation-dynamics-bound} to \(b\Phi_\lambda\) and \(b\Psi\) at time \(t/b\), with the same model constants. Using \(\tau_{t/b}^{b\Xi}=\tau_t^\Xi\) and \(J_{b\Psi}=bJ_\Psi\) gives
\[
 \|\tau_t^{\Phi_\lambda+\Psi}(\gamma_u)-\tau_t^{\Phi_\lambda}(\gamma_u)\|
 \le C_DJ_\Psi|t|(1+|t|/b)^{m_D}.
\]
Integrating against \(|k(t)|\) proves \eqref{eq:filtered-field-perturbation}.
\end{proof}

\paragraph{Local decompositions.}

We now decompose the corrections \(\mathsf F_u^{\beta,\lambda}-\mathsf F_u^{\beta,0}\) and \(\mathsf G_u^{\beta,\lambda}-\mathsf F_u^{\beta,0}\) from \eqref{eq:thermal-coordinates} and \eqref{eq:FG-coordinates} into odd local terms with the small weighted sums required by Hypothesis~\ref{hyp:perturbed-family}(ii).
In regime~\ref{ass:exponential}, we use the dyadic shell decomposition given in Lemma~\ref{lem:normalized-field-differences}(i). In regime~\ref{ass:polynomial-weight}, preserving the polynomial decay exponent requires more care. The following lemma provides the needed bound on coupling derivatives of the local representation maps.

\begin{lemma}\label{lem:rooted-parameter-derivative}
Assume Hypothesis~\ref{hyp:locality}\ref{ass:polynomial-weight}.
The local representation maps of \cite[Theorem~2.6 and Corollary~4.4]{FermionicLR} can be chosen jointly for \(\lambda\in[-1,1]\) so that, for every \(v\in\Gamma\times\{1,\ldots,2r\}\) and \(t\in\R\),
\[
 \tau^{\Phi_\lambda}_t(\gamma_v)
 =\sum_{X\ni x_v}\bigl[\mathcal T^{\Phi_\lambda}_{0,t,\{x_v\}}(\gamma_v)\bigr](X),
\]
where 
\[\bigl[\mathcal T^{\Phi_\lambda}_{0,t,\{x_v\}}(\gamma_v)\bigr](X)
 \in\A_X^-
\]
for $X\Subset\Gamma$, $x_v\in X$. The series converges absolutely in norm. For fixed \(v,t,X\), the map \(\lambda\mapsto\bigl[\mathcal T^{\Phi_\lambda}_{0,t,\{x_v\}}(\gamma_v)\bigr](X)\) is continuously differentiable. There are model constants \(C>0\) and \(m_B\in\mathbb N\) such that, for every \(t\in\R\),
\begin{equation}\label{eq:rooted-parameter-derivative}
 \sup_{|\lambda|\le1}\sup_z\sum_{X\ni z}|X|(1+\diam X)^{p_\Phi}
   \sum_{v:\,x_v\in X}
   \Bigl\|\partial_\lambda
       \Bigl[\mathcal T^{\Phi_\lambda}_{0,t,\{x_v\}}
                                      (\gamma_v)\Bigr](X)\Bigr\|
 \le C(1+|t|)^{m_B}.
\end{equation}
\end{lemma}

\begin{proof}
Let \(\jmath_x\) and \(\widehat\Phi_\lambda\) denote the isometric embeddings and extended interaction paths of \cite[Lemma~4.3]{FermionicLR}. Hypothesis~\ref{hyp:locality} and \eqref{eq:quadratic-interaction-norm-bound} give
\begin{equation}\label{eq:background-first-moment}
 \sup_{|\lambda|\le1}\|\widehat\Phi_\lambda\|^{\#}_{p_\Phi,\kappa_0}
 =\sup_{|\lambda|\le1}\|\Phi_\lambda\|^{\#}_{p_\Phi,\kappa_0}
 \le 2r\E^{2\kappa_0}C_h
       +\|\cV\|^{\#}_{p_{\cV},\kappa_0}.
\end{equation}
Using \eqref{eq:background-first-moment}, we fix the cells, scales and time cutoffs for both paths and all \(|\lambda|\le1\), as in \cite[Theorem~5.1]{FermionicLR} with \(a=p_\Phi\). Let \(Q_n(S)\) be the union of the cells indexed by \(S\) at scale \(n\), and let \(\chi_n\) be the time cutoffs.
For a finite set \(U\) of cell indices at scale \(n\), let \(\tau^{\lambda,Q_n(U)}\) be the dynamics generated by the interaction $\Phi_h(Z)+\lambda\cV(Z)$ for $Z\subseteq Q_n(U)$ and $0$ otherwise. Equations (8) and (10) of \cite{FermionicLR} give
\begin{equation}\label{eq:local-representation-formula}
 \bigl[\mathcal T^{\Phi_\lambda}_{0,t,\{x_v\}}(\gamma_v)\bigr](X)
 =\sum_{n\ge0}\chi_n(t)
   \sum_{\substack{U\subseteq S,\ Q_n(S)=X\\ x_v\in Q_n(U)}}
       (-1)^{|S|-|U|}\tau_t^{\lambda,Q_n(U)}(\gamma_v).
\end{equation}
The inner sum is over finite sets \(S,U\) of cell indices. For each pair in that sum, \(x_v\in Q_n(U)\subseteq Q_n(S)=X\). Since the dynamics restricted to \(Q_n(U)\) preserves parity,
\[
 \tau_t^{\lambda,Q_n(U)}(\gamma_v)\in\A_{Q_n(U)}^-\subseteq\A_X^-.
\]
For fixed \(t\), only finitely many \(\chi_n(t)\) are nonzero, and each finite-volume Hamiltonian is affine in \(\lambda\). The matrix exponentials defining the finite-volume evolutions are therefore real analytic in \(\lambda\). The weights in the finite sum are independent of \(\lambda\), so for fixed \(v,t,X\) the map
\[
 \lambda\longmapsto\bigl[\mathcal T^{\Phi_\lambda}_{0,t,\{x_v\}}(\gamma_v)\bigr](X)
\]
is real analytic. By \cite[Corollary~4.4]{FermionicLR},
\[
 \tau_t^{\Phi_\lambda}(\gamma_v)
 =\sum_{X\ni x_v}\bigl[\mathcal T^{\Phi_\lambda}_{0,t,\{x_v\}}(\gamma_v)\bigr](X).
\]
The series converges absolutely in norm by the bound
\[
 \sum_{X\ni x_v}
   \bigl\|\bigl[\mathcal T^{\Phi_\lambda}_{0,t,\{x_v\}}(\gamma_v)\bigr](X)\bigr\|
 \le C\E^{\kappa_0}.
\]

For each \(j\in\{1,\ldots,2r\}\), define the even on-site interaction by \(\Psi_j(\{x\}):=\jmath_x(\gamma_{x,j})\), with all other terms zero. Thus \(\|\Psi_j\|^{\#}_{p_\Phi,\kappa_0}=\E^{\kappa_0}\). Write \(\widehat\tau^{\lambda,Q_n(U)}\) for the dynamics generated by the interaction $\1_{\{Z\subseteq Q_n(U)\}}\widehat\Phi_\lambda(Z)$. For \(x\in Q_n(U)\), \cite[Lemma~4.3]{FermionicLR} gives
\[
 \widehat\tau_t^{\lambda,Q_n(U)}(\Psi_j(\{x\}))
 =\jmath_x\!\left(\tau_t^{\lambda,Q_n(U)}(\gamma_{x,j})\right).
\]
Using this identity in \eqref{eq:local-representation-formula} with \(v=(x,j)\) gives, for \(X\ni x\),
\[
 \bigl[\mathcal T^{\widehat\Phi_\lambda}_{0,t,\{x\}}(\Psi_j(\{x\}))\bigr](X)
 =\jmath_x\!\left(\bigl[\mathcal T^{\Phi_\lambda}_{0,t,\{x\}}(\gamma_{x,j})\bigr](X)\right).
\]
Since \(\jmath_x\) is independent of \(\lambda\), differentiation and its isometry give, for \(X\ni x\),
\begin{equation}\label{eq:local-derivative-norm-equality}
 \Bigl\|\partial_\lambda
   \bigl[\mathcal T^{\widehat\Phi_\lambda}_{0,t,\{x\}}
                        (\Psi_j(\{x\}))\bigr](X)\Bigr\|
 =\Bigl\|\partial_\lambda
   \bigl[\mathcal T^{\Phi_\lambda}_{0,t,\{x\}}
                        (\gamma_{x,j})\bigr](X)\Bigr\|.
\end{equation}
Since \(\partial_\lambda\widehat\Phi_\lambda(Z)=\cV(Z)\) in the enlarged algebra,
\[
 \|\partial_\lambda\widehat\Phi_\lambda\|^{\#}_{p_\Phi,\kappa_0}
 =\|\cV\|^{\#}_{p_\Phi,\kappa_0}.
\]
Apply \cite[Theorem~5.1(i), equation~(75)]{FermionicLR} to \(\widehat\Phi_\lambda\), with input \(\Psi_j\), direction \(\partial_\lambda\widehat\Phi_\lambda\), and \(n=k=1\), \(p=a=p_\Phi\). Since the only nonzero input terms are indexed by single sites, this gives
\[
 \sup_{|\lambda|\le1}\sup_z\sum_{X\ni z}|X|(1+\diam X)^{p_\Phi}
 \sum_{x\in X}\Bigl\|\partial_\lambda
   \bigl[\mathcal T^{\widehat\Phi_\lambda}_{0,t,\{x\}}(\Psi_j(\{x\}))\bigr](X)\Bigr\|\le C(1+|t|)^{m^{(1)}_{p_\Phi,1}}
      \E^{\kappa_0}\|\cV\|^{\#}_{p_\Phi,\kappa_0}.
\]
Equations (13) and (74) of \cite{FermionicLR} give
\[
 m^{(1)}_{p_\Phi,1}=1+\frac{(p_\Phi+1)(3d_{\rm vol}+p_\Phi)}{p_\Phi-d_{\rm vol}},
 \qquad m_B:=\bigl\lceil m^{(1)}_{p_\Phi,1}\bigr\rceil.
\]
Using \eqref{eq:local-derivative-norm-equality} and summing over \(j=1,\ldots,2r\) gives \eqref{eq:rooted-parameter-derivative} with \(v=(x,j)\), after absorbing \(2r\E^{\kappa_0}\|\cV\|^{\#}_{p_\Phi,\kappa_0}\) into \(C\).
\end{proof}

We use the tracial dyadic shells \(\Delta^{\rm tr}_{X,i}\) of Definition~\ref{def:tracial-shells}, with radii \(R_0=0\) and \(R_i=2^{i-1}\) for \(i\ge1\). The following lemma gives the local decompositions required by Hypothesis~\ref{hyp:perturbed-family}(ii) in both locality regimes.

\begin{lemma}\label{lem:normalized-field-differences}
Assume Hypothesis~\ref{hyp:locality}. Let \(k\) satisfy \eqref{eq:kernel-conditions}. Let \(\mathsf N=(\mathsf N_{uv})_{u,v\in\Gamma\times\{1,\ldots,2r\}}\) be a complex matrix. For every \(u\in\Gamma\times\{1,\ldots,2r\}\) and every \(\lambda\) in the range of either case below, the series
\begin{equation}\label{eq:normalized-filtered-difference}
 \Delta F_u:=\sum_{v\in\Gamma\times\{1,\ldots,2r\}} \mathsf N_{uv}
       \bigl(\gamma_v[k;\Phi_\lambda]-\gamma_v[k;\Phi_0]\bigr)
\end{equation}
converges absolutely in norm. There is a decomposition
\[
 \Delta F_u=\sum_{\substack{Z\Subset\Gamma\\ x_u\in Z}}\Delta F_{u,Z},
 \qquad \Delta F_{u,Z}\in\A_Z^-,
\]
with absolute convergence in norm and the stated bound.
\begin{enumerate}[label=(\roman*),leftmargin=*]
\item \emph{Stretched-exponential locality.}
Assume regime~\ref{ass:exponential} and \(\|\mathsf N\|^{\rm Sch}_{\mu,\theta_{\rm in}}<\infty\) for some \(\mu>0\). Suppose there are model constants \(C_*,a_*>0\) such that
\[
 |k(t)|\le C_*\E^{-a_*|t|^{1/2}},\qquad t\in\R.
\]
There is a constant \(C_A>0\) depending only on the model data, \(C_*,a_*,\mu\) and \(\|\mathsf N\|^{\rm Sch}_{\mu,\theta_{\rm in}}\). For each \(0<\theta\le\theta_{\rm in}/\sG\), there is a \(m>0\) depending only on the same quantities as \(C_A\) and on \(C_k,a_k,\sG,\theta\) so that, for every \(\lambda\) with \(|\lambda|\le1\), the local terms satisfy
\begin{equation}\label{eq:normalized-field-differences-A}
 \sup_{z\in\Gamma}\sum_{u\in\Gamma\times\{1,\ldots,2r\}}
 \sum_{\substack{Z\Subset\Gamma\\ z,x_u\in Z}}|Z|\E^{m\diam(Z)^\theta}
             \|\Delta F_{u,Z}\|\le C_A|\lambda|^{1/2}.
\end{equation}
\item \emph{Polynomial locality.}
Assume regime~\ref{ass:polynomial-weight} and \(\|\mathsf N\|^{\rm Sch,pol}_{p_\Phi}<\infty\).
Let \(m_B\) be the model integer from Lemma~\ref{lem:rooted-parameter-derivative}. There exists a model constant \(C_B>0\) such that, for every \(0<b\le1\) and every \(\lambda\in\R\) with \(b|\lambda|\le1\), the local terms can be chosen so that
\begin{equation}\label{eq:normalized-field-differences-C}
 \begin{aligned}
 &\sup_{z\in\Gamma}\sum_{u\in\Gamma\times\{1,\ldots,2r\}}
 \sum_{\substack{Z\Subset\Gamma\\ z,x_u\in Z}}|Z|(1+\diam Z)^{p_\Phi}
               \|\Delta F_{u,Z}\|\\
 &\qquad\le C_B\|\mathsf N\|^{\rm Sch,pol}_{p_\Phi}b|\lambda|
                 \int_\R|k(t)|(1+|t|/b)^{m_B}\,\dd t.
 \end{aligned}
\end{equation}
\end{enumerate}
\end{lemma}
\begin{proof}
For \(v\in\Gamma\times\{1,\ldots,2r\}\), set
\[
 \Delta\gamma_v:=\gamma_v[k;\Phi_\lambda]-\gamma_v[k;\Phi_0]
 =\int_\R k(t)(\tau_t^{\Phi_\lambda}-\tau_t^{\Phi_0})(\gamma_v)\,\dd t.
\]
For \(u\in\Gamma\times\{1,\ldots,2r\}\), \eqref{eq:normalized-filtered-difference} reads
\[
 \Delta F_u=\sum_{v\in\Gamma\times\{1,\ldots,2r\}}\mathsf N_{uv}\Delta\gamma_v.
\]
For \(x,y\in\Gamma\) and \(1\le i\le 2r\), Cauchy--Schwarz gives
\begin{equation}\label{eq:normalized-block-entry-sum}
 \sum_{j=1}^{2r}|\mathsf N_{(x,i),(y,j)}|
 \le\sqrt{2r}\,\bigl\|(\mathsf N_{(x,i'),(y,j')})_{i',j'=1}^{2r}\bigr\|.
\end{equation}
The same bound holds for summing over \(i=1,\ldots,2r\) at fixed \(j\in\{1,\ldots,2r\}\). In either locality regime, summing \eqref{eq:normalized-block-entry-sum} over \(y\in\Gamma\) gives
\[
 \sup_{u\in\Gamma\times\{1,\ldots,2r\}}
 \sum_{v\in\Gamma\times\{1,\ldots,2r\}}|\mathsf N_{uv}|
 \le \sqrt{2r}\|\mathsf N\|^{\rm Sch,pol}_{0}<\infty.
\]
By isometry and parity preservation, \(\|\Delta\gamma_v\|\le2\|k\|_1\) and \(\Delta\gamma_v\in\A^-\). Hence the series in \eqref{eq:normalized-filtered-difference} converges absolutely in norm, with \(\Delta F_u\in\A^-\).

\smallskip
\noindent\emph{(i) Spatial approximation and the dyadic decomposition.}
Assume \ref{ass:exponential} and put \(\alpha:=\theta_{\rm in}/\sG\), \(\alpha_0:=\theta_{\rm in}/2\). Apply Lemma~\ref{lem:filtered-field-locality}(iii) with \(b=1\), background \(\Phi_0\) and \(\Psi=\lambda\cV\). Using \eqref{eq:normalized-filtered-difference} and the bound on \(k\) with \(C_*,a_*\) gives
\begin{equation}\label{eq:normalized-alt-smallness}
 \sup_{u\in\Gamma\times\{1,\ldots,2r\}}\|\Delta F_u\|
 \le C_D J_{\cV}|\lambda|\sqrt{2r}\|\mathsf N\|^{\rm Sch,pol}_0
       \int_\R|k(t)|\,|t|(1+|t|)^{m_D}\,\dd t\le C_{\rm sm}|\lambda|,
\end{equation}
where \(C_{\rm sm}\) depends only on the quantities allowed for \(C_A\). For \(d(x_u,x_v)\le R/2\), the inclusion \(\ball_{R/2}(x_v)\subset\ball_R(x_u)\) implies
\[
 (\id-\Ftr{\ball_R(x_u)})\Delta\gamma_v
 =(\id-\Ftr{\ball_R(x_u)})
       (\id-\Ftr{\ball_{R/2}(x_v)})\Delta\gamma_v.
\]
For \(d(x_u,x_v)>R/2\), use \(\|\Delta\gamma_v\|\le2\|k\|_1\). Applying \(\id-\Ftr{\ball_R(x_u)}\) to \eqref{eq:normalized-filtered-difference} gives
\begin{equation}\label{eq:normalized-alt-localization}
 \begin{gathered}
 \|(\id-\Ftr{\ball_R(x_u)})\Delta F_u\| 
 \\
 \le4\|k\|_1\sum_{\substack{v\in\Gamma\times\{1,\ldots,2r\}\\d(x_u,x_v)>R/2}}|\mathsf N_{uv}|
 +2\sum_{\substack{v\in\Gamma\times\{1,\ldots,2r\}\\d(x_u,x_v)\le R/2}}|\mathsf N_{uv}|
       \|(\id-\Ftr{\ball_{R/2}(x_v)})\Delta\gamma_v\|.
 \end{gathered}
\end{equation}
Equations \eqref{eq:normalized-block-entry-sum} and \eqref{eq:Schur-norm} give
\begin{equation}\label{eq:normalized-exponential-entry-sums}
 \begin{aligned}
 \sum_{v\in\Gamma\times\{1,\ldots,2r\}}|\mathsf N_{uv}|
 &\le\sqrt{2r}\|\mathsf N\|^{\rm Sch}_{\mu,\theta_{\rm in}},\\
 \sum_{\substack{v\in\Gamma\times\{1,\ldots,2r\}\\d(x_u,x_v)>R/2}}|\mathsf N_{uv}|
 &\le\sqrt{2r}\|\mathsf N\|^{\rm Sch}_{\mu,\theta_{\rm in}}
                  \E^{-\mu(R/2)^{\theta_{\rm in}}}.
 \end{aligned}
\end{equation}
Apply Lemma~\ref{lem:filtered-field-locality}(i) at couplings \(\lambda\) and \(0\). Use the bound with \(C_*,a_*\) for \(\xi=\alpha_0\), and \eqref{eq:kernel-conditions} for \(\xi=\alpha\) if \(\alpha\ne\alpha_0\). This gives
\[
 \sup_{v\in\Gamma\times\{1,\ldots,2r\}}\|(\id-\Ftr{\ball_R(x_v)})\Delta\gamma_v\|
 \le C_\xi\E^{-c_\xi R^\xi},\qquad R\ge0.
\]
Set
\[
 \begin{gathered}
 b_\xi:=\min\{2^{-\theta_{\rm in}}\mu,2^{-\xi}c_\xi\},\\
 L_\xi:=1+\E^{b_\xi}\sqrt{2r}\|\mathsf N\|^{\rm Sch}_{\mu,\theta_{\rm in}}
             \bigl(4\|k\|_1+2C_\xi\bigr).
 \end{gathered}
\]
For \(\xi=\alpha_0\), \(b_\xi\) depends only on the model data, \(C_*,a_*\) and \(\mu\). Since \(\|k\|_1\le4C_*/a_*^2\), \(L_\xi\) depends only on these quantities and on \(\|\mathsf N\|^{\rm Sch}_{\mu,\theta_{\rm in}}\). For \(\xi=\alpha\ne\alpha_0\), both constants may also depend on \(C_k,a_k,\sG\). Substituting the bound at radius \(R/2\) and the bounds in \eqref{eq:normalized-exponential-entry-sums} into \eqref{eq:normalized-alt-localization} gives
\begin{equation}\label{eq:normalized-alt-spatial}
 \begin{gathered}
 \sup_{u\in\Gamma\times\{1,\ldots,2r\}}\|(\id-\Ftr{\ball_R(x_u)})\Delta F_u\|\\
 \le \sqrt{2r}\|\mathsf N\|^{\rm Sch}_{\mu,\theta_{\rm in}}
       \bigl(4\|k\|_1\E^{-\mu(R/2)^{\theta_{\rm in}}}
               +2C_\xi\E^{-c_\xi(R/2)^\xi}\bigr)
 \le L_\xi\E^{-b_\xi R^\xi}.
 \end{gathered}
\end{equation}
The last inequality uses \(R^{\theta_{\rm in}}\ge R^\xi\) for \(R\ge1\) and \(1\le\E^{b_\xi-b_\xi R^\xi}\) for \(0\le R\le1\).

For \(i\in\mathbb N_0\), let \(\Delta F_{u,i}:=\Delta^{\rm tr}_{\{x_u\},i}(\Delta F_u)\in\A_{\ball_{R_i}(x_u)}^-\). By Definition~\ref{def:tracial-shells},
\[
 \sum_{i=0}^{n}\Delta F_{u,i}
 =\Ftr{\ball_{R_n}(x_u)}(\Delta F_u)
 \xrightarrow[n\to\infty]{\|\cdot\|}\Delta F_u.
\]
By contractivity and \eqref{eq:normalized-alt-smallness},
\[
 \|\Delta F_{u,i}\|\le 2\|\Delta F_u\|\le 2C_{\rm sm}|\lambda|,
 \qquad i\ge0.
\]
For \(i\ge2\), \eqref{eq:tracial-shell-estimate} and \eqref{eq:normalized-alt-spatial} give
\[
 \|\Delta F_{u,i}\|
 \le\|(\id-\Ftr{\ball_{R_{i-1}}(x_u)})\Delta F_u\|
 \le L_\xi\E^{-b_\xi(R_i/2)^\xi}.
\]
Combining these bounds and including \(i=0,1\) by enlarging the constants gives
\begin{equation}\label{eq:normalized-alt-shells}
 \|\Delta F_{u,i}\|
 \le H_\xi\min\{|\lambda|,\E^{-h_\xi R_i^\xi}\},
 \qquad \xi\in\{\alpha,\alpha_0\}.
\end{equation}
Here \(h_\xi:=2^{-\xi}b_\xi\), and \(H_\xi>0\) has the same permitted dependencies as \(L_\xi,b_\xi\). The limit of the partial sums and \eqref{eq:normalized-alt-shells} give
\begin{equation}\label{eq:normalized-shell-series}
 \Delta F_u=\sum_{i=0}^\infty\Delta F_{u,i},\qquad
 \sum_{i=0}^\infty\|\Delta F_{u,i}\|
 \le H_{\alpha_0}\sum_{i=0}^\infty\E^{-h_{\alpha_0}R_i^{\alpha_0}}<\infty.
\end{equation}

\smallskip
\noindent\emph{Choosing the rate with a uniform prefactor.}
For \(\theta>0\) and \(m\ge0\), define
\[
 E_{\theta,m}(\lambda):=
 \sup_{z\in\Gamma}\sum_{u\in\Gamma\times\{1,\ldots,2r\}}
 \sum_{\substack{i\in\mathbb N_0\\z\in\ball_{R_i}(x_u)}}
       |\ball_{R_i}(x_u)|\E^{m(2R_i)^\theta}\|\Delta F_{u,i}\|.
\]
For fixed \(z\in\Gamma\) and \(i\in\mathbb N_0\), \eqref{eq:growth} gives
\[
 \sum_{\substack{u\in\Gamma\times\{1,\ldots,2r\}\\z\in\ball_{R_i}(x_u)}}|\ball_{R_i}(x_u)|
 \le 2r|\ball_{R_i}(z)|\sup_{x\in\Gamma}|\ball_{R_i}(x)|
 \le 2rC_\Gamma^2(1+R_i)^{2d_{\rm vol}}.
\]
Using \(\min\{a,b\}\le\sqrt{ab}\) in \eqref{eq:normalized-alt-shells}, we obtain
\begin{equation}\label{eq:normalized-alt-weighted-sum}
 E_{\theta,m}(\lambda)
 \le 2rC_\Gamma^2 H_\xi|\lambda|^{1/2}
       \sum_{i=0}^\infty(1+R_i)^{2d_{\rm vol}}
       \E^{m(2R_i)^\theta-h_\xi R_i^\xi/2}.
\end{equation}
Setting \(\xi=\alpha_0\) and \(m=0\) in \eqref{eq:normalized-alt-weighted-sum} gives a constant \(K_0\ge1\) that depends only on the quantities allowed for \(C_A\) such that
\[
 E_{\theta,0}(\lambda)
 \le 2rC_\Gamma^2 H_{\alpha_0}|\lambda|^{1/2}
       \sum_{i=0}^\infty(1+R_i)^{2d_{\rm vol}}
       \E^{-h_{\alpha_0}R_i^{\alpha_0}/2}
 \le K_0|\lambda|^{1/2}.
\]
For the prescribed \(0<\theta\le\alpha\), take \(\xi=\alpha\) and choose \(m_\theta>0\) with \(m_\theta2^\theta<h_\alpha/4\). For \(i\ge1\), the exponent in \eqref{eq:normalized-alt-weighted-sum} satisfies
\[
 m_\theta(2R_i)^\theta-\tfrac12h_\alpha R_i^\alpha
 \le -\tfrac14h_\alpha R_i^\alpha.
\]
Thus \(E_{\theta,m_\theta}(\lambda)\le K_\theta|\lambda|^{1/2}\) for some \(K_\theta\ge K_0\). \(K_\theta\) may depend on \(C_k,a_k,\sG,\theta\).

To obtain the prefactor \(2K_0\), we interpolate between the bounds at \(m=0\) and \(m=m_\theta\).
For \(0<\eta<1\), apply H\"older's inequality with exponents \(1/(1-\eta)\) and \(1/\eta\) to the sum at fixed \(z\in\Gamma\) in the definition of \(E_{\theta,\eta m_\theta}(\lambda)\). Taking the supremum over \(z\in\Gamma\) gives
\begin{equation}\label{eq:normalized-alt-interpolation}
 E_{\theta,\eta m_\theta}(\lambda)
 \le E_{\theta,0}(\lambda)^{1-\eta}E_{\theta,m_\theta}(\lambda)^\eta
 \le K_0^{1-\eta}K_\theta^\eta|\lambda|^{1/2}.
\end{equation}
Choose \(\eta\in(0,1)\) so that \(\eta\log(K_\theta/K_0)\le\log2\), and put \(m:=\eta m_\theta\). This choice is independent of \(\lambda\). For \(Z\Subset\Gamma\) with \(x_u\in Z\), define
\[
 \Delta F_{u,Z}:=\sum_{\substack{i\in\mathbb N_0\\\ball_{R_i}(x_u)=Z}}\Delta F_{u,i}
 \in\A_Z^-.
\]
Regrouping the absolutely convergent series in \eqref{eq:normalized-shell-series} gives
\[
 \sum_{\substack{Z\Subset\Gamma\\x_u\in Z}}\Delta F_{u,Z}
 =\sum_{i=0}^\infty\Delta F_{u,i}=\Delta F_u.
\]
Since \(\diam Z\le2R_i\) when \(Z=\ball_{R_i}(x_u)\), \eqref{eq:normalized-alt-interpolation} gives
\[
 \sup_{z\in\Gamma}\sum_{u\in\Gamma\times\{1,\ldots,2r\}}
 \sum_{\substack{Z\Subset\Gamma\\z,x_u\in Z}}
 |Z|\E^{m\diam(Z)^\theta}\|\Delta F_{u,Z}\|
 \le E_{\theta,m}(\lambda)\le 2K_0|\lambda|^{1/2}.
\]
This proves \eqref{eq:normalized-field-differences-A} with \(C_A=2K_0\).

\smallskip
\noindent\emph{(ii) Decomposing the filtered differences before applying \(\mathsf N\).}
Assume \ref{ass:polynomial-weight}. First take \(b=1\) and hence \(|\lambda|\le1\). Taking \(z=x_v\) and retaining only the terms with this \(v\) in \eqref{eq:rooted-parameter-derivative} of Lemma~\ref{lem:rooted-parameter-derivative} gives
\[
 \int_\R |k(t)|\int_{\min\{0,\lambda\}}^{\max\{0,\lambda\}}\sum_{\substack{X\Subset\Gamma\\x_v\in X}}
 \bigl\|\partial_s[\mathcal T^{\Phi_s}_{0,t,\{x_v\}}(\gamma_v)](X)\bigr\|
 \,\dd s\,\dd t
 \le C|\lambda|\int_\R |k(t)|(1+|t|)^{m_B}\,\dd t<\infty.
\]
For \(X\Subset\Gamma\) with \(x_v\in X\), Lemma~\ref{lem:rooted-parameter-derivative} gives \(\bigl[\mathcal T^{\Phi_s}_{0,t,\{x_v\}}(\gamma_v)\bigr](X)\in\A_X^-\) for \(s\in[-1,1]\). Since \(\A_X^-\) is a closed linear subspace,
\[
 \partial_s\bigl[\mathcal T^{\Phi_s}_{0,t,\{x_v\}}(\gamma_v)\bigr](X)\in\A_X^-.
\]
Define
\[
 \Delta\gamma_{v,X}:=\int_\R k(t)\int_0^\lambda
       \partial_s\bigl[\mathcal T^{\Phi_s}_{0,t,\{x_v\}}
                                     (\gamma_v)\bigr](X)\,\dd s\,\dd t
 \in\A_X^-.
\]
Taking the norms inside the integrals, Tonelli's theorem and \eqref{eq:rooted-parameter-derivative} give
\begin{equation}\label{eq:normalized-alt-QB}
 \begin{aligned}
 Q_B&:=\sup_{z\in\Gamma}\sum_{\substack{X\Subset\Gamma\\z\in X}}|X|(1+\diam X)^{p_\Phi}
 \sum_{\substack{v\in\Gamma\times\{1,\ldots,2r\}\\x_v\in X}}\|\Delta\gamma_{v,X}\|\\
 &\le\int_\R|k(t)|\int_{\min\{0,\lambda\}}^{\max\{0,\lambda\}}C(1+|t|)^{m_B}\,\dd s\,\dd t
 =C|\lambda|\int_\R|k(t)|(1+|t|)^{m_B}\,\dd t.
 \end{aligned}
\end{equation}
By absolute convergence, the fundamental theorem of calculus and Lemma~\ref{lem:rooted-parameter-derivative} give
\[
 \begin{aligned}
 \sum_{\substack{X\Subset\Gamma\\x_v\in X}}\Delta\gamma_{v,X}
 &=\int_\R k(t)\sum_{\substack{X\Subset\Gamma\\x_v\in X}}
    \bigl[(\mathcal T^{\Phi_\lambda}_{0,t,\{x_v\}}
          -\mathcal T^{\Phi_0}_{0,t,\{x_v\}})(\gamma_v)\bigr](X)\,\dd t\\
 &=\int_\R k(t)(\tau_t^{\Phi_\lambda}-\tau_t^{\Phi_0})(\gamma_v)\,\dd t
 =\Delta\gamma_v.
 \end{aligned}
\]

\smallskip
\noindent\emph{Regrouping the terms by support.}
The term \(\mathsf N_{uv}\Delta\gamma_{v,X}\) lies in \(\A_X^-\), with \(x_v\in X\). To match Hypothesis~\ref{hyp:perturbed-family}(ii), the decomposition of \(\Delta F_u\) must use sets \(Z\) containing \(x_u\). We therefore assign this term to \(Z=X\cup\{x_u\}\). For \(Z\Subset\Gamma\) with \(x_u\in Z\), set
\[
 \Delta F_{u,Z}:=
 \sum_{\substack{v\in\Gamma\times\{1,\ldots,2r\}\\
                  X\Subset\Gamma,\ x_v\in X\\X\cup\{x_u\}=Z}}
                    \mathsf N_{uv}\Delta\gamma_{v,X}\in\A_Z^-.
\]
By \eqref{eq:normalized-alt-QB} at \(z=x_v\),
\[
 \sum_{v\in\Gamma\times\{1,\ldots,2r\}}\sum_{\substack{X\Subset\Gamma\\x_v\in X}}
       |\mathsf N_{uv}|\|\Delta\gamma_{v,X}\|
 \le Q_B\sum_{v\in\Gamma\times\{1,\ldots,2r\}}|\mathsf N_{uv}|<\infty,
\]
so \eqref{eq:normalized-filtered-difference} can be regrouped as
\[
 \sum_{\substack{Z\Subset\Gamma\\x_u\in Z}}\Delta F_{u,Z}
 =\sum_{v\in\Gamma\times\{1,\ldots,2r\}}\mathsf N_{uv}
       \sum_{\substack{X\Subset\Gamma\\x_v\in X}}\Delta\gamma_{v,X}
 =\sum_{v\in\Gamma\times\{1,\ldots,2r\}}\mathsf N_{uv}\Delta\gamma_v=\Delta F_u.
\]

It remains to show \eqref{eq:normalized-field-differences-C}. Since \(x_v\in X\), we have \(|X|\ge1\), so
\[
 |X\cup\{x_u\}|\le |X|+1\le2|X|.
\]
The triangle inequality with \(x_v\in X\) also gives
\[
 (1+\diam(X\cup\{x_u\}))^{p_\Phi} \le (1+d(x_u,x_v))^{p_\Phi}(1+\diam X)^{p_\Phi}.
\]
For \(z\in\Gamma\) and \(Z=X\cup\{x_u\}\), we have \(\1_{\{z\in Z\}}\le\1_{\{z\in X\}}+\1_{\{x_u=z\}}\). Using \(|Z|\le2|X|\) and the bound above gives
\begin{equation}\label{eq:normalized-polynomial-support-split}
 \begin{aligned}
 &\sum_{u\in\Gamma\times\{1,\ldots,2r\}}\sum_{\substack{Z\Subset\Gamma\\z,x_u\in Z}}
       |Z|(1+\diam Z)^{p_\Phi}\|\Delta F_{u,Z}\|\\
 &\quad\le 2\sum_{u,v\in\Gamma\times\{1,\ldots,2r\}}\sum_{\substack{X\Subset\Gamma\\x_v\in X}}
       |X|(1+\diam X)^{p_\Phi}(1+d(x_u,x_v))^{p_\Phi}\\
 &\qquad\quad{}\times\bigl(\1_{\{z\in X\}}+\1_{\{x_u=z\}}\bigr)
       |\mathsf N_{uv}|\|\Delta\gamma_{v,X}\|.
 \end{aligned}
\end{equation}
We now bound the two contributions on the right of \eqref{eq:normalized-polynomial-support-split} separately, one from \(\1_{\{z\in X\}}\) and one from \(\1_{\{x_u=z\}}\). Equations \eqref{eq:normalized-block-entry-sum} and \eqref{eq:Schur-polynomial-norm} give the estimate used for both contributions:
\begin{equation}\label{eq:normalized-polynomial-entry-sums}
 \sup_{u\in\Gamma\times\{1,\ldots,2r\}}
 \sum_{v\in\Gamma\times\{1,\ldots,2r\}} (1+d(x_u,x_v))^{p_\Phi}|\mathsf N_{uv}|
 \le\sqrt{2r}\|\mathsf N\|^{\rm Sch,pol}_{p_\Phi}.
\end{equation}
The same bound holds with the supremum over \(v\) and the sum over \(u\). In the contribution with \(z\in X\), \(u\) ranges over all of \(\Gamma\times\{1,\ldots,2r\}\). We first sum over \(u\) using the reversed bound in \eqref{eq:normalized-polynomial-entry-sums}, then use \eqref{eq:normalized-alt-QB} for the sum over \(X\ni z\). This gives
\[
 \begin{aligned}
 &\sum_{\substack{X\Subset\Gamma\\z\in X}}|X|(1+\diam X)^{p_\Phi}
       \sum_{\substack{v\in\Gamma\times\{1,\ldots,2r\}\\x_v\in X}}\|\Delta\gamma_{v,X}\| \sum_{u\in\Gamma\times\{1,\ldots,2r\}}(1+d(x_u,x_v))^{p_\Phi}|\mathsf N_{uv}|\\
 &\qquad\le\sqrt{2r}\|\mathsf N\|^{\rm Sch,pol}_{p_\Phi}Q_B.
 \end{aligned}
\]
In the contribution with \(x_u=z\), there are only \(2r\) choices for \(u\), and \(X\) need only contain \(x_v\). Taking \(z=x_v\) in \eqref{eq:normalized-alt-QB} and retaining the term for \(v\) gives
\[
 \sum_{\substack{X\Subset\Gamma\\x_v\in X}}|X|(1+\diam X)^{p_\Phi}\|\Delta\gamma_{v,X}\|\le Q_B.
\]
Using \eqref{eq:normalized-polynomial-entry-sums} to sum over \(v\), and then summing over \(u\in\{z\}\times\{1,\ldots,2r\}\), gives
\[
 \begin{aligned}
 &\sum_{u\in\{z\}\times\{1,\ldots,2r\}} \sum_{v\in\Gamma\times\{1,\ldots,2r\}}(1+d(x_u,x_v))^{p_\Phi}|\mathsf N_{uv}| \sum_{\substack{X\Subset\Gamma\\x_v\in X}}|X|(1+\diam X)^{p_\Phi}\|\Delta\gamma_{v,X}\|\\
 &\qquad\le 2r\sqrt{2r}\|\mathsf N\|^{\rm Sch,pol}_{p_\Phi}Q_B.
 \end{aligned}
\]
Substituting the two bounds for \(z\in X\) and \(x_u=z\) into \eqref{eq:normalized-polynomial-support-split} and taking the supremum over \(z\in\Gamma\) gives
\[
 \begin{aligned}
 &\sup_{z\in\Gamma}\sum_{u\in\Gamma\times\{1,\ldots,2r\}}
 \sum_{\substack{Z\Subset\Gamma\\z,x_u\in Z}}|Z|(1+\diam Z)^{p_\Phi}
          \|\Delta F_{u,Z}\|\\
 &\le 2\bigl(\sqrt{2r}+2r\sqrt{2r}\bigr)\|\mathsf N\|^{\rm Sch,pol}_{p_\Phi}Q_B
 =2(1+2r)\sqrt{2r}\|\mathsf N\|^{\rm Sch,pol}_{p_\Phi}Q_B.
 \end{aligned}
\]
Substituting \eqref{eq:normalized-alt-QB} proves \eqref{eq:normalized-field-differences-C} for \(b=1\), with \(C_B\) depending only on the model data.

For \(0<b\le1\) and \(|b\lambda|\le1\), the substitution \(t=bs\) gives
\[
 \Delta\gamma_v
 =\int_\R b\,k(bs)
 \bigl(\tau_s^{\Phi_{bh}+(b\lambda)\cV}-\tau_s^{\Phi_{bh}}\bigr)(\gamma_v)\,\dd s.
\]
Thus we apply the case \(b=1\) of \eqref{eq:normalized-field-differences-C} with free operator \(bh\), coupling \(b\lambda\), and kernel \(s\mapsto b\,k(bs)\). The same \(C_B\) applies since \(\|bh\|^{\rm Sch,pol}_{p_h}\le bC_h\le C_h\). The resulting coupling factor is \(b|\lambda|\), and the time moment is
\[
 \int_\R b|k(bs)|(1+|s|)^{m_B}\,\dd s
 =\int_\R |k(t)|(1+|t|/b)^{m_B}\,\dd t.
\]
Substituting these two factors proves \eqref{eq:normalized-field-differences-C}.
\end{proof}

\subsection{The thermal families and comparison estimates}
\label{sec:thermal-family-proof}

We now prove Propositions~\ref{prop:thermal-family} and~\ref{prop:thermal-comparison}.

\begin{proof}[Proof of Proposition~\ref{prop:thermal-family}]
We first verify absolute norm convergence of the series in \eqref{eq:thermal-coordinates} and \eqref{eq:FG-coordinates}.
For the cutoff family under the free gap assumption, use the weight \(w\) of Lemma~\ref{lem:uniform-free-calculus} and the identity \(\Rb=\Rb P_++\Rb P_-\).
Lemma~\ref{lem:bounded-temperature-calculus} bounds \(\mathsf M\) in the Fermi-weighted case.
By Cauchy--Schwarz we find
\begin{equation}\label{eq:thermal-absolute-row-sums}
 \begin{aligned}
 \sup_{u\in\Gamma\times\{1,\ldots,2r\}}\sum_{v\in\Gamma\times\{1,\ldots,2r\}}|(\Rb)_{uv}|
 &\le\sqrt{2r}\bigl(\|\Rb P_+\|_w^{\rm Sch}
                    +\|\Rb P_-\|_w^{\rm Sch}\bigr)
 \le2\sqrt{2r}\,C_{\rm th},\\
 \sup_{u\in\Gamma\times\{1,\ldots,2r\}}\sum_{v\in\Gamma\times\{1,\ldots,2r\}}|\mathsf M_{uv}|
 &\le\sqrt{2r}\,C_{\mathsf M},
 \end{aligned}
\end{equation}
uniformly in \(\beta\).

For any \(k\in L^1(\R)\) and \(\Xi\in\mathfrak J_1\), \eqref{eq:filtered-Majorana-definition} and isometry of the dynamics give
\[
 \left\|\gamma_v[k;\Xi]\right\|
 \le\|k\|_1.
\]
Using Lemmas~\ref{lem:Gevrey} and~\ref{lem:FG-filter} in the definitions \eqref{eq:thermal-coordinates} and \eqref{eq:FG-coordinates} and applying \eqref{eq:thermal-absolute-row-sums}, we obtain absolute norm convergence of the series \eqref{eq:thermal-coordinates} and \eqref{eq:FG-coordinates} for every \(\lambda\in\R\).

\smallskip
\noindent\emph{Agreement with the free family at zero coupling.}
We now verify the identities of Equation \eqref{eq:normalized-constraints}.
The anticommutation relations \(\{\gamma_v,\gamma_w\}=2\delta_{vw}\1\) imply that
\[
 \left\|\sum_{v\in \Gamma \times \{1, \ldots, 2r\}} z_v\gamma_v\right\|
 \le\sqrt2\left(\sum_{v\in \Gamma \times \{1, \ldots, 2r\}}|z_v|^2\right)^{1/2}
\]
for every finitely supported vector \(z\in\ell^2 (\Gamma \times \{1, \ldots, 2r\})\). Fix \(u\in\Gamma\times\{1,\ldots,2r\}\) and \(k\in L^1(\R)\). Since \(h=h^*\), the operator \(\E^{-\I th}\) is unitary and
\[
 \sum_{v\in\Gamma\times\{1,\ldots,2r\}}|(\E^{-\I th})_{uv}|^2=1.
\]
By the previous estimate we find, for every finite set of labels \(J\subset\Gamma\times\{1,\ldots,2r\}\),
\[
 \begin{aligned}
 \left\|k(t)\sum_{v\in J}(\E^{-\I th})_{uv}\gamma_v\right\|
 &\le\sqrt2\,|k(t)|
       \left(\sum_{v\in J}|(\E^{-\I th})_{uv}|^2\right)^{1/2}\le\sqrt2\,|k(t)|.
 \end{aligned}
\]
By \eqref{eq:Majorana-free-dynamics}, for every \(t\in\R\),
\[
 \lim_{J\uparrow\Gamma\times\{1,\ldots,2r\}}
 \left\|k(t)\sum_{v\in J}(\E^{-\I th})_{uv}\gamma_v
       -k(t)\tau_t^{\Phi_0}(\gamma_u)\right\|=0.
\]
Dominated convergence therefore gives 
\begin{equation}\label{eq:free-filtering-identity}
 \gamma_u[k;\Phi_0]=\sum_{v\in\Gamma\times\{1,\ldots,2r\}}\left(\int_\R k(t)\E^{-\I th}\,\dd t\right)_{uv}\gamma_v =\sum_{v\in\Gamma\times\{1,\ldots,2r\}}\widehat k(-h)_{uv}\gamma_v.
\end{equation}

We first consider the cutoff family under assumption \eqref{eq:free-band}. Let \(\beta\ge0\). Applying Equation~\eqref{eq:free-filtering-identity} with \(k=f_\beta\), and using the definition~\eqref{eq:filtered-Majoranas}, we obtain, for all \(u\in\Gamma\times\{1,\ldots,2r\}\),
\begin{equation*}
 \gamma_u[f_\beta;\Phi_0]
 =\sum_{v\in\Gamma\times\{1,\ldots,2r\}}\widehat f_\beta(-h)_{uv}\gamma_v
 =\sum_{v\in\Gamma\times\{1,\ldots,2r\}}(P_++E_\beta P_-)_{uv}\gamma_v.
\end{equation*}
The last equality uses the choice of \(\widehat f\) in Section~\ref{sec:thermal-family-statement}, the definition \eqref{eq:PQE} of \(P_\pm\), and \(E_\beta=\E^{-\beta|h|/2}\).
Using \(\overline{P_+}=P_-\), \(\overline E_\beta=E_\beta\) and \(\ft(\gamma_v)=\gamma_v\), we substitute this expression and its adjoint into the definition \eqref{eq:thermal-coordinates} to obtain
\[
 \begin{aligned}
 \mathsf F_u^{\beta}[\Phi_0]
 &=\frac1{\sqrt2}\sum_{v\in\Gamma\times\{1,\ldots,2r\}}\Bigl[
    (\Rb(P_++E_\beta P_-))_{uv}\gamma_{v,1}
    +\I(\Rb(E_\beta P_++P_-))_{uv}\gamma_{v,2}\Bigr]\\
 &=\frac1{\sqrt2}\sum_{v\in\Gamma\times\{1,\ldots,2r\}}\bigl[
    a_\beta(h)_{uv}\gamma_{v,1}
    +\I a_\beta(-h)_{uv}\gamma_{v,2}\bigr]=\mathsf F_u^{\beta,0}.
 \end{aligned}
\]
For the second equality, use \(hP_\pm=\pm|h|P_\pm\) and the definitions \eqref{eq:free_normalization} and \eqref{eq:gapped-thermal-normalizer}:
\[
 a_\beta(\pm h)
 =a_\beta(|h|)P_\pm+a_\beta(-|h|)P_\mp
 =\Rb(P_\pm+E_\beta P_\mp).
\]
The last equality is the definition \eqref{eq:smooth-free-family}.

We now consider the Fermi-weighted family. Let \(\beta>0\). Since \(q_\beta\in L^1(\R)\) by Lemma~\ref{lem:FG-filter}, applying \eqref{eq:free-filtering-identity} with \(k=q_\beta\) gives
\[
 \gamma_u[q_\beta;\Phi_0]
 =\sum_{v\in\Gamma\times\{1,\ldots,2r\}}\hat q_\beta(-h)_{uv}\gamma_v.
\]
Since \(\hat q_\beta\) is real on \(\R\), \(\overline h=-h\), and \(\gamma_v^*=\gamma_v\), taking adjoints gives
\[
 (\gamma_u[q_\beta;\Phi_0])^*
 =\sum_{v\in\Gamma\times\{1,\ldots,2r\}}\hat q_\beta(h)_{uv}\gamma_v.
\]
Substituting these expressions into \eqref{eq:FG-coordinates}, using that \(\ft\) is complex-linear and fixes every Majorana, gives
\[
 \mathsf G_u^{\beta,0}
 =\sum_{v\in\Gamma\times\{1,\ldots,2r\}}\bigl[
    (\mathsf M \hat q_\beta(-h))_{uv}\gamma_{v,1}
    +\I(\mathsf M \hat q_\beta(h))_{uv}\gamma_{v,2}
   \bigr]
 =\mathsf F_u^{\beta,0},
\]
where we have used the definition \eqref{eq:smooth-free-family} and the identities
\begin{equation*}
 M_\beta(z)\hat q_\beta(-z)=\frac{a_\beta(z)}{\sqrt2},\qquad
 M_\beta(z)\hat q_\beta(z)=\frac{a_\beta(-z)}{\sqrt2}
 \qquad(z\in\R).
\end{equation*}

\par\smallskip
\noindent\emph{(i) Annihilation.}
Fix \(\Xi\in\mathfrak J_1\), \(\beta>0\), and \(\rho\in\KMS_\beta(\tau^\Xi)\). Let \(u\in\Gamma\times\{1,\ldots,2r\}\).
The state \(\rho\) is even by Proposition~\ref{prop:kms-nonempty}. Let \(\widehat\rho\) be its diagonal thermal extension, with GNS triple \((\widehat\pi,\mathcal H,\Omega)\) as in Proposition~\ref{prop:lift}.

By \eqref{eq:CD} and~\eqref{eq:FG-fields}, the first identity in Proposition~\ref{prop:lift}, applied to \(A=\gamma_u[k;\Xi]\) with \(k\in\{f_\beta,q_\beta\}\), gives
\[
 \widehat\pi\Bigl(
       \gamma_u[k;\Xi]^{(1)}
       +\I\bigl(\ft(\gamma_u[k;\Xi]^*)\bigr)^{(2)}
       \Bigr)\Omega=0.
\]

By Equation~\eqref{eq:modified-thermal-family} and norm convergence, the preceding identities give \(\widehat\pi(F_u)\Omega=0\) for either \(F_u\in\{\mathsf F_u^\beta[\Xi],\mathsf G_u^\beta[\Xi]\}\). Thus
\[
 \widehat\rho(F_u^*F_u)
 =\|\widehat\pi(F_u)\Omega\|^2=0,
\]
which proves (i).

\smallskip
\noindent\emph{(ii) Local decompositions.}
Let \(|\lambda|\le1\) for the cutoff family and \(|\eta\lambda|\le1\), with \(\eta=\beta/(1+\beta)\), for the Fermi-weighted family.
Under the free gap assumption \eqref{eq:free-band}, equation~\eqref{eq:thermal-coordinates} gives
\begin{equation}\label{eq:thermal-correction-cut}
 \begin{aligned}
 \mathsf F_u^{\beta,\lambda}-\mathsf F_u^{\beta,0}
 &=\sum_{v\in\Gamma\times\{1,\ldots,2r\}}\frac{(\Rb)_{uv}}{\sqrt2}
   \Bigl[(\gamma_v[f_\beta;\Phi_\lambda]-\gamma_v[f_\beta;\Phi_0])^{(1)}\\
 &\qquad\qquad
       +\I\bigl(\ft((\gamma_v[f_\beta;\Phi_\lambda]-\gamma_v[f_\beta;\Phi_0])^*)\bigr)^{(2)}\Bigr].
 \end{aligned}
\end{equation}
Without a free gap assumption, Equation~\eqref{eq:FG-coordinates} gives
\begin{equation}\label{eq:thermal-correction-FG}
 \begin{aligned}
 \mathsf G_u^{\beta,\lambda}-\mathsf F_u^{\beta,0}
 &=\sum_{v\in\Gamma\times\{1,\ldots,2r\}}\mathsf M_{uv}
   \Bigl[(\gamma_v[q_\beta;\Phi_\lambda]-\gamma_v[q_\beta;\Phi_0])^{(1)}\\
 &\qquad\qquad
       +\I\bigl(\ft((\gamma_v[q_\beta;\Phi_\lambda]-\gamma_v[q_\beta;\Phi_0])^*)\bigr)^{(2)}\Bigr].
 \end{aligned}
\end{equation}

Since \(\overline h=-h\), we have \(\overline{h^2}=h^2\). The definitions \eqref{eq:gapped-thermal-normalizer} and \eqref{eq:FG-normalizer} express \(\Rb\) and \(\mathsf M\) as real-valued functions of \(h^2\). Their entries are therefore real.

Using \(\Rb=\Rb P_++\Rb P_-\), Lemmas~\ref{lem:uniform-free-calculus} and~\ref{lem:bounded-temperature-calculus} provide model constants \(C_{\rm th},C_{\mathsf M}>0\) and, in regime~\ref{ass:exponential}, a model rate \(\mu_g\in(0,\mu_h)\), such that
\[
 \begin{aligned}
 \|\Rb\|^{\rm Sch}_{\mu_g,\theta_{\rm in}}
 &\le2C_{\rm th},
 &\|\mathsf M\|^{\rm Sch}_{\mu_h,\theta_{\rm in}}
 &\le C_{\mathsf M}
 &&\text{in regime~\ref{ass:exponential}},\\
 \|\Rb\|^{\rm Sch,pol}_{p_h}
 &\le2C_{\rm th},
 &\|\mathsf M\|^{\rm Sch,pol}_{p_h}
 &\le C_{\mathsf M}
 &&\text{in regime~\ref{ass:polynomial-weight}}.
 \end{aligned}
\]
The bounds on \(\Rb\) are uniform for all \(\beta\ge0\) under \eqref{eq:free-band}, and the bound on \(\mathsf M\) is uniform for \(\beta>0\).
Since \(p_\Phi\le p_h\), these estimates supply the matrix bounds required by Lemma~\ref{lem:normalized-field-differences} for the two pairs
\begin{equation*}
 (k,\mathsf N)=(f_\beta,\Rb/\sqrt2),
 \qquad (q_\beta,\mathsf M).
\end{equation*}
\par\smallskip\noindent\emph{Cutoff family.}
In regime~\ref{ass:exponential}, Lemma~\ref{lem:Gevrey} verifies the condition \eqref{eq:kernel-conditions} for \(f_\beta\) and gives model constants \(C_*,a_*>0\) such that \(|f_\beta(t)|\le C_*\E^{-a_*|t|^{1/2}}\).
Together with the bounds on \(\Rb\), Lemma~\ref{lem:normalized-field-differences}(i), applied with \(k=f_\beta\), \(\mathsf N=\Rb/\sqrt2\) and \(\theta=\theta_{\rm in}/\sG\), therefore gives \(\Delta F_{u,Z}\in\A_Z^-\), with \(Z\Subset\Gamma\) containing \(x_u\), such that
\begin{equation}\label{eq:cutoff-single-copy-decomposition}
 \sum_{v\in\Gamma\times\{1,\ldots,2r\}}\frac{(\Rb)_{uv}}{\sqrt2}
       \bigl(\gamma_v[f_\beta;\Phi_\lambda]-\gamma_v[f_\beta;\Phi_0]\bigr)
 =\sum_{\substack{Z\Subset\Gamma\\x_u\in Z}}\Delta F_{u,Z},
\end{equation}
with norm convergence. Define
\begin{equation}\label{eq:cutoff-doubled-local-terms}
 \Delta\mathsf F_{u,Z}^{\beta,\lambda}
 :=(\Delta F_{u,Z})^{(1)}
       +\I\bigl(\ft((\Delta F_{u,Z})^*)\bigr)^{(2)}
 \in(\A_Z^\st)^-.
\end{equation}
The adjoint, reversal and copy embeddings preserve norms and supports, so
\[
 \|\Delta\mathsf F_{u,Z}^{\beta,\lambda}\|
 \le2\|\Delta F_{u,Z}\|.
\]
Since \((\Rb)_{uv}\) is real, Equation~\eqref{eq:thermal-correction-cut} gives
\[
 \mathsf F_u^{\beta,\lambda}-\mathsf F_u^{\beta,0}
 =\sum_{\substack{Z\Subset\Gamma\\x_u\in Z}}\Delta\mathsf F_{u,Z}^{\beta,\lambda}
\]
with norm convergence. Taking the constant \(\mu>0\) given by Lemma~\ref{lem:normalized-field-differences}(i), which only depends on model data and $\theta$, we obtain
\[
 \begin{aligned}
 &\sup_{z\in\Gamma}\sum_{u\in\Gamma\times\{1,\ldots,2r\}}
 \sum_{\substack{Z\Subset\Gamma\\z,x_u\in Z}}|Z|\E^{\mu\diam(Z)^\theta}
       \|\Delta\mathsf F_{u,Z}^{\beta,\lambda}\|\\
 &\quad\le2 \sup_{z\in\Gamma}\sum_{u\in\Gamma\times\{1,\ldots,2r\}}
 \sum_{\substack{Z\Subset\Gamma\\z,x_u\in Z}}|Z|\E^{\mu\diam(Z)^\theta}\|\Delta F_{u,Z}\|
 \le C|\lambda|^{1/2},
 \end{aligned}
\]
where \(C\) is a model constant.

In regime~\ref{ass:polynomial-weight}, Lemma~\ref{lem:Gevrey} gives
\[
 \int_\R|f_\beta(t)|(1+|t|)^{m_B}\,\dd t\le C
\]
with a model constant \(C\).
Together with the bounds on \(\Rb\), Lemma~\ref{lem:normalized-field-differences}(ii), applied with \(k=f_\beta\), \(\mathsf N=\Rb/\sqrt2\) and \(b=1\), therefore provides terms \(\Delta F_{u,Z}\) satisfying \eqref{eq:cutoff-single-copy-decomposition}. Defining \(\Delta\mathsf F_{u,Z}^{\beta,\lambda}\) by \eqref{eq:cutoff-doubled-local-terms} gives
\[
 \begin{aligned}
 &\sup_{z\in\Gamma}\sum_{u\in\Gamma\times\{1,\ldots,2r\}}
   \sum_{\substack{Z\Subset\Gamma\\z,x_u\in Z}}|Z|(1+\diam Z)^{p_\Phi}
       \|\Delta\mathsf F_{u,Z}^{\beta,\lambda}\|\\
 &\quad\le2C_B\|\Rb/\sqrt2\|_{p_\Phi}^{\rm Sch,pol}|\lambda|
       \int_\R|f_\beta(t)|(1+|t|)^{m_B}\,\dd t
 \le C|\lambda|,
 \end{aligned}
\]
where \(C\) is a model constant.

\par\smallskip\noindent\emph{Fermi-weighted family.}
Fix \(\beta>0\).

In regime~\ref{ass:exponential}, Lemma~\ref{lem:FG-filter} gives \(\| q_\beta\|_1\le C\log(3+\beta)\), and Lemma~\ref{lem:filtered-field-locality}(ii) gives the spatial decay of the operators \(\gamma_v[q_\beta;\Phi_\lambda]\) and \(\gamma_v[q_\beta;\Phi_0]\).
We use these bounds to estimate the tracial shells of Definition~\ref{def:tracial-shells}. For \(i\ge0\), set
\[
 \Delta \mathsf G_{u,i}^{\beta,\lambda}:=\Delta^{\rm tr}_{\{x_u\},i}(\mathsf G_u^{\beta,\lambda}-\mathsf F_u^{\beta,0}).
\]
These terms are odd and supported in \(\ball_{R_i}(x_u)\). We group terms with the same support. For \(Z\Subset\Gamma\) containing \(x_u\), set
\[
 \Delta \mathsf G_{u,Z}^{\beta,\lambda}
 :=\sum_{\substack{i\in\mathbb N_0\\\ball_{R_i}(x_u)=Z}}\Delta \mathsf G_{u,i}^{\beta,\lambda}.
\]
The telescoping series for the shells gives
\[
\mathsf G_u^{\beta,\lambda}-\mathsf F_u^{\beta,0}
 =\sum_{\substack{Z\Subset\Gamma\\x_u\in Z}}\Delta \mathsf G_{u,Z}^{\beta,\lambda},
\]
where the sum converges in norm.

We first bound the spatial decay of the shell terms. For \(i\ge2\), notice that if \(d(x_u,x_v)\le R_i/4\), then \(\ball_{R_i/4}(x_v)\subset\ball_{R_i/2}(x_u)\). We therefore find, with a model constant \(c>0\),
\[
 \begin{aligned}
 \|\Delta& \mathsf G_{u,i}^{\beta,\lambda}\|
 \le\|(\id-\Ftr{\ball_{R_i/2}(x_u)})
                   (\mathsf G_u^{\beta,\lambda}-\mathsf F_u^{\beta,0})\|\\
 &\le8\| q_\beta\|_1
          \sum_{\substack{v\in\Gamma\times\{1,\ldots,2r\}\\d(x_u,x_v)>R_i/4}}|\mathsf M_{uv}|\\
 &\quad+4\sum_{\substack{v\in\Gamma\times\{1,\ldots,2r\}\\d(x_u,x_v)\le R_i/4}}|\mathsf M_{uv}|
          \|(\id-\Ftr{\ball_{R_i/4}(x_v)})
             (\gamma_v[q_\beta;\Phi_\lambda]-\gamma_v[q_\beta;\Phi_0])\|\\
 &\le C\log(3+\beta)\Bigl(
          \E^{-\mu_h(R_i/4)^{\theta_{\rm in}}}
          +\E^{-c_{\rm FG}(R_i/4)^{\theta_{\rm in}}/(1+\beta)}\Bigr)\\
 &\le C'\log(3+\beta)
          \E^{-c R_i^{\theta_{\rm in}}/(1+\beta)}.
 \end{aligned}
\]
Here we used Lemmas~\ref{lem:conditional-expectations}, \ref{lem:FG-filter},~\ref{lem:bounded-temperature-calculus} and~\ref{lem:filtered-field-locality}(ii). For \(i=0,1\), contractivity gives
\(\|\Delta \mathsf G_{u,i}^{\beta,\lambda}\|\le C\| q_\beta\|_1
\le C'\log(3+\beta)\E^{-cR_i^{\theta_{\rm in}}/(1+\beta)}\), since \(R_i\le1\).

We also need a bound proportional to \(|\eta\lambda|\). Lemma~\ref{lem:filtered-field-locality}(iii), applied with \(k= q_\beta\), \(b=\eta\), background \(\Phi_\lambda\) and perturbation \(-\lambda\cV\), gives, for every \(i\ge0\),
\[
 \begin{aligned}
 \|\Delta \mathsf G_{u,i}^{\beta,\lambda}\|
 &\le2\|\mathsf G_u^{\beta,\lambda}-\mathsf F_u^{\beta,0}\|\\
 &\le4C_D|\lambda|J_{\cV}
       \Bigl(\sup_{w\in\Gamma\times\{1,\ldots,2r\}}\sum_{v\in\Gamma\times\{1,\ldots,2r\}}|\mathsf M_{wv}|\Bigr) \int_\R| q_\beta(t)|\,|t|(1+|t|/\eta)^{m_D}\,\dd t\\
 &\le C(1+\beta)^{m_D+1}|\eta\lambda|.
 \end{aligned}
\]
Here we used Lemmas~\ref{lem:bounded-temperature-calculus} and~\ref{lem:FG-filter}, together with \(\beta/\eta=1+\beta\). Consequently, \(\log(3+\beta)\le C(1+\beta)^{m_D+1}\) yields
\[
 \|\Delta \mathsf G_{u,i}^{\beta,\lambda}\|
 \le C(1+\beta)^{m_D+1}
       \min\{|\eta\lambda|,\E^{-cR_i^{\theta_{\rm in}}/(1+\beta)}\}.
\]
It remains to sum these bounds with the required spatial weight. Choose a model constant \(\mu_{\rm FG}>0\) such that \(2^{\theta_{\rm in}}\mu_{\rm FG}\le c/4\).
Volume growth bounds both \(|\ball_{R_i}(x_u)|\) and the number of labels \(u\) with \(z\in\ball_{R_i}(x_u)\) by \(C(1+R_i)^{d_{\rm vol}}\). Using these bounds and \(\min\{a,b\}\le\sqrt{ab}\) gives
\begin{equation}\label{eq:FG-weighted-shell-sum}
 \begin{aligned}
 &\sup_{z\in\Gamma}\sum_{i=0}^\infty
       \sum_{\substack{u\in\Gamma\times\{1,\ldots,2r\}\\z\in\ball_{R_i}(x_u)}}
 |\ball_{R_i}(x_u)|\E^{\mu_{\rm FG}(2R_i)^{\theta_{\rm in}}/(1+\beta)}
            \|\Delta \mathsf G_{u,i}^{\beta,\lambda}\|\\
 &\quad\le C|\eta\lambda|^{1/2}(1+\beta)^{m_D+1}
       \sum_{i=0}^\infty(1+R_i)^{2d_{\rm vol}}
              \E^{-cR_i^{\theta_{\rm in}}/[4(1+\beta)]}.
 \end{aligned}
\end{equation}
For \(i\ge1\) we observe the elementary inequality
\[
 (1+R_i)^{2d_{\rm vol}}
       \E^{-cR_i^{\theta_{\rm in}}/[4(1+\beta)]}
 \le C\int_{R_i/2}^{R_i}r^{2d_{\rm vol}-1}
       \E^{-cr^{\theta_{\rm in}}/[4(1+\beta)]}\,\dd r.
\]
Applying this to the sum in Equation \eqref{eq:FG-weighted-shell-sum} gives 
\[
 \begin{aligned}
 \sum_{i=0}^\infty(1+R_i)^{2d_{\rm vol}}
       \E^{-cR_i^{\theta_{\rm in}}/[4(1+\beta)]}&\le C\left(1+\int_0^\infty
          r^{2d_{\rm vol}-1}
          \E^{-cr^{\theta_{\rm in}}/[4(1+\beta)]}\,\dd r\right)\\
 &\le C'(1+\beta)^{2d_{\rm vol}/\theta_{\rm in}}.
 \end{aligned}
\]
Here \(C,C'>0\) are model constants. We therefore find
\begin{equation}\label{eq:fermi-weighted-shell-sum}
 \begin{gathered}
 \sup_{z\in \Gamma}\sum_{i=0}^\infty
       \sum_{\substack{u\in\Gamma\times\{1,\ldots,2r\}\\z\in\ball_{R_i}(x_u)}}
 |\ball_{R_i}(x_u)|\E^{\mu_{\rm FG}(2R_i)^{\theta_{\rm in}}/(1+\beta)}
            \|\Delta \mathsf G_{u,i}^{\beta,\lambda}\|\\
 \le C|\eta\lambda|^{1/2}
              (1+\beta)^{m_D+1+2d_{\rm vol}/\theta_{\rm in}}.
 \end{gathered}
\end{equation}
Set \(\mu=\mu_{\rm FG}/(1+\beta)\). By the definition of \(\Delta \mathsf G_{u,Z}^{\beta,\lambda}\) and \(\diam(\ball_{R_i}(x_u))\le2R_i\), the triangle inequality gives
\[
 \sup_{z\in \Gamma}\sum_{u\in\Gamma\times\{1,\ldots,2r\}}
       \sum_{\substack{Z\Subset\Gamma\\z,x_u\in Z}}
 |Z|\E^{\mu\diam(Z)^{\theta_{\rm in}}}
             \|\Delta \mathsf G_{u,Z}^{\beta,\lambda}\|\le\sup_{z\in\Gamma}\sum_{i=0}^\infty
       \sum_{\substack{u\in\Gamma\times\{1,\ldots,2r\}\\z\in\ball_{R_i}(x_u)}}
 |\ball_{R_i}(x_u)|\E^{\mu(2R_i)^{\theta_{\rm in}}}
             \|\Delta \mathsf G_{u,i}^{\beta,\lambda}\|.
\]
Equation~\eqref{eq:fermi-weighted-shell-sum} therefore proves the asserted bound in regime~\ref{ass:exponential}.

In regime~\ref{ass:polynomial-weight}, Lemma~\ref{lem:FG-filter} and \(\beta/\eta=1+\beta\) give
\[
 \int_\R| q_\beta(t)|(1+|t|/\eta)^{m_B}\,\dd t\le C(1+\beta)^{m_B}.
\]
Together with the bound on \(\mathsf M\), Lemma~\ref{lem:normalized-field-differences}(ii), with \(k= q_\beta\), \(\mathsf N=\mathsf M\) and \(b=\eta\), gives local terms \(\Delta F_{u,Z}\in\A_Z^-\) such that, for \(u\in\Gamma\times\{1,\ldots,2r\}\),
\[
 \sum_{v\in\Gamma\times\{1,\ldots,2r\}}\mathsf M_{uv}
       (\gamma_v[q_\beta;\Phi_\lambda]-\gamma_v[q_\beta;\Phi_0])
 =\sum_{\substack{Z\Subset\Gamma\\x_u\in Z}}\Delta F_{u,Z}
\]
with absolute convergence in norm. For each \(Z\Subset\Gamma\) containing \(x_u\), define
\[
 \Delta\mathsf G_{u,Z}^{\beta,\lambda}
 :=(\Delta F_{u,Z})^{(1)}+\I\bigl(\ft((\Delta F_{u,Z})^*)\bigr)^{(2)}
 \in(\A_Z^\st)^-.
\]
Since \(\mathsf M_{uv}\) is real, \eqref{eq:thermal-correction-FG} gives
\[
 \mathsf G_u^{\beta,\lambda}-\mathsf F_u^{\beta,0}
 =\sum_{\substack{Z\Subset\Gamma\\x_u\in Z}}\Delta\mathsf G_{u,Z}^{\beta,\lambda}.
\]
The adjoint, reversal and copy embeddings preserve the norm, so \(\|\Delta\mathsf G_{u,Z}^{\beta,\lambda}\|\le2\|\Delta F_{u,Z}\|\). Thus the last series converges absolutely in norm, and Lemma~\ref{lem:normalized-field-differences}(ii) gives
\[
 \begin{aligned}
 &\sup_{z\in\Gamma}\sum_{u\in\Gamma\times\{1,\ldots,2r\}}
   \sum_{\substack{Z\Subset\Gamma\\z,x_u\in Z}}
   |Z|(1+\diam Z)^{p_\Phi}\|\Delta\mathsf G_{u,Z}^{\beta,\lambda}\|\\
 &\quad\le 2C_B\|\mathsf M\|^{\rm Sch,pol}_{p_\Phi}|\eta\lambda|
       \int_\R |q_\beta(t)|(1+|t|/\eta)^{m_B}\,\dd t
 \le C(1+\beta)^{m_B}|\eta\lambda|,
 \end{aligned}
\]
where \(C\) is a model constant.
\end{proof}

We now turn to the proof of Proposition~\ref{prop:thermal-comparison}. We first establish the following lemma, which compares filtered Majorana operators under defects and finite-volume restrictions.

\begin{lemma}\label{lem:defect-source}
Assume Hypothesis~\ref{hyp:locality}. For the coupling \(\lambda\) appearing in each estimate, set \(\Xi=\Phi_\lambda+\Psi\), \(S=\Sigma(\Psi)\), in the defect case, and \(\Xi=\Xi_{\Lambda,K}\), \(S=\Lambda^c\cup B\), in the finite-volume case as in \eqref{eq:cut-interaction}.
\begin{enumerate}[label=(\roman*),leftmargin=*]
\item \emph{General filters.} Let \(k\) satisfy \eqref{eq:kernel-conditions}.
There are constants \(C_{\rm def}>0\) and, in regime~\ref{ass:exponential}, \(c_{\rm def}>0\), depending only on the model data and the bounds on \(k\) in \eqref{eq:kernel-conditions}, such that, for every \(|\lambda|\le1\) and \(u\in\Gamma\times\{1,\ldots,2r\}\),
\begin{equation}\label{eq:filtered-defect-comparison}
 \Bigl\|\gamma_u[k;\Xi]-\gamma_u[k;\Phi_\lambda]\Bigr\|
 \le C_{\rm def}\begin{cases}
 \E^{-c_{\rm def}d(x_u,S)^{\theta_{\rm in}/\sG}},&\ref{ass:exponential},\\
 (1+d(x_u,S))^{-p_\Phi},&\ref{ass:polynomial-weight}.
 \end{cases}
\end{equation}

\item \emph{Fermi-weighted filter.}
Let \(q_\beta\) be the filter defined by \eqref{eq:FG-kernel}.
There are a model constant \(C_{\rm def,FG}>0\), a model constant \(c_{\rm def,FG}>0\) in regime~\ref{ass:exponential}, and a model constant \(m_{\rm cr}>0\) in regime~\ref{ass:polynomial-weight}, such that, for every \(\beta>0\), \(u\in\Gamma\times\{1,\ldots,2r\}\), and \(\lambda\) with \(\beta|\lambda|/(1+\beta)\le1\),
\begin{equation}\label{eq:FG-defect-bound}
 \bigl\|\gamma_u[q_\beta;\Xi]-\gamma_u[q_\beta;\Phi_\lambda]\bigr\|
 \le C_{\rm def,FG}\begin{cases}
 \log(3+\beta)
       \E^{-c_{\rm def,FG}d(x_u,S)^{\theta_{\rm in}}/(1+\beta)},&\ref{ass:exponential},\\
 (1+\beta)^{m_{\rm cr}+1}(1+d(x_u,S))^{-p_\Phi},&\ref{ass:polynomial-weight}.
 \end{cases}
\end{equation}
\end{enumerate}
The bounds hold uniformly for all admissible defects and all finite-volume restrictions, independently of the strength of \(\Psi\) and \(K\).
\end{lemma}
\begin{proof}
If \(S=\varnothing\), the two dynamics agree.
For \(d(x_u,S)\le4\), the bound \(2\|k\|_1\) gives \eqref{eq:filtered-defect-comparison}. The estimate \(\| q_\beta\|_1\le C\log(3+\beta)\) from Lemma~\ref{lem:FG-filter} gives both cases of \eqref{eq:FG-defect-bound}, using \(\log(3+\beta)\le C(1+\beta)^{m_{\rm cr}+1}\) in regime~\ref{ass:polynomial-weight}.
Suppose now that \(d(x_u,S)>4\), and put
\[
 U:=\ball_{d(x_u,S)/2}(x_u).
\]
Then \(U\subset S^c\) and \(d(x_u,U^c)\ge d(x_u,S)/2\). Write \(\tau^{\lambda,U}\) for the dynamics generated by \(\sum_{Z\subset U}\Phi_\lambda(Z)\), which preserves \(\A_U\).
For a defect, every nonzero \(\Psi(Z)\) is even and supported in \(S\subset U^c\), so it commutes with \(\A_U\). In the finite-volume case, \(S^c\subset\Lambda\), and \(K\) is also supported in $S$ and therefore commutes with \(\A_U\).  Hence, for \(A\in\A_U\),
\[
 \delta_{\Xi_{\Lambda,K}}(A)-\sum_{Z\subset U}[\Phi_\lambda(Z),A]
 =\sum_{\substack{Z\subset\Lambda\\ Z\cap U\ne\varnothing\\ Z\cap U^c\ne\varnothing}}
       [\Phi_\lambda(Z),A].
\]
By the triangle inequality we have
\begin{equation}\label{eq:defect-alt-generator}
 \Bigl\|\delta_\Xi(A)-\sum_{Z\subset U}[\Phi_\lambda(Z),A]\Bigr\|
 \le\sum_{\substack{Z\cap U\ne\varnothing\\ Z\cap U^c\ne\varnothing}}\|[\Phi_\lambda(Z),A]\|,
\end{equation}
for all $A \in \A_U$. The sum is bounded by \(2|U|\|A\|\|\Phi_\lambda\|_1\), and hence well-defined. By Proposition~\ref{prop:weak-dynamics}, \(\A_U\) lies in the domain of \(\delta_\Xi\). Differentiating \(s\mapsto\tau_{t-s}^\Xi(\tau_s^{\lambda,U}(\gamma_u))\) in norm, integrating, and using \eqref{eq:defect-alt-generator} gives
\[
 \|\tau_t^\Xi(\gamma_u)-\tau_t^{\lambda,U}(\gamma_u)\|
 \le\int_{\min\{0,t\}}^{\max\{0,t\}}
       \sum_{\substack{Z\cap U\ne\varnothing\\ Z\cap U^c\ne\varnothing}}
       \|[\Phi_\lambda(Z),\tau_s^{\lambda,U}(\gamma_u)]\|\,\dd s.
\]
The same bound holds with \(\Phi_\lambda\) in place of $\Xi$. The triangle inequality and isometry therefore give
\begin{equation}\label{eq:defect-alt-dynamics-comparison}
\|\tau_t^\Xi(\gamma_u)-\tau_t^{\Phi_\lambda}(\gamma_u)\|\le\min\left\{2,\,
 2\int_{\min\{0,t\}}^{\max\{0,t\}}
       \sum_{\substack{Z\cap U\ne\varnothing\\ Z\cap U^c\ne\varnothing}}
       \|[\Phi_\lambda(Z),\tau_s^{\lambda,U}(\gamma_u)]\|\,\dd s\right\}.
\end{equation}
The right-hand side of \eqref{eq:defect-alt-dynamics-comparison} involves only \(\Phi_\lambda\), so its bounds are independent of the norms of \(\Psi\) and \(K\).

\par\smallskip\noindent\emph{(i) General filters.}
In regime~\ref{ass:exponential}, the interaction norm \(\|\Phi_\lambda\|_{\mu,\theta_{\rm in}}\), with \(\mu=\min\{\mu_h,\mu_{\cV}\}\), is uniformly bounded for \(|\lambda|\le1\). Apply \cite[Lemma~6.4]{LPPLV3} with interaction \(\Phi_\lambda\), region \(U\), and observable \(\gamma_u\). Since \(d(x_u,U^c)\ge d(x_u,S)/2\), it gives model constants \(C,a,v>0\) such that, for every \(s\in\R\),
\begin{equation}\label{eq:defect-commutator-A}
 \sum_{\substack{Z\cap U\ne\varnothing\\ Z\cap U^c\ne\varnothing}}
       \|[\Phi_\lambda(Z),\tau_s^{\lambda,U}(\gamma_u)]\|
 \le C\E^{v|s|-a d(x_u,S)^{\theta_{\rm in}}}.
\end{equation}
The conclusion on filtered dynamics of the same lemma, applied to \eqref{eq:defect-alt-dynamics-comparison} with filter \(k\), proves part~(i) in regime~\ref{ass:exponential}, with the stated dependence on the bounds in \eqref{eq:kernel-conditions}.

In regime~\ref{ass:polynomial-weight}, use \eqref{eq:background-first-moment} to bound \(\|\Phi_\lambda\|^{\#}_{p_\Phi,\kappa_0}\) uniformly for \(|\lambda|\le1\). Apply \cite[Proposition~5.3(ii)]{FermionicLR} with interaction \(\Phi_\lambda\), spatial parameters \(a=p=p_\Phi\), region \(U\), observable \(\gamma_u\in\A_{\{x_u\}}\), and filter \(k\). The proposition applies to the odd observable \(\gamma_u\). It gives model constants \(C>0\) and 
\[
 m_{\rm cr}:=m_{p_\Phi,0}
 =\frac{(p_\Phi+1)(d_{\rm vol}+p_\Phi)}{p_\Phi-d_{\rm vol}},
\]
such that, for every \(s\in\R\),
\begin{equation}\label{eq:defect-commutator-B}
 \sum_{\substack{Z\cap U\ne\varnothing\\ Z\cap U^c\ne\varnothing}}
       \|[\Phi_\lambda(Z),\tau_s^{\lambda,U}(\gamma_u)]\|
 \le\frac{C(1+|s|)^{m_{\rm cr}}}{(1+d(x_u,U^c))^{p_\Phi}}.
\end{equation}
Integrating \eqref{eq:defect-alt-dynamics-comparison} against \(|k(t)|\), applying that estimate, and using \(d(x_u,U^c)\ge\tfrac12d(x_u,S)\) gives
\begin{align*}
 \Bigl\|\gamma_u[k;\Xi]-\gamma_u[k;\Phi_\lambda]\Bigr\|
 &\le C(1+d(x_u,S))^{-p_\Phi}
       \int_\R|k(t)|\,|t|(1+|t|)^{m_{\rm cr}}\,\dd t\\
 &\le C_{\rm def}(1+d(x_u,S))^{-p_\Phi}.
\end{align*}
Here $C_{\rm def} > 0$ is a constant, with the stated dependence, by \eqref{eq:kernel-conditions}. This completes the proof of part~(i).

\par\smallskip\noindent\emph{(ii) Fermi-weighted filter.}
For part~(ii), put \(\eta:=\beta/(1+\beta)\). We apply \eqref{eq:defect-alt-dynamics-comparison} with background interaction \(\eta\Phi_\lambda\) and perturbed interaction \(\eta\Xi\), at time \(t/\eta\). Since \(0<\eta\le1\) and \(|\eta\lambda|\le1\), the locality norms of \(\eta\Phi_\lambda=\Phi_{\eta h}+(\eta\lambda)\cV\) are bounded by model constants. The support \(S\) is unchanged, and \(\tau_{t/\eta}^{\eta\Xi}=\tau_t^\Xi\). Thus the boundary estimates below apply at time \(t/\eta\), uniformly in the temperature and coupling ranges of~(ii).

In regime~\ref{ass:exponential}, inserting \eqref{eq:defect-commutator-A} for the rescaled interaction into \eqref{eq:defect-alt-dynamics-comparison} gives
\[
 \|\tau_t^\Xi(\gamma_u)-\tau_t^{\Phi_\lambda}(\gamma_u)\|
 \le\min\{2,C\E^{v|t|/\eta-a d(x_u,S)^{\theta_{\rm in}}}\}.
\]
Split the integral against \( q_\beta\) at \(T=\eta a d(x_u,S)^{\theta_{\rm in}}/(2v)\). The pointwise and \(L^1\) bounds in Lemma~\ref{lem:FG-filter}, together with \(\beta/\eta=1+\beta\), give
\[
 \bigl\|\gamma_u[q_\beta;\Xi]-\gamma_u[q_\beta;\Phi_\lambda]\bigr\|
 \le C\log(3+\beta)\left(
       \E^{-a d(x_u,S)^{\theta_{\rm in}}/2}
       +\E^{-a d(x_u,S)^{\theta_{\rm in}}/[8v(1+\beta)]}\right).
\]
This proves the first case of \eqref{eq:FG-defect-bound} with \(c_{\rm def,FG}:=\min\{a/2,a/(8v)\}\).

In regime~\ref{ass:polynomial-weight}, use \eqref{eq:defect-commutator-B} for the rescaled interaction in \eqref{eq:defect-alt-dynamics-comparison}. For \(|\eta\lambda|\le1\), this gives
\[
 \begin{aligned}
\bigl\|\gamma_u[q_\beta;\Xi]-\gamma_u[q_\beta;\Phi_\lambda]\bigr\|&\le C(1+d(x_u,S))^{-p_\Phi}
       \int_\R| q_\beta(t)|\,(|t|/\eta)(1+|t|/\eta)^{m_{\rm cr}}\,\dd t\\
 &\le C(1+d(x_u,S))^{-p_\Phi}(1+\beta)^{m_{\rm cr}+1},
 \end{aligned}
\]
where the last inequality follows from Lemma~\ref{lem:FG-filter} and \(\beta/\eta=1+\beta\). This proves the second case of \eqref{eq:FG-defect-bound}.
\end{proof}

We now prove Proposition~\ref{prop:thermal-comparison}. We combine the bounds on the matrices \(\Rb\) and \(\mathsf M\) with Lemma~\ref{lem:filtered-field-locality}(iii) for part~(i) and Lemma~\ref{lem:defect-source} for part~(ii).

\begin{proof}[Proof of Proposition~\ref{prop:thermal-comparison}]
\noindent\emph{(i) Perturbation bound.}
Put \(\eta:=\beta/(1+\beta)\). Using \(\Rb=\Rb P_++\Rb P_-\), Lemmas~\ref{lem:uniform-free-calculus} and~\ref{lem:bounded-temperature-calculus} give the unweighted Schur bounds
\[
 \|\Rb\|_0^{\rm Sch,pol}\le2C_{\rm th},
 \qquad
 \|\mathsf M\|_0^{\rm Sch,pol}\le C_{\mathsf M}.
\]
Together with the kernel bounds in Lemmas~\ref{lem:Gevrey} and~\ref{lem:FG-filter}, these estimates give absolute norm convergence of the series defining \(\mathsf F_u^{\beta,\lambda},\mathsf G_u^{\beta,\lambda}\) and \(\mathsf F_u^\beta[\Xi],\mathsf G_u^\beta[\Xi]\), for each \(u\in\Gamma\times\{1,\ldots,2r\}\).
Let \(\Psi\in\mathfrak J_1\) and \(\Xi:=\Phi_\lambda+\Psi\), and let \(\mathsf F_u^\beta[\Xi],\mathsf G_u^\beta[\Xi]\) be defined by \eqref{eq:modified-thermal-family} for \(\Xi\).
For the cutoff family, the definitions \eqref{eq:thermal-coordinates} and \eqref{eq:modified-thermal-family} give a model constant $C > 0$ such that
\begin{equation*}
 \begin{aligned}
 \|\mathsf F_u^{\beta,\lambda}-\mathsf F_u^\beta[\Xi]\|
 &\le\sqrt2\sum_v|(\Rb)_{uv}|
       \|\gamma_v[f_\beta;\Phi_\lambda]-\gamma_v[f_\beta;\Xi]\|\\
 &\le\sqrt2\,C_DJ_\Psi\Bigl(\sum_v|(\Rb)_{uv}|\Bigr)
       \int_\R|f_\beta(t)|
                 |t|(1+|t|)^{m_D}\,\dd t\le CJ_\Psi,
 \end{aligned}
\end{equation*}
where we used Lemma~\ref{lem:filtered-field-locality}(iii) with \(b=1\), Lemma~\ref{lem:Gevrey}, and~Lemma~\ref{lem:uniform-free-calculus}. 
For the Fermi-weighted family, the definition \eqref{eq:modified-thermal-family} gives a model constant \(C>0\) such that
\begin{equation*}
 \begin{aligned}
 \|\mathsf G_u^{\beta,\lambda}-\mathsf G_u^\beta[\Xi]\|
 &\le2\sum_v|\mathsf M_{uv}|
       \|\gamma_v[q_\beta;\Phi_\lambda]-\gamma_v[q_\beta;\Phi_\lambda+\Psi]\|\\
 &\le2C_DJ_\Psi\Bigl(\sum_v|\mathsf M_{uv}|\Bigr)
       \int_\R| q_\beta(t)|\,|t|(1+|t|/\eta)^{m_D}\,\dd t\\
 &\le CJ_\Psi\bigl(\beta+\eta^{-m_D}\beta^{m_D+1}\bigr)
 \le C'\eta(1+\beta)^{m_D+1}J_\Psi,
 \end{aligned}
\end{equation*}
where we used Lemma~\ref{lem:majorana-reversal}, Lemma~\ref{lem:filtered-field-locality}(iii) with \(b=\eta\) and Lemmas~\ref{lem:bounded-temperature-calculus} and~\ref{lem:FG-filter}.
The last inequality follows from \(\beta/\eta=1+\beta\).
Since the bounds are uniform in \(u\), they prove \eqref{eq:thermal-family-perturbation}.

\smallskip
\noindent\emph{(ii) Spatial comparison.}
Let \((\Xi,S)\) be either of the two choices in part~(ii) of the proposition, and use the families \((\mathsf F_u^\beta[\Xi])_{u\in\Gamma\times\{1,\ldots,2r\}}\) and \((\mathsf G_u^\beta[\Xi])_{u\in\Gamma\times\{1,\ldots,2r\}}\) defined in \eqref{eq:modified-thermal-family}.
We first choose the weights and bound the weighted sums of the entries of \(\Rb\) and \(\mathsf M\), using Lemmas~\ref{lem:uniform-free-calculus} and~\ref{lem:bounded-temperature-calculus}. We then prove the spatial comparison for the cutoff family and the Fermi-weighted family in turn.

In regime~\ref{ass:exponential}, choose \(c_{\rm def}>0\) given by Lemma~\ref{lem:defect-source} for the kernel \(f_\beta\). This constant only depends on the model data and the cutoff. For the cutoff family, take \(\mu:=\min\{c_{\rm def},\mu_g\}/2\), where \(\mu_g\) is the model constant from Lemma~\ref{lem:uniform-free-calculus}. For the Fermi-weighted family, let \(c_{\rm def,FG}>0\) be the model constant provided by Lemma~\ref{lem:defect-source}(ii), put \(\mu_{\rm def,FG}:=\min\{c_{\rm def,FG},\mu_h\}/2\) and take \(\mu:=\mu_{\rm def,FG}/(1+\beta)\).
Set \(W(R)=\exp (\mu R^\theta)\) in regime~\ref{ass:exponential}, with \(\theta=\theta_{\rm in}/\sG\) for the cutoff family and \(\theta=\theta_{\rm in}\) for the Fermi-weighted family. In regime~\ref{ass:polynomial-weight}, set \(W(R)=(1+R)^{p_\Phi}\).
For the cutoff family in regime~\ref{ass:exponential}, \(\mu<\mu_g\) and \(\sG>1\) give
\[
 W(R)=\E^{\mu R^{\theta_{\rm in}/\sG}}
 \le\E^{\mu_g}\E^{\mu_gR^{\theta_{\rm in}}},\qquad R\ge0.
\]
For the Fermi-weighted family in regime~\ref{ass:exponential}, \(\mu\le\mu_h\) gives \(W(R)\le\E^{\mu_hR^{\theta_{\rm in}}}\).
In regime~\ref{ass:polynomial-weight}, \(p_\Phi\le p_h\) gives \(W(R)\le(1+R)^{p_h}\) for both families.
The Schur norms use operator norms of the \(2r\times2r\) blocks connecting pairs of sites. The sums below run over individual Majorana labels. For the block \(\mathsf M_{xy}\) connecting sites \(x\) and \(y\), Cauchy--Schwarz gives, for each \(i\in\{1,\ldots,2r\}\),
\[
 \sum_{j=1}^{2r}|(\mathsf M_{xy})_{ij}|
 \le\sqrt{2r}\left(\sum_{j=1}^{2r}|(\mathsf M_{xy})_{ij}|^2\right)^{1/2}
 \le\sqrt{2r}\,\|\mathsf M_{xy}\|.
\]
The same estimate applies to the blocks of \(\Rb\).
The matrix bounds from these lemmas, using \(\Rb=\Rb P_++\Rb P_-\), therefore give a model constant \(C_R>0\) such that
\[
 \begin{aligned}
 \sup_{\beta>0}\sup_{u\in\Gamma\times\{1,\ldots,2r\}}
 \sum_{v\in\Gamma\times\{1,\ldots,2r\}}|(\Rb)_{uv}|W(d(x_u,x_v))
 &\le C_R&&\text{for the cutoff family},\\
 \sup_{u\in\Gamma\times\{1,\ldots,2r\}}
 \sum_{v\in\Gamma\times\{1,\ldots,2r\}}|\mathsf M_{uv}|W(d(x_u,x_v))
 &\le\sqrt{2r}\,C_{\mathsf M}&&\text{for the Fermi-weighted family}.
 \end{aligned}
\]
For \(\mathsf F^{\beta,\lambda}\), the bounds in Lemma~\ref{lem:Gevrey} and Lemma~\ref{lem:defect-source}(i) give
\[
 \Bigl\|\int_\R f_\beta(t)(\tau_t^{\Phi_\lambda}-\tau_t^\Xi)(\gamma_v)\,\dd t\Bigr\|
 \le\frac{C_{\rm def}}{W(d(x_v,S))},
\]
with \(C_{\rm def}\) depending only on the model data and, in regime~\ref{ass:exponential}, on \(\theta\). Inserting the definition \eqref{eq:thermal-coordinates}, using isometry of adjoints, reversal and copy embeddings, and applying Lemma~\ref{lem:weighted-Schur-properties} gives
\[
 \|\mathsf F_u^{\beta,\lambda}-\mathsf F_u^\beta[\Xi]\|
\le\sqrt2\,C_{\rm def}
       \sum_v\frac{|(\Rb)_{uv}|}{W(d(x_v,S))} \le\frac{\sqrt2\,C_{\rm def}}{W(d(x_u,S))}
       \sum_v|(\Rb)_{uv}|W(d(x_u,x_v)).
\]
Bounding $\Rb$ with Lemma~\ref{lem:uniform-free-calculus} therefore proves part~(ii) for the cutoff family.

For \(\mathsf G^{\beta,\lambda}\) in regime~\ref{ass:exponential}, Lemma~\ref{lem:defect-source}(ii) bounds \(\|\gamma_v[q_\beta;\Phi_\lambda]-\gamma_v[q_\beta;\Xi]\|\) by \(C_{\rm def,FG}\log(3+\beta)/W(d(x_v,S))\), since \(\mu\le c_{\rm def,FG}/(1+\beta)\). In regime~\ref{ass:polynomial-weight}, Lemma~\ref{lem:defect-source}(ii) bounds the same difference by \(C_{\rm def,FG}(1+\beta)^{m_{\rm cr}+1}/W(d(x_v,S))\), uniformly for \(\beta|\lambda|/(1+\beta)\le1\). We use the definition \eqref{eq:FG-coordinates}, apply these bounds and use Lemma~\ref{lem:weighted-Schur-properties} to find in the same way
\[
\|\mathsf G_u^{\beta,\lambda}-\mathsf G_u^\beta[\Xi]\| \le\frac{2 C_{\rm def,FG}}{W(d(x_u,S))}
       \left(\sum_v|\mathsf M_{uv}|W(d(x_u,x_v))\right)
       \begin{cases}
       \log(3+\beta),&\ref{ass:exponential},\\
       (1+\beta)^{m_{\rm cr}+1},&\ref{ass:polynomial-weight}.
       \end{cases}
\]
Bounding $\mathsf M$ with Lemma~\ref{lem:bounded-temperature-calculus} therefore proves part~(ii) for the Fermi-weighted family, with a model prefactor. In regime~\ref{ass:polynomial-weight}, we may take \(M=m_{\rm cr}+1\).
\end{proof}

\subsection{Ground-state families}
\label{sec:ground-family}

We now apply the preceding estimates to prove Lemma~\ref{lem:ground-free-family} and Propositions~\ref{prop:ground-family} and~\ref{prop:ground-comparison}.

\begin{proof}[Proof of Lemma~\ref{lem:ground-free-family}]
We first verify the Parseval condition for the displayed coefficients. Since \(P_+ + P_- = \1\) and \(\overline{P_+}=P_-\), we have \(2\RealPart P_\pm=\1\). The sum in \eqref{eq:parseval-condition} converges absolutely by the Cauchy--Schwarz inequality, since the columns of \(P_\pm\) are square summable. For \(v,w\in\Gamma\times\{1,\ldots,2r\}\), substituting the displayed coefficients gives
\[
 \left(\sum_{u\in\Gamma\times\{1,\ldots,2r\}}\overline{\sa_{u,(v,j)}}\sa_{u,(w,k)}\right)_{j,k=1}^2
 =2\begin{pmatrix}
 (P_+^*P_+)_{vw}&\I(P_+P_-)_{vw}\\
 -\I(P_-P_+)_{vw}&(P_-^*P_-)_{vw}
 \end{pmatrix}=2\begin{pmatrix}
 (P_+)_{vw}&0\\
 0&(P_-)_{vw}
 \end{pmatrix}.
\]
Taking real parts and using \(2\RealPart P_\pm=\1\) proves the Parseval condition of Definition~\ref{def:parseval-family}. Lemma~\ref{lem:parseval-reconstruction} therefore implies that the series in \eqref{eq:ground-free-family} defining each \(\mathsf F_u^{\infty,0}\) converges in operator norm.

With \(w\) as in Lemma~\ref{lem:uniform-free-calculus}, Cauchy--Schwarz on each row and column of the \(2r\times2r\) site blocks gives
\[
 \bigl\|(|(P_\pm)_{uv}|)_{u,v\in\Gamma\times\{1,\ldots,2r\}}\bigr\|_w^{\rm Sch}
 \le\sqrt{2r}\,\|P_\pm\|_w^{\rm Sch}
 \le\sqrt{2r}\,C_{\rm th},
\]
where the last inequality follows from that lemma. Inserting the coefficients \(\sa_{u,(v,1)}=\sqrt2(P_+)_{uv}\) and \(\sa_{u,(v,2)}=\sqrt2\I(P_-)_{uv}\) gives the bound in Equation~\eqref{eq:ground-free-coefficients}, with \(\mu_1=\mu_g\) in regime~\ref{ass:exponential} and \(C_0=\max\{1,4\sqrt{r}\,C_{\rm th}\}\) in both regimes. Since \(w\ge1\), this bound also gives absolute norm convergence of \eqref{eq:ground-free-family}.
\end{proof}

\begin{lemma}\label{lem:linear-KMS-filter}
Let \(\tau\) be a strongly continuous dynamics on \(\A^\st\), let \(c>0\), and let \(q\in C_c^\infty(\R)\) have support in \(( -\infty,-c)\). Let \(\check q\) be its inverse Fourier transform and \(A\in\A^\st\). Set
\[
 A_q:=\int_\R\check q(t)\tau_t(A)\,\dd t.
\]
Then \(A_q\) is entire analytic for \(\tau\), and the following hold.
\begin{enumerate}[label=(\roman*),leftmargin=*]
\item Let \(\omega\) be a ground state of \(\tau\), with GNS representation \((\pi,\mathcal H,\Omega)\) and nonnegative GNS Hamiltonian \(L\) implementing \(\tau\). Then
\[
 \pi(A_q)\1_{[0,c]}(L)=0.
\]
\item There is a constant \(C>0\), depending only on \(q\) and \(c\), such that, for every \(\beta>0\), \(\rho\in\KMS_\beta(\tau)\) and \(E,G\in\A^\st\),
\[
 |\rho(A_q^*E)|+|\rho(GA_q)|
 \le C\E^{-\beta c}\|A\|\max\{\|E\|,\|G\|\}.
\]
If \(\supp q\subset(-\infty,-c-\delta]\) for some \(\delta>0\), then \(C\) can be chosen to depend only on \(\delta\) and an upper bound for \(\|q\|_1+2\|q'\|_1+\|q''\|_1\).
\end{enumerate}
\end{lemma}
\begin{proof}
Lemma~\ref{lem:analytic-filtering} gives analyticity of \(A_q\) and the identity
\begin{equation}\label{eq:linear-filter-complex-time}
 \tau_z(A_q)=\int_\R \check q(t-z)\tau_t(A)\,\dd t.
\end{equation}
To show assertion (i) we notice that smoothness and the support condition of \(q\) imply that it vanishes on \([-c,\infty)\). Since \(\pi\) is bounded and intertwines the dynamics, we have
\[
 \pi(A_q)=\int_\R\check q(t)\E^{\I tL}\pi(A)\E^{-\I tL}\,\dd t.
\]
The statement therefore follows from \cite[Lemma~6.2]{LPPLV3}, with \(k=\check q\), \(B=\pi(A)\) and \(E=\varepsilon=c\).

To prove (ii), we use Equation~\eqref{eq:linear-filter-complex-time} at \(z=-\I\beta\). The corresponding kernel is
\[
 \check q(t+\I\beta)
 =\frac1{2\pi}\int_\R
 \E^{\beta\omega}q(\omega)\E^{-\I t\omega}\,\dd\omega.
\]
Thus the Fourier transform of the function \(t\mapsto\check q(t+\I\beta)\) is \(\omega\mapsto\E^{\beta\omega}q(\omega)\). Since \(q\) is smooth and compactly supported, two integrations by parts yield
\[
 t^2\check q(t+\I\beta)
 =-\frac1{2\pi}\int_\R
 \E^{\beta\omega}\bigl(\beta^2q(\omega)+2\beta q'(\omega)+q''(\omega)\bigr)
 \E^{-\I t\omega}\,\dd\omega.
\]
Choose \(\delta>0\) such that \(\supp q\subset(-\infty,-c-\delta]\). The first integral representation gives a bound independent of \(t\), while the integration-by-parts identity gives a bound proportional to \(t^{-2}\). We use the first bound for \(|t|\le1\) and the second for \(|t|>1\). We find
\[
 \begin{aligned}
 \int_\R|\check q(t+\I\beta)|\,\dd t
 &\le\frac1\pi\Bigl(\|\E^{\beta\omega}q\|_1
 +\|\E^{\beta\omega}(\beta^2q+2\beta q'+q'')\|_1\Bigr)\\
 &\le(1+\beta)^2\sup_{\omega\in\supp q}\E^{\beta\omega}
 \bigl(\|q\|_1+2\|q'\|_1+\|q''\|_1\bigr)\le C\E^{-\beta c}.
 \end{aligned}
\]
The last inequality follows from \(\sup_{\beta\ge0}(1+\beta)^2\E^{-\beta\delta}<\infty\), with the stated dependence of \(C\). It follows that
\begin{equation}\label{eq:norm-bound-fourier}
 \left\|\int_\R \check q(t+\I\beta)\tau_t(A)\,\dd t\right\|
 \le\int_\R|\check q(t+\I\beta)|\,\dd t\,\|A\|
 \le C\E^{-\beta c}\|A\|.
\end{equation}
Equation~\eqref{eq:linear-filter-complex-time} identifies the operator in Equation~\eqref{eq:norm-bound-fourier} with \(\tau_{-\I\beta}(A_q)\). The KMS condition, extended to arbitrary \(G,E\in\A^\st\) by norm continuity, gives
\[
 \rho(GA_q)=\rho(\tau_{-\I\beta}(A_q)G),
 \qquad
 \rho(A_q^*E)=\overline{\rho(\tau_{-\I\beta}(A_q)E^*)}.
\]
Taking absolute values and applying Equation~\eqref{eq:norm-bound-fourier} proves (ii).
\end{proof}

We now prove Proposition~\ref{prop:ground-family}. 

\begin{proof}[Proof of Proposition~\ref{prop:ground-family}]
\noindent\emph{(i) Ground-state and KMS estimates.}
By Lemmas~\ref{lem:graded-ground} and~\ref{lem:ground-state-equivalence}, \(\omega_\st\) is a ground state of \(\tau^{\st,\lambda,\Upsilon}=\tau^{(\Phi_\lambda+\Upsilon)^\st}\). Since \(\supp\widehat f\subset[-2C_h,-7g/8]\subset(-\infty,-g_*)\), Lemma~\ref{lem:linear-KMS-filter}(i), applied with \(q=\widehat f\), \(A=(\gamma_{u,1}+\I\gamma_{u,2})/\sqrt2\) and \(c=g_*\), gives, by \eqref{eq:ground-doubled-family}, \(\pi(\mathsf F_u^{\infty}[\Phi_\lambda+\Upsilon])\1_{[0,g_*]}(L)=0\), where \((\pi,\mathcal H,\Omega)\) and \(L\) are its GNS representation and nonnegative implementing Hamiltonian. In particular, \(\mathsf F_u^{\infty}[\Phi_\lambda+\Upsilon]\) annihilates the GNS ground-state vector. Thus
\[
 \omega_\st\bigl((\mathsf F_u^{\infty}[\Phi_\lambda+\Upsilon])^*\mathsf F_u^{\infty}[\Phi_\lambda+\Upsilon]\bigr)
 =0.
\]

To prove \eqref{eq:ground-family-kms}, note that Lemma~\ref{lem:graded-KMS} gives \(\rho_{\beta,\st}\in\KMS_\beta(\tau^{\st,\lambda,\Psi})\). The derivative estimates in Lemma~\ref{lem:gevrey-cutoff} give
\[
 \|\widehat f\|_1+2\|\widehat f'\|_1+\|\widehat f''\|_1\le C
\]
uniformly in \(\sG\). Since \(\supp\widehat f\subset[-2C_h,-7g/8]\) and \(7g/8-g_*=g/8>0\), Lemma~\ref{lem:linear-KMS-filter}(ii), applied with \(q=\widehat f\), \(A=(\gamma_{u,1}+\I\gamma_{u,2})/\sqrt2\) and \(c=g_*\), gives the following bound with a model constant \(C\):
\[
 \begin{aligned}
 |\rho_{\beta,\st}((\mathsf F_u^{\infty}[\Phi_\lambda+\Psi])^*E)|
 +|\rho_{\beta,\st}(G\mathsf F_u^{\infty}[\Phi_\lambda+\Psi])|&\le C\E^{-\beta g_*}\left\|\frac{\gamma_{u,1}+\I\gamma_{u,2}}{\sqrt2}\right\|
              \max\{\|E\|,\|G\|\}\\
 &=\sqrt2C\E^{-\beta g_*}\max\{\|E\|,\|G\|\},
 \end{aligned}
\]
where we used that the CAR give \(\|(\gamma_{u,1}+\I\gamma_{u,2})/\sqrt2\|=\sqrt2\).

\smallskip
\noindent\emph{(ii) Local decompositions.}
We first note that Lemma~\ref{lem:Gevrey} gives
\[
 |f(t)|\le C\E^{-a|t|^{1/\sG}},\qquad
 |f(t)|\le C_*\E^{-a_*|t|^{1/2}}
\]
for some \(a>0\). The first bound shows that \(f\) satisfies \eqref{eq:kernel-conditions} in both regimes. The second bound, with \(C_*,a_*>0\) independent of \(\sG\), verifies the additional kernel hypothesis of Lemma~\ref{lem:normalized-field-differences}(i). For \(\mathsf N=\1\) and any $\mu >0$ it holds that \(\|\1\|^{\rm Sch}_{\mu,\theta_{\rm in}}=\|\1\|^{\rm Sch,pol}_{p_\Phi}=1\). We can therefore apply Lemma~\ref{lem:normalized-field-differences} with \(k=f\), \(\mathsf N=\1\), and \(b=1\) in regime~\ref{ass:polynomial-weight}. Using \eqref{eq:filtered-Majorana-definition}, we obtain the norm-convergent decomposition
\[
 \gamma_u[f;\Phi_\lambda]-\gamma_u[f;\Phi_0]
 =\sum_{Z\ni x_u}\Delta\gamma_{u,Z}^\lambda,
\]
where $\Delta\gamma_{u,Z}^\lambda\in\A_Z^-$. Define
\[
 \Delta \mathsf F_{u,Z}^{\infty,\lambda}:=\frac1{\sqrt2}
 \bigl[(\Delta\gamma_{u,Z}^\lambda)^{(1)}
 +\I(\ft((\Delta\gamma_{u,Z}^\lambda)^*))^{(2)}\bigr].
\]
Reversal preserves supports and norms. Hence, it follows that $\Delta \mathsf F_{u,Z}^{\infty,\lambda}\in(\A_Z^\st)^-$ and
\[
 \|\Delta \mathsf F_{u,Z}^{\infty,\lambda}\|\le\sqrt2\|\Delta\gamma_{u,Z}^\lambda\|.
\]
Definition~\eqref{eq:ground-filtered-family} and \(\mathsf F_u^{\infty}[\Phi_0]=\mathsf F_u^{\infty,0}\) therefore give
\[
 \mathsf F_u^{\infty,\lambda}-\mathsf F_u^{\infty,0}=\sum_{Z\ni x_u}\Delta \mathsf F_{u,Z}^{\infty,\lambda}.
\]
In \ref{ass:exponential}, choose \(\mu>0\) less than or equal to the rate supplied by Lemma~\ref{lem:normalized-field-differences}(i). The bounds of Lemma \ref{lem:normalized-field-differences} then imply \eqref{eq:ground-family-correction}, where \(C_{\rm gs}\) is a model constant.
\end{proof}

We now prove Proposition~\ref{prop:ground-comparison}.

\begin{proof}[Proof of Proposition~\ref{prop:ground-comparison}]
Lemma~\ref{lem:Gevrey} verifies the kernel assumptions of Lemma~\ref{lem:defect-source}(i) for \(k=f\). In regime~\ref{ass:exponential}, choose \(0<\mu\le c_{\rm def}\), where \(c_{\rm def}\) is supplied by Lemma~\ref{lem:defect-source}(i) with \(k=f\). Put \(W(R)=\exp(\mu R^\theta)\). In regime~\ref{ass:polynomial-weight}, use \(W(R)=(1+R)^{p_\Phi}\).

If \(\Psi=0\), the dynamics agree and \(\mathsf F_u^{\infty}[\Phi_\lambda+\Psi]=\mathsf F_u^{\infty,\lambda}\).
Definition~\eqref{eq:ground-filtered-family}, isometry of \(\ft\), and the triangle inequality, followed by Lemma~\ref{lem:defect-source}(i), yield
\[
 \|\mathsf F_u^{\infty,\lambda}-\mathsf F_u^{\infty}[\Phi_\lambda+\Psi]\|
 \le\sqrt2\left\|\int_\R f(t)
 (\tau_t^{\Phi_\lambda}-\tau_t^{\Phi_\lambda+\Psi})(\gamma_u)\,\dd t\right\|\le\frac{C}{W(d(x_u,\Sigma(\Psi)))}.
\]
This proves \eqref{eq:ground-family-defect}. The constant depends only on the bounds in Lemmas~\ref{lem:Gevrey} and~\ref{lem:defect-source}, and hence on the stated data.
\end{proof}

\appendix

\section{Graded CAR algebras}
\label{app:graded}

In this appendix we collect the properties of graded CAR algebras used in the proofs. We first discuss the doubled algebra and products of states. We use the local algebras, Majorana labels \(\cJ\), and graded derivatives of Section~\ref{sec:comparison-setting}. These estimates apply to both \(\A\) and \(\A^\st\).

\subsection{The doubled algebra and products of states}
\label{sec:doubled-notation}

\begin{definition}\label{def:stacked-algebra}
Let $\Hsp$ be a one-particle Hilbert space and \(\Hsp_\aux\) a second copy. We set \(\A_\aux=\operatorname{CAR}(\Hsp_\aux)\). The doubled CAR algebra is
\[
 \A^\st:=\operatorname{CAR}(\Hsp\oplus\Hsp_\aux)
 \cong\A\widehat\otimes\A_\aux.
\]
We denote the copy embeddings by \(A^{(1)}:=A\widehat\otimes\1\) and \(B^{(2)}:=\1\widehat\otimes B\), and set \(\A_X^\st:=\A_X\widehat\otimes\A_{\aux,X}\).
For homogeneous elements, the graded tensor-product operations are
\[
 (A_1\widehat\otimes B_1)(A_2\widehat\otimes B_2)
 =(-1)^{|B_1||A_2|}A_1A_2\widehat\otimes B_1B_2
\]
and
\[
 (A\widehat\otimes B)^*=(-1)^{|A||B|}A^*\widehat\otimes B^*.
\]
Here \(|A|\in\{0,1\}\) is determined by \(\Theta(A)=(-1)^{|A|}A\).
\end{definition}
Thus \(A^{(1)}B^{(2)}=A\widehat\otimes B\), with the conventions of Section~\ref{sec:canonical-extension} and \cite[Appendix~A.1]{StabilityBdG26}.

\begin{definition}\label{def:majorana-reversal}
The reversal \(\ft\) on local Majorana polynomials is the complex-linear map defined by $\ft(\1)=\1$ and
\[
 \qquad \ft(\gamma_{u_1}\cdots\gamma_{u_k})
 =\gamma_{u_k}\cdots\gamma_{u_1},
\]
for any \(k\in\mathbb N\) and \(u_1,\ldots,u_k\in\Gamma\times\{1,\ldots,2r\}\).
\end{definition}

\begin{lemma}\label{lem:majorana-reversal}
The reversal of Definition~\ref{def:majorana-reversal} extends uniquely to a complex-linear isometric map \(\ft:\A\to\A\) such that, for all \(A,B\in\A\), it holds that $\ft(AB)=\ft(B)\ft(A)$ and $\ft(A^*)=\ft(A)^*$. Moreover the reversal satisfies \(\ft(\1)=\1\), \(\ft^2=\id\), $\ft\circ\Theta=\Theta\circ\ft$ and $\ft(\A_X) = \A_X$ for all $X \subset \Gamma$.
In particular, \(\ft\) is a \(*\)-antiautomorphism.
\end{lemma}
\begin{proof}
Fix some \(X\Subset\Gamma\). We first check that the prescribed reversal is compatible with the defining relations of \(\A_X\). This then shows that $\ft$ is well-defined on $\A_X$. To this end, notice that reversing $\gamma_u\gamma_v+\gamma_v\gamma_u=2\delta_{uv}\1$ gives
\[
 \gamma_v\gamma_u+\gamma_u\gamma_v=2\delta_{uv}\1,
\]
and reversing \(\gamma_u^*=\gamma_u\) leaves the identity unchanged. Thus polynomial expressions representing the same operator have the same reversal. Hence $\ft$ defines a complex-linear map on \(\A_X\).

For \(A=\gamma_{u_1}\cdots\gamma_{u_k}\) and \(B=\gamma_{w_1}\cdots\gamma_{w_\ell}\), we find 
\[
 \ft(AB)
 =\gamma_{w_\ell}\cdots\gamma_{w_1}\gamma_{u_k}\cdots\gamma_{u_1}
 =\ft(B)\ft(A).
\]
Since the generators are self-adjoint we also have
\[
 \ft(A^*)=\gamma_{u_1}\cdots\gamma_{u_k}
 =\ft(A)^*.
\]
Linearity extends the first identity, and conjugate-linearity of the adjoint extends the second, to all \(A,B\in\A_X\). If \(B\) is invertible, then
\[
 \ft(B)\ft(B^{-1})
 =\ft(B^{-1}B)=\1
 =\ft(BB^{-1})
 =\ft(B^{-1})\ft(B).
\]
Hence, $\ft$ preserves invertibility, which shows that \(\spec(\ft(B))=\spec(B)\). Since \(\ft(AA^*)=\ft(A)^*\ft(A)\), we obtain
\[
 \|\ft(A)\|^2
 =\max\spec\bigl(\ft(A)^*\ft(A)\bigr)
 =\max\spec(AA^*)=\|A\|^2.
\]
Thus \(\ft\) is isometric on \(\A_\loc\) and extends uniquely to \(\A\) by norm density. The displayed identities pass to the extension by continuity.

Reversal keeps every label in a monomial, so \(\ft(\A_X)=\A_X\), first for finite \(X\) and then for all \(X\subset\Gamma\) by local norm density. Finally,
\[
 \Theta\bigl(\ft(\gamma_{u_1}\cdots\gamma_{u_k})\bigr)
 =(-1)^k\gamma_{u_k}\cdots\gamma_{u_1}
 =\ft\bigl(\Theta(\gamma_{u_1}\cdots\gamma_{u_k})\bigr),
\]
which proves commutation with parity by linearity and continuity.
\end{proof}

\begin{definition}\label{def:graded-product-state}
For states \(\rho\) on \(\A\) and \(\eta\) on \(\A_\aux\), their graded product \(\rho\widehat\otimes\eta\), when it exists, is the state on \(\A^\st\) satisfying
\[
 (\rho\widehat\otimes\eta)(A^{(1)}B^{(2)})=\rho(A)\eta(B),
\]
for all $A\in\A$ and $B\in\A_\aux$.
\end{definition}

\begin{lemma}[{\cite[Theorem~1(1)]{ArakiMoriya03}}]\label{lem:graded-product}
Let \(\rho\) be a state and \(\eta\) an even state on CAR algebras over two disjoint sets of fermionic modes, respectively. Their graded product exists and is unique. If \(\rho\) is also even, then \(\rho\widehat\otimes\eta\) is even.
\end{lemma}

\subsection{Averaging and localization}
\label{app:localization}

We work on the CAR algebra \(\cC\) throughout this subsection. A state \(\omega\) on \(\cC\) is tracial if
\[
 \omega(AB)=\omega(BA),\qquad A,B\in\cC.
\]
This state on \(\cC\) exists and is unique (see \cite[Definition~4.1]{ArakiMoriyaGibbs03}); we shall denote it by \(\omega^{\rm tr}\).  Its restriction to each finite local algebra is the normalized matrix trace.

\begin{definition}\label{def:majorana-averaging}
For \(v\in\cJ\) and $A \in \cC$ we define the linear operator
\[
 \mathcal P_v(A):=\tfrac12(A+\gamma_v\Theta(A)\gamma_v).
\]
For \(Z\Subset\Gamma\) we enumerate its Majorana labels as \(v_1,\ldots,v_N\) and set
\[
 \mathbb E_Z:=\mathcal P_{v_N}\circ\cdots\circ\mathcal P_{v_1}.
\]
For \(Z=\varnothing\), set \(\mathbb E_Z=\id\).
\end{definition}

\begin{lemma}\label{lem:majorana-averaging}
For all \(v,w\in\cJ\), the maps \(\mathcal P_v:\cC\to\cC\) are unital completely positive contractions and satisfy \(\mathcal P_v\mathcal P_w=\mathcal P_w\mathcal P_v\).
For every \(Z\Subset\Gamma\), the map \(\mathbb E_Z\) is independent of the enumeration in Definition~\ref{def:majorana-averaging} and is a unital completely positive contraction satisfying $\mathbb E_Z(\cC)=\cC_{\Gamma\setminus Z}$ and $\mathbb E_Z(A)=A$ for all $A\in\cC_{\Gamma\setminus Z}$. For every \(A\in\cC\) it holds that
\begin{equation}\label{eq:majorana-oscillation}
 \|A-\mathbb E_Z(A)\|\le\tfrac12\sum_{v\in\cJ:\,x_v\in Z}\|\partial_vA\|.
\end{equation}
Moreover, along every increasing sequence of finite sets \(Z_m\) with \(\bigcup_mZ_m=\Gamma\),
\[
 \lim_{m\to\infty}\|\mathbb E_{Z_m}(A)-\omega^{\rm tr}(A)\1\|=0,
 \qquad A\in\cC.
\]
\end{lemma}
\begin{proof}
We first prove unitality, complete positivity and contractivity. Since \(\gamma_v\) is a self-adjoint unitary, the map \(A\mapsto\gamma_v\Theta(A)\gamma_v\) is a unital \(*\)-automorphism. Thus \(\mathcal P_v\), as the average of two unital \(*\)-automorphisms, is unital and completely positive, and
\[
 \|\mathcal P_v(A)\|
 \le\tfrac12\bigl(\|A\|+\|\gamma_v\Theta(A)\gamma_v\|\bigr)
 =\|A\|.
\]
These properties pass to the finite composition \(\mathbb E_Z\).

To show the commutativity assertion we first notice that the anticommutation relations of the Majoranas give for pairwise distinct labels \(w_1,\ldots,w_k\),
\[
 \mathcal P_v(\gamma_{w_1}\cdots\gamma_{w_k})
 =\begin{cases}
 0,&v\in\{w_1,\ldots,w_k\},\\
 \gamma_{w_1}\cdots\gamma_{w_k},&v\notin\{w_1,\ldots,w_k\}.
 \end{cases}
\]
Thus \(\mathcal P_v\mathcal P_w=\mathcal P_w\mathcal P_v\) on these products and on \(\1\), hence on \(\cC\) by linearity and norm density. This proves independence of $\mathbb E_Z$ from the enumeration for all $Z \Subset \Gamma$ and gives
\[
 \mathbb E_Z(\gamma_{w_1}\cdots\gamma_{w_k})
 =\begin{cases}
 \gamma_{w_1}\cdots\gamma_{w_k},&x_{w_j}\notin Z\text{ for every }j,\\
 0,&x_{w_j}\in Z\text{ for some }j.
 \end{cases}
\]
Since these products and \(\1\) span each local algebra, for every \(K\Subset\Gamma\), we have $\mathbb E_Z(\cC_K)=\cC_{K\setminus Z}$ and $\mathbb E_Z(B)=B$ for all $B\in\cC_{K\setminus Z}$. Continuity and density of local operators therefore yield $\mathbb E_Z\big|_{\cC_{\Gamma\setminus Z}}=\id$ and $\mathbb E_Z(\cC) = \cC_{\Gamma\setminus Z}$.

To prove Equation~\eqref{eq:majorana-oscillation}, we first observe that
\[
 (\id-\mathcal P_v)(A)
 =\tfrac12\bigl(A-\gamma_v\Theta(A)\gamma_v\bigr)
 =\tfrac12\gamma_v\partial_vA.
\]
Enumerating the Majorana labels in $Z$ and using the commutativity of the $\mathcal P_{v_i}$ we obtain the identity
\[
 A-\mathbb E_Z(A)
 =\sum_{j=1}^N\mathcal P_{v_{j-1}}\cdots\mathcal P_{v_1}
       (\id-\mathcal P_{v_j})(A)
 =\tfrac12\sum_{j=1}^N\mathcal P_{v_{j-1}}\cdots\mathcal P_{v_1}
       (\gamma_{v_j}\partial_{v_j}A).
\]
Contractivity then gives
\[
 \|A-\mathbb E_Z(A)\|
 \le\tfrac12\sum_{j=1}^N\|\gamma_{v_j}\partial_{v_j}A\|
 =\tfrac12\sum_{v\in\cJ:\,x_v\in Z}\|\partial_vA\|.
\]

Finally, uniqueness of the tracial state implies its invariance under every \(*\)-automorphism, so
\[
 \omega^{\rm tr}(\mathcal P_v(A))
 =\tfrac12\bigl(\omega^{\rm tr}(A)+\omega^{\rm tr}(\gamma_v\Theta(A)\gamma_v)\bigr)
 =\omega^{\rm tr}(A).
\]
Consequently, \(\omega^{\rm tr}\circ\mathbb E_Z=\omega^{\rm tr}\). For finite \(X\subset Z\) and \(B\in\cC_X\), we have $\mathbb E_Z(B)\in\cC_\varnothing=\C\1$ and $\mathbb E_Z(B)=\omega^{\rm tr}(B)\1$.
For \(A\in\cC\) and local \(B\in\cC_X\), we have for large enough $m$
\[
 \|\mathbb E_{Z_m}(A)-\omega^{\rm tr}(A)\1\|
 =\|\mathbb E_{Z_m}(A-B)-\omega^{\rm tr}(A-B)\1\|\le2\|A-B\|.
\]
Taking the limit superior in \(m\) and then the infimum over \(B\in\cC_\loc\) proves norm convergence to \(\omega^{\rm tr}(A)\1\).
\end{proof}

\begin{lemma}\label{lem:local-graded-commutator}
Let \(Z\Subset\Gamma\), let \(B\in\cC_Z\) be homogeneous, and let \(A\in\cC\). Then
\begin{equation*}
 \|BA-\Theta^{|B|}(A)B\|
 \le\|B\|\sum_{v\in\cJ:\,x_v\in Z}\|\partial_vA\|.
\end{equation*}
\end{lemma}
\begin{proof}
Let \(\mathbb E_Z\) be the averaging map of Definition~\ref{def:majorana-averaging}. Since \(\mathbb E_Z(A)\in\cC_{\Gamma\setminus Z}\) we have
\[
 B\mathbb E_Z(A)=\Theta^{|B|}(\mathbb E_Z(A))B.
\]
Subtracting this identity and applying Lemma~\ref{lem:majorana-averaging} yields
\[
 \begin{aligned}
 \|BA-\Theta^{|B|}(A)B\|
 &=\|B(A-\mathbb E_Z(A))
       -\Theta^{|B|}(A-\mathbb E_Z(A))B\|\\
 &\le2\|B\|\|A-\mathbb E_Z(A)\|\\
 &\le\|B\|\sum_{v\in\cJ:\,x_v\in Z}\|\partial_vA\|,
 \end{aligned}
\]
which proves the lemma.
\end{proof}

\begin{definition}\label{def:tracial-localization}
Let \(X\Subset\Gamma\). The tracial conditional expectation \(\Ftr X:\cC\to\cC_X\) is the unique conditional expectation satisfying
\[
 \omega^{\rm tr}\bigl(\Ftr X(D)B\bigr)=\omega^{\rm tr}(DB)
\]
for all \(D\in\cC\) and \(B\in\cC_X\). Its existence and uniqueness follow from \cite[Theorem~4.7]{ArakiMoriyaGibbs03}.
\end{definition}

\begin{lemma}\label{lem:conditional-expectations}
For every \(X\Subset\Gamma\), the map \(\Ftr X\) is a parity-preserving unital completely positive contraction onto \(\cC_X\). For \(A,B\in\cC_X\) and \(D\in\cC\), it satisfies
\[
 \Ftr X(ADB)=A\,\Ftr X(D)B.
\]
For every \(Z\Subset\Gamma\), we have \(\Ftr X(\cC_Z)\subset\cC_{X\cap Z}\). If \(X\subset Z\), then \(\Ftr X\Ftr Z=\Ftr X\). Moreover, for every \(D\in\cC\),
\begin{equation}\label{eq:expectation-approximation}
 \|D-\Ftr X(D)\|\le2\inf_{B\in\cC_X}\|D-B\|.
\end{equation}
\end{lemma}
\begin{proof}
The range, support and composition properties follow from \cite[Theorem~4.7]{ArakiMoriyaGibbs03}. Unitality, contractivity and the identity \(\Ftr X(ADB)=A\Ftr X(D)B\) are given in \cite[Lemma~2.2]{ArakiMoriyaGibbs03}, and parity preservation in \cite[Corollary~4.8]{ArakiMoriyaGibbs03}. Complete positivity is a general property of conditional expectations. The cited results apply after relabeling the countable set of fermionic modes and do not require lattice geometry.
For the remaining estimate, let \(D\in\cC\) and \(B\in\cC_X\). Since \(\Ftr X(B)=B\), contractivity gives
\[
 \|D-\Ftr X(D)\|
 =\|(\id-\Ftr X)(D-B)\|
 \le2\|D-B\|.
\]
Taking the infimum over \(B\in\cC_X\) proves Equation~\eqref{eq:expectation-approximation}.
\end{proof}

\begin{definition}\label{def:tracial-shells}
For finite nonempty \(X\subset\Gamma\), set \(R_0=0\) and \(R_i=2^{i-1}\) for \(i\ge1\). The tracial dyadic shells are the maps
\[
 \Delta^{\rm tr}_{X,0}:=\Ftr X,
\]
and, for \(i\ge1\),
\[
 \Delta^{\rm tr}_{X,i}:=\Ftr{X^{(R_i)}}-\Ftr{X^{(R_{i-1})}}.
\]
\end{definition}
This is the decomposition used in \cite[Corollary~5.5]{FermionicLR} and \cite[Section~4.1]{LPPLV3}. Its partial sums satisfy
\[
 \sum_{i=0}^m\Delta^{\rm tr}_{X,i}(A)=\Ftr{X^{(R_m)}}(A)\longrightarrow A.
\]
The convergence follows from local norm density and contractivity. The composition property in Lemma~\ref{lem:conditional-expectations} gives, for \(i\ge1\),
\begin{equation}\label{eq:tracial-shell-estimate}
 \|\Delta^{\rm tr}_{X,i}(A)\|
 =\|\Ftr{X^{(R_i)}}(A-\Ftr{X^{(R_{i-1})}}(A))\|\le\|A-\Ftr{X^{(R_{i-1})}}(A)\|.
\end{equation}

\section{Dynamics and equilibrium states}
\label{app:dynamics}

In this appendix we collect facts about dynamics, ground states and KMS states of CAR algebras. The statements apply to CAR algebras with finitely many modes per site, including \(\A\), \(\A_\aux\) and \(\A^\st\).

\subsection{Dynamics}

\begin{proposition}\label{prop:weak-dynamics}
Let \(\Xi\in\mathfrak J_1\) be an interaction on \(\cC\), and set \(H_\Lambda:=\sum_{Z\subset\Lambda}\Xi(Z)\) for \(\Lambda\Subset\Gamma\). There is a strongly continuous group \((\tau_t^\Xi)_{t\in\R}\) of parity-preserving \(*\)-automorphisms of \(\cC\) such that, for every increasing finite exhaustion \(\Lambda_n\uparrow\Gamma\), every \(A\in\cC\) and every \(T>0\),
\[
 \lim_{n\to\infty}\sup_{|t|\le T}
 \bigl\|\E^{\I tH_{\Lambda_n}}A\E^{-\I tH_{\Lambda_n}}-\tau_t^\Xi(A)\bigr\|=0.
\]
Its infinitesimal generator is \(\I\overline\delta_\Xi\), where \(\overline\delta_\Xi\) is the closure of $\delta_\Xi$.

Moreover, let \(\Xi_1,\Xi_2\in\mathfrak J_1\) be interactions on CAR algebras \(\cC_1,\cC_2\), respectively. On \(\cC_1\widehat\otimes\cC_2\), we have
\[
 \tau^{\Xi_1^{(1)}+\Xi_2^{(2)}}_t(A\widehat\otimes B)
 =\tau^{\Xi_1}_t(A)\widehat\otimes\tau^{\Xi_2}_t(B)
\]
for all \(A\in\cC_1\), \(B\in\cC_2\) and \(t\in\R\).
\end{proposition}
\begin{proof}
The first assertion is \cite[Proposition~2.1]{LPPLV3}, whose proof applies to \(\|\Xi\|_1<\infty\) and finitely many modes per site. See also \cite[Section~4]{BruPedra17} and \cite{NSY18,NSY19}.
For the product interaction, the finite-volume Hamiltonians of the two factors are even, so
\[
 [H_{1,\Lambda}^{(1)},H_{2,\Lambda}^{(2)}]=0
\]
for any $\Lambda \Subset \Gamma$. Consequently,
\[
 \begin{aligned}
 &\E^{\I t(H_{1,\Lambda}^{(1)}+H_{2,\Lambda}^{(2)})}
 A^{(1)}B^{(2)}
 \E^{-\I t(H_{1,\Lambda}^{(1)}+H_{2,\Lambda}^{(2)})}=
 \bigl(\E^{\I tH_{1,\Lambda}}A\E^{-\I tH_{1,\Lambda}}\bigr)^{(1)}
 \bigl(\E^{\I tH_{2,\Lambda}}B\E^{-\I tH_{2,\Lambda}}\bigr)^{(2)}.
 \end{aligned}
\]
Taking the norm limit proves the asserted identity.
\end{proof}

\begin{lemma}\label{lem:reversed-interactions}
Let \(\Xi\in\mathfrak J_1\) be an interaction on \(\A\), and define \((\ft\Xi)(Z):=\ft(\Xi(Z))\) for \(Z\Subset\Gamma\). Then \(\ft\Xi\in\mathfrak J_1\), with \(\|\ft\Xi\|_1=\|\Xi\|_1\), and, for every \(t\in\R\),
\[
 \ft\circ\tau_t^\Xi
 =\tau_{-t}^{\ft\Xi}\circ\ft.
\]
Moreover, for states \(\omega,\rho\) on \(\A\) and every \(\beta>0\),
\[
 \begin{aligned}
 \omega\in\GS(\tau^\Xi)
 &\Longleftrightarrow\omega\circ\ft\in\GS(\tau^{\ft\Xi}),\\
 \rho\in\KMS_\beta(\tau^\Xi)
 &\Longleftrightarrow\rho\circ\ft\in\KMS_\beta(\tau^{\ft\Xi}).
 \end{aligned}
\]
\end{lemma}
\begin{proof}
By Lemma~\ref{lem:majorana-reversal}, for every \(Z\Subset\Gamma\) it holds that \((\ft\Xi)(Z)^*=(\ft\Xi)(Z)\in\A_Z^+\) and \(\|(\ft\Xi)(Z)\|=\|\Xi(Z)\|\). Thus \(\ft\Xi\in\mathfrak J_1\) with \(\|\ft\Xi\|_1=\|\Xi\|_1\).

By Lemma~\ref{lem:majorana-reversal} we have for any \(A\in\A_\loc\)
\begin{equation}\label{eq:reversed-derivation}
 \ft(\delta_\Xi(A))
 =\sum_{Z\Subset\Gamma}[\ft(A),\ft(\Xi(Z))]
 =-\delta_{\ft\Xi}(\ft(A)).
\end{equation}
Since \(\ft(\A_\loc)=\A_\loc\) and \(\A_\loc\) is a core for both derivations by Proposition~\ref{prop:weak-dynamics}, taking closures gives \(\ft\bigl(\operatorname{Dom}(\overline\delta_\Xi)\bigr)
 =\operatorname{Dom}(\overline\delta_{\ft\Xi})\) and
\[
 \begin{aligned}
 \ft(\I\overline\delta_\Xi)\ft
 &=-\I\overline\delta_{\ft\Xi}.
 \end{aligned}
\]
By Equation~\eqref{eq:dynamics-convention}, the groups \(\ft\circ\tau_t^\Xi\circ\ft\) and \(\tau_{-t}^{\ft\Xi}\) therefore have the same generator \(-\I\overline\delta_{\ft\Xi}\). Uniqueness therefore proves the asserted identity for the dynamics.

Let \(\omega\in\GS(\tau^\Xi)\) and \(A\in\A_\loc\). Lemma~\ref{lem:ground-state-equivalence} gives \(\omega(\delta_\Xi(\ft(A)\ft(A)^*))=0\). Hence Equation~\eqref{eq:reversed-derivation} and Definition~\ref{def:equilibrium}(ii) give
\[
 \begin{aligned}
 \RealPart(\omega\circ\ft)(A^*\delta_{\ft\Xi}(A))
 &=-\RealPart\omega\bigl(\delta_\Xi(\ft(A))\ft(A)^*\bigr)=\RealPart\omega\bigl(\ft(A)\delta_\Xi(\ft(A)^*)\bigr)\ge0.
 \end{aligned}
\]

Let \(\rho\in\KMS_\beta(\tau^\Xi)\) and \(A,B\in\A_\an(\tau^{\ft\Xi})\). Since \(\ft\) is bounded and linear (Lemma~\ref{lem:majorana-reversal}), the function \(z\mapsto\ft(\tau_{-z}^{\ft\Xi}(A))\) is norm-entire. For \(t\in\R\), we have \(\ft(\tau_{-t}^{\ft\Xi}(A))=\tau_t^\Xi(\ft(A))\). Thus \(\ft(A)\in\A_\an(\tau^\Xi)\), with
\[
 \tau_z^\Xi(\ft(A))
 =\ft(\tau_{-z}^{\ft\Xi}(A)),
 \qquad z\in\C.
\]
The same argument applies to \(B\). Definition~\ref{def:equilibrium}(i) therefore gives
\[
 \begin{aligned}
 (\rho\circ\ft)(B\tau_{\I\beta}^{\ft\Xi}(A))
 &=\rho\bigl(\tau_{-\I\beta}^\Xi(\ft(A))\ft(B)\bigr)=\rho(\ft(B)\ft(A))
 =(\rho\circ\ft)(AB).
 \end{aligned}
\]
The converse implications follow from \(\ft^2=\id\) (Lemma~\ref{lem:majorana-reversal}).
\end{proof}

\subsection{Ground states}
\label{app:state-conditions}

\begin{lemma}\label{lem:ground-state-equivalence}
Let \(\Xi\in\mathfrak J_1\) be an interaction on \(\cC\). Let \(\rho\) be a state on $\cC$ with GNS triple \((\pi,\mathcal H,\Omega)\). The following are equivalent:
\begin{enumerate}[label=(\roman*),leftmargin=*]
\item For every \(B\in\cC_\loc\),
\[
 \RealPart\rho(B^*\delta_\Xi(B))\ge0.
\]
\item The state \(\rho\) is \(\tau^\Xi\)-invariant, and its GNS Hamiltonian \(L_\rho\), determined by
\[
 \E^{\I tL_\rho}\pi(B)\Omega=\pi(\tau^\Xi_t(B))\Omega,
\]
is nonnegative.
\end{enumerate}
In this case \(L_\rho\Omega=0\), and for every \(B\in\cC\),
\[
 \pi(\tau^\Xi_t(B))=\E^{\I tL_\rho}\pi(B)\E^{-\I tL_\rho}.
\]
\end{lemma}
\begin{proof}
By \cite[Lemma~2.2]{LPPLV3}, condition (i) is equivalent to the algebraic ground-state condition on \(\operatorname{Dom}(\overline\delta_\Xi)\). The equivalence with (ii) and the properties of the GNS Hamiltonian then follow from \cite[Proposition~5.3.19]{BR97}, with infinitesimal generator \(\I\overline\delta_\Xi\).
\end{proof}

\begin{lemma}[{\cite[Section~2.1]{LPPLV3}}]\label{lem:ground-state-existence}
Every interaction \(\Xi\in\mathfrak J_1\) admits a ground state.
\end{lemma}

\subsection{KMS states}

\begin{proposition}\label{prop:kms-nonempty}
Let \(\Xi\in\mathfrak J_1\) be an interaction on \(\cC\), and let \(\beta>0\). Then \(\KMS_\beta(\tau^\Xi)\) is nonempty, and all its states are even.
\end{proposition}
\begin{proof}
Existence follows from \cite[Proposition~5.3.25]{BR97} and Proposition~\ref{prop:weak-dynamics}, since any weak-\(*\) cluster point of the finite-volume Gibbs states, extended by the trace outside the volume, is a \(\beta\)-KMS state for \(\tau^\Xi\).

Evenness is immediate if \(\Gamma\) is finite: then \(\cC\) is a matrix algebra, the only \(\beta\)-KMS state is the Gibbs state of the even Hamiltonian \(\sum_Z\Xi(Z)\), and its density matrix is even. Let \(\Gamma\) be infinite.

For evenness, we verify the hypotheses of the evenness criterion \cite[Proposition~4.1 and Lemma~4.4]{Kishimoto01}, see also \cite{FVV84}. The criterion applies to an automorphism commuting with the dynamics when it is a point-norm limit of inner automorphisms implemented by unitaries in the generator domain with uniformly bounded generator norms. We apply it to \(\Theta\) and \(\tau^\Xi\), whose generator is \(\I\overline\delta_\Xi\).

Since every interaction term is even, Proposition~\ref{prop:weak-dynamics} gives
\[
 \Theta\tau^\Xi_t=\tau^\Xi_t\Theta
\]
for all \(t\in\R\). Choose a sequence of sites \(x_n\) leaving every finite set. The Majoranas \(\gamma_{x_n,1}\) are self-adjoint unitaries and belong to \(\cC_\loc\subset\operatorname{Dom}(\overline\delta_\Xi)\). For \(B\in\cC_X\) and \(x_n\notin X\), the commutation relations give
\[
 \gamma_{x_n,1}B\gamma_{x_n,1}=\Theta(B).
\]
Thus, for \(A\in\cC\) and such a local \(B\),
\[
\|\gamma_{x_n,1}A\gamma_{x_n,1}-\Theta(A)\|
 =\|\gamma_{x_n,1}(A-B)\gamma_{x_n,1}-\Theta(A-B)\|\le2\|A-B\|.
\]
Local norm density proves point-norm convergence of \(\operatorname{Ad}(\gamma_{x_n,1})\) to \(\Theta\).
Finally, interaction terms supported away from \(x_n\) commute with \(\gamma_{x_n,1}\), so
\[ \|\I\overline\delta_\Xi(\gamma_{x_n,1})\|
 =\left\|\sum_{Z\ni x_n}[\Xi(Z),\gamma_{x_n,1}]\right\| \le2\sum_{Z\ni x_n}\|\Xi(Z)\|
  \le2\|\Xi\|_1.
\]
The series is absolutely convergent and the bound is uniform in \(n\). All hypotheses of the criterion are therefore satisfied, and it yields \(\rho\circ\Theta=\rho\) for every \(\rho\in\KMS_\beta(\tau^\Xi)\).
\end{proof}

\subsection{Products of equilibrium states}
\label{app:product-dynamics}

\begin{lemma}\label{lem:graded-ground}
Let \(\Xi_1,\Xi_2\in\mathfrak J_1\) be interactions on \(\A\) and \(\A_\aux\), respectively. If \(\omega_1\) and \(\omega_2\) are ground states of these interactions and \(\omega_2\) is even, then
\[
 \omega_1\widehat\otimes\omega_2\in\GS(\tau^{\Xi_1^{(1)}+\Xi_2^{(2)}}).
\]
\end{lemma}
\begin{proof}
The product exists by Lemma~\ref{lem:graded-product}. Since the interaction terms are even, we have
\[
 \delta_{\Xi_1^{(1)}+\Xi_2^{(2)}}(A^{(1)}B^{(2)})
 =\delta_{\Xi_1}(A)^{(1)}B^{(2)}+A^{(1)}\delta_{\Xi_2}(B)^{(2)}
\]
for local \(A,B\). The assertion is therefore \cite[Lemma~8.2]{LPPLV3}, applied with \(U=\Gamma\), \(\delta_1=\delta_{\Xi_1}\) and \(\delta_2=\delta_{\Xi_2}\).
\end{proof}

\begin{lemma}\label{lem:graded-KMS}
Let \(\tau^{(1)}\) and \(\tau^{(2)}\) be strongly continuous parity-preserving dynamics on CAR algebras \(\cC_1\) and \(\cC_2\) over two disjoint sets of fermionic modes, respectively. Write \(A^{(1)}\) and \(B^{(2)}\) for the canonical embeddings into \(\cC_1\widehat\otimes\cC_2\). Their product dynamics is defined by
\[
 (\tau^{(1)}\widehat\otimes\tau^{(2)})_t(A^{(1)}B^{(2)})
 :=\tau^{(1)}_t(A)^{(1)}\tau^{(2)}_t(B)^{(2)}.
\]
For every \(\beta>0\), if \(\rho_i\in\KMS_\beta(\tau^{(i)})\), \(i=1,2\), are even, then
\[
 \rho_1\widehat\otimes\rho_2\in\KMS_\beta(\tau^{(1)}\widehat\otimes\tau^{(2)}).
\]
\end{lemma}
\begin{proof}
Set \(\tau:=\tau^{(1)}\widehat\otimes\tau^{(2)}\). The two automorphisms preserve the canonical anticommutation relations within each factor. For homogeneous \(A\in\cC_1\) and \(B\in\cC_2\) we find the relation
\[
 \tau^{(2)}_t(B)^{(2)}\tau^{(1)}_t(A)^{(1)}
 =(-1)^{|A||B|}\tau^{(1)}_t(A)^{(1)}\tau^{(2)}_t(B)^{(2)}.
\]
Thus the universal property of the CAR algebra defines \(\tau_t\) on \(\cC_1\widehat\otimes\cC_2\). The group law follows from those of the two dynamics, and \(\tau_{-t}\) is its inverse. Strong continuity holds on products \(A^{(1)}B^{(2)}\) by continuity in each factor, and extends to the whole algebra by norm density and isometry.

The state \(\rho_1\widehat\otimes\rho_2\) exists by Lemma~\ref{lem:graded-product}. For homogeneous entire analytic \(A_i,B_i\in\cC_i\), \(i=1,2\), set \(A=A_1^{(1)}A_2^{(2)}\) and \(B=B_1^{(1)}B_2^{(2)}\). The product \(B\) is entire analytic, with
\[
 \tau_z(B)=\tau^{(1)}_z(B_1)^{(1)}\tau^{(2)}_z(B_2)^{(2)},
 \qquad z\in\C.
\]
Graded multiplication and the two KMS identities give
\[
 \begin{aligned}
 (\rho_1\widehat\otimes\rho_2)(A\tau_{\I\beta}(B))
 &=(-1)^{|A_2||B_1|}
   \rho_1(A_1\tau^{(1)}_{\I\beta}(B_1))
   \rho_2(A_2\tau^{(2)}_{\I\beta}(B_2))\\
 &=(-1)^{|A_2||B_1|}\rho_1(B_1A_1)\rho_2(B_2A_2),\\
 (\rho_1\widehat\otimes\rho_2)(BA)
 &=(-1)^{|B_2||A_1|}\rho_1(B_1A_1)\rho_2(B_2A_2).
 \end{aligned}
\]
Evenness gives \(\rho_i(B_iA_i)=0\) whenever \(|A_i|\ne|B_i|\). Thus both expressions vanish unless \(|A_i|=|B_i|\) for \(i=1,2\), in which case the signs agree. This proves the KMS identity on these products. Homogeneous entire analytic elements span a norm-dense subalgebra in each factor, since the parity projections commute with the dynamics. Their products span a norm-dense invariant \(*\)-subalgebra of entire analytic elements in \(\cC_1\widehat\otimes\cC_2\), so bilinearity and \cite[Theorem~21]{Pillet06} yield the assertion.
\end{proof}

\subsection{Analytic filtering and Gevrey cutoffs}
\label{app:gevrey}

\begin{lemma}\label{lem:analytic-filtering}
Let \(\tau\) be a strongly continuous dynamics on a \(C^*\)-algebra \(\cC\), and let \(q\in C^2(\R)\) satisfy
\[
 \sum_{j=0}^2\int_\R\E^{b|\omega|}|q^{(j)}(\omega)|\,\dd\omega<\infty
 \qquad\text{for every }b>0.
\]
Then \(\int_\R\check q(t)\tau_t(A)\,\dd t\) is entire analytic for every \(A\in\cC\), and
\begin{equation}\label{eq:analytic-filter-shift}
 \tau_z\left(\int_\R\check q(t)\tau_t(A)\,\dd t\right)
 =\int_\R\check q(t-z)\tau_t(A)\,\dd t,
 \qquad z\in\C.
\end{equation}
\end{lemma}
\begin{proof}
For a compact \(K\subset\C\), two integrations by parts give, for \(\ell=0,1\),
\[
 \sup_{z\in K}(1+t^2)|\partial_z^\ell\check q(t-z)|
 \le\frac1{2\pi}\sup_{z\in K}\sum_{j\in\{0,2\}}
 \left\|\partial_\omega^j\bigl((\I\omega)^\ell\E^{\I z\omega}q(\omega)\bigr)\right\|_1
 <\infty.
\]
The assumed exponential integrability justifies the Fourier integrals and their derivatives. Since \(\|\tau_t(A)\|=\|A\|\), the bound permits norm differentiation under the time integral. Its value at real \(z\) is the translated orbit by a change of variables, proving \eqref{eq:analytic-filter-shift}.
\end{proof}

\begin{lemma}\label{lem:gevrey-cutoff}
Let \(I\subset\R\) be a nonempty compact interval and let \(\delta>0\).
There is a constant \(C_*>0\), depending only on \(I\) and \(\delta\), with the following property.
For every \(\sG>1\), there are a real-valued function \(q\in C_c^\infty(\R)\) and a constant \(C>0\), depending only on \(I,\delta,\sG\), such that
\[
 0\le q\le1,\qquad q=1\text{ on }I,\qquad
 \supp q\subset I+[-\delta,\delta],
\]
and, for every \(n\ge0\),
\[
 \begin{aligned}
 \|q^{(n)}\|_\infty+\|q^{(n)}\|_1&\le C^{n+1}(n!)^{\sG},\\
 \|q^{(n)}\|_\infty+\|q^{(n)}\|_1&\le C_*^{n+1}(n!)^2.
 \end{aligned}
\]
\end{lemma}

\begin{proof}
Choose nonnegative mollifiers \(\rho_2,\rho_{\sG}\) of Gevrey orders \(2\) and \(\sG\), respectively, both supported in \([-\delta/4,\delta/4]\) and of integral one.
Their existence follows from the standard construction in \cite[Example~1.4.9 and Proposition~1.4.6]{Rodino93}; see also \cite[Section~6.1]{LPPLV3}.
Fix \(\rho_2\) independently of \(\sG\), and set
\[
 q:=\mathbf1_{I+[-\delta/2,\delta/2]}*\rho_2*\rho_{\sG}.
\]
Then \(q\) is real and smooth, takes values in \([0,1]\), equals one on \(I\), and has support in \(I+[-\delta,\delta]\).
Differentiating either mollifier and using Young's inequality gives, for \(\sigma\in\{2,\sG\}\),
\[
 \begin{aligned}
 \|q^{(n)}\|_\infty+\|q^{(n)}\|_1
 &\le\|\mathbf1_{I+[-\delta/2,\delta/2]}\|_1
       \bigl(\|\rho_\sigma^{(n)}\|_\infty+\|\rho_\sigma^{(n)}\|_1\bigr)\\
 &\le C_\sigma^{n+1}(n!)^\sigma.
 \end{aligned}
\]
Here we used that \(\|\rho_2\|_1=\|\rho_{\sG}\|_1=1\).
Taking \(C=C_{\sG}\) and \(C_*=C_2\) proves both bounds, with \(C_*\) independent of \(\sG\).
\end{proof}

\section{Proofs of lemmas from Section~\ref{sec:setting}}
\label{app:decay-profile-bounds}

\begin{proof}[Proof of Lemma~\ref{lem:site-sum-bounds}]
The first bound follows directly from the monotonicity of \(G\):
\[
 \sum_{y\in Y}G(d(y,S))\le\sum_{y\in Y}G(d(Y,S))=|Y|G(d(Y,S)).
\]
For the boundary estimate, choose for each \(y\in Y\) a nearest point \(s(y)\in S\) and then a nearest point \(z(y)\in Y\) to \(s(y)\). By the definition of the metric boundary it holds that \(z(y)\in\partial Y\). We have
\[
 d(y,z(y))\le d(y,s(y))+d(s(y),z(y))\le2d(y,S).
\]
Together with the estimate \(d(y,S)\ge d(Y,S)\) this yields
\[
 \begin{aligned}
 \cS(Y,S)
 &=\sum_{z\in\partial Y}\sum_{\substack{y\in Y\\z(y)=z}}G(d(y,S))\\
 &\le\sum_{z\in\partial Y}\sum_{\substack{y\in Y\\z(y)=z}}
       G\bigl(\max\{d(Y,S),d(y,z)/2\}\bigr)\\
 &\le |\partial Y|\sup_{z\in\Gamma}
       \sum_{y\in\Gamma}G\bigl(\max\{d(Y,S),d(y,z)/2\}\bigr).
 \end{aligned}
\]
We estimate this sum separately in the two regimes. In the stretched-exponential regime \ref{ass:exponential}, the volume-growth condition gives
\[
 \begin{aligned}
 \sup_{z\in\Gamma}\sum_{y\in\Gamma}
       \E^{-m\max\{d(Y,S),d(y,z)/2\}^\theta}
 &\le\E^{-m'd(Y,S)^\theta}\sup_{z\in\Gamma}\sum_{y\in\Gamma}
       \E^{-(m-m')(d(y,z)/2)^\theta}\\
 &\le C_\Gamma\E^{-m'd(Y,S)^\theta}
       \sum_{j\ge0}(j+2)^{d_{\rm vol}}\E^{-(m-m')(j/2)^\theta}\\
 &\le C\E^{-m'd(Y,S)^\theta}.
 \end{aligned}
\]
In the polynomial regime \ref{ass:polynomial-weight}, we first sum over sites at distance at most \(2d(Y,S)\) from \(z\), and then over regions of successively doubled radius. By volume growth, we find that
\[
 \begin{aligned}
 \sup_{z\in\Gamma}\sum_{y\in\Gamma}
       (1+\max\{d(Y,S),d(y,z)/2\})^{-p_\Phi}
 &\le\sup_{z\in\Gamma}\left[
       \frac{|\ball_{2d(Y,S)}(z)|}{(1+d(Y,S))^{p_\Phi}}
       +\sum_{k\ge0}
       \frac{|\ball_{2^{k+2}(1+d(Y,S))}(z)|}
            {\bigl(2^k(1+d(Y,S))\bigr)^{p_\Phi}}
       \right]\\
 &\le C_\Gamma5^{d_{\rm vol}}(1+d(Y,S))^{d_{\rm vol}-p_\Phi}
       \left(1+\sum_{k\ge0}2^{k(d_{\rm vol}-p_\Phi)}\right)\\
 &\le C(1+d(Y,S))^{d_{\rm vol}-p_\Phi},
 \end{aligned}
\]
since \(p_\Phi>d_{\rm vol}\). Combining these bounds proves the claim.
\end{proof}

\section*{Acknowledgments}

We thank Tom Wessel and Marius Wesle for helpful discussions. S.T.\ thanks Domenico Monaco and Gianluca Panati at Sapienza University of Rome and Massimo Moscolari at Politecnico di Milano for their hospitality during his two-month sabbatical visits to their respective institutions.
This work was supported by the Deutsche Forschungsgemeinschaft (DFG, German Research Foundation) -- 470903074 and 465199066.

\section*{Statements and Declarations}

The authors have no competing interests to declare.

\section*{AI methodology}

During the preparation of this manuscript, we used OpenAI's ChatGPT Pro (GPT 5.6 Sol and GPT 6 Astra) to assist in exploring and critically examining mathematical arguments and literature, and in preparing the exposition. We critically reviewed and revised all AI-assisted material. We take full responsibility for the correctness, originality, and content of the entire manuscript.

\end{document}